\documentclass[11pt,a4paper]{article}
\pdfoutput = 1
\usepackage[utf8]{inputenc}
\usepackage{color,graphicx}
\usepackage{epsfig}
\usepackage{amsmath}
\usepackage{verbatim}
\usepackage{amssymb}
\usepackage[mathcal]{eucal}
\usepackage{stmaryrd}
\usepackage{esint}
\usepackage{braket}
\usepackage{physics}
\usepackage{amsfonts}
\usepackage{cite}
\usepackage{array}
\usepackage{setspace}
\usepackage{braket}
\usepackage{float}
\usepackage{url}
\usepackage{mathtools} 
\usepackage{slashed}
\usepackage{tikz}
\usetikzlibrary{decorations.markings,arrows.meta}
\usepackage{graphicx}
\usepackage{epstopdf}
\usepackage[labelfont=bf,font={sf}]{caption}
\usepackage{pgfplots}
\usepgfplotslibrary{colormaps}
\usepackage{mathrsfs}
\usepackage{fancybox}
\usepackage{eurosym}
\usepackage{tcolorbox}  
\usepackage{tensor}
\usepackage{bbm}
\usepackage{dsfont}
\allowdisplaybreaks
\usepackage{subcaption}
\usepackage{float}
\usepackage{enumitem}
\usetikzlibrary{hobby}
\usetikzlibrary {arrows.meta}
\usepackage[percent]{overpic}
\usepackage{nicefrac}
\usepackage[normalem]{ulem}
\usepackage{xcolor}

\usepackage[margin = 2.2cm]{geometry}
\usepackage[ragged]{footmisc}
\usepackage[bookmarks=true,bookmarksnumbered=false,
colorlinks=true, hyperindex=true,bookmarksopen=true,hyperfigures=true,linkcolor=black,citecolor=black,urlcolor=black,breaklinks]{hyperref}
\definecolor{webgreen}{rgb}{0, 0.5, 0}
\definecolor{webblue}{rgb}{0, 0, 0.5}
\definecolor{webred}{rgb}{0.5, 0, 0}
\definecolor{darkgreen}{rgb}{0,0.5,0}

\newcommand{\normord}[1]{:\mathrel{#1}:}

\newcommand{\floor}[1]{\lfloor #1 \rfloor}

\def\lowerj{{\cal J}^-}

\def\ben{\begin{equation}}
\def\een{\end{equation}}

     \let\r=v

\def\be{\begin{equation}}
\def\ee{\end{equation}}
\def\ba{\begin{array}}
\def\ea{\end{array}}

\def\ie{\rm i.e.\ }
\def\dalemb#1#2{{\vbox{\hrule height .#2pt
       \hbox{\vrule width.#2pt height#1pt \kern#1pt
               \vrule width.#2pt}
       \hrule height.#2pt}}}

\newcommand{\bea}{\begin{eqnarray}}
\newcommand{\eea}{\end{eqnarray}}

\let\tilde=\widetilde

\def\diag{{\rm diag}}

\def\tt{tt^{\ast}}

\newcommand{\1}{\mathbbm{1}}

\newcommand{\cM}{\mathcal{M}}
\newcommand{\cN}{\mathcal{N}}

\newcommand{\cU}{\mathcal{U}}

\newcommand{\bZ}{\mathbb{Z}}

\numberwithin{equation}{section}

\usepackage{amsthm}

\newtheoremstyle{indentedbold}%
  {6pt}   % Space above
  {6pt}   % Space below
  {\normalfont}  % Body font
  {2em}   % Indent amount
  {\bfseries} % Head font
  {.}     % Punctuation after head
  {.5em}  % Space after head
  {}      % Head spec

\theoremstyle{indentedbold}
\newtheorem{theorem}{Theorem}
\newtheorem{lemma}{Lemma}

\makeatletter
\renewenvironment{proof}[1][\proofname]{\par
  \pushQED{\qed}%
  \normalfont
  \topsep6\p@\@plus6\p@\relax
  \trivlist
  \item\relax
  {\itshape #1\@addpunct{.}}\hspace{0.5em}\ignorespaces
}{%
  \popQED\endtrivlist\@endpefalse
}
\makeatother

\title{}

\pgfplotsset{compat=1.18} 
\begin{document}

\begin{flushright}
CERN-TH-2026-072\\
SISSA 06/2026/FISI\\
\end{flushright}

\thispagestyle{empty}
\begin{center}
    ~\vspace{5mm}

     {\LARGE \bf 
   Chaos of Berry curvature for BPS microstates \\
   }
    
   \vspace{0.4in}
    
    {\bf Yiming Chen$^1$, Sean Colin-Ellerin$^{2,3}$, Ohad Mamroud$^{4,5}$, Kyriakos Papadodimas$^{3}$}

    \vspace{0.4in}
   {$^1$ Leinweber Institute for Theoretical Physics, Stanford University, Stanford, CA 94305, USA \\
   $^2$ Leinweber Institute for Theoretical Physics and Department of Physics,
   University of California, Berkeley, CA 94720, USA \\
   $^3$ Theoretical Physics Department, CERN, CH-1211 Geneva 23, Switzerland \\
   $^4$ SISSA, Via Bonomea 265, 34127 Trieste, Italy\\
   $^5$ INFN, Sezione di Trieste, Via Valerio 2, I-34127 Trieste, Italy\\}
    \vspace{0.1in}
    \texttt{ymchen.phys@gmail.com,  sean.julian.colin-ellerin@cern.ch, omamroud@sissa.it, kyriakos.papadodimas@cern.ch}
    
\end{center}

\vspace{0.4in}

\begin{abstract}
\noindent 

We expect black hole microstates to differ in their chaotic properties from states associated with other geometries. For supersymmetric black holes, ordinary level statistics cannot diagnose this distinction, since their energy levels are exactly degenerate. We propose that there is an intrinsic probe of chaos, encoded in the mixing of the microstates under changes in the couplings of the theory, as determined by the non-Abelian Berry curvature of the BPS states under certain deformations. For states dual to horizonless geometries in holographic systems, such as 1/2-BPS states in the D1/D5 CFT and 1/4-BPS states in $\mathcal{N}=4$ SYM, we find that the Berry curvature for marginal deformations is non-random and often exactly zero at generic couplings. By contrast, for states dual to supersymmetric black holes, we show through computations in $\mathcal{N}=2$ super-JT gravity and explicit numerics in the $\mathcal{N}=2$ SYK model that the Berry curvature resembles a random matrix. We also uncover interesting topological features of the 
$\mathcal{N}=2$ SYK moduli space, as probed by Chern numbers.  These results suggest that the Berry curvature sharply distinguishes black hole microstates from smooth horizonless states and provides a robust diagnostic of chaos in supersymmetric sectors.

\end{abstract}

\pagebreak
\setcounter{page}{1}
\tableofcontents

\newpage

%%%%%%%%%%%
% SECTION %
%%%%%%%%%%%
\section{Introduction}

A simple measure of chaos in a quantum system is the extent to which the repulsion of its energy levels is consistent with that of a random matrix. Black holes are very chaotic objects so one expects that the quantum Hamiltonian describing a black hole should exhibit such eigenvalue statistics. Indeed, there is evidence that this is the case for ordinary black holes in anti-de Sitter (AdS) spacetimes. For example, the SYK model \cite{kitaev2015, Maldacena:2016hyu}, a microscopic low-dimensional toy model for black holes, exhibits such repulsion \cite{Cotler:2016fpe}. For higher dimensional black holes, apart from the expectation that the dual strongly-coupled CFT is chaotic, evidence also comes from the gravitational path integral, as the eigenvalue repulsion is exhibited in the spectral form factor via the contribution of a saddle known as the double-cone wormhole \cite{Saad:2018bqo}. However, this repulsion cannot be present for supersymmetric black holes because they are exactly degenerate: all of the energy levels of the black hole microstates with the same charges saturate the same bound, namely the Bogomol'nyi–Prasad–Sommerfield (BPS) bound. This leads to the basic puzzle \cite{Lin:2022zxd, Chen:2024oqv}: how does one detect chaos for supersymmetric black holes?

One proposed solution to this puzzle, due to Lin, Maldacena, Rozenberg, and Shan (LMRS) \cite{Lin:2022zxd}, is to consider a simple operator $O$, project it onto the supersymmetric states to obtain $\widehat{O}$, and analyze the eigenvalue statistics of $\widehat{O}$. It is conjectured that the eigenvalue spacings of $\widehat{O}$ will exhibit random matrix theory (RMT) statistics, independent of which simple $O$ one chooses, thereby probing the chaos of the supersymmetric black hole. The implication is that there is no preferred basis to the black hole microstates. Those authors found this to be true for the two-dimensional $\mathcal{N}=2$ super-JT gravity, which describes the near-horizon region of certain higher-dimensional supersymmetric black holes \cite{Heydeman:2020hhw,Boruch:2022tno,Heydeman:2024ezi,Heydeman:2024fgk}. One can roughly think of this diagnostic as a supersymmetric version of the eigenstate thermalization hypothesis, even though supersymmetric black holes have zero temperature. We will call this the LMRS criterion, following \cite{Chen:2024oqv}. Armed with a diagnostic, one can ask a more fine-grained question: can the eigenvalues of $\widehat{O}$ distinguish between supersymmetric states that are microstates of macroscopic black holes and those that are microstates of smooth, horizonless geometries in supergravity? 

It is certainly possible that the non-black hole states are also chaotic, but one expects them to be less so than the black hole microstates. This can be quantified by the Thouless time $t_{\rm Th}$ of the system, and its dependence on the number of degrees of freedom in the system $N$: if it grows as $t_{\rm Th} \sim O(N^0)$, such that the eigenvalue repulsion extends throughout the entire spectrum, we call the system strongly chaotic, while if $t_{\rm Th} \sim O(N^{\# >0})$ and the repulsion only extends to a small part of the spectrum, we say that the system is weakly chaotic. One conjectures then that black holes are strongly chaotic, while other geometries are at most weakly chaotic, if at all.

In the AdS/CFT correspondence, there is hope of answering this fine-grained question because, given a supersymmetric black hole in AdS with some energy and charges, one can try to find all of the supersymmetric states in the dual CFT with that energy and those charges. Recently, a conjecture has been made by Chang and Lin for how to differentiate the black hole microstates from the non-black hole ones based on their behavior as a function of the number of degrees of freedom of the CFT \cite{Chang:2024zqi}. They observed that while some supersymmetric states, dubbed monotonous, exist for all $N$, there are also other states, dubbed fortuitous, that appear only for small enough $N$ as they require trace (or other) relations to satisfy the supersymmetry conditions. Together, the fortuitous and monotonous states comprise all of the BPS states. Since the monotonous states alone cannot account for the black hole entropy \cite{Kinney:2005ej,Chang:2013fba}, they deduced that typical black hole microstates must be fortuitous, while smooth, horizonless microstates are monotonous. Further evidence for the latter part of the conjecture comes from matching the smooth supergravity solutions with their dual monotone states \cite{Lunin:2001fv,Lin:2004nb}.

One can then ask if the monotonous states in the CFT are less chaotic than the fortuitous ones. An investigation of this question using the LMRS criterion was carried out in \cite{Chen:2024oqv}, who found that the 1/2-BPS and 1/4-BPS states in $\mathcal{N}=4$ SYM and some 1/2-BPS states in D1/D5, all of which are monotones, can in principle exhibit random matrix universality in short-range level correlations. However, they found that the Thouless time $t_{\mathrm{Th}}$, grows with the system size as $t_{\mathrm{Th}} \sim N^{\# > 0}$ so $\widehat{O}$ is only weakly chaotic in these sectors. In contrast, the Thouless time for supersymmetric black hole microstates was computed in $\mathcal{N}=2$ super-JT gravity and found to be $t_{\mathrm{Th}} \sim O(N^{0})$, and hence they exhibit strong chaos \cite{Lin:2022zxd}.

In this work, we propose a different diagnostic for the chaos of supersymmetric black holes: the mixing of their microstates as one varies the couplings of the theory. Consider a family of quantum mechanical systems described by Hamiltonians $H(\lambda)$ parametrized by a set of moduli $\mathcal{M} = \{\lambda_{\mu}\}$. Pick a point $\lambda^{(0)} \in \mathcal{M}$ and consider a degenerate energy eigenspace $\mathbb{H}_{n}(\lambda^{(0)})$ of the Hamiltonian $H(\lambda^{(0)})$ with energy $E_{n}$ and dimension $D$. If we now evolve the system adiabatically by slowly varying the couplings in time, a state $\ket{a} \in \mathbb{H}_{n}(\lambda^{(0)})$ will mix under evolution with all the other states in this degenerate subspace, but not with the states outside it. Schr\"odinger's equation for such an evolution then implies that the mixing matrix $U$ is a unitary matrix determined by a $D \times D$ Hermitian matrix-valued connection $A$ for $\mathbb{H}_{n}(\lambda)$ over the moduli space $\mathcal{M}$, known as the Berry connection \cite{Berry:1984jv, PhysRevLett.51.2167} (or Wilczek-Zee connection in the degenerate case \cite{PhysRevLett.52.2111}). There is a corresponding curvature for this connection, $F = dA - i A \wedge A$, known as the Berry curvature. In quantum mechanics, the curvature can be computed in perturbation theory around any point, and it reads
\begin{equation}
    \label{eq:Berry curvature QM}
    \bigl(F_{\mu\nu}\bigr)_{ab} = \sum_{m \not \in \mathbb{H}_{n}}\frac{i\bra{b}\partial_{[\mu}H\ket{m}\bra{m}\partial_{\nu]}H\ket{a}}{(E_{n}-E_{m})^{2}} \,.
\end{equation}

\begin{figure}[t]
\begin{center}
\includegraphics[scale=1]{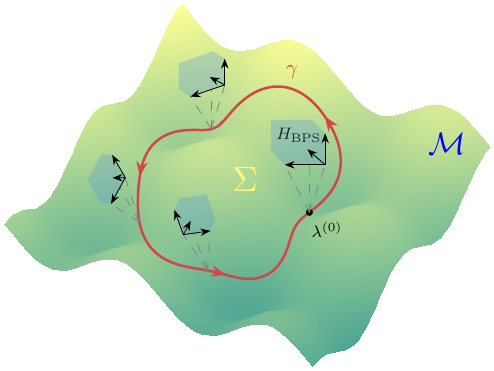}
\end{center}
\caption{Loop $\gamma$ in moduli space $\mathcal{M}$ based at $\lambda^{(0)}$ along which the system is evolved. The fibres $\mathbb{H}_{\mathrm{BPS}}$ twist along the curve and the holonomy is the exponential of the integral of the Berry curvature on $\Sigma$ with $\gamma = \partial\Sigma$.}
\label{fig:loopinM}
\end{figure}

If the evolution is along a loop $\gamma$, illustrated in Figure \ref{fig:loopinM}, then the mixing matrix $U$ is the holonomy of this connection (up to the usual dynamical factor), which can be written in terms of an integral of the curvature on a 2-dimensional surface $\Sigma$ such that $\gamma=\partial \Sigma$. When the energy level is non-degenerate, this holonomy is simply a phase -- the famous Berry phase \cite{Berry:1984jv}. Under a local change of basis for $\mathbb{H}_{n}(\lambda)$, the curvature $F$ transforms as the field strength of a non-Abelian $U(D)$ gauge field. It is a gauge redundancy of the curvature $F$ and the gauge-invariant information is encoded in its eigenvalues. A more detailed discussion of the basics of the Berry curvature can be found in Section \ref{sec:general}.

The Berry curvature plays an important role in a wide variety of physical phenomena, including quantum computation \cite{Zanardi:1999gv, Pachos:1999ve, ZHANG20231}, the anomalous Hall effect \cite{PhysRevLett.93.206602, Gradhand_2012}, topological phases in condensed matter systems \cite{RevModPhys.83.1057, Bradlyn_2022}, and quantum field theory \cite{Kapustin:2020eby,Hsin:2020cgg,Cordova:2019jnf,Cordova:2019uob}, as well as in atomic, molecular, and optical physics \cite{ZYGELMAN1987476, resta2000manifestations}.  It was also studied in the context of conformal field theory, where a natural choice of Hilbert space is that of the local operators in the theory, given by radial quantization, and natural deformations are exactly marginal operators $\mathcal{O}_{\mu}$, $\mathcal{O}_\nu$. The Berry connection (and curvature) then describes how the operators in the theory $\phi_{a}$, $\phi_b$ mix as one moves along a conformal manifold \cite{Seiberg:1988pf,Kutasov:1988xb, Ranganathan:1992nb, Ranganathan:1993vj},
\begin{equation}
    \label{eq:Berry curv CFT}
    (F_{\mu\nu})_{ab} = {i \over \mathcal{S}_d}\left[ \int_{|x|\leq 1} d^dx \int_{|y|\leq1}d^dy\,\,\langle \phi_b^\dagger(\infty) \,\,\mathcal{O}_{[\mu}(x) \,\,\mathcal{O}_{\nu]}(y)\,\,\phi_a(0)\rangle\right]_{\rm renormalized} \,,
\end{equation}
where $\mathcal{S}_{d}$ is a constant, with a prescription that deals with the UV-divergences at coincident points \cite{Ranganathan:1992nb,Ranganathan:1993vj,Baggio:2017aww}, as elaborated upon in Section \ref{sec:regularization}. When $\phi_{a,b}$ are chosen to be the marginal operators themselves, this is nothing but the curvature of the Zamolodchikov metric on the conformal manifold.

Closer to the subject of this paper, the Berry curvature also has a storied history in supersymmetric quantum field theory, especially in relation to the $\tt$ equations for the curvature of chiral primaries in topologically twisted $\mathcal{N}=(2,2)$ superconformal field theories in two dimensions \cite{Cecotti:1991me}. It has since been computed for $\mathcal{N}=(4,4)$ SCFTs in two dimensions \cite{deBoer:2008ss} and four-dimensional SCFTs with supersymmetry $\mathcal{N} \geq 2$ \cite{Papadodimas:2009eu, Baggio:2014ioa, Baggio:2017aww, Niarchos:2018mvl}, as well as other developments \cite{Pedder:2007ff, Pedder:2007wp, Pedder:2008je, Sonner:2008fi, Dedushenko_2023}.

We will be interested in the Berry curvature of supersymmetric states in holographic SCFTs in various dimensions. Given that the BPS states at fixed charges form an energy eigenspace, their Berry curvature matrix is a natural object to consider. It determines how much the states in the BPS subspace get mixed up after we perform evolution along a loop in moduli space. In particular, for a flat Berry connection $F=0$, the evolution matrix $U$ along homotopically trivial loops $\gamma$
is the identity operator and no mixing occurs. On the other hand, if $F$ is a random matrix then the states---in a generic basis---will get highly mixed after the evolution because $U$ will act as a random unitary (but not necessarily a Haar random one).

The question we investigate in this work is whether the Berry curvature of BPS states has different chaotic properties for black hole microstates compared to non-black hole microstates.\footnote{A closely related object known as the adiabatic gauge potential (AGP) has been used to characterize integrable vs. chaotic systems \cite{Pandey:2020tru} (see also \cite{Sharipov:2024lah}), but in that context it is the scaling of the AGP with system size (polynomial vs. exponential) that diagnoses chaos, which is very different from the random matrix theory behavior of the Berry curvature matrix for a degenerate subspace that we study in this work.} For the former, we analyze the $\mathcal{N}=2$ SYK model \cite{Fu:2016vas} numerically and $\mathcal{N}=2$ super-JT gravity analytically. These are toy models for supersymmetric black holes where all the BPS states are fortuitous \cite{Chang:2024lxt}, and we find that the Berry curvature matrix $F$ has RMT statistics. Based on these results, we conjecture that this will hold true for any BPS black hole microstates, including the 1/16-BPS black hole microstates in $\mathcal{N}=4$ super-Yang Mills theory (SYM) and the 1/4-BPS black hole microstates in the D1/D5 CFT. On the other hand, for all of the monotones that we examine, we show that $F$ is not a random matrix. These consist of the monotones in variants of the $\mathcal{N}=2$ SYK model, the 1/2-BPS states and some monotone 1/4-BPS states in the D1/D5 theory, and the 1/2-BPS and 1/4-BPS states in $\mathcal{N}=4$ SYM theory all at generic coupling. In fact, in many (but not all) of these monotone cases, we find a flat connection $F=0$. This indicates that there is a sharp distinction between the eigenvalue statistics of the curvature for the black hole microstates compared to the non-black hole microstates.\footnote{Admittedly, our analysis is not exhaustive as it does not include the 1/8-BPS states in $\mathcal{N}=4$ SYM theory, which are conjectured to all be monotones \cite{Chang:2023ywj}, the 1/16-BPS monotone states in $\mathcal{N}=4$ SYM theory, or some of the 1/4-BPS monotones in the D1/D5 CFT. We remain agnostic about whether we expect that their Berry curvature matrix is random, but we make some comments on them in Sections \ref{sec:1/8and1/16} and \ref{sec:D1/D5_1/4BPS}, respectively. We expect that even if they exhibit random matrix statistics, there should still be sharp distinctions from the black hole cases in terms of the Thouless time.}

\noindent Given that the paper is rather lengthy, we provide below a summary of our results.

\subsection{Summary of results}

We begin in Section \ref{sec:general} with an introduction to the Berry curvature. For the Hilbert bundle of states over the moduli space $\mathcal{M}$, we explain in Section \ref{sec:basics} how the Berry connection follows from treating Schr\"odinger's equation as a parallel transport equation for states, and derive \eqref{eq:Berry curvature QM}, which makes manifest the singularities in the curvature when level crossing occurs. These singularities give non-trivial topology and can be thought of as ``sources" for the curvature whose flux is measured by Chern numbers \eqref{eqn:Chernnumbers}. Then, in systems with supersymmetry, we introduce in Section \ref{sec:conj} a special class of deformations, known as conjugation deformations, under which the number of BPS states is preserved. They are obtained by conjugating the supercharge by a general invertible matrix $M$. When $M=e^{\lambda\Lambda}$ with $\Lambda = \pm \Lambda^\dagger$, we derive a simplified formula for the Berry curvature \eqref{eqn:Fconjdef} that looks similar to an LMRS-observable, and proves to be useful in the lower-dimensional examples we study.\footnote{A generalization of \eqref{eqn:Fconjdef} exists for generic deformations \eqref{eqn:curvformula2}, and it would be interesting to understand the properties of generic deformations as well.}  Finally, in Section \ref{sec:regularization}, we review how to regulate the UV divergences that appear in the Berry curvature for QFTs and, in the case of CFTs, we explain \eqref{eq:Berry curv CFT}.

In Section \ref{sec:D1/D5}, we study our first example: the D1/D5 system of $\mathcal{N}=(4,4)$ SCFTs for $M=K3$ and $M=T^{4}$, dual to Type IIB string theory in $\mathrm{AdS}_{3} \times S^{3} \times M$ whose moduli space is $\mathcal{M} = SO_{0}(4,n)/(SO(4) \times SO(n))$. We review in Section \ref{sec:D1/D5_1/2BPS} previous results for the Berry curvature of 1/2-BPS states in $\mathcal{N}=(4,4)$ SCFTs and extend them to include the $T^{4}$ case whose superconformal algebra (SCA) is the medium (large contracted) $\mathcal{N}=(4,4)$ one. The final result is the $\tt$ equations
\begin{equation}\label{eqn:D1/D5curv_intro}
\left(F_{\mu\nu}\right)_{KL} = i\delta_{ab}\left(\left(\delta_{ij}-\mathcal{F}_{ijKL}\right)-[C_{i},C_{j}^{\dagger}]_{KL}\right)
\end{equation}
where $\mu = (a,i)$, $\nu = (b,j)$ are moduli directions, $K,L$ label chiral primaries, $\mathcal{F}_{ijKL}$ is a function of the global symmetry charges of the marginal operators and external states, and $C_{iK}^{M}$ are chiral ring coefficients. Equation \eqref{eqn:D1/D5curv_intro} holds for any $N$ and anywhere on moduli space. Using the fact that these coefficients are covariantly constant on $\mathcal{M}$, one immediately deduces that the curvature is also covariantly constant. Since $\mathcal{M}$ is a symmetric space, a mathematical theorem implies that $F$ is fixed by representations of the holonomy group $SO(4) \times SO(n)$, which allows us to use tools from representation theory to prove that $F$ can never be a random matrix in Section \ref{sec:D1/D5_non-random}. We then discuss in Section \ref{sec:symT4} the various subtleties that arise in the $T^4$ case due to extra global symmetries and an enlarged moduli space coming from double-trace currents, as well as some added complication at the orbifold point $\mathrm{Sym}^{N}(T^4)$ in Section \ref{sec:orbifold}. Next, Section \ref{ccdagertt} uses the $\tt$ equations to show that the chiral ring coefficients, when viewed suitably as a matrix, also cannot be a random matrix, allowing us in Section \ref{sec:LMRSD1D5} to make contact with and improve the BPS chaos analysis \cite{Chen:2024oqv}, who found evidence for weak chaos in a smaller subspace than all of the 1/2-BPS states at fixed charge. Finally, we explain in Section \ref{sec:D1/D5_1/4BPS} why our techniques no longer work for general 1/4-BPS states due to their lesser supersymmetry, but succeed in showing that some 1/4-BPS monotones obtained from superconformal generators, including some superstrata and singleton microstates, have the same curvature as the 1/2-BPS states, and hence it is not a random matrix.

The subject of Section \ref{sec:SYM} is the Berry curvature for BPS states in $\mathcal{N}=4$ SYM, whose moduli space\footnote{In this paper by moduli space we are referring to the conformal manifold, not the moduli space of vacua.} is the upper half-plane parametrized by the complexified gauge coupling $\tau={\theta \over 2\pi} + i {4\pi \over g_{\rm YM}^2}$, which we compute for any $N$ and $\tau$. We first review its 1/2- and 1/4- BPS states and its marginal operators in Section \ref{sec:SYM_review}. Then, in Section \ref{sec:SYM_1/2BPS}, we review the previously known fact that the Berry curvature vanishes for 1/2-BPS states, provide a new proof using Ward identities and crossing symmetry, and check it explicitly using the fermion description of the 1/2-BPS sector. We then show that the Berry curvature also vanishes for the 1/4-BPS states in Section \ref{sec:curv14}, using superconformal Ward identities to localize the integrated four-point function to boundary terms and carefully analyzing the resulting OPEs. Finally, Section \ref{sec:1/8and1/16} explains how our methods fail for the 1/8- and 1/16-BPS states due to the lack of localization to boundary terms, and we speculate on their curvature.

Section \ref{sec:SYK} contains our first example of the Berry curvature for fortuitous states. We consider the $\mathcal{N}=2$ supersymmetric SYK model for fixed couplings $C_{ijk}$ whose BPS states are all fortuitous generically and can be obtained by performing exact diagonalization of the Hamiltonian numerically. We study the Berry curvature due to two different types of deformations. In Section \ref{sec:SYKdirect} we change the couplings $C_{ijk}$, while in Section \ref{sec:SYKconj} we study a conjugation deformation originating from two operators $\Lambda_\mu$, $\Lambda_\nu$ that commute in the UV, in which case the Berry curvature reduces to the commutator of two LMRS-observables $F_{\mu\nu}=i[\widehat{\Lambda}_{\mu},\widehat{\Lambda}_{\nu}]$. In both cases, we find the nearest-neighbor eigenvalue distribution agrees with the Wigner surmise for the GUE Wigner-Dyson ensemble of random matrices, and the Thouless time to be small, indicating long-range level repulsion and strong chaos. Finally, Section \ref{sec:SYKmonotone} analyzes two variants of the $\mathcal{N}=2$ SYK model containing monotones: (1) a two-flavor model, which has monotonous and fortuitous states, but the former are coupling independent so have trivially vanishing Berry curvature and (2) an integrable model where the couplings are restricted to be a wedge product of three one-forms, which has only monotonous states. The Berry curvature can be analytically computed and has few eigenvalues with large multiplicities, hence it is not a random matrix.

We turn in Section \ref{sec:JTgravity} to the $\mathcal{N}=2$ super-JT gravity. The theory describes the near-horizon dynamics of the 1/16-BPS Gutowski-Reall black holes in $\mathrm{AdS}_{5} \times S^{5}$, and so it can be used to probe the amount of chaos in the Berry curvature of 1/16-BPS states in the dual $\cN = 4$ SYM. We start in Section \ref{sec:review JT} with a review of $\mathcal{N}=2$ super-JT gravity and of the LMRS computation for the statistics of a simple operator projected to the BPS subspace. It involves computation of correlation functions using Witten diagrams. For heavy operators whose conformal dimension $\Delta\to\infty$ the computation simplifies considerably, and becomes a combinatorial problem of counting non-crossing chord diagrams on the relevant topology. Next, in Section \ref{sec:JT Berry comp} we move to the Berry curvature for conjugation deformations by two heavy operators, which again reduces to the commutator of two LMRS-observables $F_{\mu\nu}=i[\widehat{\Lambda}_{\mu},\widehat{\Lambda}_{\nu}]$. We find that it has the same spectrum as the free commutator of two GUE random matrices, then derive a novel formula for the second order cumulant of the free commutator, $\bigl\langle \Tr\bigl(F_{\mu\nu}^m\bigr)\Tr\bigl(F_{\mu\nu}^n\bigr) \big\rangle$, and use it to show that the spectral form factor of $F_{\mu\nu}$ has a linear ramp with an $O(1)$ Thouless time, thus exhibiting strong chaos.

In Section \ref{sec:topo}, we explore the non-trivial topology of the Hilbert bundle of BPS states in the $\mathcal{N}=2$ SYK model, which comes from singularities at loci in moduli space where extra BPS states appear and leads to non-zero, quantized Chern numbers on $2$-spheres. Although this is somewhat tangential to the main theme of the paper, it has the interesting consequence that, despite $F$ being well approximated by a random matrix at every generic point in the moduli space, it nevertheless satisfies global quantization conditions. Section \ref{sec:specialloci} provides some simple examples of such special loci that are present for any $N$ and gives a general---albeit very complicated---set of equations for these loci for even $N$. A subtlety related to redundancies on subspaces of the moduli space whose action must be quotiented out, and the resulting stratified spaces, is addressed in Section \ref{sec:enhancedsym}. We then perform an explicit analysis of the moduli space for $4 \leq N \leq 6$ in Section \ref{sec:modulismallN}, where all singular loci and the result of stratification can be completely understood. Finally, Section \ref{sec:Chernnumbers} contains numerical computations of the first Chern number of $S^{2}$ for various $N$ from which we extrapolate a general $N$ formula \eqref{eqn:ChernnumbergeneralN}. The Chern number displays exponential growth in $N$, which can be interpreted as associated to the singularity at the origin $C_{ijk}=0$ for all $i,j,k$, where all states become BPS. 
        
We conclude in Section \ref{sec:disc} with some discussion and various open questions. In Appendix \ref{sec:quantummmetric}, we discuss the quantum metric tensor---the symmetrized cousin of the Berry curvature---and use it to derive an upper bound on the first Chern number by a ``quantum volume". The details of our analysis for the D1/D5 case are provided in Appendix \ref{sec:D1/D5details}, including an in-depth analysis of the subtleties in the $T^{4}$ case, and some explicit checks around the orbifold point. Appendix \ref{sec:SYMdetails} contains our new proof that the Berry curvature for 1/2-BPS states vanishes in $\mathcal{N}=4$ SYM and the details of our proof that it also vanishes for the 1/4-BPS states. In Appendix \ref{sec:SYKdetails}, we provide some technical results for the $\mathcal{N}=2$ SYK model, prove a theorem about the topology of its moduli space, and compute the Berry curvature for the integrable variant of the model.

%%%%%%%%%%%
% SECTION %
%%%%%%%%%%%
\section{Berry curvature: general discussion}
\label{sec:general}

We begin with a basic introduction to the Berry curvature and some of its properties. The curvature for a special type of deformation, known as conjugation deformation, is then presented. This allows us to make contact with the LMRS criterion and will be used in Sections \ref{sec:SYK} and \ref{sec:JTgravity}. Finally, we discuss the issue of divergences that appear for the curvature in quantum field theories. For a nice textbook on Berry curvature containing the mathematical prerequisites along with many examples and applications, see \cite{chruscinski2004geometric}.

%%%%%%%%%%%
% SECTION %
%%%%%%%%%%%
\subsection{Basics}
\label{sec:basics}
 
Consider a family of quantum mechanical systems, each with a Hilbert space $\mathbb{H}$ and a Hamiltonian $H(\lambda)$ that depends on some parameters $\{\lambda_{\mu}\}$. The parameters can be regarded as coordinates on a manifold  $\mathcal{M}$, and then this defines a Hilbert bundle with base space $\mathcal{M}$ whose fibres are $\mathbb{H}$. Let 
\begin{equation}\label{eqn:Espace}
    \mathbb{H}_{n} = \bigl\{\,\ket{a(\lambda)}\;\big|\; a=1,\ldots,D\,\bigr\} \; \subset \; \mathbb{H}
\end{equation}
be an eigenspace of the Hamiltonian containing all the states with energy $E_{n}(\lambda)$ and degeneracy $D$, i.e., the fibre of a sub-bundle.\footnote{Note that there is an associated $U(D)$ principal bundle since every element of $\mathbb{H}_{n}$ can be obtained from an element of $U(D)$ acting on a reference state.} For every point $\lambda \in \mathcal{M}$, we choose an arbitrary basis for the fibre. To relate the different fibres, we need to define a connection on the bundle so that we can parallel transport states. As we shall see, Schr\"odinger's equation in the adiabatic limit gives a natural choice of connection on this sub-bundle \cite{PhysRevLett.51.2167}.

Suppose we evolve the system by moving along some curve $\{\lambda(t')\,|\,0 \leq t' \leq t\}$ in $\mathcal{M}$ such that at any point the states remain degenerate\footnote{The requirement for exact degeneracy can be somewhat relaxed as long as there's a separation of scales between the splitting and the gap to the other states.} and separated from the other states. If the evolution is performed adiabatically, then a state in $\mathbb{H}_{n}$ will remain in this subspace under evolution \cite{BornFock1928, 1950JPSJ....5..435K}. Starting with a basis element $\ket{a(\lambda(t=0))}$, one finds the solution of Schr\"odinger's equation to be
\begin{equation}\label{eqn:evolvedstate}
    \ket{\psi(t)} = e^{-\frac{i}{\hbar}\int_{0}^{t}dt'\, E_{n}(t')}\sum_{b=1}^{D}U_{ab}^{}(t)\ket{b(\lambda(t))} \,,
\end{equation}
which has the expected dynamical phase along with a unitary transformation $U(t) \in U(D)$. In the non-degenerate case $K=1$, $U(t)$ is simply the Berry phase \cite{Berry:1984jv}.

Plugging \eqref{eqn:evolvedstate} into Schr\"odinger's equation and taking the inner product with some basis element $\bra{c(\lambda)}$, we find
\begin{equation}\label{eqn:Ueqn}
\frac{d}{dt}U_{ac}+\sum_{b=1}^{D}U_{ab}\bra{c}\frac{d}{dt}\ket{b} = 0\,.
\end{equation}
We recognize this as the parallel transport equation for the following connection on the Hilbert bundle\footnote{\label{foo:transposeA}We warn the reader that in this convention, locally the connection is written with the transpose of the matrix defined above, \ie $A = (A^T)_{ba} |b\rangle\langle a|$. This will cause various transposes to appear when we write the connection (and its curvature) in terms of other operators without explicit indices.} 
\begin{equation}\label{eqn:Adef}
    A_{ab} = -i\left(U^{-1}\frac{d}{dt}U\right)_{ab} = i \bra{b}\mathrm{d}\ket{a} \,,
\end{equation}
where $\mathrm{d}$ is the exterior derivative on $\mathcal{M}$, known as the Wilczek-Zee connection \cite{PhysRevLett.52.2111}. Under a local change of basis 
\begin{equation}
    \bigl\{\ket{a(\lambda)}\bigr\} \to \bigl\{\ket{a'(\lambda)}\bigr\} = \bigl\{V(\lambda)\ket{a(\lambda)}\bigr\}  \,,
\end{equation}
for some unitary $V(\lambda)$, the connection transforms as
\begin{equation}
    A \to VAV^{\dagger}+i\,\mathrm{d}VV^{\dagger}
    \,,
\end{equation}
and so it defines a $U(D)$ gauge field on $\cM$, with gauge transformations being arbitrary changes of basis at any point on the moduli space $\cM$.
We can now solve the equation of motion for $U$ in \eqref{eqn:Ueqn} in terms of this connection to find
\begin{equation}
    U(t) = P\,e^{i \int_{\lambda(0)}^{\lambda(t)} A_{\mu}(\lambda)d\lambda^{\mu}} \,,
\end{equation}
where $P$ denotes the path-ordered exponential. In sum, we have found that Schrödinger's equation in the adiabatic limit is the parallel transport equation for the Wilczek-Zee connection.

It is natural to study the curvature of this connection
\begin{equation}
    F = \mathrm{d} A-i A \wedge A \,,
\end{equation}
known as the Berry curvature. It transforms under a local change of basis as $F \to VFV^{\dagger}$, and hence its moments $\Tr(F^k)$, or eigenvalues, provide gauge-invariant data for the geometry of the Hilbert bundle. Another useful gauge-invariant object arises when the deformation takes us around a loop $\gamma$ in moduli space, \ie $\lambda(0) = \lambda(t)$. Despite the fact that we start and end with the same Hamiltonian, our subspace $\mathbb{H}_n$ can mix non-trivially with itself in a way encoded by $U(\gamma)$, the holonomy of the connection around the loop. When our subspace is one-dimensional, the holonomy is simply the Berry phase. More generally, it is given by 
\begin{equation}
    U = Pe^{i \oint_{\gamma}A_{\mu}dx^{\mu}} = \mathcal{P}e^{i \int_{\Sigma}\widehat{F}_{\mu\nu}dx^{\mu}\wedge dx^{\nu}} \,,
\end{equation}
where $\gamma=\partial\Sigma$ and we have used the non-Abelian Stokes's Theorem \cite{Schles, Arefeva, Hirayama:1997ct, Karp:1999xw} so $\widehat{F}$ is the twisted curvature obtained from the curvature $F$ by dressing it with Wilson lines anchored to $\gamma$, $\mathcal{P}$ is a two-dimensional generalization of the usual path-ordering, and the holonomy is independent of the choice of $\Sigma$.

The Berry curvature will be the primary object of study in this work so we will derive various formulae for it and analyze their properties. To start, using \eqref{eqn:Adef} the curvature can be written explicitly as
\begin{equation}\label{eqn:Fexplicit}
    (F_{\mu\nu})_{ab} = i\big(\langle\partial_{\mu}b|\partial_{\nu}a\rangle-\langle\partial_{\nu}b|\partial_{\mu}a\rangle+\langle c|\partial_{\mu}a\rangle\langle b|\partial_{\nu}c\rangle-\langle c|\partial_{\nu}a\rangle\langle b|\partial_{\mu}c\rangle\big) \,.
\end{equation}
By inserting a resolution of the identity in terms of energy eigenstates, it is straightforward to rewrite this as
\begin{equation}\label{eqn:curvformula}
    \left(F_{\mu\nu}\right)_{ab} = \sum_{m \not \in \mathbb{H}_{n}}\frac{i}{(E_{n}-E_{m})^{2}}\bra{b}\partial_{\mu}H\ket{m}\bra{m}\partial_{\nu}H\ket{a} - (\mu \leftrightarrow \nu) \,.
\end{equation}
This reformulation reveals two important properties of the Berry curvature. First, observe that whenever the theory has a global symmetry that is preserved by the deformation, so $\partial_{\mu}H$ is neutral, the curvature is block-diagonal with respect to the corresponding charge sectors and the sum over intermediate states truncates to only those with the same charges as the external states. Use of this fact is pervasive throughout our analysis, and we will always work in subspaces of fixed global charges. Second, \eqref{eqn:curvformula} makes manifest that the Berry curvature diverges in regions of moduli space where level-crossing occurs because the corresponding energy differences in the denominator vanish. These are singular regions of the Hilbert bundle and must be removed from the moduli space $\mathcal{M}$ in order to have a well-defined bundle. By the von Neumann-Wigner theorem, such loci generically have real codimension greater than or equal to $3$ \cite{1929PhyZ...30..467V}. Cutting them out will in general lead to a non-trivial topology beyond the intrinsic topology of $\mathcal{M}$, which will be explored in detail for $\mathcal{N}=2$ SYK in Section \ref{sec:topo}.

This non-trivial topology has interesting consequences. As the curvature of a fibre bundle, $F$ must have quantized Chern numbers \cite{Eguchi:1980jx}
\begin{equation}\label{eqn:Chernnumbers}
    c_{k} = \frac{1}{(2 \pi)^{k}}\int_{\sigma_{2k}} s_{k}\left(F\right) \in \mathbb{Z} \,,
\end{equation}
where $s_{k}(F)$ is the $k$-th symmetric polynomial in the eigenvalues of $F$, which can always be written in terms of traces and wedge products of $F$, and $\sigma_{2k} \subset \mathcal{M}$ is a $2k$-cycle.\footnote{In practice, we choose this $2k$-cycle to be a sphere $\sigma_{2k}=S^{2k}$.} For instance, the first Chern number is simply
\begin{equation}\label{eqn:firstChernnumber}
    c_{1} = \frac{1}{2 \pi}\int_{\sigma_{2}} \Tr F \in \mathbb{Z} \,.
\end{equation}
These integers can be non-zero when $\sigma_{2k}$ is non-contractible.\footnote{In Appendix \ref{sec:quantummmetric} we explain an interesting upper bound on the first Chern number by the ``quantum volume" of the corresponding $2$-sphere.} One way this can occur is that $\sigma_{2k}$ encloses one of the level-crossing regions. Therefore, these level-crossings can be thought of as sources creating some non-trivial flux through $\sigma_{2k}$. In fact, this source picture can be made precise for the simple example of a spin coupled to a magnetic field in three spatial dimensions. The moduli space $\mathcal{M} = \mathbb{R}^{3}-\{0\}$ is the value of the magnetic field with the origin removed, as it is a point of level crossing where all states have zero energy. The Berry curvature is the same as the electromagnetic field strength due to a magnetic monopole at the origin \cite{Berry:1984jv}.

We note in passing a curious formula for the curvature, derived by Berry and Robbins \cite{berrychaos} in the Abelian case and generalized here to the non-Abelian one.  Writing the denominator of \eqref{eqn:curvformula} as 
\begin{equation}\label{eqn:denom_rewrite}
    \frac{1}{(E_{n}-E_{m})^{2}} = -\lim_{\epsilon \to 0}\int_{0}^{\infty}du\,u e^{-\epsilon u}\cos\bigl((E_{n}-E_{m})u\bigr) \,,
\end{equation}
and denoting\footnote{We define $O_{\mu}^T$ with a transpose for reasons explained in Footnote~\ref{foo:transposeA}. One also needs to be careful about it to get the correct sign in \eqref{eqn:curvformula2}.} $O_{\mu} = (\partial_{\mu}H)^T$ and its projection to the $n^{\rm th}$ energy eigenspace by $\widehat{O}_{\mu} = P_{n}O_{\mu}P_{n}$, the curvature takes the form
\begin{equation}\label{eqn:curvformula2}
    F_{\mu\nu} = \frac{i}{2}\lim_{\epsilon \to 0}\int_{0}^{\infty}du\,u e^{-\epsilon u} \Bigl(P_n \left[O_{[\mu}(u), O_{\nu]}(0)\right] P_n
    - \bigl[\widehat{O}_{[\mu},\widehat O_{\nu]}\bigr]\Bigr) \,,
\end{equation}
where we defined\footnote{Since $\mathbb{H}_n$ is degenerate, $\widehat O$ commutes with the Hamiltonian and we can drop its time dependence.} $O(u) = e^{iHu}O(0)e^{-iHu}$. The integrand in \eqref{eqn:curvformula2} almost looks like an LMRS-type operator, and might allow for convenient path integral computations as in Section~\ref{sec:JTgravity}. In particular, one could use this to compute all of the moments of $F_{\mu\nu}$, and for the BPS subspace these are thermal correlation functions in the limit of zero temperature, which may be computable from the dual gravity picture. Below we will see that for certain types of deformations, we find an even more simplified version of this formula.

%%%%%%%%%%%
% SECTION %
%%%%%%%%%%%
\subsection{Conjugation deformations}\label{sec:conj}

For any supersymmetric quantum mechanical theory with at least two supercharges $Q$ and $Q^\dagger$, there exists a special type of deformation called conjugation deformation. It has the nice property that it preserves the number of BPS states,\footnote{In this section by BPS states we mean supersymmetric ground states obeying $Q|a\rangle=Q^\dagger|a\rangle=0$.} not just the Witten index \cite{WITTEN1982253}. Hence, level crossing cannot occur for the BPS subspace under such deformations, and the corresponding Berry curvature does not diverge. We will now show that it is given by a simple formula.

Let $Q(\lambda)$ and $Q^{\dagger}(\lambda)$ be a complex supercharge and its Hermitian conjugate at some point in moduli space $\lambda \in \mathcal{M}$. These give rise to the Hamiltonian
\begin{equation}
    H(\lambda) = \bigl\{Q(\lambda),Q^{\dagger}(\lambda)\bigr\} \,,
\end{equation}
whose null space is the BPS subspace. Now, suppose we change the supercharges via conjugation by an invertible matrix $M$
\begin{equation}\label{eqn:conjdeform}
    Q \to \tilde{Q} = M^{-1}QM, \qquad Q^{\dagger} \to \tilde{Q}^{\dagger} = M^{\dagger}Q^\dagger (M^{-1})^{\dagger} \,,
\end{equation}
which preserves the nilpotency of the supercharges $\tilde{Q}^{2}=(\tilde{Q}^{\dagger})^{2}=0$. The Hamiltonian becomes
\begin{equation}
    \tilde{H} = \bigl\{\tilde{Q}(\lambda),\tilde{Q}^{\dagger}(\lambda)\bigr\} = M^{-1}QMM^{\dagger}Q^\dagger (M^{-1})^{\dagger}+M^{\dagger}Q^\dagger (M^{-1})^{\dagger}M^{-1}QM\,.
\end{equation}
If we take $M$ to be unitary, then the deformation is rather trivial,\footnote{We remind the reader that one can have a non-trivial holonomy even for unitary transformations. For example, one finds a non-trivial Berry phase for a spin coupled to a magnetic field under rotations of that field.} $\tilde{H} = M^\dagger HM$. However, for more general $M$ this gives non-trivial deformations, changing the non-zero energy eigenvalues while preserving the number of ground states \cite{WITTEN1982253}, as they map $Q$-cohomology representatives to $Q$-cohomology representatives. Notice that they are not enough for finding the BPS states of the deformed theory given knowledge of the BPS states in the undeformed theory.\footnote{Observe that conjugation deformations can preserve fortuity in the following way. Given the projection $\pi_{N}$ from the $N = \infty$ Hilbert space to the finite-$N$ Hilbert space, if one considers a sequence of invertible maps $M_{N}$ that commute with the projection, namely $\pi_{N}M_{\infty}^{-1} = M_{N}^{-1}\pi_{N}$, then $M_{N}^{-1}\psi$ is fortuitous whenever $\psi$ is fortuitous. To see this, suppose by way of contradiction that $\psi$ is fortuitous, but $M_{N}^{-1}\psi$ is not. Then there exists a $\tilde{Q}_{\infty}$-closed $M_{\infty}^{-1}\tilde{\psi}_{\infty}$ in the $N=\infty$ Hilbert space such that $\pi_{N}M_{\infty}^{-1}\tilde{\psi}_{\infty} = M_{N}^{-1}\psi$, but this implies that $\pi_{N}\tilde{\psi}_{\infty}=\psi$, contradicting the fact that $\psi$ is fortuitous. }

Let us parametrize a family of such deformations using a fixed complex operator $\tilde\Lambda$ by 
\begin{equation}
    M(\lambda) = e^{\lambda \tilde\Lambda} \,,
\end{equation}
i.e., starting at $\lambda_{0}$, we have a one-parameter family of supercharges $Q(\lambda_{0}+\lambda) = M^{-1}(\lambda)Q(\lambda_{0})M(\lambda)$ for $\lambda \in \mathbb{R}$. In general, $\tilde\Lambda$ can be decomposed into a Hermitian and anti-Hermitian part. If $\tilde\Lambda$ is anti-Hermitian, then $M$ is unitary. It is interesting to instead consider Hermitian $\tilde\Lambda$. Under an infinitesimal deformation, $\partial_{\lambda} Q = [Q,\tilde\Lambda]$ and hence the Hamiltonian changes as
\begin{equation}\label{eqn:deltaHconjdeform}
    \partial_{\lambda}H = \bigl\{Q^{\dagger},[Q,\tilde\Lambda]\bigr\} + \bigl\{Q,[\tilde\Lambda,Q^{\dagger}]\bigr\} \,.
\end{equation}
Given two such different deformations $\tilde\Lambda_{\mu}$, $\tilde\Lambda_{\nu}$, a convenient trick to derive the Berry curvature of the BPS subspace is to first define the auxiliary quantity 
\begin{equation}\label{eqn:auxcurv}
    \left(F_{\mu\nu}\right)_{ab}(x) = i\bra{b}\partial_{\mu}H\frac{1}{(H-x)^{2}}\partial_{\nu}H\ket{a} - (\mu \leftrightarrow \nu) \,,
\end{equation}
whose limit as $x\rightarrow 0$ gives the Berry curvature in \eqref{eqn:curvformula},
\begin{equation}
    F_{\mu\nu} = \lim_{x \to 0}F_{\mu\nu}(x)\,.
\end{equation}
Plugging (\ref{eqn:deltaHconjdeform}) into \eqref{eqn:auxcurv}, and since $H$ commutes with the supercharges and the supercharges annihilate the external states, one arrives at a simple expression
\begin{equation}\label{eqn:curvconjdeformx}
    \left(F_{\mu\nu}\right)_{ab}(x) = i\bra{b}\tilde\Lambda_{\mu}\frac{H^{2}}{(H-x)^{2}}\tilde\Lambda_{\nu}\ket{a} - (\mu \leftrightarrow \nu) \,.
\end{equation}
For any observable $O$ on the full Hilbert space, define the BPS-projected LMRS-type operator \cite{Lin:2022zxd}
\be
\label{projectedBPS}
\widehat{O} = P_{\mathrm{BPS}}\,O\,P_{\mathrm{BPS}}\,.
\ee Then, adding and subtracting appropriate terms from \eqref{eqn:curvconjdeformx} and taking $x \to 0$, we arrive at
\begin{equation}
\label{eqn:Fconjdef}
    F_{\mu\nu} = i\left([\widehat{\Lambda}_{\mu},\widehat{\Lambda}_{\nu}] - \widehat{[\Lambda_{\mu},\Lambda_{\nu}]}\right) \,,
\end{equation}
where we defined\footnote{The transpose is due to reasons explained in Footnote~\ref{foo:transposeA}. One also needs to be careful about it to get the correct sign in \eqref{eqn:Fconjdef}.} $\Lambda_{\mu,\nu} = \tilde\Lambda^T_{\mu,\nu}$. When $\Lambda_{\mu}$ is a simple operator, $\widehat{\Lambda}_{\mu}$ is an LMRS observable so we have managed to write the curvature purely in terms of LMRS-type observables. Note the simplification compared to \eqref{eqn:curvformula2}, which applies for general deformations. 

Thus, conjugation deformations allow us to make contact with the LMRS criterion for chaos. However, we emphasize that conjugation deformations only comprise a very special subspace in the space of all deformations and in quantum field theories there can be obstacles to performing such deformations. For instance, in seminal work by Witten \cite{WITTEN1982253}, it was shown that in non-Abelian gauge theories one cannot change both the Yang-Mills coupling and the theta angle by such deformations because it maps normalizable states to non-normalizable states. So we are not aware of a way to apply conjugation deformations for computing Berry curvature in $\mathcal{N}=4$ super-Yang Mills theory where these are the only two possible directions in moduli space that preserve the full $\mathcal{N}=4$ supersymmetry. However, we will analyze them in cases where they are well-defined, including $\mathcal{N}=2$ SYK and $\mathcal{N}=2$ JT gravity in Sections \ref{sec:SYK} and \ref{sec:JTgravity}, respectively.

%%%%%%%%%%%
% SECTION %
%%%%%%%%%%%
\subsection{Subtleties in QFT and renormalization scheme}
\label{sec:regularization}

In the rest of the paper, we will compute the Berry curvature of BPS states in CFTs in various dimensions as a function of the marginal deformations of the theory. The computation of Berry curvature in QFT is not straightforward due to the infinite number of degrees of freedom, coming both from the IR and the UV. By placing the theory in compact volume, as we will do for CFTs by placing them on $S^{d-1}\times {\mathbb R}$, we can overcome IR-related issues but we may still encounter difficulties from the UV. The difficulty comes from the fact that under a marginal deformation we are modifying the state at all scales and the vacuum state of the deformed theory may not be realizable as a normalizable state in the Hilbert space of the undeformed theory. Technically, in a direct computation of the Berry curvature using \eqref{eqn:curvformula}, we may encounter UV divergences from the infinite sum over intermediate states.

On the other hand, we expect that these divergences, coming from the deep UV of the theory, are state-independent. Hence one might hope that there is a natural regularization and renormalization procedure where we compute the ``vacuum-subtracted" Berry phase of excited states, i.e. a scheme where we renormalize the Berry phase of the vacuum to be zero (assuming the vacuum is unique) and then associate a finite Berry phase to excited states. It would be interesting to investigate this more systematically but we will not do it in this paper.

Instead, and since we will be considering the Berry curvature along marginal deformations in CFTs, we will adopt a complementary approach which will allow us to associate a UV-finite Berry curvature to states of the CFT. The state-operator map provides a natural identification between states of the CFT on $S^{d-1}\times {\mathbb R}$ and local operators on the plane. This mapping then provides a relation of the Berry connection for states to the connection characterizing the  mixing of local operators over the conformal manifold of CFTs. The latter has been discussed in various earlier works and provides an elegant way to renormalize UV divergences encountered in the computation.

A connection on the space of local operators, which describes the mixing of operators with the same quantum numbers under marginal deformations, has been discussed in several previous works \cite{Kutasov:1988xb, Ranganathan:1992nb, Ranganathan:1993vj}. The idea is that in the framework of conformal perturbation theory it is natural to think of local operators as being associated to vector bundles over the conformal manifold. There is a natural metric on the fibres, which is the analogue of the Zamolodchikov metric. For any two operators $\phi_a$, $\phi_b$ of conformal dimension $\Delta$ that belong to a fibre, the metric at a point on the conformal manifold is
\begin{equation}\label{Zmetric}
    g_{ab} = \lim_{z\to\infty} z^{2\Delta} \langle \phi_b^{\dagger}(z)\,\phi_a(0)\rangle \,.
\end{equation}
There is an integrable, metric-compatible connection on these bundles. The curvature of this connection can be computed in conformal perturbation theory by considering a doubly-integrated antisymmetrized 4-point function
\be
\label{berry4point}
(F_{\mu\nu})_{ab} = {i\over \mathcal{S}_d} \left[ \int_{|x|\leq 1} d^dx \int_{|y|\leq1}d^dy\,\,\langle \phi_b^\dagger(\infty) \,\,\mathcal{O}_{[\mu}(x) \,\,\mathcal{O}_{\nu]}(y)\,\,\phi_a(0)\rangle\right]_{\rm renormalized} \,,
\ee
where $\mathcal{O}_{\mu}$, $\mathcal{O}_{\nu}$ are the marginal operators along which the curvature is computed. The overall constant $\mathcal{S}_d$, related to the normalization conventions of marginal operators, is selected so as to simplify certain formulae below. In 2d we take $\mathcal{S}_2=\pi^2$ and in 4d $\mathcal{S}_4= (2\pi)^4$. The UV divergences of the Berry phase anticipated in QFT make their appearance in \eqref{berry4point} as short-distance singularities in the integration domain when operators approach each other. Conformal perturbation theory suggests a natural way of renormalizing these divergences, as discussed\footnote{See also \cite{Friedan:2012hi, Balthazar:2022hzb} for a slightly different regularization scheme that involves tracking possible contact terms, and for application of CFT inversion formulas to write the curvature in a way reminiscent of \eqref{eq:Berry curvature QM}.} in detail in \cite{Ranganathan:1993vj} for the 2d case, see also \cite{deBoer:2008ss, Baggio:2017aww}. This allows us to define a finite curvature for the bundle of operators on the conformal manifold. We then adopt the renormalization scheme that UV divergences encountered when applying \eqref{eqn:curvformula} for states $|a\rangle, |b\rangle$ of the CFT on $S^{d-1}\times {\mathbb R}$ are renormalized by mapping the computation to \eqref{berry4point} for the dual local operators $\phi_a,\phi_b$.\footnote{For a demonstration that for continuous families of CFTs the Berry curvature \eqref{eqn:curvformula} for states on $S^{d-1}\times R$ under marginal deformations can be naturally related to the curvature on the space of local operators \eqref{berry4point} computed by conformal perturbation theory, see \cite{Baggio:2017aww}.} The details of the renormalization procedure for \eqref{berry4point} can be found in the references mentioned above. They amount to excluding small balls of radius $\epsilon$ around the insertions of the various operators from the integration region, so the $y$ integral is over $\{ \epsilon \le |y| \le 1 \} \cap \{ |x-y| \ge \epsilon \}$ and the $x$ integral over $\{ \epsilon \le |x| \le 1 \}$. The next step is to evaluate the inner integral in the limit $\epsilon \to 0$ while using minimal subtraction, keeping only its finite piece in $\epsilon$. Last, one evaluates the outer integral in the same limit, keeping again only the finite piece.

%%%%%%%%%%%
% SECTION %
%%%%%%%%%%%
\section{Monotones in D1/D5 CFT}
\label{sec:D1/D5}

The theory of Type IIB strings in $\mathrm{AdS}_{3} \times S^{3} \times M$ for $M=K3$ or $T^4$ is obtained from the low-energy limit of $N_{1}$ D1-branes and $N_{5}$ D5-branes in $\mathbb{R}^{10}$ compactified on $M \times S^{1}$ with semiclassicality reached at large $N=N_{1}N_{5}$. The moduli space of this so-called D1/D5 system is a conformal manifold of $\mathcal{N}=(4,4)$ SCFTs\footnote{There are actually two different $\mathcal{N}=(4,4)$ superconformal algebras at play here because $M=K3$ has the small $\mathcal{N}=(4,4)$ superconformal algebra, while $M=T^4$ has a larger $\mathcal{N}=(4,4)$ superconformal algebra, known as the medium or (contracted) large algebra. See Appendix \ref{sec:N=(4,4)SCA} for the details.} that is given by (up to discrete quotients)\footnote{These discrete quotients do not affect our discussion of the curvature because, away from fixed points of the quotient, the curvature of the true moduli space will be the same as that of \eqref{eqn:D1/D5modulispace_ST}.}
\begin{equation}\label{eqn:D1/D5modulispace_ST}
     \mathcal{M} = \frac{SO_{0}(4,n)}{SO(4) \times SO(n)} \,,
\end{equation}
where $SO_{0}(4,n)$ is the identity component of $SO(4,n)$, and $n$ is equal to the number of chiral primaries in the theory with conformal dimensions $(h,\widetilde{h})=(\frac{1}{2},\frac{1}{2})$ ($n=5$ for $T^{4}$ and $n=21$ for $K3$).\footnote{In the case of $T^4$, the counting $n=5$ corresponds to $({1\over 2},{1\over 2})$ chiral primaries which are also primaries under the enhanced symmetry algebra, see Section \ref{sec:symT4}.} Certain superconformal descendants of these operators give the marginal operators generating flows along $\mathcal{M}$ \cite{Larsen:1999uk,Seiberg:1999xz}. In the case of $T^4$, the superconformal algebra of the theory is somewhat larger. This leads to extra marginal deformations of $J\widetilde{J}$ form and the conformal manifold contains an extra factor in addition to \eqref{eqn:D1/D5modulispace_ST}. This extra factor will be discussed in more detail in Section \ref{sec:symT4}.

It is expected that there exists a ``strongly-coupled" region dual to supergravity in $\mathrm{AdS}_{3} \times S^{3} \times M$ and a ``free" locus in $\mathcal{M}$ given by the symmetric product orbifold of the non-linear sigma model CFT with target space the symmetric product of $N$ copies of $M$, denoted $\mathrm{Sym}^{N}(M)$, which is dual to tensionless strings.\footnote{It has been argued that tensionless string theory in $\mathrm{AdS}_{3} \times S^{3} \times M$ is actually dual to a grand canonical ensemble of $\mathrm{Sym}^{N}(M)$ theories (with an extra factor of $T^{4}$ in the $T^{4}$ case) \cite{Kim:2015gak, Eberhardt:2021jvj, Aharony:2024fid}.}

The 1/2-BPS states of the theory are chiral primaries for both the left- and the right-moving sectors
\begin{equation}\label{eqn:D1/D5_1/2BPS}
    \bigg\{1/2 \mathrm{-BPS\;states}\bigg\} = \{\ket{\mathrm{chiral\;primary}}\ket{\mathrm{chiral\;primary}}\} \,,
\end{equation}
as they preserve $8$ of the $16$ supercharges.  The total number of such states is constant on the moduli space \eqref{eqn:D1/D5modulispace_ST} in the case $M=K3$ \cite{Aharony:1999ti}, since all states are bosonic and cannot recombine into long multiplets,  and is strongly expected to be constant for $M=T^4$ as well.

There is mixing of these states as one moves along the moduli space leading to a non-trivial Hilbert bundle and we will see that this bundle has non-zero curvature.

All of the 1/2-BPS states are monotones because they can be embedded from some fixed $N=N_{\ast}$ Hilbert space into any larger $N$ Hilbert space by appending trivial cycles and symmetrizing. In supergravity, which becomes semiclassical in the large $N$ limit, the states with energies $E \sim O(N^{0})$ are dual to BPS supergravitons \cite{Larsen:1998xm, Deger:1998nm, deBoer:1998kjm} and the states whose energies scale as $E \sim O(N)$ are dual to smooth, horizonless, two-charge Lunin-Mathur (LM) geometries \cite{Lunin:2001fv}.

The 1/4-BPS states correspond to states with excitations on the left-moving sector (or right-moving), thereby breaking all supersymmetries for that chiral half, viz.,
\begin{equation}\label{eqn:D1/D5_1/4BPS}
    \bigg\{1/4 \mathrm{-BPS\;states}\bigg\} = \{\ket{\mathrm{anything}}\ket{\mathrm{chiral\;primary}}\} \,.
\end{equation}
Most of these states are fortuitous \cite{Chang:2025wgo} and include the black hole microstates in $\mathrm{AdS}_{3} \times S^{3} \times M$ at the supergravity point \cite{Strominger:1996sh}, while some are monotonous \cite{Chang:2025rqy}. There is a subset of the monotonous states that have known smooth, horizonless supergravity duals, such as singletons \cite{Hughes:2025car} and superstrata \cite{Bena:2015bea, Bena:2016agb, Shigemori:2020yuo}. A lightning review of the D1/D5 moduli space and summary of its BPS states can be found in \cite{Shigemori:2020yuo}, see also the review \cite{David:2002wn}.

We first consider the Berry curvature of the chiral primary states in \eqref{eqn:D1/D5_1/2BPS}. In Section \ref{sec:D1/D5_1/2BPS}, we will review the computations in \cite{deBoer:2008ss, deBoer:2008qe, Baggio:2017aww} where a formula for the curvature was obtained in terms of the chiral ring coefficients, which is the ${\cal N}=(4,4)$ analogue of the $\tt$ equations \cite{Cecotti:1991me}, and generalize the result slightly to allow for the ${\rm Sym}^N(T^{4})$ case. We then review the arguments in \cite{deBoer:2008ss} that the curvature is determined by representations of the holonomy group of $\mathcal{M}$. In Section \ref{sec:D1/D5_non-random}, we prove that this implies the curvature can never have RMT statistics. Next, in Section \ref{sec:symT4}, we discuss the ${\rm Sym}^N(T^{4})$  case, which is slightly more complicated than the ${\rm Sym}^N(K3)$ case, due to the enlarged symmetry, but allows for explicit computations at the orbifold point discussed in Section \ref{sec:orbifold}. In Section \ref{ccdagertt} we demonstrate how the $\tt$ equations provide an alternative way to determine the spectrum of BPS three-point functions computed in \cite{Chen:2024oqv} and in Section \ref{sec:LMRSD1D5}, we explain the implication for the BPS chaos discussion in \cite{Chen:2024oqv}. In particular, we find that upon including the full 1/2-BPS subspace of $\textrm{Sym}^N (T^4)$ into the computation, the weak chaos found in \cite{Chen:2024oqv} is completely removed. Finally, in Section \ref{sec:D1/D5_1/4BPS}, we compute the Berry curvature for the 1/4-BPS states where $\ket{\mathrm{anything}}$ in \eqref{eqn:D1/D5_1/4BPS} is obtained by elements of the superconformal algebra acting on chiral primary states, which includes some of the singletons and superstrata, and find similar results to the 1/2-BPS case. 

We will use the following notation in this section.

\paragraph{Notation.}
\begin{itemize}
    \item $i,j$ label $(\frac{1}{2},\frac{1}{2})$ chiral primaries
    \item $a,b$ label $SO(4)$ supercurrent directions
    \item $K,L$ label general chiral primaries
    \item $r,\tilde{r}$ are left- and right-moving $SU(2)$ R-charges
    \item $\alpha,\beta$ label complex fermions
    \item $c = 6\hat{k}$ where $c$ is the central charge and $\hat{k}$ is the Kac-Moody level of the $SU(2)$ R-symmetry
\end{itemize}

%%%%%%%%%%%
% SECTION %
%%%%%%%%%%%
\subsection{Berry curvature of 1/2-BPS states in ${\cal N}=(4,4)$ theories}
\label{sec:D1/D5_1/2BPS}

\noindent Let us begin by reviewing the Berry curvature for 1/2-BPS chiral primary states. This was originally computed for ${\cal N}=(2,2)$ theories by Cecotti and Vafa in  \cite{Cecotti:1991me} and generalized to ${\cal N}=(4,4)$ theories in \cite{deBoer:2008ss}. Strictly speaking, the analysis of \cite{deBoer:2008ss} only holds for theories with the small ${\cal N}=4$ superconformal algebra, such as the $\textrm{Sym}^N (K3)$ CFT, but we will present a small generalization to include the $\textrm{Sym}^N (T^4)$ case in Section \ref{sec:symT4}.

\paragraph{Background.} Consider a two-dimensional superconformal field theory with the (small) ${\cal N}=(4,4)$ supersymmetry in the Neveu-Schwarz (NS) sector. The left-moving symmetry currents include the stress tensor $T(z)$, 4 supercurrents $G^{aA}(z)$, $a,A=\pm$, and an $SU(2)$ triplet of R-currents $J^{\pm}(z),J^3(z)$. The supercurrents transform as a doublet of the $SU(2)$ R-symmetry with respect to the index $a$. The index $A$ corresponds to a doublet under an outer\footnote{This means that there is no corresponding conserved current, or conserved charge, generating this transformation.} $SU(2)$ automorphism of the algebra. The right moving symmetry currents are analogous.

Superconformal primary states of the theory are characterized by representations of the $SU(2)_{\rm left}\times SU(2)_{\rm right}$ R-symmetry and we denote these representations by two non-negative (half)-integers $(r,\widetilde{r})$. A unitarity bound implies that the conformal dimension $(h,\widetilde{h})$ of the superconformal primary must obey
\be
\label{boundsn4}
h\geq r \qquad,\qquad \widetilde{h}\geq \widetilde{r} \,.
\ee
If both bounds are saturated then we have a 1/2-BPS multiplet, if only one of the two is saturated then a 1/4-BPS multiplet. Multiplets which saturate these bounds are shorter than the typical multiplet. Let us consider a superconformal primary state\footnote{In what follows we often switch between referring to states and local operators. Their representation theory is equivalent via the state-operator map.} $|\varphi\rangle$ which is also a highest weight state with respect to the global part of the $SU(2)_{\rm left}\times SU(2)_{\rm right}$ R-symmetry. The latter condition means that it is aligned so that the eigenvalues of $J^3$, $\widetilde{J}^3$ obey $r_3 = r$, $\widetilde{r}_3=\widetilde{r}$. For such a state we have that
\be
\label{shortn4}
G_{-{1 \over 2}}^{+A}|\varphi\rangle = 0 \qquad \Leftrightarrow \qquad h=r \,.
\ee
Notice that the first condition necessarily holds (or fails) simultaneously for both possible values of $A$. Because the state $|\varphi\rangle$ is annihilated by the supercharge raising operators the corresponding super-multiplet is short. In addition to \eqref{shortn4}, we also have the following conditions from the fact that the state is a superconformal primary
\begin{equation}\label{eqn:2dsuperconfprimary}
     G_{n-\frac{1}{2}}^{a A}\ket{\varphi} = 0\,, \qquad n>0 \,,
\end{equation}
from which $L_{n}\ket{\varphi} = J_n^{3}\ket{\varphi} = 0$ for $n>0$ follows from the SCA \eqref{eqn:smallN=(4,4)SCA}.\footnote{\label{foo:currentprimary}There is an extended notion of primary by imposing that the state is also a current primary meaning $J_n^{\pm}\ket{\varphi} = 0$ for $n>0$. These have a stricter bound on their allowed $R$-charge of $r \leq \hat{k}/2$. Unless otherwise stated, we will not use this extended notion.}  These 1/2-BPS superconformal primaries, known as chiral primaries, have an upper bound on their $R$-charge given by $r \leq \hat{k}$. From now on, we focus on 1/2-BPS multiplets which saturate the unitarity bound \eqref{boundsn4} on both left- and right-moving sector. The non-highest weight states in these multiplets are all superconformal primaries, and we will refer to these states as ``chiral multiplet states".

The chiral multiplet states $\ket{\varphi_K}$ at fixed R-symmetry representation $(r,\widetilde{r})$ form a sub-bundle of the Hilbert bundle with fibres $\mathbb{H}_{\mathrm{chiral}}^{(r,\widetilde{r})}$. The non-Abelian Berry connection $(A_{\mu}^{(r,\widetilde{r})})_{KL}$ defined in \eqref{eqn:Adef} gives a connection on this sub-bundle with respect to which states can be parallel-transported. There is also a natural metric on the sub-bundle given by the two-point function on the sphere
\begin{equation}
\label{2pt}
    g_{KL} = \lim_{z \to \infty}z^{2h}\bar{z}^{2\widetilde{h}}\langle{\varphi}^\dagger_{L}(z)\,\varphi_{K}(0)\rangle \,,
\end{equation}
which is the analogue of the usual Zamolodchikov metric on the tangent bundle of $\mathcal{M}$. 

Furthermore, one can define multiplication of elements in different sub-bundles as follows. Consider the OPE between two chiral multiplet operators $\varphi_K$, $\varphi_L$ corresponding to R-symmetry representations $(r_K,\widetilde{r}_K)$ and $(r_L,\widetilde{r}_L)$, respectively. These operators need to be further characterized by their $J^3_0,\widetilde{J}^3_0$ eigenvalues. On the right hand side of the OPE of these operators there will generally be short and long operators. We can find which chiral multiplet operators $\varphi_M$ appear on the RHS of the OPE by considering the relevant 3-point functions between two such operators, which have the general form
\begin{align}
\label{n43pt}
\langle \varphi_{K; (m_K,\widetilde{m}_K)}(\infty) & \,\,\varphi_{L;(m_L,\widetilde{m}_L)}(z) \,\,\varphi_{M;(m_M,\widetilde{m}_M)}(0)\rangle = \cr &={C_{KLM} \over z^{(r_L+r_M-r_K)}\,\, \overline{z}^{(\widetilde{r}_L+\widetilde{r}_M-\widetilde{r}_K)}}\,\,\begin{pmatrix} r_K & r_L & r_M \\ m_K & m_L & m_M \end{pmatrix}
\begin{pmatrix} \widetilde{r}_K & \widetilde{r}_L & \widetilde{r}_M \\ \widetilde{m}_K & \widetilde{m}_L & \widetilde{m}_M \end{pmatrix} \,.
\end{align}
The constant $C_{KLM}$ does not depend on the alignment of the quantum numbers $(m,\widetilde{m})$ of the operators, which is fixed by R-symmetry and expressed by the 3-j symbols. The chiral ring OPE gives rise to a product  for elements of different sub-bundles at the same point $p \in \mathcal{M}$
\begin{equation}
C_{KL}^{M} : \quad \mathbb{H}_{\mathrm{chiral}}^{(r_K,\widetilde{r}_K)} \otimes \mathbb{H}_{\mathrm{chiral}}^{(r_L,\widetilde{r}_L)} \to \mathbb{H}_{\mathrm{chiral}}^{(r_M,\widetilde{r}_M)}\,.
\end{equation}
Notice that the R-symmetry selection rules imply that for an element in a chiral multiplet to appear on the RHS of the OPE of two other operators in chiral multiplets it is necessary that $|r_K-r_L|\leq r_M\leq r_K+r_L$ and similarly for the right movers.\footnote{The general form \eqref{n43pt} of the 3-point function of chiral primaries does not capture all selection rules coming from the ${\cal N}=4$ SCA. For example, using the ``pseudospin" $SU(2)$ generated by $J_{-1}^+, J_0^3-{\hat{k}\over 2}, J_{+1}^-$, one can show that $C_{KLM}=0$ for certain combination of multiplets $K,L,M$, even though these multiplets satisfy the $J_0^+,J_0^-,J_0^3$ $SU(2)$ selection rule mentioned in the text.} This means that the chiral ring OPE in ${\cal N}=(4,4)$ SCFTs can in general contain singular terms, unlike what happens in ${\cal N}=(2,2)$ SCFTs. 
In what follows we will focus on the part of the chiral ring which corresponds to ``extremal" OPE multiplication, where we have $r_M=r_K +r_L$ and $\widetilde{r}_M = \widetilde{r}_K + \widetilde{r}_L$ and where we focus on the OPE between highest-weight states in each representation, i.e we restrict to only chiral primaries. This provides a close analogue of the chiral ring of ${\cal N}=(2,2)$ theories \cite{Lerche:1989uy}. In this case, the OPE
\begin{equation}
    \varphi_{K}(x) \cdot \varphi_{L}(0) = C_{KL}^{M}\,\varphi_{M}(0) + \ldots 
\end{equation}
is regular.

For theories with only $\mathcal{N}=(2,2)$ supersymmetry, the chiral ring coefficients will generally be non-trivial functions of the moduli. However, when there exists $\mathcal{N}=(4,4)$ supersymmetry, it was proven in \cite{deBoer:2008ss, Baggio:2012rr} that they are covariantly constant
\begin{equation}\label{eqn:covconstCRC}
    \nabla_{\mu}C_{KL}^{M} = 0 \,,
\end{equation}
where the covariant derivative is defined with respect to the Berry connection. The three-point function coefficients \eqref{n43pt} are also covariantly constant. This property of the chiral ring coefficients will prove to be very useful for determining the Berry curvature.

\paragraph{Moduli space.} To compute the curvature, we need to first understand better the structure of the moduli space $\mathcal{M}$. Marginal operators preserving supersymmetry are superconformal descendants of chiral primaries $\varphi_i$ of dimension $({1\over 2},{1\over 2})$. As reviewed in Appendix \ref{app:realchiral}, it is convenient to go to a basis of orthonormal $({1\over 2},{1\over 2})$ chiral primaries, where they obey
\be
\label{hermbasis}
[J_0^-,[\widetilde{J}_0^-,\varphi_i]] = \varphi_i^\dagger \,.
\ee
Then from each such chiral primary we can get 4 Hermitian marginal operators which are mutually orthogonal
\be
\label{marginal1}
{\cal O}_{a,i} = {1\over \sqrt{2}}\sigma_{a,{A\dot{B}}} \{G_{-{1\over2}}^{-A},[\widetilde{G}_{-{1\over 2}}^{-\dot{B}},\varphi_i]\} \,,\qquad a =1,2,3,4 \,,
\ee
where the sigma matrices are $\sigma_a=(\sigma_1,\sigma_2,\sigma_3,i {\mathbbm{1}})$.

Notice that using \eqref{hermbasis} and the ${\cal N}=4$ algebra we have the following  property\footnote{This can be seen as follows, where we are using short-hand notation and not write explicitly the nested (anti)-commutators: $G_{-{1\over 2}}^{-A}\widetilde{G}_{-{1\over 2}}^{-\dot{B}}\varphi_i=G_{-{1\over 2}}^{-A}\widetilde{G}_{-{1\over 2}}^{-\dot{B}}J_0^+ \widetilde{J}_0^+\varphi_i^\dagger =G_{-{1\over 2}}^{-A}J_0^+\widetilde{G}_{-{1\over 2}}^{-\dot{B}} \widetilde{J}_0^+\varphi_i^\dagger =- G_{-{1\over 2}}^{-A}J_0^+\widetilde{G}_{-{1\over 2}}^{+\dot{B}} \varphi_i^\dagger $ where in the last equality we used that $[\widetilde{J}_0^+,\widetilde{G}_{-{1\over 2}}^{-\dot{B}}]=\widetilde{G}_{-{1\over 2}}^{+\dot{B}}$ and that $\widetilde{G}_{-{1\over 2}}^{-\dot{B}}\varphi_i^\dagger=0$. Proceeding with an exactly equivalent argument for the left-moving generators, we arrive at the desired \eqref{specialn4}.}
\be
\label{specialn4}
\{G^{-A}_{-{1\over 2}},[\widetilde{G}^{-\dot{B}}_{-{1\over 2}},\varphi_i]\} =\{G_{-{1\over2}}^{+A},[\widetilde{G}_{-{1\over 2}}^{+\dot{B}},\varphi_i^\dagger]\} \,.
\ee
This allows us to rewrite the same set of operators \eqref{marginal1} as
\be
\label{marginal2}
{\cal O}_{a,i} = {1\over \sqrt{2}}\sigma_{a,{A\dot{B}}} \{G_{-{1\over2}}^{+A},[\widetilde{G}_{-{1\over 2}}^{+\dot{B}},\varphi_i^\dagger]\} \,,\qquad a =1,2,3,4 \,.
\ee
In other words, in ${\cal N}=(4,4)$ SCFTs a given marginal operator can be written either as a descendant of what would be a chiral primary in ${\cal N}=(2,2)$ language, or a descendant of an anti-chiral primary. This is not possible in ${\cal N}=(2,2)$ SCFTs.

Notice that the operators ${\cal O}$ above are mutually orthonormal, if $\varphi_i$ are selected to be orthonormal. This can be seen from the 2-point function
\be
\langle {\cal O}_{a,i}(x)\,\,{\cal O}_{b,j}(y)\rangle = \delta_{ab} \,\,\partial_x \partial_{\overline{x}}\, \langle \varphi_i(x) \,\,\varphi_j^\dagger(y)\rangle \,,
\ee
which follows from superconformal Ward identities.

The marginal operators \eqref{marginal1}, labeled by the two indices $a=1,\ldots,4$ and $i=1,\ldots,n$, span the tangent space of the conformal manifold \eqref{eqn:D1/D5modulispace_ST} and can be thought of as sections of a vector bundle which is isomorphic to the tangent bundle of ${\cal M}$. The holonomy of the latter is generally $SO(4n)$. Since the marginal operators have the form \eqref{marginal1}, it means that the holonomy group is reduced and factorizes into a tensor product of the holonomy of the supercurrents and that of the chiral primaries $\varphi_i$.\footnote{See Appendix \ref{sec:desccurv} for a CFT justification of the multiplicative nature of the holonomies.} In other words, the holonomy of the tangent manifold reduces from $SO(4n)$ down to $SO(4)\times SO(n)$. This completely fixes $\mathcal{M}$ to be locally of the form \eqref{eqn:D1/D5modulispace_ST} \cite{Seiberg:1988pf,Cecotti:1990kz}.

We also notice the following useful formula, which follows from superconformal Ward identities
\be
\label{n4ward}
\langle \varphi_L^\dagger(\infty) \,{\cal O}_{a,i}(x) \,\, {\cal O}_{b,j}(y)\,\, \varphi_K(0)\rangle = \delta_{ab}\,\,\partial_x \partial_{\overline{x}}\, \langle \varphi_L^\dagger(\infty) \,\varphi_i(x)\,\, \varphi_j^\dagger(y) \,\,\varphi_K(0)\rangle \,,
\ee
where $\varphi_K,\varphi_L$ are general chiral primaries. 
To prove this, we write each of the marginal operators as in \eqref{marginal2}. Then we move the supercharges away from $y$ using Ward identities. We do not get any contributions from $\varphi_K(0)$, since the operator is annihilated by the relevant supercharges and no contribution from $\varphi_L^\dagger$ since the operator is ``at infinity" (see \cite{deBoer:2008ss} for further details). Then we end up with 4 supercharges acting on $\varphi_i(x)$. To simplify that combination, we use the following identity\footnote{From the ${\cal N}=4$ SCA, we can write $G_{-{1\over 2}}^{+B} = -[L_{-1},G_{+{1\over 2}}^{+B}]$. Then we have $G^{+A}_{-{1\over 2}}G^{+B}_{-{1\over 2}}\varphi_i^\dagger = - G^{+A}_{-{1\over 2}}[L_{-1},G_{+{1\over 2}}^{+B}]\varphi_i^\dagger = G^{+A}_{-{1\over 2}} G^{+B}_{+{1\over 2}} L_{-1}\varphi_i^\dagger $, where in the last equality we used that $\varphi_i^\dagger$ is superconformal primary. To proceed, we use the anticommutator $\{G_{-{1\over 2}}^{+A}, G_{+{1\over 2}}^{+B}\} = \epsilon^{AB}J^+$ which leads to \eqref{anotherid} after some obvious steps.} 
\be
\label{anotherid}
\{G^{+A}_{-{1\over 2}},[G^{+B}_{-{1\over 2}} ,\varphi_i^\dagger]\} =\epsilon^{AB}[L_{-1}, \, [J_0^+,\varphi_i^\dagger]] \,,
\ee
as well as a similar one for the right-movers. Then \eqref{n4ward} follows.

\paragraph{Berry curvature.} With these preliminaries sorted out, there are two equivalent ways to compute the Berry curvature for the 1/2-BPS states: (1) use the general expression for the curvature \eqref{berry4point} and simplify the computation by using the identity \eqref{n4ward} to integrate by parts and localize the integrals to boundary terms, resulting in the $\tt$ equations \eqref{eqn:D1D5_1/2BPS_tildeF_final_N=4}, as described in \cite{deBoer:2008ss}, or (2) use the Hamiltonian formalism in radial quantization, i.e. on the cylinder \cite{Cecotti:1992vy}, \cite{Baggio:2017aww} in which case the Berry curvature is given by \eqref{eqn:curvformula}.\footnote{There is a small caveat: regularizing the divergences of the Berry curvature in the Hamiltonian formalism requires some insight from the conformal perturbation theory method.} We will review method 2, as it will be the easier method when we analyze the 1/4-BPS states in Section \ref{sec:D1/D5_1/4BPS}.

We consider two marginal directions which we denote as $\mu = (a,i)$ and $\nu=(b,j)$. For technical convenience we express the first marginal deformation using \eqref{marginal1} and the second marginal deformation using \eqref{marginal2}.
The variation of the Hamiltonian along these marginal directions is\footnote{In the version of the computation in \cite{Baggio:2017aww} they used a different basis for operators. The reason is that if one starts in Euclidean radial quantization and transforms to Lorentzian signature, then the dilatation operator on the plane, corresponding to the cylinder Hamiltonian $H$, no longer has real positive eigenvalues, rather they become complex. Relatedly, eigenstates are not normalizable anymore. To obtain positive, unitary representations, one can switch to the conformal Hamiltonian $H_{\mathrm{conf}} = -\frac{1}{2}(P_{0}+K_{0})$, which is related to the dilatation operator by a change of basis. Implementing this basis change on all operators leads to a sensible Berry curvature in Lorentzian signature. We choose to instead work directly in Euclidean radial quantization.}
\begin{equation}\label{eqn:D1D5partialH}
    \partial_{\mu}H = \frac{1}{\pi}\int_{0}^{2\pi}d\theta\,{1\over \sqrt{2}}\sigma_{a,A\dot{B}}\,G_{+\frac{1}{2}}^{-A}\tilde{G}_{+\frac{1}{2}}^{-\dot{B}} \cdot \varphi_{i}\,, \qquad \partial_{\nu}H = \frac{1}{\pi}\int_{0}^{2\pi}d\theta\,{1\over \sqrt{2}}\sigma_{b,C\dot{D}}\,G_{-\frac{1}{2}}^{+C}\tilde{G}_{-\frac{1}{2}}^{+\dot{D}} \cdot \varphi_j^\dagger \,,
\end{equation}
where we used the shorthand notation for nested (anti-)commutators $G_{\frac{1}{2}}^{-A}\tilde{G}_{\frac{1}{2}}^{-\dot{B}} \cdot \varphi_{i} \equiv \{G_{\frac{1}{2}}^{-A},[\tilde{G}_{\frac{1}{2}}^{-\dot{B}},\varphi_{i}(w)]\}$ and the integral is the angular integral on the cylinder.\footnote{Notice that $G_{+{1\over 2}}$ does not annihilate the superconformal primary $\varphi$ since it is not located at the origin.} Let us clarify a potentially confusing point. The standard form of the marginal operator in \eqref{marginal1} uses $G_{-\frac{1}{2}}^{-A}$ and $\tilde{G}_{-\frac{1}{2}}^{-\dot{B}}$ acting on $\varphi_{i}$, but \eqref{eqn:D1D5partialH} uses the opposite moding. The resolution is that when integrated on the unit circle the two are equivalent, viz., 
\begin{equation}
    \left[\tilde{G}_{+\frac{1}{2}}^{-\dot{B}},\varphi(\bar{w})\right] = \oint_{\bar{w}}\frac{d\bar{z}}{2\pi i}\,\bar{z}\tilde{G}^{-\dot{B}}(\bar{z})\varphi(\bar{w}) =  \oint_{\bar{w}}\frac{d\bar{z}}{2\pi i}\,\frac{\bar{z}}{\bar{z}-\bar{w}}\left(\tilde{G}_{-\frac{1}{2}}^{-\dot{B}}\varphi\right)(\bar{w}) = \bar{w}\left(\tilde{G}_{-\frac{1}{2}}^{-\dot{B}}\varphi\right)(\bar{w}) \,,
\end{equation}
where in the second equality we used the OPE in eq. (B.1) of \cite{deBoer:2008ss}. Taking the anti-commutator with $G_{+\frac{1}{2}}^{-A}$ gives \eqref{marginal1} times $|w|^2$ so on the unit circle the two are the same. It will turn out to be easier to work with the form in \eqref{eqn:D1D5partialH}.

The curvature can be computed using the same trick as in Section \ref{sec:conj}. Namely, for a chiral primary in a fixed charge sector $(r,\widetilde{r})$ with energy $E=r+\widetilde{r}$, we define the auxiliary quantity
\begin{equation}
\label{curvaturef}
    (F_{\mu\nu})_{KL}(x) = \bra{L}\partial_{\mu}H\frac{1}{(H-R-x)^{2}}\partial_{\nu}H\ket{K} - (\mu \leftrightarrow \nu) \,,
\end{equation}
where $H=L_0+\widetilde{L}_0$ and $R= J_{0}^3 + \widetilde{J}_{0}^3$. Here we used that the intermediary state $\ket{m}$ in \eqref{eqn:curvformula} must have the same $R$-charges $(r,\tilde{r})$ as the external states, since the variations $\partial H$ are neutral, so the denominator can be rewritten as $(E-E_{n})^{2}=((E-(r+\widetilde{r})-(E_{n}-(r+\widetilde{r})))^{2} = (E_{n}-(r+\widetilde{r}))^{2}$. The Berry curvature is then recovered from the limit
\begin{equation}
    F_{\mu\nu} = \lim_{x \to 0}F_{\mu\nu}(x) \,.
\end{equation}

The crux of this trick is to manipulate the numerator in \eqref{curvaturef} to obtain an expression in terms of $(H-R)^2$ which can then cancel the denominator in the limit $x\rightarrow 0$. To do this, we use that the external states are superconformal primaries annihilated by various supercharges \eqref{eqn:2dsuperconfprimary} and we use the ${\cal N}=4$ SCA in \eqref{eqn:smallN=(4,4)SCA} to deduce
\begin{equation}
    \left[G_{\mp\frac{1}{2}}^{\pm A},H-R\right]=0 \,.
\end{equation}
Armed with these two facts, and moving the supercharges from $\partial_{\nu}H$ to the left gives 
\begin{align}
\begin{split}
\label{eqn:D1D5_Fsimplified}
    (F_{\mu\nu})_{KL}(x) &= \frac{i}{2\pi^{2}}(\sigma_{a})_{A\dot{B}}(\sigma_{b})_{C\dot{D}}\bra{L}\left[\left\{G_{+\frac{1}{2}}^{-A},G_{-\frac{1}{2}}^{+C}\right\}\left\{\tilde{G}_{+\frac{1}{2}}^{-\dot{B}},\tilde{G}_{-\frac{1}{2}}^{+\dot{D}}\right\},\int \varphi_{i}\right]\frac{1}{(H-R-x)^{2}}\int \varphi_j^\dagger\ket{K}
    \\  &\qquad - (\mu \leftrightarrow \nu) \,,
\end{split}
\end{align}
where we used that $[G_{-\frac{1}{2}}^{+},\varphi_{i}(w)] = \oint_{w}\frac{dz}{2\pi i}\,G^{+}(z)\varphi_{i}(w) = 0$ since the OPE is regular, and the same for the right-movers. From the SCA algebra \eqref{eqn:smallN=(4,4)SCA}, we find
\begin{align}
\begin{split}
    \left\{G_{+\frac{1}{2}}^{-A},G_{-\frac{1}{2}}^{+C}\right\}\left\{\widetilde{G}_{+\frac{1}{2}}^{-\dot{B}},\widetilde{G}_{-\frac{1}{2}}^{+\dot{D}}\right\} &= \epsilon^{AC}\epsilon^{\dot{B}\dot{D}}\left(L_{0}-J_{0}^3\right)\left(\widetilde{L}_{0}-\widetilde{J}^3_{0}\right) 
    \\  &= \frac{1}{4}\epsilon^{AC}\epsilon^{\dot{B}\dot{D}}\left(\left(H-R\right)^{2}-\left(L_{0}-\widetilde{L}_{0}-(J^3_{0}-\widetilde{J}^3_{0})\right)^{2}\right) \,.
\end{split}
\end{align}
Observe that $L_{0}-\widetilde{L}_{0}$ is the angular momentum generator which commutes with the zero-mode $\int \varphi_{i}$ and $J^3_{0}-\widetilde{J}^3_{0}$ also commutes with this chiral primary operator since $r_{i} = \widetilde{r}_{i} = \frac{1}{2}$. Thus, we have arrived at the desired simplification
\begin{equation}
    (F_{\mu\nu})_{KL}(x) = -\frac{i}{(2\pi)^{2}}\delta_{ab}\bra{L}\int \varphi_{i}\frac{(H-R)^{2}}{(H-R-x)^{2}}\int \varphi_j^\dagger \ket{K} - (\mu \leftrightarrow \nu) \,,
\end{equation}
so taking the limit $x \to 0$ gives\footnote{Our result differs by a factor of $\frac{1}{4}$ compared to \cite{Baggio:2017aww} due to our choice of normalization of the supercharges, as can be seen from the SCA in \eqref{eqn:smallN=(4,4)SCA}.}
\begin{align}\label{eqn:D1/D5curv}
\begin{split}
    (F_{\mu\nu})_{KL} &= -\frac{i}{(2\pi)^{2}}\delta_{ab}\sum_{n \not \in \mathbb{H}_{\mathrm{chiral}}^{(r,\widetilde{r})}}\bra{L}\int \varphi_{i}\ket{n}\bra{n}\int \varphi_j^\dagger\ket{K} - (\mu \leftrightarrow \nu)
     \\  &= \frac{i}{(2\pi)^{2}}\delta_{ab}\Bigg[{-}\underbrace{\bra{L} [\int\varphi_{i},\int \varphi_j^\dagger ]\ket{K}}_{\mathrm{current\;term}} 
     \\ &\qquad + \sum_{M \in \mathbb{H}_{\mathrm{chiral}}^{(r,\widetilde{r})}}\left(\bra{\bar{L}}\int \varphi_{i}\ket{M}\bra{M}\int \varphi_j^\dagger\ket{K} - \bra{L}\int \varphi_j^\dagger\ket{M}\bra{M}\int \varphi_{i} \ket{K}\right)\Bigg] \,,
\end{split}
\end{align}
where the sum in the first equality is over all orthonormal eigenstates of the Hamiltonian on the cylinder with $R$-charges $(r,\widetilde{r})$ that are not chiral primaries.  We note in passing that this result is very reminiscent of the curvature formula for conjugation deformations in \eqref{eqn:Fconjdef}, where $\int \varphi_i$ plays the role of $\Lambda$ there.

The ``current term" on the second line of (\ref{eqn:D1/D5curv}) is naively zero since we are in Euclidean signature, but it is actually divergent and needs to be suitably regularized, leading to a finite, non-zero answer. The computation of this term can be found in Appendix E of \cite{Baggio:2017aww}, using the renormalization scheme discussed in Section \ref{sec:regularization}, i.e. by mapping the computation to the equivalent conformal perturbation theory computation \cite{deBoer:2008ss}. It receives contributions from all operators in the $\varphi_{i} \times \varphi_{j}^\dagger$ OPE with $h+\widetilde{h} \leq 1$. These include the identity operator, $R$-currents and possibly other conserved currents with $(h,\widetilde{h})=(1,0)$ or $(0,1)$ which may be present in the theory. We will denote the contribution from these conserved currents by $\mathcal{F}_{ijKL}(\{\mathcal{Q},\bar{\mathcal{Q}}\})$ where $\mathcal{Q}$ is a collective label for all of the charges under the corresponding global symmetries of the external states and the chiral primaries used to deform the theory. When the only such current is the $R$-symmetry currents $J_{R}^{3}$,  $\widetilde{J}_{R}^{3}$, which is the case when $M=K3$, one finds $\mathcal{F}_{ijKL}(\{\mathcal{Q},\bar{\mathcal{Q}}\}) = \frac{6}{c}(r+\widetilde{r})\delta_{ij}\delta_{KL}$.\footnote{The factor of 2 difference compared to \cite{deBoer:2008ss} follows from different conventions in the normalization of the $SU(2)$ R-charge, with $r_{\rm here} = {1\over 2} q_{\rm there}$.} However, as we shall see in Section \ref{sec:T4}, for the case of $M=T^4$, $\mathcal{F}_{ijKL}(\{\mathcal{Q},\bar{\mathcal{Q}}\})$ is more complicated due to enhanced global symmetries.

The other term in the curvature \eqref{eqn:D1/D5curv} is the commutator of chiral ring coefficients $[C,C^{\dagger}]$ so we arrive at the final result
\begin{equation}\label{eqn:D1D5_1/2BPS_tildeF_final_N=4}
\left(F_{\mu\nu}\right)_{KL} = i\delta_{ab}\Big(\left(\delta_{ij}\delta_{KL}-\mathcal{F}_{ijKL}(\{\mathcal{Q},\bar{\mathcal{Q}}\})\right)- [C_i,C_j^\dagger]_{KL}\Big) \,.
\end{equation}
Here we use the notation $\mu=(a,i)$, $\nu=(b,j)$ for the marginal operators, where $a,b=1,\ldots,4$ is the $SO(4)$ index \eqref{marginal1} and $i,j=1,\ldots,n$ labels one of the $n$ chiral primaries of dimension $({1\over 2},{1\over 2})$. The indices $K,L$ label general chiral primaries, with left and right $SU(2)$ R-symmetry charges\footnote{The curvature is diagonal in R-charge, so $K,L$ must have the same R-charges.} $r,\widetilde{r}$ in ${\cal N}=(4,4)$ conventions. Here $C_i$ is the map acting on the entire chiral ring corresponding to ``multiplication with $\varphi_i$" and with dagger we denote the adjoint map. This means that the last term in the equation above has the explicit form
\be
\label{comexplain}
[C_i,C_j^\dagger]_{KL} = C_{iK}^{M} \delta_{MM'} (C_{j L}^{M'})^*- \delta_{KK'}(C_{j N'}^{K'})^* \delta^{N'N}C_{iN'}^{L'}\delta_{L'L} \,,
\ee
where in this formula the indices $M,M'$ are summed over chiral primaries of R-charge $(r_K+{1\over 2} , \widetilde{r}_K+{1\over 2})$ while the indices $N,N'$ over chiral primaries of R-charge  $(r_K-{1\over 2} , \widetilde{r}_K-{1\over 2})$. Here we assumed that we work in an orthonormal basis of chiral primaries, otherwise the $\delta$'s above need to replaced by the matrices of 2-point functions, i.e. the (generalized) Zamolodchikov metric \eqref{2pt}.

Stripping off $\delta_{ab}$, this is the curvature in general $\mathcal{N}=(2,2)$ theories, which corresponds to the $\tt$ structure discovered by Cecotti and Vafa for the Ramond sector of topologically twisted $\mathcal{N}=(2,2)$ theories\footnote{Up to a small difference in the current term contribution due to the difference between NS and R sectors.} \cite{Cecotti:1991me}.  In what follows we will refer to \eqref{eqn:D1D5_1/2BPS_tildeF_final_N=4} as the $\tt$ equations.

Observe that the result in \eqref{eqn:D1D5_1/2BPS_tildeF_final_N=4} is not manifestly antisymmetric under the exchange $\mu \leftrightarrow \nu$. In fact, the first term is symmetric so the $[C_i,C_j^{\dagger}]_{KL}$ contribution must have a symmetric part that exactly cancels this first term. We prove in Appendix \ref{sec:CRCidentity}, using Ward identities, that for two marginal operators $\mu=(a,i)$, $\nu=(b,j)$, with $i=j$, the curvature actually vanishes so indeed such a cancellation occurs. Furthermore, when $i \neq j$ so the first contribution vanishes, we see in explicit examples in Appendix \ref{app:orbifold} that $[C_i,C_j^{\dagger}]_{KL}$ is indeed antisymmetric.

For $\mathcal{N}=(4,4)$ theories, we can now use that the chiral ring coefficients are covariantly constant \eqref{eqn:covconstCRC} to find that the curvature is covariantly constant
\begin{equation}
    \nabla_{\rho}F_{\mu\nu} = 0 \,.
\end{equation}
This turns out to be very powerful due to the fact that the base manifold $\mathcal{M}$ is a symmetric space \eqref{eqn:D1/D5modulispace_ST}. In particular, there exists a mathematical theorem that for vector bundles with a symmetric base space, if the curvature of the bundle is covariantly constant, then it is obtained by composing the curvature of the tangent bundle for the base space $R^{\mathcal{M}}$ with a representation $\rho$ of the holonomy group of the base space
\begin{equation}
    \mathrm{Hol}(\mathcal{M}) \cong SO(4) \times SO(n) \,,
\end{equation}
on the fibre $\mathbb{H}_{\mathrm{chiral}}^{(r,\widetilde{r})}$, i.e., $\rho : SO(4) \times SO(n) \to SO(m)$ where $m = \dim \mathbb{H}_{\mathrm{chiral}}^{(r,\widetilde{r})}$. These are known as homogeneous bundles. It is clear from \eqref{eqn:D1D5_1/2BPS_tildeF_final_N=4} that the representation of the $SO(4)$ part of the holonomy algebra is always trivial. One can thus derive an alternate formula for the curvature given the representation $\mathcal{R}$ of $SO(n)$ in which $\mathbb{H}_{\mathrm{chiral}}^{(r,\widetilde{r})}$ lives
\begin{equation}\label{eqn:D1D5_1/2BPS_tildeF_final_N=4_repformula}
\left(F_{\mu\nu}\right)_{KL} = {1\over \hat{k}}i\delta_{ab}(\Sigma_{ij}^{\mathcal{R}})_{KL} \,,
\end{equation}
where we remind that $c=6\hat{k}$ and the prefactor in the equation above sets the overall scale of the moduli space, and $\Sigma_{ij}^{\mathcal{R}}$ are the representation matrices for the $ij$-th generator of $\mathfrak{so}(n)$. Here and henceforth, we are rather sloppy about Lie group versus Lie algebra representations since there is a unique mapping between the two, but for the curvature it is the holonomy algebra that is indeed the correct mathematical object.\footnote{Note that one can actually prove a stronger statement for vector bundles with covariantly constant curvature than just the curvature being fixed by representations. It is proven in \cite{guijarro2001parallel} that for any rank-$k$ vector bundle $E$ over $\mathcal{M}=G/H$ simply-connected with fibre $V$, whose curvature is covariantly constant, there exists a representation $\rho: \mathrm{Hol}(\mathcal{M}) \to SO(k)$ such that $E \cong \mathcal{M} \times_{\rho} V$ where subscript $\rho$ means the set of equivalence classes $(gh,v) \sim (g,\rho(h)v)$ for $g \in G$, $h \in H$, and $v \in V$. Here it is the Lie group representation $\rho$ that plays the important role.} All of the necessary mathematical details along with a derivation of \eqref{eqn:D1D5_1/2BPS_tildeF_final_N=4_repformula} can be found in Appendix \ref{sec:D1/D5math}. We want to mention that ${\rm Sym}^N(T^4)$ is very similar with a shift $\hat k \to \hat k -1$, as in \eqref{eq:curv_T4_rep}.

The significance of this result is that the curvature \eqref{eqn:D1D5_1/2BPS_tildeF_final_N=4} looks like it depends sensitively on the OPE coefficients of the chiral primaries, but it is actually just fixed by the geometry of the moduli space. This is what will allow us to show in the next subsection that the Berry curvature matrix does not have RMT statistics for any $N$ and any charge sector. 

\paragraph{Closing remarks.}
Before we close this subsection, notice that combining \eqref{eqn:D1D5_1/2BPS_tildeF_final_N=4} and \eqref{eqn:D1D5_1/2BPS_tildeF_final_N=4_repformula}, we get the following constraints for the chiral ring coefficients of any SCFT with the small ${\cal N}=(4,4)$ algebra, such as the case of ${\rm Sym}^N(K3)$
 \be
 \label{generalid}
 [C_i,C_j^\dagger]_{KL} = \delta_{ij}\delta_{KL}\left(1-{r + \widetilde{r} \over \hat{k}}\right) - \frac{1}{\hat{k}}(\Sigma_{ij}^\mathcal{R})_{KL} \,,
 \ee
 where the last term is a (generally reducible) representation of $SO(n)$ and we use the notation $c=6\hat{k}$. In this equation $i,j$ are chiral primaries of dimension (${1\over 2},{1\over 2}$) while $K,L$ are general chiral primaries. Notice that if we consider the case where $i=j$ the RHS simplifies, as the second term vanishes. We will come back to this special case in the Subsection \ref{ccdagertt}.

For the special case where we take the external states $K,L$ to be chiral primaries of dimension (${1\over 2},{1\over 2}$) --- and hence denote them as $k,l$ --- then we expect that the curvature corresponds to the $SO(n)$ part of the holonomy of the conformal manifold \eqref{eqn:D1/D5modulispace_ST} and we have
\be
(F_{\mu\nu})_{kl} = {i \over \hat{k}}\delta_{ab} (\Sigma_{ij}^{\rm fund})_{kl} \,,
\ee
where $\Sigma_{ij}^{\rm fund}$ is the fundamental representation of $SO(n)$ \eqref{eqn:so(n)gens}. In that case we can work in a basis where the chiral ring coefficients are real, as explained in Appendix \ref{app:realchiral}, and  the general equation  \eqref{generalid} simplifies to
\be
\label{chiralringidentity}
C_{ik}^M C_{jl}^M = \left(1-{1\over \hat{k}}\right)\left(\delta_{ij}\delta_{kl}+\delta_{il}\delta_{jk}\right) +{1\over \hat{k}} \delta_{ik}\delta_{jl} \,,
\ee
which was also derived in \cite{Gomis:2016sab}. Here the indices $i,j,k,l$ run over $({1\over 2},{1\over 2})$ chiral primaries, while the index $M$ is summed over $(1,1)$ chiral primaries. Notice that the second term in the $[C,C^\dagger]$ commutator \eqref{comexplain}, with the $(0,0)$ internal state, was moved to the RHS of the equation \eqref{chiralringidentity} and is the term $\delta_{jk}\delta_{il}$. We provide some checks of this identity in the ${\rm Sym}^N(T^4)$ theory in Appendix \ref{app:orbifold}, see equation \eqref{identitycrt4}.  

For ${\rm Sym}^N(K3)$, this is the end of the story. In the case of ${\rm Sym}^N(T^4)$, the theory has enlarged symmetry and there are some minor differences which are discussed in more detail in Section \ref{sec:symT4}, though the essential part of the story remains the same.

%%%%%%%%%%%
% SECTION %
%%%%%%%%%%%
\subsection{Non-randomness of 1/2-BPS Berry curvature}
\label{sec:D1/D5_non-random}

The goal of the current section is to determine whether or not the Berry curvature matrix for the 1/2-BPS states can ever exhibit eigenvalue repulsion consistent with the Wigner surmise for a random matrix ensemble. Of particular interest are the chiral primaries at large charge $r,\widetilde{r} \sim c$ whose coherent superpositions are dual to the Lunin-Mathur geometries. This seems very hard to answer directly from the formulation in terms of the chiral ring coefficients in \eqref{eqn:D1D5_1/2BPS_tildeF_final_N=4}. Thankfully, the alternate formula \eqref{eqn:D1D5_1/2BPS_tildeF_final_N=4_repformula} in terms of representations of $\mathfrak{so}(n)$ means that we can use tools from the representation theory of Lie algebras to answer this question.

First, observe that the representation $\mathcal{R}^{(r,\widetilde{r})}$ of $SO(n)$ for the chiral primaries at charge $(r,\widetilde{r})$ is, in general, a reducible representation. When $\mathcal{R}$ is reducible, the curvature $F_{\mu\nu}$ will be block-diagonal according to the decomposition of $\mathcal{R}^{(r,\tilde{r})}$ into irreps
\begin{equation}
     \left(F_{\mu\nu}\right)_{KL} = \frac{1}{\hat{k}}i\delta_{ab} \left(\begin{matrix} \Sigma_{ij}^{\mathcal{R}_{1}} & & & & \\ & \Sigma_{ij}^{\mathcal{R}_{2}} & & & \\ & & \ddots & & \\ & & & &  \Sigma_{ij}^{\mathcal{R}_{k}}   \end{matrix}\right)\,.
\end{equation}
It is a standard result in random matrix theory that block diagonal matrices with independent blocks, each of which is a random matrix, cannot have RMT statistics \cite{GUHR1998189}. The reason is that even if the eigenvalues of each block repel, there is no repulsion between eigenvalues in different blocks. Hence naively $F$ is not a random matrix. Some readers might be confused as usually one should first focus on a specific block and only then compare with RMT. We emphasize that the difference here is that $SO(n)$ is not a symmetry of the full physical theory, and therefore the existence of such a decomposition here is already reflecting the special features of the $1/2$-BPS states. However, one might worry that a single irrep of $SO(n)$ in the decomposition has dimension that grows as a function of $r,\widetilde{r}$ while the other irreps do not, making $F_{\mu\nu}$ dominated by that single block in the large $c$ limit. We will now argue against that scenario, and then show that even if it were to happen, any block (which is an irrep of $\mathfrak{so}(n)$) is far from a random matrix, as it has large degeneracies and no eigenvalue repulsion.

Some evidence that this concern is not realized can be seen for ${\rm Sym}^N(K3)$. The authors of \cite{Bourget:2015ffp} obtained a generating function for the characters of the representation of $SO(21)$ in which the $(r,\widetilde{r})$ chiral primaries transform, which can be decomposed into characters of irreps. The results for small charges are displayed in Table \ref{tab:so(21)repdecomp} below. We see that the ratio of the dimension of the largest irrep in the decomposition to the dimension of the full representation actually decreases as we increase the charge, indicating that a single irrep does not dominate.

\vspace{3mm}

\begin{table}[h]
\centering
\caption{Decomposition of $SO(21)$ representations for $K3$ chiral primaries}
\label{tab:so(21)repdecomp}
{\renewcommand{\arraystretch}{1.4}
\begin{tabular}{|c|c|c|c|}
\hline
$(r,\widetilde{r})$ & Representation & Decomposition & $\frac{\mathrm{Dim\;of\;largest\;irrep}}{\mathrm{Dim\;of\;rep}}$ \\
\hline
$(\frac{1}{2},\frac{1}{2})$ & $\mathbf{21}$ & $\mathbf{21}$ & $1$ \\
\hline
$(1,1)$ & $254$ & $\mathbf{230}+\mathbf{21}+3 \times \mathbf{1}$ & 0.9 \\
\hline
$(\frac{3}{2},\frac{3}{2})$ & $2278$ & $\mathbf{1750}+\mathbf{230}+\mathbf{210}+4 \times \mathbf{21}+4 \times \mathbf{1}$ & 0.77 \\
\hline
$(2,2)$ & $16744$ & 
$\begin{aligned} & \quad\quad\quad\mathbf{10395}+\mathbf{3059}+\mathbf{1750}\\& +5 \times \mathbf{230} +\mathbf{210}+8 \times \mathbf{21}+12 \times \mathbf{1} \end{aligned} $ & 0.62 \\
\hline
\end{tabular}}
\end{table}

\vspace{3mm}

However, these results are only for small charges and it is possible that this pattern does not continue beyond some charges $(r_{\ast},\widetilde{r}_{\ast})$. Therefore, to obtain a watertight argument, we will prove that even if a single irrep of $\mathfrak{so}(n)$ dominates the decomposition, the curvature still cannot have random matrix statistics.\footnote{Even though the eigenvalues of the representation matrices are all integers, after unfolding, the probability distribution of eigenvalue spacings will depend on the size of the matrix, and it could in principle resemble the Wigner surmise. In this section, we prove that this does not happen and so the Berry curvature matrix can be easily told apart from a random matrix in practice.} 

\paragraph{Some example irreps.} Before proceeding to the proof, let us look at some simple examples to build some intuition. The simplest case, which is universal to both ${\rm Sym}^N(K3)$ and ${\rm Sym}^N(T^4)$, is the $n$ single-trace chiral primaries with charges $(\frac{1}{2},\frac{1}{2})$ whose descendants give the marginal operators. These states transform in the vector representation $\mathrm{fund}$ of $\mathfrak{so}(n)$ so $\Sigma_{ij}^{\mathrm{fund}}$ are the standard $\mathfrak{so}(n)$ generators
\begin{equation}\label{eqn:so(n)gens}
    (\Sigma_{ij}^{\mathrm{fund}})_{kl} = \delta_{il}\delta_{jk}-\delta_{ik}\delta_{jl} \,.
\end{equation}
These are clearly not random matrices since their eigenvalues are $\pm i$ with multiplicity $1$ and $0$ with multiplicity $n-2$. Even arbitrary combinations of these matrices that arise from  marginal deformations in some directions $\vec{v}$ and $\vec{w}$ still have large degeneracy. The curvature along them is
\begin{equation}
F^{\mathrm{fund},\;\mathrm{arbitrary}} \equiv F_{(a,i)(b,j)}^\mathrm{fund}v^{i}w^{j} = {-}\frac{1}{\hat{k}}i\delta_{ab}(vw^{T}-wv^{T}) \,,
\end{equation}
which is not a random matrix since it has rank $2$ (its kernel is $\{\mathrm{Span}(v,w)\}^{\perp}$) with eigenvalue $0$ with multiplicity $n-2$ and eigenvalues $\pm\lambda \equiv \pm \sqrt{v^2w^2-(v\cdot w)^{2}}$ with multiplicity $1$. 

For general tensor representations, it is convenient to still write the fundamental representation in the basis $V = U \oplus U^{\perp}$ where $U = \mathrm{span}(v,w)$. The $k$-th anti-symmetric representation decomposes in terms of $U$ as
\begin{equation}
    \Lambda^{k}V = \left(\Lambda^{2}U \wedge \Lambda^{k-2}U^{\perp}\right) \oplus \left(U \wedge \Lambda^{k-1}U^{\perp}\right) \oplus \Lambda^{k}U^{\perp},
\end{equation}
where $\Lambda^{k}V$ is the $k$th antisymmetric product. Then any $k$-form from the first term $\Lambda^{2}U \wedge \Lambda^{k-2}U^{\perp}$ has zero eigenvalue since $T$ maps $w \to \lambda v$ and $v \to -\lambda w$. Also, any $k$-form from the last term $\Lambda^{k}U^{\perp}$ gives zero eigenvalue. So we have ${n-2 \choose k-2} + {n-2 \choose k}$ many zero eigenvalues. The $k$-forms in the middle term $U \wedge \Lambda^{k-1}U^{\perp}$ give eigenvectors whose eigenvalues are $\pm \lambda$ with multiplicity ${n-2 \choose k-1}$.

Next, consider the $\ell$-th symmetric traceless representation $[\ell]$. The decomposition is simplest to work out for $\ell=2$ where it is given by
\begin{equation}
    ST^{2}V = ST^2 U \oplus (U \otimes U^{\perp})_{\mathrm{sym}} \oplus ST^2 U^{\perp} \oplus \mathrm{span}\left((n-2)\delta_{\mu\nu}^{U}-2\delta_{\mu\nu}^{U^{\perp}}\right),
\end{equation}
where $ST^{k}V$ is the symmetric traceless product, and $\mathrm{sym}$ means symmetrization. The last term takes into account the symmetric $2$-tensor whose traces on $U$ and on $U^{\perp}$ are non-zero, but their sum is zero. Then tensors in $ST^2 U^{\perp}$ and $\mathrm{span}\left((n-2)\delta_{\mu\nu}^{U}-2\delta_{\mu\nu}^{U^{\perp}}\right)$ have zero eigenvalue giving multiplicity ${n-1 \choose 2}$. The eigenvalues for the subspace $ST^2 U$ are $\pm 2 \lambda$ with multiplicity $1$ and for $(U \otimes U^{\perp})_{\mathrm{sym}}$ the eigenvalues are $\pm \lambda$ with multiplicity $(n-2)$ each.

The main takeaway from these examples is that there are only a few distinct eigenvalues each of which has a large multiplicity. Nevertheless, one could complain that these examples are too simple. Since tensor representations of $\mathfrak{so}(n)$ are determined by Young tableaux, it is possible that for very complicated choices of Young tableaux that become increasingly complicated as the dimension of the irrep grows, RMT statistics emerges.  We will now prove that this does not happen and the large multiplicities are a feature of any irrep of $\mathfrak{so}(2k+1)$ so there cannot be any eigenvalue repulsion.

\paragraph{General proof for any irrep of $\mathfrak{so}(2k+1)$.}

Given an irrep $\mathcal{V}$ of $\mathfrak{so}(2k+1)$, we want to understand the Berry curvature in an arbitrary direction
\begin{equation}
    F^\mathbf{\mathcal{V},\;\mathrm{arbitrary}} \equiv F_{(a,i) (b,j)}^\mathcal{V}v^{i}w^{j} = \frac{1}{\hat{k}}i\delta_{ab}\rho^{\mathcal{V}}(\mathcal{T}_{ij})v^{i}w^{j} = \frac{1}{\hat{k}}i\delta_{ab}\rho^{\mathcal{V}}(\mathcal{T}_{ij}v^{i}w^{j})
\end{equation}
where $\mathcal{T}_{\mu\nu}$ is a generator of $\mathfrak{so}(2k+1)$ and in the last equality we used linearity of the representation map $\rho^{\mathcal{V}}$. Recall that the inner automorphism group of $\mathfrak{so}(n)$ is $SO(n)$ which acts by conjugation. Since this is an automorphism, acting on $\mathfrak{g} \in \mathfrak{so}(n)$ with any $g \in SO(n)$ changes $\rho^{\mathcal{V}}(\mathfrak{g})$ by a similarity transformation, and hence preserves the eigenvalues. We saw above that there exists an $SO(n)$ change of basis matrix that takes $\mathcal{T}_{ij}v^{i}w^{j} = (vw^{T}-wv^{T})$ to $\lambda \mathcal{T}_{12}$ so for the eigenvalues of $F^\mathbf{\mathcal{V},\;\mathrm{arbitrary}}$, we only need to study $\rho^{\mathcal{V}}(\lambda \mathcal{T}_{12})=\lambda\rho^{\mathcal{V}}(\mathcal{T}_{12})$. Observe that $\mathcal{T}_{12}$ is one of the elements of the canonical choice of Cartan subalgebra for $\mathfrak{so}(n)$ given by $\mathfrak{C} = \{\mathcal{T}_{(2i+1)(2i+2)}|\,i \in 0,\ldots,\floor{\frac{n}{2}}-1\}$. So the problem has reduced to studying eigenvalues of the representation of a Cartan element. But, these are simply the elements of the weights for this Cartan element, that is, there are $\mathrm{dim}\,\mathcal{V}$-many weights $\mu_{i}$ and we need to compute $\mu_{i}(\mathcal{T}_{12})$. Using this fact, we are able to prove the desired theorem:

\begin{theorem}
For any irrep $\mathcal{V}$ of $\mathfrak{so}(2k+1)$ for $k \geq 2$, the eigenvalues of the matrix representation of a Cartan basis element never exhibit eigenvalue repulsion. In particular, the distribution of nearest-neighbor spacings behaves as $P(s) \to \delta(s)$ as $\dim \mathcal{V} \to \infty$.
\end{theorem}

\begin{proof}
Let $\mathfrak{g}=\mathfrak{so}(2k+1)$ with $k\ge2$, and choose the standard orthogonal basis $\epsilon_{1},\dots,\epsilon_{k}$ of the weight space. Write the highest weight of the irrep $V_{\lambda}$ as
\begin{equation} 
\lambda=\sum_{i=1}^{k}\ell_{i}\epsilon_{i},\qquad\ell_{1}\ge\ell_{2}\ge\cdots\ge\ell_{k}\ge0\,.
\end{equation}
The $\ell_i$'s are related to the usual Dynkin indices by 
\begin{equation}
    \ell_i = \sum_{j=i}^{k-1} \Lambda_i + \frac{\Lambda_k}{2} \,,
\end{equation}
and so whenever any Dynkin label diverges, $\Lambda_i \to \infty$, so does $\ell_1$. Now fix a Cartan basis element $H_{a}$ dual to $\epsilon_{a}$, so that a weight $\mu=\sum_{i}\mu_{i}\epsilon_{i}$ gives the eigenvalue $\mu(H_{a})=\mu_{a}$. The weights of $V_{\lambda}$ lie in the convex hull of the Weyl orbit $W\cdot\lambda$ \cite{Hall2015}, and for  $\mathfrak{so}(2k+1)$ the Weyl group acts by signed permutations of the $\epsilon_i$. In a such a convex hull, every weight $\mu$ satisfies $|\mu_{a}|\le\ell_{1}$. Moreover, every weight is (half-)integer. Therefore the number of distinct eigenvalues of $H_{a}$ on $V_{\lambda}$ is at most $2\ell_{1}+1$.

Now use the Weyl dimension formula in weight form. For $\mathfrak{so}(2k+1)$, the positive roots are $\epsilon_{i}$ and $\epsilon_{i}\pm\epsilon_{j}$ for $i<j$, together with $\rho=\sum_{i=1}^{k}\rho_{i}\,\epsilon_{i}$ where $\rho_{i}=k-i+\tfrac{1}{2}$. Thus 
\begin{equation}
\dim V_{\lambda} = \prod_{\alpha>0}\frac{(\lambda+\rho,\alpha)}{(\rho,\alpha)} = \prod_{i=1}^{k}\frac{\ell_{i}+\rho_{i}}{\rho_{i}}\prod_{1\le i<j\le k}\frac{\ell_{i}-\ell_{j}+\rho_{i}-\rho_{j}}{\rho_{i}-\rho_{j}}\prod_{1\le i<j\le k}\frac{\ell_{i}+\ell_{j}+\rho_{i}+\rho_{j}}{\rho_{i}+\rho_{j}}\,.
\end{equation} 
Since $\ell_{1}\ge\cdots\ge l_{k}\ge0$, every factor in a numerator is greater or equal to 1, while the denominators are independent of the $\ell$'s. Keeping only terms in the first and third product that depend on $\ell_1$, we obtain 
\begin{equation} \dim V_{\lambda}\ge c_{k}\ell_{1}^{k}\,,
\end{equation} 
for some constant $c_{k}>0$ that depends only on $k$.

Now sort the eigenvalues of $H_a$ by size. Since they have $2\ell_{1}+1$ allowed values, there are at most $2\ell_1$ non-zero gaps between them, while overall there are at least $\dim\left(V_{\lambda}\right)-1\ge O(\ell_{1}^{k})$ gaps, and so the probability of finding a non-zero gap vanishes when any Dynkin weight $\Lambda_i \to \infty$,
\begin{equation}
    \frac{\#\text{ of non-zero spacings}}{\#\mathrm{\;of\;spacings}} \le  \frac{2\ell_1}{\dim(V_\lambda)-1} \le \frac{2}{c_k \ell_1^{k-1}-\frac{1}{\ell_{1}}} \xrightarrow{\ell_1 \to \infty} 0 \,.
\end{equation}
In other words, the probability distribution for the spacings $s$ behaves as $P(s) \to \delta(s)$ for any of the $\Lambda_{i} \to \infty$ so there is no eigenvalue repulsion.
\end{proof}

%%%%%%%%%%%
% SECTION %
%%%%%%%%%%%
\subsection{Case of ${\rm Sym}^N(T^{4})$}
\label{sec:symT4}
In this section we discuss some subtleties for the case of ${\rm Sym}^N(T^4)$. At the generic point on the conformal manifold, the ${\rm Sym}^N(T^4)$ CFT has more symmetries than the ${\rm Sym}^N(K3)$ CFT. Because of this the structure of the conformal manifold, the chiral ring and the Berry curvature of 1/2-BPS states are somewhat modified. At the orbifold point, the ${\rm Sym}^N(T^4)$ theory also has accidental symmetries, which need to be taken into account in certain comparisons with orbifold computations, this will be discussed in Subsection \ref{sec:orbifold}.\footnote{\label{d1d5vssym}In this section we focus on the curvature of chiral primaries over the moduli space of the ${\rm Sym}^N(T^4)$ CFT. The D1/D5 CFT, with $M=T^4$, is somewhat larger and contains additional marginal deformations \cite{Larsen:1999uk,Aharony:2024fid}. However, there is no curvature of the chiral primaries along those directions since they are of $J\widetilde{J}$ form and the chiral primaries are neutral under these extra currents.}

\paragraph{The symmetry algebra.} 

The superconformal algebra (SCA) of the ${\rm Sym}^N(T^4)$ CFT, at a generic point of its conformal manifold, is larger than the standard small ${\cal N}=4$ SCA. It is sometimes called contracted large ${\cal N}=4$ SCA or medium ${\cal N}=4$ SCA \cite{Ali:2003aa}. In addition to the usual ${\cal N}=4$ generators $(T,G^{aA},J^i)$ it contains 4 left-moving fermions, as well as 4 $U(1)$ bosonic currents, and the explicit algebra can be found in Appendix \ref{sec:mediumN=(4,4)SCA}. We combine the 4 real $(\frac{1}{2},0)$ fermionic operators into two complex combinations that we denote as $\Psi^\alpha(z),\,\alpha=1,2$ --- and similarly for the right movers. At the orbifold point, the free fermions have the form
\be
\Psi^\alpha(z) = {1\over \sqrt{N}}\sum_n \psi^\alpha_n(z) \,,\qquad \alpha=1,2 \,,
\ee
where $n$ runs over the $N$ strands of the orbifold theory. We emphasize that we continue to have $({1\over 2},0)$ free fermions $\Psi^\alpha$ everywhere on the conformal manifold. By taking products of these fermions we can  construct an $SO(4)$ worth of ``double-trace" $(1,0)$ conserved currents. For reasons that will become clear later, we factorize this as $SO(4)=SU(2)^D_R\times SU(2)^D_{\rm flavor}$, where we denote the generators of the first $SU(2)$ factor as  the ``double-trace R-symmetry currents"
\be
\label{doubletracer}
J_{R}^{ D,+} = -\Psi^1 \Psi^2 \,, \qquad J_{R}^{ D,-} = \Psi^{1\dagger} \Psi^{2\dagger} \,, \qquad J_{R}^{ D,3} = {1\over 2}\left( \normord{\Psi^1 \Psi^{1\dagger}}+\normord{\Psi^2\Psi^{2\dagger}}\right) \,,
\ee
and another combination gives the second $SU(2)$, which we call  ``double-trace flavor symmetry"  generators
\be
\label{doubletracef}
J_{{\rm flavor}}^{ D,+} = -\Psi^1 \Psi^{2\dagger} \,, \qquad J_{{\rm flavor}}^{ D,-} = \Psi^{1\dagger} \Psi^{2} \,, \qquad J_{\rm flavor}^{ D,3} = {1\over 2}\left( - \normord{\Psi^1 \Psi^{1\dagger}}+\normord{\Psi^2\Psi^{2\dagger}}\right) \,.
\ee
Both $SU(2)^D_{\rm R}$ and $SU(2)^{D}_{\rm flavor}$ remain good symmetries everywhere on the conformal manifold since the corresponding currents do not receive anomalous dimensions.\footnote{Since the fermions $\Psi^i$ have dimension $(\frac{1}{2},0)$ everywhere on the conformal manifold, correlators of these fermions can be fixed by holomorphicity and Ward identities. By examining the conformal block expansion of these correlators we can deduce the existence of the conserved currents \eqref{doubletracer}, \eqref{doubletracef} at the generic point of the conformal manifold.} We have the entirely equivalent story for the right-moving sector.

Notice that the full ${\cal N}=4$ symmetry generators $J_{R}^+,J_{R}^-,J_{R}^3$ (which at the orbifold point are single-trace) are not orthogonal to the double trace currents mentioned above. It is convenient to consider orthogonal linear combinations. In particular, we define
\be
\label{shiftedcurrent}
\widehat{J}_{R}^3 = J_{R}^3-J_{R}^{D,3}.
\ee
We also define the corresponding shifted SU(2) R-charges
\be
\label{shiftedcharges}
\widehat{r} \equiv r - r^D \,,
\ee
where $r$ is the standard R-charge (eigenvalue of $J_{R,0}^3$) and $r^D$ is the corresponding charge for the double-trace R-current $J^{D,3}_{R,0}$ in \eqref{doubletracer}. Notice that if the 2-point function $\langle J_{R}^3(z) J_{R}^3(0)\rangle ={\hat{k}\over z^2}$, with $\hat{k}=c/6$, then we have 
\be
\langle \widehat{J}_{R}^3(z) \,\widehat{J}_{R}^3(0)\rangle = {\hat{k}-1 \over z^2} \,.
\ee
We can proceed using similar arguments to remove from each medium ${\cal N}=4$ SCA generator its double-trace component.

\paragraph{The chiral ring.}

The 1/2-BPS states of the theory are counted by the Göttsche formula \cite{goettsche1990hilbert} as follows.
First, for a given $N$ we define the partition function over 1/2-BPS states as\footnote{Here we are working in the NS sector and the 1/2-BPS states are chiral primaries on both sides, i.e. $h=r,\widetilde{h}=\widetilde{r}$.}
\be
\label{bpspartdef}
Z_{N}(y,\widetilde{y}) = {\rm Tr}_{{1\over 2}-\rm BPS}[y^{2J_{R,0}^3}\, \widetilde{y}^{2\widetilde{J}_{R,0}^3}] \,,
\ee
then the grand-canonical generating function of these partition functions is \cite{goettsche1990hilbert,Dijkgraaf:1996xw},
\be
\label{goettsche}
\sum_{N=0}^\infty Q^N Z_{N}(y,\widetilde{y}) = \prod_{m=1}^\infty \prod_{p,q=0}^2 (1+(-1)^{p+q+1} Q^m y^{p+m-1}\widetilde{y}^{q+m-1})^{(-1)^{p+q+1} h^{p,q}} \,,
\ee
where the Hodge numbers of $T^4$ are $h^{0,0}=h^{2,0}=h^{0,2}=h^{2,2}=1$, $h^{1,0}=h^{0,1}=h^{1,2}=h^{2,1}=2$ and $h^{1,1}=4$.

The chiral primaries in the case of ${\rm Sym}^N(T^4)$ can be further classified into representations of the free-fermion algebra \cite{Maldacena:1999bp,Chang:2025rqy}. We can define chiral primaries which are also ``primaries" under this fermionic algebra by the condition that
\be
\label{primaryfermionic}
\oint dz \,\Psi^{\alpha \dagger}(z) \,\varphi(0) = 0\,, \qquad \alpha=1,2 \,,
\ee
which means that there is no first-order pole in the OPE between $\Psi^{\alpha,\dagger}$ and $\varphi(0)$. We have a similar condition for the right-moving fermions. Let us call such chiral primaries {\it primitive}. 
Such primaries can be decorated by acting with fermions and form Clifford multiplets.\footnote{For example, consider the correlator
\be
\langle \varphi(x)^\dagger \Psi^\alpha(z) \Psi^{\alpha\dagger}(w) \varphi(y)\rangle
=\langle \varphi(x)^\dagger \varphi(y)\rangle\,{1\over z-w}
\ee
where we assumed that $\varphi(x)$ is a chiral primary, which is also a primary of the fermionic algebra \eqref{primaryfermionic} and the RHS of the equation was determined by holomorphicity of $\Psi^\alpha(z)$ and analysis of all possible poles. Taking the limit $z\rightarrow y, w\rightarrow x$ we conclude that there must exist a chiral primary, which can be thought of as the composite $\Psi^\alpha(x) \varphi(x)$, with dimension $(h_\varphi+{1\over 2},\widetilde{h}_\varphi)$. Recursively, we conclude that starting with a given fermionic-primary chiral primary $\varphi$ we can decorate it with any element of the Hodge diamond of a single $T^4$. In other words, the chiral primaries of ${\rm Sym}^N(T^4)$ can be decomposed into representations of this fermionic algebra \cite{Chang:2025rqy, Maldacena:1999bp}.} Notice that chiral ring multiplication respects the structure of these representations, i.e. the OPE of two primitive chiral-primaries is also primitive.

At the level of 1/2-BPS state counting, the aforementioned decomposition into representations of the fermionic algebra can be seen from the fact that the 1/2-BPS partition function of ${\rm Sym}^N(T^4)$, for any $N$, is divisible by the 1/2-BPS partition function $Z_1(y,\widetilde{y})$ of a single $T^4$ \cite{Maldacena:1999bp},  that is
\be
\label{bpsfactor}
Z_N(y,\widetilde{y})= Z_1(y,\widetilde{y}) \cdot \hat{Z}_N(y,\widetilde{y}) \,,
\ee
where 
\be
Z_1(y,\widetilde{y})=(1+y)^2(1+\widetilde{y})^2 \,,
\ee
and $\hat{Z}_N(y,\widetilde{y})$ is a polynomial in $y,\widetilde{y}$ with {\it positive, integer} coefficients.  In Appendix \ref{app:factorization} we provide an elementary proof of \eqref{bpsfactor}. The reduced 1/2-BPS partition function $\hat{Z}_N(y,\widetilde{y})$ only counts primitive chiral primaries.

\paragraph{The conformal manifold.}

As discussed in Subsection \ref{sec:D1/D5_1/2BPS}, marginal deformations preserving the ${\cal N}=(4,4)$ supersymmetry are super-descendants of chiral primaries of dimension $({1\over2}, {1\over 2})$. From \eqref{goettsche} we find that for ${\rm  Sym}^N(T^4)$ with $N>1$ there are 9 such states. Using the factorization \eqref{bpsfactor} we have 5 primitive chiral primaries. The remaining four correspond to double-trace chiral primaries made out of the free fermions
\be
\label{doubletraceprimaries}
\Psi^\alpha \widetilde{\Psi}^\beta\,, \qquad \alpha,\beta=1,2 \,.
\ee
We make sure that we select the 5 primitive chiral primaries to be orthogonal to the double-trace ones.\footnote{At the orbifold point these 5 chiral primaries correspond to single-trace chiral primaries (slightly shifted by double-trace admixture so as to be orthogonal) and the twist-$2$ chiral primary $\Sigma_2$.}

The conformal manifold factorizes locally into two factors\footnote{We remind the reader that here we are talking strictly about the conformal manifold of the ${\rm Sym}^N(T^4)$ SCFT, see also Footnote \ref{d1d5vssym}.} as 
\be
\label{confmant4}
{\cal M} = {SO_{0}(4,5) \over SO(4) \times SO(5)} \times {SO_{0}(4,4) \over SO(4)\times SO(4)} = {\cal M}_5 \times {\cal M}_4 \,.
\ee
The factor ${\cal M}_4$ is spanned by marginal operators which are descendants of the double-trace chiral primaries \eqref{doubletraceprimaries}. These marginal operators are of $J\widetilde{J}$ form, where the currents are selected from the left and right $U(1)^4$ bosonic currents. On the other hand, the factor ${\cal M}_5$ is generated by marginal operators which are descendants of the 5 primitive chiral primaries.

One can confirm the factorized structure \eqref{confmant4} by explicitly computing the curvature along various directions, using \eqref{berry4point} for external states being marginal operators. Marginal operators along ${\cal M}_4$ are of $J\widetilde{J}$ form so the relevant 4-point function for the computation of the curvature is fixed by Ward identities \cite{Kutasov:1988xb}. Moreover,  using Ward identities one can see that the cross-terms of the curvature vanish. Finally, as will be explained in more detail below, the geometry of the ${\cal M}_5$ factor follows from the usual reasoning for ${\cal N}=(4,4)$ theories with small ${\cal N}=4$ SCA \cite{Seiberg:1988pf, Cecotti:1990kz}, with  $n=5$ corresponding to the 5 primitive chiral primaries of weight $({1\over 2},{1\over 2})$ and with a shift of $\hat{k} \rightarrow \hat{k}-1$ in fixing the overall scale of ${\cal M}_5$.\footnote{As will be discussed below, there is no Berry-connection mixing between double-trace and primitive chiral primaries of weight $({1\over 2},{1\over 2})$.}

\paragraph{The curvature of chiral primaries.}

Now we discuss a minor modification of the $\tt$ equations for the ${\rm Sym}^N(T^4)$ CFT, due to its enhanced symmetry algebra, relative to theories with small ${\cal N}=4$ SCA, such as the ${\rm Sym}^N(K3)$. The point is that the ``current term" ${\cal F}_{ijKL}$ in equation \eqref{eqn:D1D5_1/2BPS_tildeF_final_N=4} receives contribution from $(1,0)$ and $(0,1)$ currents. To be general, suppose we have a set of $(1,0)$ conserved currents $J_A$ normalized so that
\be
\label{current2pt}
\langle J_A(z) J_B(0)\rangle = {\hat{k}_A \delta_{AB} \over z^2} \,,
\ee
and similarly for the right-movers. These currents may appear in the OPE of chiral primaries. We define the OPE coefficient 
\be
\varphi_i(z) \cdot \varphi_j^\dagger(0) = \ldots+X_{i\overline{j}}^{A} {J_A(0) \over z^{h_i + h_j-1}\overline{z}^{\widetilde{h}_i + \widetilde{h}_j}} + \ldots \,,
\ee
and the 3-point function coefficient
\be
\langle \varphi_K (\infty) \varphi_L^\dagger(1) J_A(0) \rangle = X_{K \overline{L} A} \,.
\ee
We denote the analogue coefficients from the right movers by $Y$. 
Then the curvature formula \eqref{eqn:D1D5_1/2BPS_tildeF_final_N=4} becomes
\be
\label{ttgenj}
(F_{\mu\nu})_{KL} =i \delta_{ab}\left[\left(\delta_{ij}-\sum_A {X_{i\overline{j}}^A X_{K\overline{L}A} \over \hat{k}_A}-\sum_{\widetilde{A}} {Y_{i\overline{j}}^{\widetilde{A}} Y_{K\overline{L}\widetilde{A}} \over \widetilde{\hat{k}}_{\widetilde{A}}}\right)-[C_i,C_j^\dagger]_{KL}\right]\,,
\ee
where we remind the notation for the marginal operators $\mu=(a,i)$, $\nu=(b,j)$. In the case where we only have the small ${\cal N}=4$ SCA, and working in the usual real basis for the chiral ring, then the two sums in the bracket collapse to $-{1\over \hat{k}}(r+\widetilde{r})\delta_{ij}\delta_{KL}$, with $\hat{k}=c/6$. In this formula it is important to include only currents which are conserved in an open neighborhood of the point at which we are computing the curvature.\footnote{In the derivation of the curvature outlined in Section \ref{sec:D1/D5_1/2BPS} a spin 1 current can contribute only if it has dimension {\it exactly} equal to $(1,0)$ or $(0,1)$ \cite{deBoer:2008ss}. Suppose $P$ is a point where we have accidental currents. We are usually interested in the curvature at point $P$ defined as the limiting value of the curvature as we approach the point $P$, hence accidental currents will not contribute.}

When we are at the generic point of the ${\rm Sym}^N(T^4)$ conformal manifold the currents that contribute are the $J_{R}^3$ R-current and the double-trace currents $J_{R}^D$
and $J_{{\rm flavor}}^D$ in equations \eqref{doubletracer} and \eqref{doubletracef}, respectively. When applying the formula \eqref{ttgenj} it is important to work in a basis where the currents are orthogonal, as in \eqref{current2pt}, which requires a change of basis as in \eqref{shiftedcurrent}.

In practice, it may be more convenient to isolate the total contribution from the currents by considering the OPE expansion of the 4-point function  $\langle \varphi^\dagger_L(\infty) \, \varphi_i(x) \, \varphi^\dagger_j(y) \, \varphi_K(0)\rangle$ in the $x\rightarrow y$ channel
\be\label{eqn:4ptfn_currentscontr}
\langle \varphi^\dagger_L(\infty) \, \varphi_i(x) \, \varphi^\dagger_j(y) \, \varphi_K(0)\rangle = \ldots+ a{1\over (\overline{x}-\overline{y})y}  + \widetilde{a} {1\over (x-y)\overline{y}} +\ldots \,,
\ee
then we can directly read off the desired sums by noticing that
\be
\label{eqn:readoffa}
a= \sum_A {X_{i\overline{j}}^A X_{K\overline{L}A} \over \hat{k}_A}\,, \qquad \widetilde{a}=\sum_A {Y_{i\overline{j}}^{\widetilde{A}} Y_{K\overline{L}\widetilde{A}} \over \widetilde{\hat{k}}_{\widetilde{A}}} \,.
\ee
However, if the correlator \eqref{eqn:4ptfn_currentscontr} is evaluated at a point where there are accidental currents, it is important that their contribution is removed by hand from \eqref{eqn:readoffa} before plugging the sums into \eqref{ttgenj}.

Now we discuss the curvature of the chiral primaries over the conformal manifold \eqref{confmant4}, as predicted by \eqref{ttgenj}. We summarize the final result and postpone details of the derivation to Appendix \ref{sec:detailssymt4}:
\begin{enumerate}
\item The curvature of chiral primaries along ${\cal M}_4$ vanishes.
\item The curvature of chiral primaries along mixed directions on ${\cal M}_5\times {\cal M}_4$ vanishes.
\item The curvature of chiral primaries along ${\cal M}_5$ is non-trivial. The curvature is the same for all elements in a given fermionic-multiplet and the curvature takes the form
\end{enumerate}
\be
\label{curvaturet4main}
(F_{\mu \nu})_{KL} = i \delta_{ab}\left[\delta_{ij}\delta_{KL}\left(1 -{1\over \hat{k}-1} (r + \widetilde{r})\right)-[C_i, C_j^\dagger]_{KL}\right] \,,
\ee
where $r,\widetilde{r}$ are the R-symmetry charges of the primitive chiral primary inside a given fermionic multiplet.

To summarize, we notice that once we factor out the extra free-fermion sector, in the sense that: i) we focus on primitive chiral primaries and ii) we focus  on directions of the conformal manifold spanned by descendants of primitive $({1\over 2}, {1\over 2})$ chiral primaries, the story for ${\rm Sym}^N(T^4)$ is exactly the same as that of ${\rm Sym}^N(K3)$ with an effective shift of the central charge $c\rightarrow c-6$ in the $\tt$ equations.

In particular, since the 3-point functions are covariantly constant, we find that the curvature \eqref{curvaturet4main} is also covariantly constant.  So, we have homogeneous bundles over ${\cal M}_5$ (which are flat along ${\cal M}_4$). As discussed before, this implies that the  chiral primaries fall into representations of $SO(5)$ and the curvature of these bundles have the form
\be
\label{eq:curv_T4_rep}
(F_{\mu\nu})_{KL} = {1\over \hat{k}-1}i\delta_{ab}(\Sigma_{ij}^{\mathcal{R}})_{KL} \,,
\ee
where now the matrices on the RHS correspond to representations of $SO(5)$.

Notice that the case of $N=1$, i.e. the SCFT of a single $T^4$, is somewhat special and we analyze it in detail in Appendix \ref{sec:T4}.

\subsection{Comparison with orbifold computations}
\label{sec:orbifold}

In order to compare with symmetric orbifold computations, and focusing for simplicity on the case ${\rm Sym}^N(T^4)$, we need to study the curvature of the chiral primaries near the orbifold locus ${\cal M}_0$. This is a submanifold of the full conformal manifold, which can be generated by starting with the naive symmetric product of $N$ copies of $T^4$ (with particular values of the $T^4$ moduli) and deforming by all possible untwisted-sector operators. There are $4 \times 8 =32$ of those.  There are 4 directions perpendicular to ${\cal M}_0$ which are generated by descendants of the twist-$2$ chiral primary $\Sigma_2$, i.e. this deformation takes us toward strong-coupling. The 8 untwisted-sector $({1\over 2}, {1\over 2})$ chiral primaries, whose superconformal descendant marginal operators span ${\cal M}_0$, can be further subdivided into 4 primitive chiral primaries and 4 which are fermionic descendants (under the free fermion algebra) of the identity operator. The latter are double-trace operators and the corresponding marginal deformations are of the $J\widetilde{J}$ form. In terms of the decomposition \eqref{confmant4}, we can think of ${\cal M}_0$ as covering the ${\cal M}_4$ part of the conformal manifold times another 16-dimensional slice inside ${\cal M}_5$.

On ${\cal M}_0$ there are additional, ``accidental" conserved currents, the single-trace version of the flavor symmetry currents \eqref{doubletracef}. Under deformation by twist-$2$ operators moving perpendicular to ${\cal M}_0$, these currents acquire anomalous dimensions \cite{Benjamin:2021zkn}. For the computation of the curvature of 1/2-BPS states near ${\cal M}_0$ it is important {\it not} to include the contribution of the accidental conserved currents, when applying  equation \eqref{ttgenj}, since it is no longer conserved at any small (finite) distance away from ${\cal M}_0$.

In Appendix \ref{app:orbifold} we present a few checks, by explicit computation in the orbifold CFT, that the curvature of chiral primaries in the neighborhood of ${\cal M}_0$ has the structure described in, and above equation \eqref{curvaturet4main}.

%%%%%%%%%%%
% SECTION %
%%%%%%%%%%%
\subsection{Computing $CC^\dagger$ eigenvalues via consistency with $\tt$ equations}
\label{ccdagertt}

In Subsection \ref{sec:D1/D5_non-random}, we found that the Berry curvature of 1/2-BPS states is not chaotic. We will now show that the chiral ring OPE functions $(C_i)_{K}^L$ are also not chaotic, where $i$ labels chiral primaries of dimension $({1\over 2},{1\over 2}$) and $KL$ are general chiral primaries. In particular, we will show that the $\tt$ equations predict the eigenvalues of the matrices $C_i C_i^\dagger$ and $C_i^\dagger C_i$. First, we consider the case of ${\rm Sym}^N(K3)$, where we do not have the extra fermions and then discuss the ${\rm Sym}^N(T^4)$ case.

Our starting point is the identity\footnote{We remind the reader that \eqref{generalid}, and hence \eqref{identitychiral}, follows from combining the formula for the curvature of the chiral ring with the non-renormalization theorem of the chiral ring for ${\cal N}=(4,4)$ SCFTs \eqref{eqn:covconstCRC}. In Appendix \ref{sec:CRCidentity} we provide a more direct proof of \eqref{identitychiral} based on crossing symmetry.} \eqref{generalid}. Evaluating it for $i=j$ we get
\be
\label{identitychiral}
[C_i,C_i^\dagger]_{KL} = \delta_{KL} \left(1-{r+\widetilde{r}\over N}\right) \,,
\ee
where we used that $\hat{k}=N$ and
where $i$ can be any of the dimension $(\frac{1}{2},\frac{1}{2})$ chiral primaries. This identity will allow us to fully determine the distribution of eigenvalues of $C_i C_i^\dagger$ and of $C_i^\dagger C_i$ individually, just by group theory and knowledge of the 1/2-BPS partition function. Notice that the result we will present is valid anywhere on the conformal manifold since the argument does not assume weak coupling. Indeed, from the non-renormalization theorem of chiral ring 3-point functions we expect that the spectrum of these matrices does not depend on the coupling. 

The basic point is that \eqref{identitychiral} implies that the 1/2-BPS chiral primaries organize themselves in $SU(2)$ representations for which $C_i,C_i^\dagger$ plays the role of $J_+, J_-$. In particular, the eigenvalues of $C_i C_i^\dagger$ are the same as those of $J_+ J_-$ in the corresponding representation of $SU(2)$. Notice that this $SU(2)$ should not be confused with the R-symmetry (or outer automorphism) of the ${\cal N}=4$ SCA, but is rather analogous to the $SU(2)$ Lefschetz action in cohomology.

Each representation of this $SU(2)$ starts with a lowest weight state, which is defined by the condition that it is annihilated by the action of $C_i^\dagger$ (within the chiral ring). Suppose that this lowest weight chiral primary is $\varphi_0$ and has dimension $(r_0,\widetilde{r}_0)$. The rest of the  representation is generated by acting on $\varphi_0$ with $C_i$, giving us the vectors
\be
\label{lefrep}
(C_i)^n\cdot \varphi_0 \,, \qquad n=0,1,\ldots,n_{\rm max}\,,\qquad n_{\rm max} =2N-2(r_0+\widetilde{r}_0) \,,
\ee
where the notation $(C_i)^n\cdot\varphi_0$ means chiral ring multiplication by $\varphi_i$ $n$-times, within the chiral ring. Notice that this chain terminates, as $(C_i)^{n_{\rm max}+1}\cdot \varphi_0=0$ within the chiral ring. The set of vectors \eqref{lefrep} form an $SU(2)$ representation of spin $N-(r_0+\widetilde{r}_0)$, where $C_i$ plays the role of $J_+$ and $C_i^\dagger$ of $J_-$. 

Within this representation $C_i^\dagger C_i$ has now eigenvalues
\be
\lambda_n^+=(n+1)\left(1-{r_0 +\widetilde{r}_0\over N} -{n \over 2N}\right) \,,
\ee
and $C_i C_i^\dagger$ has eigenvalues
\be
\lambda_n^-=n\left(1-{r_0 +\widetilde{r}_0\over N} -{(n-1) \over 2N}\right) \,.
\ee
In fact, taking the difference of the two matrices in this representation, we find that $C_i C_i^\dagger - C_i^\dagger C_i$ has diagonal elements as required by the $\tt$ equations \eqref{identitychiral}.

It is easy to see that the entire chiral ring decomposes into the direct sum of representations of the form above. Since each of these chains are orthogonal, the eigenvalues of the full $CC^\dagger$ matrix acting on the full\footnote{We are still focusing on the ``extremal" part of the chiral ring, as discussed in Section \ref{sec:D1/D5_1/2BPS}. It would be interesting to also study more general chiral ring 3-point functions allowed by the $SU(2)$ selection rules. } chiral ring will simply be the union of the eigenvalues of each $SU(2)$ chain. Then, we start with the 1/2-BPS partition function $Z_N(y,\widetilde{y})$, evaluated at $\widetilde{y}=y$ since we only care about the sum $r_0+\widetilde{r}_0$, and decompose it into $SU(2)$ characters as follows
\be
\label{bpspartsu2}
Z_N(y,y) = \sum_{s\leq N}\,\,{\cal N}_{s}\,\,P_N^{s}(y) \,,
\ee
where
\be
P_N^{s} (y)= y^{2s} (1+ y^2 +y^4 +\ldots+ (y^2)^{2N-2s}) \,,
\ee
where ${\cal N}_{s}$ are non-negative integers and $s\equiv r_0+\widetilde{r}_0$ is (half)-integer. Having determined the integers ${\cal N}_{s}$, we can immediately get the full spectrum of all eigenvalues of $C_i C_i^\dagger$: for each $s=r_0+\widetilde{r}_0$  we have ${\cal N}_{s}$ copies of the eigenvalues
\be
\lambda_n^+=(n+1)\left(1-{r_0+\widetilde{r}_0\over N}-{n\over 2N}\right) \,, \qquad n=0,1,\ldots,2N-2(r_0+\widetilde{r}_0) \,.
\ee
This holds for any $({1\over 2},{1\over 2})$ chiral primary $\varphi_i$ in the ${\rm Sym}^N(K3)$ theory.

In the case of ${\rm Sym}^N(T^4)$ we first notice that fermionic decorations do not change the eigenvalues of $C_i C_i^\dagger$ and $C_i^\dagger C_i$, as long as $i$ is one of the 5 primitive chiral primaries. Then to determine the eigenvalues of $C_i C_i^\dagger$ we need: i) restrict to the partition function of primitive chiral primaries, which we denoted by $\widehat{Z}_N$ in the decomposition \eqref{bpsfactor}, ii) follow the decomposition into $SU(2)$ characters \eqref{bpspartsu2} applied to $\widehat{Z}_N$ and finally iii) multiply the number of eigenvalues by 16 due to the fermionic decorations. 

It is straightforward to compare the above results with explicit orbifold computations. We take $i$ to be the twist-$2$ chiral primary $\Sigma_2$, which is one of the 5 primitive chiral primaries of dimension $({1\over 2},{1\over 2})$. Then, for instance, for $N=4$ we find the reduced partition function
\be
\hat{Z}_{N=4}=1 + 7 y^2 + 8 y^3 + 36 y^4 + 56 y^5 + 113 y^6 + 56 y^7 + 36 y^8 +
 8 y^9 + 7 y^{10} + y^{12} \,.
\ee
This decomposes as
\be
\hat{Z}_{N=4}= P^0_3(y)+6 P^1_3(y)+8 P^{3/2}_3(y)+29 P^2_3(y)+48 P^{5/2}_3(y)+77 P^3_3(y) \,.
\ee
Using the equations above, we find the following eigenvalues 
\be
(0,169),\quad \left(\frac{1}{6},48\right),\quad \left(\frac{1}{3},58\right), \quad \left(\frac{1}{2},16\right), \quad \left(\frac{2}{3},20\right), \quad (1,14), \quad \left(\frac{5}{3},2\right), \quad (2,2) ,
\ee
where the first number of each pair is the eigenvalue and the second is the degeneracy. 
Multiplying the degeneracies by 16 to account for the fermionic decorations, we find precise agreement with an explicit symmetric orbifold computation of the eigenvalues of the matrix $C_{\Sigma_2}C_{\Sigma_2}^\dagger$ using the action of $\Sigma_2$ on chiral primaries as derived in \cite{Avery:2010hs,Carson:2014ena}.

\subsection{Connection to the LMRS BPS chaos discussion}
\label{sec:LMRSD1D5}

In \cite{Chen:2024oqv}, a closely related computation was discussed. In our language, they analyzed the chaotic property of an operator 
\begin{equation}\label{Ohat}
    \widehat{O} =P_{r=\tilde{r} = \frac{N}{2} + \frac{k}{2}}  (C + C^\dagger) P_{r=\tilde{r} = \frac{N}{2} + \frac{k}{2}} \,,
\end{equation}
where $C$ and $C^\dagger$ are the chiral ring coefficients, viewed as matrices, for the twist-$2$ (${1\over 2},{1\over 2}$) chiral/anti-chiral primaries, while $P_{r=\tilde{r} = \frac{N}{2} + \frac{k}{2}}$ is a further projection into a subspace of $1/2$-BPS states whose number of strands $k$ is related to the $R$-charges via $r=\tilde{r} = \frac{N}{2} + \frac{k}{2}$. Notice that the $R$-charges of a $1/2$-BPS state with $k$ strands always satisfies the inequality
\begin{equation}\label{Rchargeinequality}
 \frac{N}{2} - \frac{k}{2} \leq   r, \tilde{r} \leq \frac{N}{2} + \frac{k}{2} \,,
\end{equation}
since a single strand of length $w$ can have $r,\tilde{r} = \frac{w-1}{2}, \frac{w}{2}, \frac{w+1}{2}$. Therefore, $P_{r=\tilde{r} = \frac{N}{2} + \frac{k}{2}}$ projects into a subspace of states which saturates the upper end of the inequality in (\ref{Rchargeinequality}) (these are the ``all plus" states in the notation of \cite{Chen:2024oqv}). In \cite{Chen:2024oqv}, through a combination of analytic and numerical explorations, it was found that $\widehat{O}$ exhibits weak chaos, with a Thouless time scaling polynomially in $N$.

However, following the discussion in Section \ref{ccdagertt}, we can now understand that if we remove the projection $P_{r=\tilde{r} = \frac{N}{2} + \frac{k}{2}}$ in (\ref{Ohat}), there will be no chaos at all! This is because as we discussed earlier, $1/2$-BPS chiral primaries organize into $SU(2)$ representations in which $C, C^\dagger, [C,C^\dagger]$ (up to overall normalization) acts as the role of $J_+ , J_-, J_3$. Therefore, $C + C^\dagger$ acts as the $J_+ + J_- = J_1$ operator, which up to normalization factors must have exactly the same eigenvalues as $[C, C^\dagger]$. This is consistent with the weak chaos result in \cite{Chen:2024oqv} since the projection $P_{r=\tilde{r} = \frac{N}{2} + \frac{k}{2}}$ does not commute with $C+C^\dagger$, so the eigenvalues of $\widehat{O}$ are not directly related to those of $C+C^\dagger$. 

Note that due to the non-zero Berry curvature, states with different cycle structures do mix under marginal deformations. Therefore, it is unclear how one can extend the projection $P_{r=\tilde{r} = \frac{N}{2} + \frac{k}{2}}$ unambiguously away from the orbifold limit. On the other hand, the computation without this further projection is well-defined at arbitrary couplings, including the supergravity limit. 

One might ask why the further projection $P_{r=\tilde{r} = \frac{N}{2} + \frac{k}{2}}$ does not produce strong chaos. Indeed, a random further projection should produce a genuine random matrix. The answer to this question is presumably that the subspace with $r=\tilde{r} = \frac{N}{2} + \frac{k}{2}$ isn't just an arbitrary subspace in the full $1/2$-BPS, but rather retains certain simple structures as the original twist-$2$ operators. Most importantly, $\widehat{O}$ has the same ``banded" structure as $C+C^\dagger$ in a basis of states ordered by the number of strands, a key property that was used to provide a bound for chaos in \cite{Chen:2024oqv}. It would be  satisfying if one can combine the information from $P_{r=\tilde{r} = \frac{N}{2} + \frac{k}{2}}$ and $C+C^\dagger$ to explain the features of the spectrum of $\widehat{O}$ observed in \cite{Chen:2024oqv}, such as the clustering of eigenvalues, but we have not managed to do so. 

For other choices of simple probe operators, such as chiral primaries which are heavier, one does not expect similar simplifications to occur for their associated chiral ring coefficients. In particular, the proof in Appendix \ref{sec:CRCidentity} would no longer hold. We still expect that generically (such as by taking simple linear combinations of heavier chiral primaries), the LMRS observables should exhibit weak chaos.

%%%%%%%%%%%
% SECTION %
%%%%%%%%%%%
\subsection{Some comments on 1/4-BPS states}
\label{sec:D1/D5_1/4BPS}

Recall that the 1/4-BPS states are chiral primaries for the right- (or the left-) moving sector, while the state on the other sector can be anything \eqref{eqn:D1/D5_1/4BPS}. Observe that if we try to compute the Berry curvature using the same method as for the 1/2-BPS states in Section \ref{sec:D1/D5_1/2BPS}, it does not simplify to the form in \eqref{eqn:D1D5_Fsimplified} because $G_{-\frac{1}{2}}^{+C}$ no longer annihilates the ket and $G_{+\frac{1}{2}}^{-A}$ no longer annihilates the bra. Similarly, if one tried to use the double-integrated four-point function in \eqref{berry4point} instead, as was done in \cite{deBoer:2008ss}, one finds that superconformal Ward identities only localize one of the two integrals so one is still left with an integrated correlator. Nevertheless, we now consider some simple examples of 1/4-BPS states where one can compute the curvature.

Starting with a 1/2-BPS state, the simplest way to get a 1/4-BPS state is to act with the superconformal generators on the left-moving part of the state. Given that the 1/2-BPS states do not lift, the superconformal algebra implies that such states also cannot lift. Known as singletons, these states have the simple bulk interpretation of adding excitations of boundary gravitons, $SU(2)$ Chern-Simons gauge fields, and their superpartners on top of the LM geometry. A more sophisticated class of 1/4-BPS states can be obtained, at the orbifold point, by acting with the generators $\{L_{-1},G_{-\frac{1}{2}}^{-A},J_{0}^{-}\}$ on each strand individually. Such states could in principle lift as one moves away from the orbifold point, but some do not because they contribute to the index \cite{Shigemori:2019orj}. It has been conjectured that at the supergravity point, they are dual to smooth, horizonless supergravity solutions known as superstrata \cite{Bena:2015bea, Shigemori:2020yuo}.\footnote{The superstrata are actually more general than those discussed here. A set of such geometries dual to fractional modes of single strand generators have also been constructed \cite{Bena:2016agb}.} All of the aforementioned states have been shown to be monotones \cite{Hughes:2025car, Chang:2025rqy}.

We have not been able to obtain a simple result for the Berry curvature for all of these 1/4-BPS monotones, but we did succeed in computing it for the subset obtained by acting with arbitrarily many $SU(2)$ current modes either on the full state (singletons) or on a strand (superstrata), and for a single supercharge mode or Virasoro mode. We will see that the curvature for these states is the same or closely related to that of the underlying 1/2-BPS states from which they are constructed, akin to what happens in $\mathcal{N}=4$ SYM \cite{Niarchos:2021iax}. Note that we will only compute the matrix elements of the curvature between different monotones, but they could in principle also mix with the fortuitous states (see Section \ref{sec:disc} for more comments on this). 

Consider a 1/2-BPS state $\ket{K}$, which is also a current primary (see Footnote \ref{foo:currentprimary}), with $R$-charges $(r_{k},\tilde{r}_{k})$ in any $\mathcal{N}=(4,4)$ SCFT theory, from which we obtain a family of 1/4-BPS states
\begin{equation}\label{eqn:D1/D51/4-BPScurrentdesc}
    \ket{K;\{n_{m,a}\}} = \frac{1}{\sqrt{\mathcal{N}_{\{n_{m,a}\}}}}\prod_{a=\pm,3}\prod_{m=0}^{\infty}(J_{-m}^{a})^{n_{m,a}}\ket{K} \,.
\end{equation}
where $\mathcal{N}_{\{n_{m,a}\}}$ is a normalization factor whose precise form is unimportant for us (these correspond to the singletons). These states have conformal dimension and $R$-charge
\begin{equation}
    h=r_{k}+\sum_{a=\pm,3}\sum_{m=0}^{\infty}n_{m,a}m \,, \qquad r = r_{k}+\sum_{m=0}^{\infty}\left(n_{m,+}-n_{m,-}\right) \,,
\end{equation}
For fixed $(h,r)$, we form an orthonormal basis $\{\ket{K;(h,r)_{1}},\ldots,\ket{K;(h,r)_{D}}\}$ which will be linear combinations of the ``elemental" states in \eqref{eqn:D1/D51/4-BPScurrentdesc}. To compute the Berry curvature of these orthonormal basis states, we use the same trick as for the 1/2-BPS states by considering the $x \to 0$ limit of the following quantity
\begin{equation}\label{eqn:D1/D51/4-BPS_curv}
    (F_{\mu\nu})_{(K;\mathfrak{a})(L;\mathfrak{b})}(x) = \bra{L;(h,r)_{\mathfrak{b}}}\partial_{\mu}H\frac{i}{(H-R-(h-r)-x)^{2}}\partial_{\nu}H\ket{K;(h,r)_{\mathfrak{a}}} - (\mu \leftrightarrow \nu) \,.
\end{equation}
Note that the marginal operators $\partial_{\mu}H$ are not only neutral under $SU(2)$, but in fact have a regular OPE with all of the $SU(2)$ currents (see Appendix G of \cite{deBoer:2008ss}), implying that $[J_{n}^{a},\partial_{\mu}H] = 0$ for all $n \in \mathbb{Z}$. We also need the commutators $[J_{n}^{\pm},H-R] = (n \pm 1)J_{n}^{\pm}$ and $[J_{n}^{3},H-R] = nJ_{n}^{3}$ obtained from \eqref{eqn:smallN=(4,4)SCA}, along with the following fact about commutators: if $[A,B] = pA$ for some $p\in\mathbb{C}$ then $AB^{m}=(B+p)^{m}A$, easily proven by induction. Now \eqref{eqn:D1/D51/4-BPS_curv} can be simplified by considering the case where the bra and ket are ``elemental" states. We can move all of the $(J_{m}^{a})^{n}$ from the bra to the right, through the denominator which shifts it to $(H-R-x)^{-2}$, and finally arriving at the ket. Combining all of the contributions from the different ``elemental" states appearing in $\ket{K;(h,r)_{\mathfrak{a}}}$ and $\ket{L;(h,r)_{\mathfrak{b}}}$ one simply obtains orthonormality (with the normalization factor $\mathcal{N}_{\{n_{m,a}\}}$ cancelling), so we find
\begin{equation}
    (F_{\mu\nu})_{(K;\mathfrak{a})(L;\mathfrak{b})}(x) = \delta_{\mathfrak{a}\mathfrak{b}}(F_{\mu\nu})_{KL}(x)\,.
\end{equation}
Taking $x \to 0$ gives the same curvature as the 1/2-BPS state. 

This general argument immediately applies to the singleton states. For the superstrata states, since the $SU(2)$ currents only act on individual strands, it is not obvious that $[(J_{0}^{-})_{\mathrm{strand}},\partial_{\mu}H] = 0$, and indeed this is only true when $\partial_{\mu}H$ is one of the untwisted marginal deformations. Thus, the argument only holds for untwisted deformations of superstrata states, and not for the twist-$2$ deformation that moves us towards strong coupling.

Next, consider the 1/4-BPS states obtained by acting once with any of the superconformal generators $G_{-r}^{-A}$ or $L_{-n}$ on a 1/2-BPS state. We show in Appendix \ref{sec:desccurv} that the curvature of these states is the sum of the curvature of the symmetry current bundle plus the curvature of the 1/2-BPS state. It is demonstrated there that the curvature of the stress-tensor bundle vanishes and the only symmetry current whose bundle has non-trivial curvature is the supercharge \cite{Gomis:2016sab}, but it is only two-dimensional.\footnote{One can also consider states obtained by acting with a single $J_{-m}^{a}$, which is a special case of what was done above. In this case, we show in Appendix \ref{sec:desccurv} that the curvature of the $SU(2)$ $R$-symmetry bundle is vanishing, which is consistent with what we found in the more general case, namely the curvature is the same as for the chiral primary.} Therefore, the eigenvalues of the full curvature are equal to the eigenvalues of $(F_{\mu\nu})_{KL}$ plus one of the two eigenvalues of this two-dimensional supercharge curvature, which cannot have random matrix statistics since $(F_{\mu\nu})_{KL}$ does not.\footnote{More explicitly, let $F^{(s/p)}$ denote the curvature of the supercharge/chiral primary. The full curvature is $F = F^{(s)} \otimes \mathds{1}^{(p)}+\mathds{1}^{(s)} \otimes F^{(p)}$ whose eigenvalues are $\{\lambda^{(s)}+\lambda^{(p)}\,|\,\lambda^{(s/p)} = \mathrm{eigenvalue\;of\;}F^{(s/p)}\}$.} Again, this argument applies directly for the singleton states, however, for the superstrata it does not apply. The reason is that superstrata states are not the product of a symmetry operator with a 1/2-BPS state. For example, in the $\mathrm{Sym}^{2}(T^{4})$ theory the $S_{2}$-invariant operator $G_{-\frac{1}{2}}^{-(1)}\Phi \otimes \chi + \chi \otimes G_{-\frac{1}{2}}^{-(2)}\Phi$, with $\Phi$ and $\chi$ chiral primaries on a single copy, is not a product of a symmetry generator and a 1/2-BPS operator.

%%%%%%%%%%%
% SECTION %
%%%%%%%%%%%
\section{Monotones in $\mathcal{N}=4$ SYM}
\label{sec:SYM}

In this section, we consider the Berry curvature of short multiplets in the ${\cal N}=4$ SYM on its conformal manifold, parametrized by the Yang-Mills coupling $g_{\rm YM}$ and the $\theta$ angle. As usual, we consider the complexified coupling
\be
\tau={\theta \over 2\pi} + i {4\pi \over g_{\rm YM}^2}\,.
\ee
The Berry curvature is given by the general formula
\be
\label{integralcurv}
(F_{\mu\nu})_{KL} = \frac{i}{(2\pi)^4} \int_{|x|<1} d^4x \int_{|y|<1} d^4y\,\,\langle \phi_L^-(\infty) \, {\cal O}_{[\mu} (x) \, {\cal O}_{\nu]} (y) \, \phi_K^+(0)\rangle \,,
\ee
where ${\cal O}_{\mu,\nu}$ are the marginal operators and $\phi^+/\phi^-$ are $SU(4)$ highest/lowest weight states of the corresponding short multiplets labeled by $K,L$. Brackets denote anti-symmetrization and one has to be careful about regularization issues, as discussed in Section \ref{sec:regularization}. As explained there, we use interchangeably the notion of local operators and states on $S^3\times {\mathbb R}$.

Since the conformal manifold is 1-complex dimensional we have $F_{\tau \tau}=F_{\overline{\tau}\overline{\tau}}=0$, which can be seen from the vanishing of the anti-symmetrized correlator in \eqref{integralcurv} and the Berry curvature is captured by $(F_{\tau \overline{\tau}})_{KL}$.

It is known \cite{Baggio:2014ioa}, \cite{Niarchos:2018mvl} that the Berry curvature for 1/2-BPS states in ${\cal N}=4$ SYM vanishes. We will show that this is also true for 1/4-BPS states. This result follows from superconformal Ward identities and is exact for all values of $\tau$ and any $N$. For 1/8- and 1/16-BPS states the superconformal Ward identities are not powerful enough and in order to compute the Berry curvature one would need to know the full 4-point function in the integrand of \eqref{integralcurv}. At large $N$, the heavy 1/2-BPS, 1/4-BPS, and 1/8-BPS states correspond to horizonless geometries of the bubbling AdS type \cite{Lin:2004nb,Chen:2007du}, while the heavy 1/16-BPS states correspond also to black holes. When we are in the black-hole regime, we expect from the bulk that 1/16-BPS states will exhibit strongly chaotic Berry curvature. 

Crucial to the vanishing of the Berry curvature for 1/2- and 1/4-BPS states are the enhanced symmetries of ${\cal N}=4$ SYM. For example, in 4d ${\cal N}=2$ SCFTs the Berry curvature of Coulomb-branch chiral primaries is non-trivial \cite{Baggio:2014ioa}.\footnote{The Higgs-branch chiral primaries in 4d ${\cal N}=2$ SCFTs have vanishing Berry curvature \cite{Niarchos:2018mvl}.}

\subsection{Introduction and review of BPS multiplets} 
\label{sec:SYM_review}

We start with 1/2-BPS operators in the ${\cal N}=4$ SYM. The superconformal primaries of the multiplet are Lorentz scalars in the $[0,p,0]$ representation of $SU(4)$ and have dimension $\Delta=p$. In general we denote the $SU(4)$ highest weight vector of the representation by $\phi^+$ and the lowest one by $\phi^-$. If we define $Z=\Phi^1 + i \Phi^2$, we can think of $\phi^+$ as being generally a  multi-trace operator, which contains in total $p$ copies of the letter $Z$.

The highest weight vector of a 1/2-BPS state is annihilated by the following set of supercharges\footnote{We are using notation $Q_{i,a}$, where $i=1,\ldots,4$ is $SU(4)$ index and $a=1,2$ the Lorentz spinor index.}
\be
\label{halfp}
[Q_{1,a}, \phi^+] = [Q_{2,a},\phi^+] = [\overline{Q}_{3,{\dot{a}}},\phi^+] = [\overline{Q}_{4,\dot{a}},\phi^+]=0 \,,
\ee
and the lowest weight vector by
\be
\label{halfm}
[Q_{3,a}, \phi^-] = [Q_{4,a},\phi^-] = [\overline{Q}_{1,{\dot{a}}},\phi^-] = [\overline{Q}_{2,\dot{a}},\phi^-]=0 \,.
\ee
These equations hold for all values of the Lorentz spinor indices $a=1,2,\dot{a}=1,2$.

Then we consider  1/4-BPS multiplets. The corresponding superconformal primaries are Lorentz scalars in the $SU(4)$ representation $[k,p,k]$ with dimension $\Delta=p+2k$. At weak coupling they can be constructed out of two scalars $Z$, $X$, where $X=\Phi^3+i\Phi^4$. In the free theory, any operator composed out of only $X$ and $Z$ is 1/4-BPS, but as we turn on the coupling, most of them are lifted, and it can be viewed as level crossing that happens at zero coupling. The operators that remain 1/4-BPS at generic coupling are heuristically those that do not contain any term of the form $[X,Z]$. Despite this level crossing, we will see below that the Berry curvature still vanishes. The highest weight vector of a 1/4-BPS multiplet is annihilated by
\be
\label{quartp}
[Q_{1,a},\phi^+]=[\overline{Q}_{4,\dot{a}},\phi^+]=0 \,,
\ee
and the lowest weight by
\be
\label{quartm}
[Q_{4,a},\phi^-]= [\overline{Q}_{1,\dot{a}},\phi^-]=0 \,.
\ee

We will not discuss the structure of 1/8-BPS or 1/16-BPS multiplets of the ${\cal N}=4$ theory, since we are not able to compute their Berry curvature without having knowledge of the full 4-point function \eqref{integralcurv} at general values of the coupling, as explained in Section \ref{sec:1/8and1/16}. 

Finally, we describe the marginal operators of the theory. We have two ${\cal N}=4$ preserving marginal operators, corresponding to changing $g_{\rm YM}$ and $\theta$. We combine them into (anti)-holomorphic combinations ${\cal O}_\tau$, ${\cal O}_\tau^\dagger$. These are scalar operators with $\Delta=4$ which are neutral under $SU(4)$. They are contained in the $[0,2,0]$ supermultiplet, i.e. they are super-descendants of ${\rm Tr}(Z^2)$.

Thinking in 4d ${\cal N}=2$ language one would write
\be
\label{marginaln41}
{\cal O}_\tau = {1\over 4} \{Q_{4,1},[Q_{4,2},\{Q_{3,1},[Q_{3,2},{\rm Tr}(Z^2)]\}]\} \,,
\ee
\be
\label{marginaln42}
{\cal O}_\tau^\dagger = {1\over 4}\{\overline{Q}_{4,1}, [\overline{Q}_{4,2}, \{\overline{Q}_{3,1}, [\overline{Q}_{3,2} ,{\rm \Tr}(\overline{Z}^2)]\}]\} \,,
\ee
In ${\cal N}=2$ language ${\cal O}_{\tau}$ would be a descendant of a  chiral primary and ${\cal O}_\tau^\dagger$ a descendant of an anti-chiral primary.
However, due to the enhanced ${\cal N}=4$ symmetry, it is actually possible to write both marginal operators as descendants of the highest weight state ${\rm Tr}(Z^2)$ or the lowest weight state ${\rm Tr}(\overline{Z}^2)$. This follows from the underlying $SU(4)$ R-symmetry which can rotate ${\rm Tr}(Z^2)$ to ${\rm Tr}(\overline{Z}^2)$.\footnote{This property is crucial in proving various non-renormalization theorems for BPS correlators in this theory \cite{Baggio:2012rr} and is the analogue of similar properties in ${\cal N}=(4,4)$ SCFTs in 2d, as discussed in Section \ref{sec:D1/D5}.}
For example, we can represent the marginal operators as
\be
\label{eq:marginal O Q_1}
{\cal O}_\tau ={1\over 4} \{Q_{1,1},[Q_{1,2},\{Q_{2,1},[Q_{2,2},{\rm Tr}(\overline{Z}^2)]\}]\} \,,
\ee
\be
{\cal O}_\tau^\dagger = {1\over 4}\{\overline{Q}_{4,1}, [\overline{Q}_{4,2}, \{\overline{Q}_{3,1}, [\overline{Q}_{3,2} ,{\rm \Tr}(\overline{Z}^2)]\}]\} \,,
\ee
so now they are both written as descendants of ${\rm Tr}(\overline{Z}^2)$. Similarly, we can write both as descendants of ${\rm Tr}(Z^2)$.

 In what follows we will examine the curvature of superconformal primaries of 1/2- and 1/4-BPS multiplets. Notice that the curvature of conformal primaries which are super-descendants is given as the sum of the curvature of the superconformal primary in the same short supermultiplet plus curvature contributions from the supercharges acting on it. The Berry curvature of the supercharges is characterized by a $U(1)$ bundle over the conformal manifold, corresponding to a phase rotations automorphism  of the form $Q_{i,a} \rightarrow e^{i \theta} Q_{i,a}$ and $\overline{Q}_{i,\dot{a}}\rightarrow e^{-i\theta} \overline{Q}_{i,\dot{a}}$. The curvature of this bundle is proportional to the K\"ahler form of the Zamolodchikov metric \cite{Papadodimas:2009eu, Niarchos:2021iax}.

\subsection{Vanishing of curvature for 1/2-BPS states}
\label{sec:SYM_1/2BPS}

We now review the vanishing of the Berry curvature of 1/2-BPS operators in ${\cal N}=4$ SYM, as computed from \eqref{integralcurv} using superconformal Ward identities. We only present the results and refer the reader to original references for more details. We start with a more general case, that of computing the Berry curvature of chiral primaries (Coulomb branch operators) in 4d ${\cal N}=2$ SCFTs, whose R-symmetry is $SU(2)\times U(1)$. This was computed in \cite{Papadodimas:2009eu}, resulting in a 4d analogue of the $\tt$ equations.
\be
\label{maintt4d}
(F_{\tau \overline{\tau}})_{KL}=i\left[\delta_{KL}\left(1+{p\over 2c}\right)-[C_2, C_2^\dagger]_{KL} \right] \,,
\ee
where $K,L$ denote chiral primary states and $p$ is their $U(1)$ R-charge.\footnote{We write the formula in conventions where the ${\cal N}=2$ Coulomb branch BPS bound is $\Delta=p$.} This formula also applies to 1/2-BPS operators in ${\cal N}=4$ SYM, where $p$ labels the $SU(4)$ R-charge $[0,p,0]$.\footnote{\label{foo:1/2BPS_normalization}Here we assume that we are working in an orthonormal basis of the states $K,L$. In a more general basis the $\delta_{KL}$ would have to be replaced by the matrix of 2-point functions $g_{KL}$.} Here $(C_2)_{K}^M$ denotes the chiral ring OPE coefficient of ${\rm Tr}(Z^2)$ acting on $\phi_K$, ${\rm Tr}(Z^2) \, \phi_K \sim \phi_M$, and  $C_2^{\dagger}$ the conjugate. Finally $c$ is the usual $c$-anomaly coefficient of the theory.  

The vanishing of \eqref{maintt4d} for 1/2-BPS states in ${\cal N}=4$ SYM was shown in \cite{Baggio:2014ioa}. An alternative proof was given in \cite{Niarchos:2018mvl}. In Appendix \ref{app:4dcrossing}, we give yet another proof, by providing an elementary argument based on crossing symmetry that 
\be
\label{identitychiral4d}
[C_2, C_2^\dagger]_{KL} = \delta_{KL}\left(1+{p\over 2c}\right) \,,
\ee
for all values of the coupling, which does not require making use of the non-renormalization theorem of chiral primary 3-point functions.

\paragraph{Comments on eigenvalues of $[C_2,C_2^\dagger]$ for 1/2-BPS.} We can also perform an explicit computation to verify (\ref{identitychiral4d}) and therefore the vanishing of the Berry curvature in the $\mathcal{N}=4$ SYM theory. The computation can be generalized to compute properties of other structure constants of 1/2-BPS operators, and not necessarily $\Tr(Z^2)$.

For simplicity, let's first consider the case where the gauge group is $U(N)$, then we will describe the modifications needed for the $SU(N)$ case. The 1/2-BPS sector has a description in terms of $N$ fermions in a harmonic potential \cite{Berenstein:2004kk}. The creation and annihilation operators for each level of the harmonic potential are given by 
\begin{equation}
    \{c_n , c_m^\dagger\} = \delta_{nm}, \quad \{c_n , c_m\} = \{c_n^\dagger, c_m^\dagger \} = 0,\quad n,m = 1,\ldots,\infty \,.
\end{equation}
After projecting into the 1/2-BPS sector, we have \cite{Takayama:2005yq}
\begin{equation}\label{Z2Zbar2}
\begin{aligned}
  &  \widehat{ \textrm{Tr}(Z^2)}  \quad \rightarrow \quad \frac{\alpha}{2} \sum_{\ell = 0}^\infty \sqrt{(\ell+1) (\ell+2)} c_{\ell+2}^\dagger c_{\ell}   \\
  & \widehat{\textrm{Tr}(\bar{Z}^2 )} \quad \rightarrow \quad \frac{\alpha}{2}  \sum_{\ell = 0}^\infty \sqrt{(\ell+1) (\ell+2)} c_{\ell}^\dagger c_{\ell+2} \,,
\end{aligned}
\end{equation}
where $\alpha$ is a normalization constant we will fix later.  The two operators in (\ref{Z2Zbar2}) are $C_2^\dagger$ and $C_2$ in (\ref{identitychiral4d}), respectively. It is straightforward to find that 
\begin{equation}\label{commZ2}
    [ \widehat{\textrm{Tr}(\bar{Z}^2)} , \widehat{\textrm{Tr}(Z^2)} ] = \alpha^2 \sum_{\ell=0}^\infty \left(\ell + \frac{1}{2} \right) c_{\ell}^\dagger c_{\ell} = \alpha^2 \left( \frac{N^2}{2} + p \right) \,,
\end{equation}
which is indeed an identity operator in the sector with fixed $R$-charge $p$. 

In the normalization described in Footnote \ref{foo:1/2BPS_normalization}, the left-hand side of (\ref{commZ2}) evaluated in the vacuum state should be equal to one, so we find $\alpha = \frac{\sqrt{2}}{N}$
and therefore
\begin{equation}
      [ \widehat{\textrm{Tr}(\bar{Z}^2)} , \widehat{\textrm{Tr}(Z^2)} ] = 1 + \frac{p}{N^2/2} =  1+ \frac{p}{2c_{U(N)}} \,,
\end{equation}
where $c_{U(N)} =\frac{N^2}{4}$. Therefore we have recovered (\ref{identitychiral4d}).

In the $SU(N)$ theory the complex scalar is traceless, so we should work with
\begin{equation}
    \widetilde Z \equiv Z - \frac{1}{N}\textrm{Tr}(Z) \,,
\end{equation}
such that $\textrm{Tr}(\widetilde Z ) = 0$. In the fermion language we should also project out the $U(1)$ mode. This can be done by introducing
\begin{equation}
    J_1 \equiv \sum_{\ell=0}^{\infty}\sqrt{\ell+1}\,c_{\ell+1}^\dagger c_\ell,
    \qquad
    J_1^\dagger \equiv \sum_{\ell=0}^{\infty}\sqrt{\ell+1}\,c_{\ell}^\dagger c_{\ell+1} \,,
\end{equation}
which represent $\textrm{Tr} (Z), \textrm{Tr} (\bar{Z})$ (up to an overall normalization) and satisfies $ [J_1^\dagger,J_1] = N$.
In the Hilbert space of $N$ fermions, the $SU(N)$ 1/2-BPS Hilbert space is obtained by projecting to the $U(1)$ vacuum, i.e. imposing
\begin{equation}\label{fermionphysical}
    J_1^\dagger \ket{\psi} =0\,.
\end{equation}
Now, for the projected operators $\widehat{\textrm{Tr}(\tilde{Z}^2)}$ and $\widehat{\textrm{Tr}(\tilde{\bar{Z}}^2)}$, the maps into the fermion language are given by
\begin{equation}
\begin{aligned}
   \widehat{ \textrm{Tr}(\tilde{Z}^2) }=  \widehat{\textrm{Tr}(Z^2)} - \frac{1}{N}\widehat{\textrm{Tr}(Z)}^2  \quad & \rightarrow \quad \tilde{\alpha} \left( \frac{1}{2} \sum_{\ell = 0}^\infty \sqrt{(\ell+1) (\ell+2)} c_{\ell+2}^\dagger c_{\ell}  - \frac{1}{2N} J_1^2\right) \\
     \widehat{ \textrm{Tr}(\tilde{\bar{Z}}^2)} = \widehat{ \textrm{Tr}(\bar{Z}^2)} - \frac{1}{N}\widehat{\textrm{Tr}(\bar{Z})}^2  \quad & \rightarrow \quad \tilde{\alpha} \left( \frac{1}{2} \sum_{\ell = 0}^\infty \sqrt{(\ell+1) (\ell+2)} c_{\ell}^\dagger c_{\ell+2}  - \frac{1}{2N} J_1^{\dagger 2}\right) \,.
\end{aligned}
\end{equation}
One can check that they indeed have the correct property of preserving (\ref{fermionphysical}). Again, we have introduced an overall normalization factor $\tilde{\alpha}$ to be fixed later. Now, in terms of the fermion operators it is straightforward to repeat the analysis as in the $U(N)$ case, where one finds
\begin{equation}
      [ \widehat{\textrm{Tr}(\tilde{\bar{Z}}^2)} , \widehat{\textrm{Tr}(\tilde{Z}^2)} ] =  1+ \frac{p}{2c_{SU(N)}} \,,
\end{equation}
where $c_{SU(N)} = \frac{N^2-1}{4}$.

The explicit computation above can be generalized to cases of $\textrm{Tr}(Z^k), k>2$ easily, though in such cases the commutator is no longer an identity operator in a sector with fixed $R$-charge and degeneracies are generically broken. In fact, by considering operators that are linear combinations of different 1/2-BPS operators with the same $R$-charge, the spectrum of the commutator could exhibit weak chaos, in a similar fashion as in \cite{Chen:2024oqv}. Note however, in these other cases it is not clear whether $[C,C^\dagger]$ is related to a notion of curvature.

\subsection{Vanishing of curvature for 1/4-BPS states}
\label{sec:curv14}

We then move on to 1/4-BPS states and consider the mixed component $F_{\tau \overline{\tau}}$ of the curvature. For that we consider the 4-point function
\be
\label{again4pointf}
\langle \phi_L^-(\infty) \,\,{\cal O}_{\tau} (x)\,\, {\cal O}_\tau^\dagger(y)\,\, \phi_K^+(0)\rangle \,,
\ee
where $\phi_{K,L}$ are 1/4-BPS multiplets. We take advantage of the fact  that we are dealing with a marginal operator in a ${\cal N}=4$ theory and, as discussed above, we  represent the marginal operators as
\be
{\cal O}_\tau = {1\over 4}\{Q_{1,1},[Q_{1,2},\{Q_{2,1},[Q_{2,2},{\rm Tr}(\overline{Z}^2)]\}]\}
\ee
\be
{\cal O}_\tau^\dagger = {1\over 4} \{\overline{Q}_{4,1}, [\overline{Q}_{4,2}, \{\overline{Q}_{3,1}, [\overline{Q}_{3,2} ,{\rm \Tr}(\overline{Z}^2)]\}]\} \,.
\ee
We go back to the 4-point function \eqref{again4pointf} and we exploit the fact that the 1/4-BPS operator $\phi_K^+(0)$ is annihilated by both $Q_{1,a}$ and $\overline{Q}_{4,\dot{a}}$ \eqref{quartp} to move those supercharges around using Ward identities. We do not get any contribution from $\phi_L^-(\infty)$ since it is located at infinity, see for example \cite{Papadodimas:2009eu}. Then \eqref{again4pointf} is proportional to
\be
\big\langle \phi_L^-(\infty)\, \Big(\!\{\overline{Q}_{4,1}, [\overline{Q}_{4,2}, \{Q_{2,1}, [Q_{2,2} ,{\rm \Tr}(\overline{Z}^2)]\}]\}\!\Big)\!(x)  \,
\Big(\!\{Q_{1,1}, [Q_{1,2}, \{\overline{Q}_{3,1}, [\overline{Q}_{3,2} ,{\rm \Tr}(\overline{Z}^2)]\}]\}\!\Big)\!(y) \, \phi_K^+(0)\big\rangle \,.
\ee

Now, each of the two operators located at $x$ and $y$ in the expression above have the property that they are i) Lorentz scalars ii) have $\Delta=4$ and have equal number of $Q$'s and $\overline{Q}$'s. Examining the structure of the $[0,2,0]$ multiplet, one can show that
\be
\label{identity2}
\{Q_{1,1}, [Q_{1,2}, \{\overline{Q}_{3,1}, [\overline{Q}_{3,2} ,{\rm \Tr}(\overline{Z}^2)]\}]\}(y)  = 4\, \Box_y {\rm Tr}(Y^2) \,,
\ee
where $Y= \Phi^5+i \Phi^6$. The relation above is true at any value of the coupling: by ${\rm Tr}(Z^2)$ we denote the highest weight state of the $[0,2,0]$ multiplet and similarly by ${\rm Tr}(Y^2)$ another element in the same $[0,2,0]$ representation\footnote{In the free theory we have the three complex scalars $Z=\Phi^1+i \Phi^2\,,\,X=\Phi^3+i \Phi^4\,,\,Y=\Phi^5+i \Phi^6$. Then starting with the highest weight state $Z$ of $[0,1,0]$ we get the rest of the states as $E_2^\dagger Z = X$, $E_1^\dagger X= \overline{Y}$, $E_3^\dagger X = Y, E_1^\dagger Y = -\overline{X}, E_3^\dagger \overline{Y} = -\overline{X}, E_2^\dagger \overline{X} = -\overline{Z}.$} which can be written as
\be
\label{lambdaresult}
{\rm Tr}(Y^2) =\kappa [E_3^\dagger,[E_2^\dagger,[E_3^\dagger,[E_2^\dagger,{\rm Tr}(Z^2)]]]] \,,
\ee
where $E_i$ are the simple positive roots of $SU(4)$ and the constant $\kappa$ is selected so that the 2-point function of ${\rm Tr(Y^2)}$ is equal to that of ${\rm Tr(Z^2)}$.
Similarly, we find
\be
\label{identity1}
\{\overline{Q}_{4,1}, [\overline{Q}_{4,2}, \{Q_{2,1}, [Q_{2,2} ,{\rm \Tr}(\overline{Z}^2)]\}]\}(x) =4
\,\Box_x {\rm Tr}(\overline{Y}^2) (x) \,.
\ee

All in all we have
\be
\label{finalder}
\langle \phi_L^-(\infty) \,{\cal O}_{\tau} (x) \,{\cal O}_\tau^\dagger(y) \,\phi_K^+(0)\rangle\,\, = \,\,\Box_x \Box_y \,\, \langle \phi_L^-(\infty) \,\,{\rm Tr}(\overline{Y}^2)(x) \,\,{\rm Tr}(Y^2)(y) \,\,\phi_K^+(0)\rangle \,.
\ee
Then the Berry curvature \eqref{integralcurv} takes the form
\be
\label{integralcurv2}
(F_{\tau \overline{\tau}})_{KL} = {i\over (2\pi)^4}\int_{|x|<1} d^4x \int_{|y|<1} d^4y\,\,\Box_x \Box_y \,\, \langle \phi_L^-(\infty) \,\,{\rm Tr}(\overline{Y}^2)(x) \,\,{\rm Tr}(Y^2)(y) \,\,\phi_K^+(0)\rangle  - \left(Y \leftrightarrow \overline{Y}\right)\,.
\ee
Now it is clear that the two Laplacians reduce the double integral to boundary terms, for which we need to analyze the OPE expansion of the correlator in the corresponding limits. First, by examining the $x\rightarrow y$ OPE for $y$ away from the boundary, we notice that there is no non-integrable singularity after anti-symmetrization, and therefore we do not get any boundary terms from the diagonal $x=y$. We then have two types of boundary terms, one as $x,y\rightarrow 0$ and another when $|x|,|y|\rightarrow 1$. The analysis is somewhat technical and we present it in appendices \ref{app:n4bd1} and \ref{app:n4bd2}, respectively. There we show, using the structure of the OPE between 1/2- and 1/4-BPS multiplets, that both boundary terms vanish. Hence we find that for  1/4-BPS operators the Berry curvature vanishes
\be
(F_{\tau \overline{\tau}})_{KL}=0 \,.
\ee

\subsection{Comments on 1/8- and 1/16-BPS operators}
\label{sec:1/8and1/16}

For 1/8-BPS and 1/16-BPS operators it is generally not possible to use superconformal Ward identities to rewrite the 4-point function entering in \eqref{integralcurv} as a total derivative which would reduce the corresponding integrals to boundary terms. In the previous cases, this was a crucial step that allowed us to concentrate on the leading terms in the OPE of the operators in the correlator. Without this step, the naive expectation is that all operators in the OPE can contribute; hence, in general, one needs to know the full 4-point function to compute the Berry curvature, or to introduce a new technique. 

For 1/8-BPS operators annihilated by $Q_{1,a}$, $a=1,2$, it is possible to partially simplify the formula so that a Laplacian with respect to $x$ appears.\footnote{There are three different 1/8-BPS sectors in $\mathcal{N}=4$ SYM and here we are considering the $\mathfrak{su}(2|3)$ sector \cite{Chang:2023ywj}.} One starts by representing the marginal operators as in \eqref{eq:marginal O Q_1} and its Hermitian conjugate. Using standard superconformal Ward identities one can move $\overline{Q}_{1,1}$ and $\overline{Q}_{1,2}$ away from $y$ and onto $x$ (exploiting that they annihilate $\phi_K^+(0)$ and that $\phi_L^-$ does not contribute since it is located at infinity), so \eqref{again4pointf} is equal to 
\begin{equation}
\Box_x\,\, \langle \phi_L^-(\infty) \; \Big(\{Q_{2,1},[Q_{2,2},{\rm Tr}(\overline{Z}^2)]\}\Big)(x)  \; \Big( \{\overline{Q}_{2,1}, [\overline{Q}_{2,2} ,{\rm \Tr}(Z^2)]\}\Big)(y) \; \phi_K^+(0)\rangle \,.
\end{equation}
Because of the Laplacian of $x$, we can reduce one of the two integrations in the computation of the curvature \eqref{integralcurv} to boundary terms, and the two operators in the correlator will have dimension $3$. For 1/16-BPS operators $\phi^+_K$ the picture is even less promising. The operators $\phi^+_K(0)$ preserves only $Q_{1,1}$. Moving that supercharge from one operator in the correlator to the other will bring out an $x$-derivative along a specific direction, instead of a Laplacian, with the remaining operators being fermionic ones of dimension $\frac72$. These operations remove one of the integrals, but the remaining integral over $y$ cannot be simplified further in any obvious way, and seems to require knowledge of the full 4-point function.

Nevertheless, the expectation is that the Berry curvature would behave differently for these two classes. Like their more supersymmetric counterparts, 1/8-BPS operators correspond to horizonless geometries, rather than to black holes \cite{Chen:2007du, Chang:2023ywj}. On the field theory side, they can be described in terms of coherent states of commuting matrices, much like the 1/4-BPS operators \cite{Berenstein:2022srd}, and according to the arguments in section 4.4 of \cite{Chen:2024oqv} we would expect correlation functions involving them, in particular the Berry curvature, to exhibit at most weak chaos.

The 1/16-BPS operators, on the other hand, describe also black hole microstates in the theory, and no coherent state picture applies for the vast majority of them. We expect the Berry curvature to exhibit strong chaos in their case, even at weak coupling. One reason for such an expectation is the domination of the Schwarzian mode over other stringy corrections at low temperature \cite{Maldacena:2016hyu}. We could try to use holography in order to compute \eqref{integralcurv}, but as we do not have access to matrix elements between individual black hole microstates from the gravity side, we could at most try to compute some statistical properties of the Berry curvature, such as $\Tr((F_{\tau\bar\tau})^k)$, by computing Euclidean correlation functions in a supersymmetric black hole background. They involve projections to the BPS sector, i.e to the lowest energy states in a given charge sector, and so when computed holographically they should be sensitive to the near horizon dynamics. But the near horizon region of these black hole backgrounds is described by $\mathcal{N}=2$ super-JT gravity \cite{Boruch:2022tno}. In Section~\ref{sec:JTgravity} we compute the Berry curvature associated with certain deformations of JT gravity\footnote{These deformations differ slightly from the ones computed here, as they come from a natural deformation in the dual quantum mechanical system.} and show it is strongly chaotic.

%%%%%%%%%%%
% SECTION %
%%%%%%%%%%%
\section{$\mathcal{N}=2$ SYK: chaotic Berry curvature for fortuitous states}
\label{sec:SYK}

In Section \ref{sec:D1/D5} and Section \ref{sec:SYM}, we have demonstrated with several examples that the Berry curvature for monotone states in holographic theories does not exhibit quantum chaos. For the most part, our discussion applies for general $N$ and generic coupling. However, it relied on the large amount of supersymmetry of these states. For states with less supersymmetry, such as the fortuitous states that are supposed to describe black hole microstates, these techniques fail. There however exists a nice toy model, the $\mathcal{N}=2$ supersymmetric SYK model \cite{Fu:2016vas}, which allows one to access a relatively large number of fortuitous states numerically at moderate values of $N$ and generic couplings \cite{Chang:2024lxt}, and provides an understanding of the large $N$ limit via disorder average.\footnote{The ground state in the $\mathcal{N}=1$ SYK model has a degeneracy of at most $4$, and is not separated from the rest of the spectrum by a gap \cite{Fu:2016vas}, making it irrelevant for our discussion of random matrices which requires large degeneracy.} 

The model consists of $N$ complex fermions satisfying the Dirac algebra
\begin{equation}\label{eqn:anticomms}
\{\psi_{i},\psi_{j}\} = \{\bar{\psi}_{i},\bar{\psi}_{j}\} = 0, \qquad \{\psi_{i},\bar{\psi}_{j}\} = \delta_{ij}\,.
\end{equation}
The supercharge and the Hamiltonian of the model are given by\footnote{We will focus on the $q=3$ version of the model for concreteness (namely, the supercharge is cubic in the fermions), though most of our discussion can be easily generalized to higher $q$.}
\begin{equation}
    Q = \sum_{1 \leq i < j < k \leq N}C_{ijk}\psi_{i}\psi_{j}\psi_{k}, \quad\quad H = \{Q,Q^{\dagger}\} \,,
\end{equation}
for some antisymmetric tensor $(C_3)_{ijk} = C_{ijk}$. We can think of $C_{3}$ as a three-form and the states as $p$-forms built out of the fermions \cite{Chang:2024lxt}.
This theory has a large number of BPS states (ground states) satisfying
\begin{equation}\label{eqn:BPSdef}
    Q\ket{\textrm{BPS}}=Q^{\dagger}\ket{\textrm{BPS}}=0 \,.
\end{equation}
It has a $U(1)$ $R$-charge, which is simply the fermion number operator
\begin{equation}
R = N_{\psi} = \sum_{i=1}^{N}\psi_{i} \bar{\psi}_i \,,
\end{equation}
which satisfies $[R,Q]=3Q$. We denote the eigenvalues of $R$ as $r$. The model also has a particle-hole symmetry implemented by an anti-unitary operator $\mathcal{P}$ that maps \cite{Kanazawa:2017dpd}
\begin{equation}
    \label{eq:particle-hole sym}
    \mathcal{P} \psi_i \mathcal{P} = \eta\bar\psi_i \,, \qquad \mathcal{P} \bar\psi_i \mathcal{P} = \eta\psi_i  \,,\qquad \mathcal{P}^2 = (-1)^{\lfloor\frac{N}{2}\rfloor} \,, \qquad \eta = (-1)^{\lfloor \frac{N-1}{2} \rfloor} \,.
\end{equation}
This implies an isomorphism of $R$-charge sectors $\mathbb{H}_{r} \cong \mathbb{H}_{N-r}$, and in particular a symmetry of the $r = \frac{N}{2}$ sector when $N$ is even.

For generic couplings, the BPS states in this model exhibit \textit{R-charge concentration}: only in the spaces with $r = \frac{N}{2} ,\frac{N}{2} \pm 1$ does one find an exponentially large number of BPS states, with the explicit values displayed in Table \ref{tab:N=2SYK_BPSdegeneracy}. As explained in \cite{Chang:2024lxt}, all the BPS states in this model are fortuitous. 
\begin{table}[h]
\centering
\caption{Degeneracy of BPS states in $\mathcal{N}=2$ SYK for different $R$-charge sectors \cite{Fu:2016vas}.}
\label{tab:N=2SYK_BPSdegeneracy}
\renewcommand{\arraystretch}{1.6}

\begin{minipage}{0.45\textwidth}
\centering
\begin{tabular}{|c | c|}
\multicolumn{2}{c}{\textbf{$N$ even}} \\
\hline
\textbf{$R$-charge} & \textbf{BPS degeneracy} \\
\hline
$\frac{N}{2} \pm 1$ & $3^{\frac{N}{2}-1}$ \\
\hline
$\frac{N}{2}$ & $2 \times 3^{\frac{N}{2}-1}$ \\
\hline
\end{tabular}
\end{minipage}
\hfill
\begin{minipage}{0.50\textwidth}
\centering
\begin{tabular}{|c|c|c|}
\multicolumn{3}{c}{\textbf{$N$ odd}} \\

\cline{2-3}
\multicolumn{1}{c}{} & \multicolumn{2}{|c|}{\textbf{BPS degeneracy}} \\

\cline{2-3}
\cline{1-1}   % horizontal line over R-charge

\textbf{$R$-charge} 
& \textbf{$N \equiv 1 \pmod{4}$} 
& \textbf{$N \equiv 3 \pmod{4}$} \\

\hline
$\frac{N \pm 1}{2}$ 
& $3^{\frac{(N-1)}{2}}$ 
& $3^{\frac{(N-1)}{2}}$ \\
\hline

$\frac{N \pm 3}{2}$ 
& $1$ or $3$
& $0$ \\
\hline
\end{tabular}
\end{minipage}
\end{table}

The moduli space of the $\mathcal{N}=2$ SYK model is parameterized by the couplings $C_3$ and is of dimension $O(N^3)$, which is much larger than in our field theory examples.
The analytic treatment of the model usually involves averaging over the couplings $C_3$. However, averaging over couplings could ``wash out" interesting topological features to be discussed in Section \ref{sec:topo} associated with special subspaces of the moduli space that are of measure zero. Unless otherwise mentioned, we do not consider averaging over couplings and will study the model numerically through exact diagonalization.

Before we move on to discuss the Berry curvature for the fortuitous states in this model, it is worth mentioning that in \cite{Chang:2024lxt}, some generalizations of the $\mathcal{N}=2$ SYK model which also contain monotone states were proposed and studied. These models have the property that the monotone states are completely coupling independent and therefore the Berry curvature is identically zero. In Section \ref{sec:SYKmonotone} we discuss a different toy model that has monotone states which are coupling dependent.

\subsection{Berry curvature for directly varying individual components of $C_{3}$}\label{sec:SYKdirect}

We will analyze the Berry curvature of the BPS sector on the moduli space $\mathcal{M}=\{C_3\}=\mathbb{C}^{\binom{N}{3}}/U(1) \cong \mathbb{C}\mathbb{P}^{\binom{N}{3}-1} \times [0,\infty)$. The quotient by $U(1)$ is due to the fact that the Hamiltonian is invariant under an overall phase rotation $C_{3} \to e^{i\theta}C_{3}$.  We have not taken into account the global structures of the moduli space, such as the identification of the three-form $C_3$ under permutation of fermions. These global structures are not important for our current discussion since we will focus on the local properties of the Berry curvature. We also have not taken into account the existence of subspaces where the BPS spectrum differs, which we should carve out in the moduli space. Such subspaces will be studied in Section \ref{sec:topo}.

We can put the general formula (\ref{eqn:curvformula}) in a more explicit form using the specific structures of the SYK model. First, using $H = \{Q,Q^\dagger\}$ and that the BPS states are annihilated by both $Q$ and $Q^\dagger$, we have
\begin{equation}\label{curvSYK}
     \left(F_{\mu\nu}\right)_{ab} = \sum_{n \notin \mathbb{H}_{\textrm{BPS}}}\frac{i}{E_n^2}\bra{b} ((\partial_{\mu}Q )Q^\dagger + (\partial_{\mu}Q^\dagger )Q )\ket{n} \bra{n} (Q (\partial_{\nu}Q^\dagger ) + Q^\dagger (\partial_{\nu}Q ) )\ket{a} - (\mu \leftrightarrow \nu)\,.
\end{equation}
Equation (\ref{curvSYK}) applies to general deformations labelled by $\mu,\nu$. Now, 
let's specify $\mu$ and $\nu$ to be variations with respect to two specific couplings $C_{ijk}$ and $C_{lmn}$, and $\bar{\mu}, \bar{\nu}$ for their complex conjugates. Since $Q$ ($Q^\dagger$) is holomorphic (anti-holomorphic) in the couplings, we find 
\begin{equation}
    F_{\mu\nu} = F_{\bar{\mu}\bar{\nu}} = 0\,.
\end{equation}
For $F_{\mu\bar{\nu}}$, we get a nonzero expression\footnote{In terms of holomorphic and anti-holomorphic couplings, we have $(F_{\mu\bar{\nu}})^\dagger = - F_{\nu \bar{\mu}} = F_{\bar{\mu} \nu }.$ In particular $F_{\mu\bar{\mu}}$ is anti-Hermitian. \label{footnoteholomorphic} }
\begin{equation}\label{curvSYK2}
   (F_{\mu\bar{\nu}} )_{ab} = \sum_{n \notin \mathbb{H}_{\textrm{BPS}}}\frac{i}{E_n^2} \left[ \bra{b} (\partial_{\mu}Q )Q^\dagger \ket{n} \bra{n} Q (\partial_{\bar{\nu}}Q^\dagger )\ket{a} - \bra{b} (\partial_{\bar{\nu}}Q^\dagger )Q \ket{n} \bra{n} Q^\dagger (\partial_{\mu}Q )\ket{a}  \right]\,.
\end{equation}
Using $\partial_\mu Q = \psi_i \psi_j \psi_k$ and $\partial_{\bar{\nu}} Q^\dagger = \bar{\psi}_n \bar{\psi}_m \bar{\psi}_l$, (\ref{curvSYK2}) becomes a completely explicit formula that is amenable to numerical analysis through exact diagonalization. 

Similar to the discussion in previous sections, we can cast the Berry curvature in a more compact form by introducing an auxiliary quantity
\begin{equation}
   \left( F_{\mu \bar{\nu}} \right)_{ab} (x) =  \bra{b} (\partial_{\mu}Q )Q^\dagger  \frac{i}{(H-x)^2} Q (\partial_{\bar{\nu}}Q^\dagger )\ket{a} -  \bra{b} (\partial_{\bar{\nu}}Q^\dagger )Q \frac{i}{(H-x)^2} Q^\dagger (\partial_{\mu}Q )\ket{a} \,.
\end{equation}
Using $[H,Q] = [H, Q^\dagger] = 0$, we arrive at
\begin{equation}\label{FxSYK}
    \left( F_{\mu \bar{\nu}} \right)_{ab} (x)  =  i\bra{b} \psi_i \psi_j \psi_k   \frac{H}{(H-x)^2} \bar{\psi}_n \bar{\psi}_m \bar{\psi}_l \ket{a} -  i\bra{b}  \bar{\psi}_n \bar{\psi}_m \bar{\psi}_l    \frac{H}{(H-x)^2} \psi_i \psi_j \psi_k \ket{a} \,,
\end{equation}
from which we can recover $  (F_{\mu\bar{\nu}} )_{ab}$ through
\begin{equation}
    (F_{\mu\bar{\nu}} )_{ab} = \lim_{x\rightarrow 0}  \left( F_{\mu \bar{\nu}} \right)_{ab} (x)\,. 
\end{equation}
We restrict attention to $i(F_{\mu\bar{\mu}})_{ab}$ since it is a Hermitian matrix (but the Hermitian combinations $F_{\mu\bar{\nu}}+F_{\bar{\mu}\nu}$ with $\mu \neq \nu$ behaves similarly). Recall that the Berry curvature is block-diagonal with blocks labeled by their $R$-charges so we must restrict to a fixed $R$-charge sector.

\paragraph{Berry curvature spectrum.}

\begin{figure}
    \centering
    \includegraphics[width=0.4\linewidth]{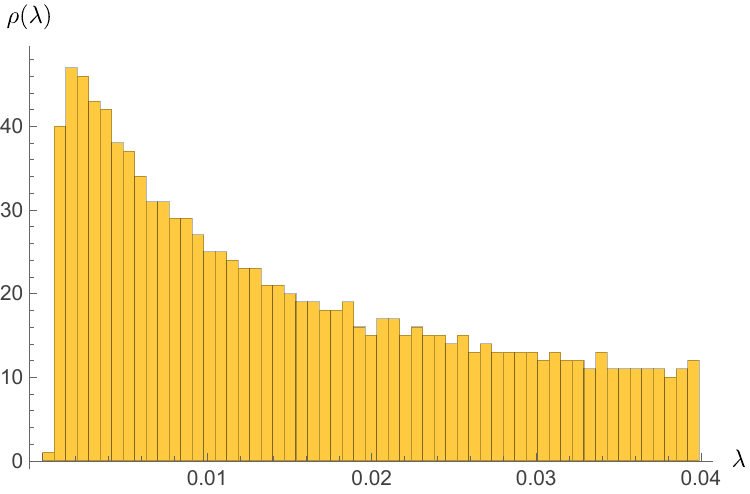}
    \caption{Spectral density $\rho(\lambda)$ for the low end of the positive spectrum (before unfolding) for the Berry matrix with $N=18$, $r=9$, exhibiting a hard edge. As $N$ increases, the edge approaches zero.}
    \label{fig:Berry_C_spectrum}
\end{figure}

The spectrum of the Berry curvature has some interesting features, which depend on the $R$-charge. For the middle sector, $r = \frac{N}{2}$ for even $N$, the particle-hole symmetry $\mathcal{P}$ maps $F_{\mu\bar\mu} \to -F_{\mu\bar\mu}$, and so implies that the eigenvalues come in pairs of opposite sign. The matrix is then of class D (for $N \equiv 0$ mod $4$) or C (for $N \equiv 2$ mod 4) in the Altland-Zirnbauer (AZ) classification. Moreover, \eqref{curvSYK2} is the difference between two terms. For $\mu = \nu$ each term is of the form $WW^\dagger$, where $W = (\partial_\mu Q) Q^\dagger H^{-1}$ or $W = (\partial_{\bar\mu}Q^\dagger)Q H^{-1}$ is a rectangular matrix from $\mathbb{H}_{\text{BPS}}$ to its complement in a given charged sector, similar to a Wishart matrix. The spectra of Wishart matrices exhibits a hard edge close to zero, which we also observe for the Berry matrices, as shown in Figure~\ref{fig:Berry_C_spectrum}, even though it does not necessarily follow from it. One caveat is that for small $N$, we observe that the Berry curvature matrix has many zero eigenvalues, but the fraction of zero eigenvalues decreases as $N$ is increased, and we expect them to disappear at large enough $N$, as in Appendix G of \cite{Lin:2022zxd}, so we will always remove them from the spectrum.

The Berry curvature in the sectors $r = \frac{N}{2}\pm 1$ for even $N$ is not symmetric and so its eigenvalues do not come in pairs. For $N=18$, we do not find any zero eigenvalues.

\paragraph{Eigenvalue repulsion.}

\begin{figure}[ht]
\begin{center}
\includegraphics[width=0.45\textwidth]{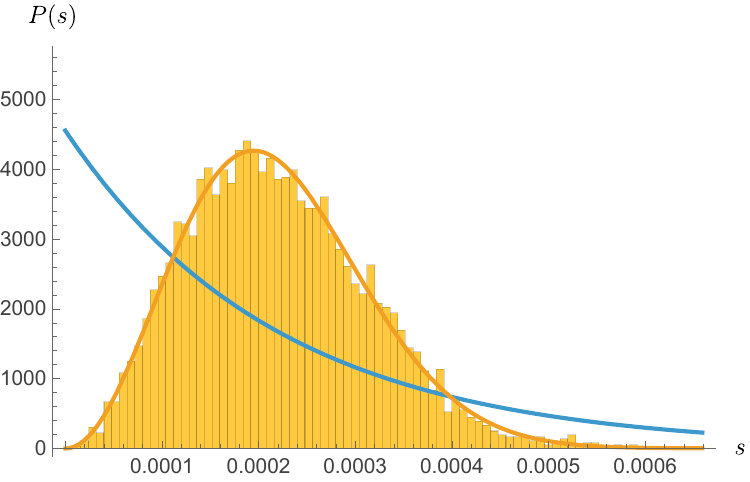}
\end{center}
\caption{Histogram of nearest-neighbor eigenvalue differences (after unfolding) for the Berry curvature $i(F_{\mu\overline{\mu}})_{ab}$ for $N=18$ and $r=9$ compared with Poisson statistics $P_{\mathrm{Poisson}}(s) = De^{-sD}$ (blue curve) and the GUE Wigner surmise $P_{\mathrm{GUE}}(s) = \frac{32}{\pi^{2}}D^{3}s^{2}e^{-\frac{4}{\pi}(sD)^{2}}$ (orange curve), where $D = 4551$ is the number of non-zero, positive eigenvalues of $i(F_{\mu\overline{\mu}})_{ab}$.}
\label{fig:Wignersurmise}
\end{figure}

To determine whether the curvature matrix $(F_{\mu\bar{\nu}})_{ab}$ is chaotic, we compute it numerically and compare the repulsion of its eigenvalues to those of a random matrix of the Wigner-Dyson GUE type, which is the appropriate one for the these type of matrices. For the short-range level repulsion, we compute the probability distribution of nearest-neighbor eigenvalue spacings $P(s)$ where $s = E_{i+1}-E_{i}$, and $\{E_i\}$ are the sorted, unfolded, eigenvalues of $F_{\mu\bar{\nu}}$ after unfolding the spectrum to make the mean level spacing equal to $1/D$, where $D$ is the size of the matrix. In this normalization the range of the spectrum is fixed to $O(D^0)$.\footnote{The RMT literature often uses different units, where the mean eigenvalue separation after unfolding is $1$. Our convention implies that strongly chaotic systems have a Thouless time of $O(1)$, while in the other units it is $O(1/D)$.} We find for $N=18$, $r=9$ that the spacings $s$ for the Berry curvature display very good agreement with the Wigner surmise prediction for a GUE ensemble, as shown in Figure \ref{fig:Wignersurmise}. We found the same to be true for $N=18$, $r=8$, although we do not show it here.

In strongly chaotic systems the eigenvalue repulsion is long-ranged, extends throughout a finite range of the spectrum, and results in spectral rigidity. One measure of this repulsion is the Spectral Form Factor (SFF),
\begin{equation}
    \text{SFF}(t) = \bigl\langle \Tr\bigl(e^{iF_{\mu\nu}t}\bigr) \Tr\bigl(e^{-iF_{\mu\nu}t}\bigr) \bigr\rangle \,,
\end{equation}
where we take $F_{\mu\nu}$ to be the matrix after unfolding. The imprint of the eigenvalue repulsion is a linear ramp of the SFF. The time scale in which this ramp starts for the connected part of the SFF is the Thouless time, $t_{\text{Th}}$. In a strongly chaotic system the Thouless time is independent of the system size, and is $O(1)$, corresponding to an eigenvalue repulsion that extends throughout the entire spectrum. The ramp should persist until late times, proportional to the dimension of the Hilbert space, where the SFF plateaus. Without unfolding, the ramp is still evident, but there could be oscillations around it. This will be relevant in Section~\ref{sec:JTgravity}.

Note, however, that the spectral form factor may exhibit several features that mask the ramp at early time. First, it might exhibit oscillations at early times, since the unfolded spectrum has compact support. In addition, and especially in sparse systems such as SYK, it might exhibit correlations due to fluctuations in the range of the spectrum between different realizations \cite{Gharibyan:2018jrp,Jia:2019orl,Berkooz:2020fvm,Caceres:2022kyr,Altland:2024ubs}. While it would be interesting to incorporate these effects in our analysis, here we will negate their effect by weighting the SFF with a Gaussian window around the middle of the (positive) spectrum, by computing $|Y_{E_0,\Delta E}|^2$, defined by \cite{Gharibyan:2018jrp}
\begin{equation}
    Y_{E_0,\Delta E} = \sum_E e^{-\frac{(E-E_0)^2}{2\Delta E^2} - i E t} \,,
\end{equation}
where $E$ are the unfolded energies, $E_0$ is chosen to be the middle of the (positive) spectrum, and $\Delta E$ will be an $O(1)$ fraction of the range of the spectrum. In such a way, we are still sensitive to a finite fraction of the spectrum.

Another standard measure of the long-range level repulsion and its resulting spectral rigidity is the level number variance \cite{GUHR1998189} defined by 
\begin{equation}
\Sigma^{2}(L) = \langle \hat{\eta}(L,\xi_{s})^{2}\rangle-\langle \hat{\eta}(L,\xi_{s})\rangle^{2}\,.
\end{equation} 
where $\hat{\eta}(L,\xi_{s})$ is the number of unfolded eigenvalues in the window $[\xi_{s}-L/2,\xi_{s}+L/2]$ and the average is over window positions $\xi_{s}$. The behavior of the level number variance is very different for a GUE random matrix compared to Poisson statistics due to stiffness of the spectrum in the GUE ensemble, which behaves at large $DL$ as
\begin{align}\label{eqn:LNV_GUEprediction}
\begin{split}
    \Sigma_{\mathrm{GUE}}^{2}(L) &= \frac{1}{\pi^{2}}\left(\log(2\pi D L)+\gamma+1\right) + O(1/D,1/L) \,,
    \\ \Sigma_{\mathrm{Poisson}}^{2}(L) &= DL \,,
\end{split}
\end{align}
where $\gamma$ is the Euler-Mascheroni constant and $D$ is the dimension of the Hilbert space. The value of $L$ at which the eigenvalues start to deviate from the RMT prediction is the Thouless energy $E_{\mathrm{Th}}$, inversely proportional to the Thouless time $t_{\mathrm{Th}}$. For the same reasons discussed above for the Gaussian weighted SFF, we truncate to the center of the spectrum when computing $\Sigma^{2}(L)$. The resulting Hilbert space is rather small, and finite $D$ effects become important. We thus compare the level number variance to that of GUE matrices acting on the same (truncated) Hilbert space.

Both the spectral form factor and the level number variance contain similar data, as they are functions of the connected 2-level correlation function, $R_{2,c}(s)$, written as function of the distance between the levels,
\begin{equation}
    \Sigma^2(L) = D L + 2\int_0^{L}ds\,\bigl(L-s\bigr)R_{2,c}(s) \,, \qquad \text{SFF}_c(t) = D + \int_{-\infty}^\infty ds \,R_{2,c}(s)e^{- i t s} \,,
\end{equation}
For the GUE, $R_{2,c}(s)$ is determined by the sine kernel to be  $-\left(\frac{\sin(D \pi s)}{\pi s}\right)^2$, which implies the formula \eqref{eqn:LNV_GUEprediction} and the linear ramp and plateau of the SFF.

\begin{figure}[ht]
\centering
\begin{subfigure}[t]{0.45\linewidth}
    \includegraphics[width=1\linewidth]{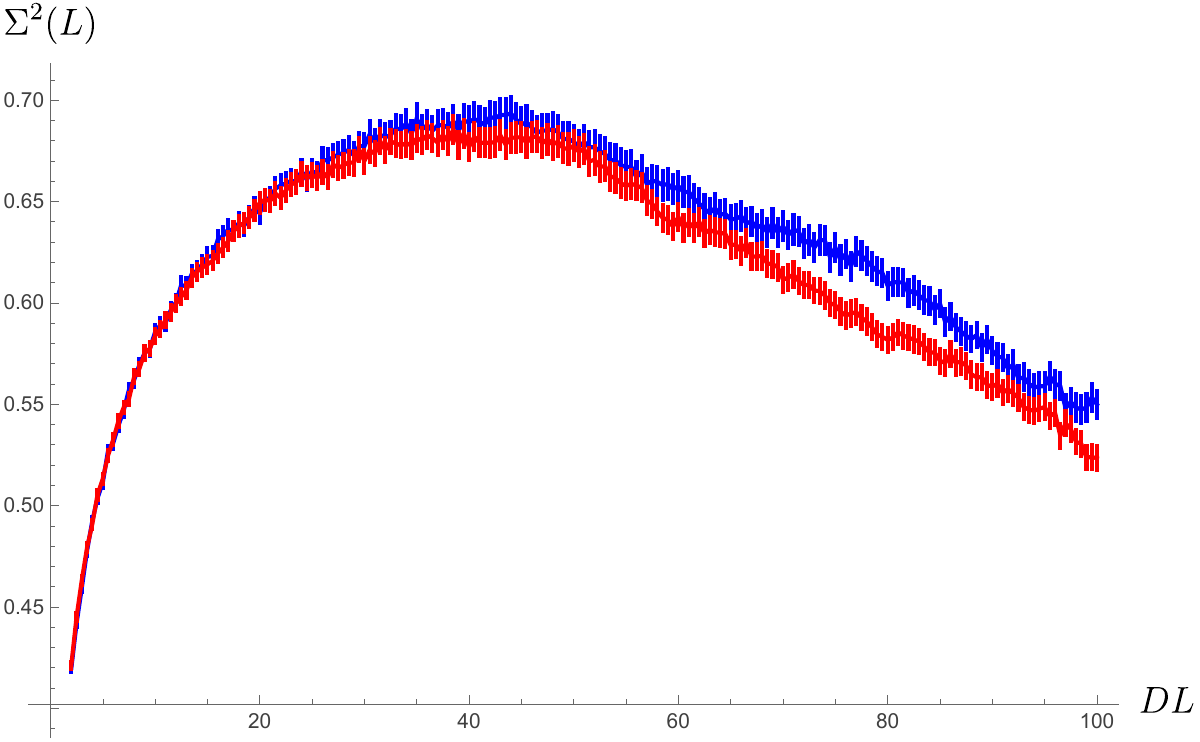}
    \caption{Level number variance $\Sigma^{2}(L)$ for the middle 10$\%$ of unfolded eigenvalues of $i(F_{\mu\overline{\mu}})_{ab}$ for $N=16$, $r=8$ averaged over an ensemble of $1000$ samples (blue), compared with an ensemble of GUE matrices of the same size (red). $D=1209$ is the total number of positive, non-zero eigenvalues and the error bars are the standard error of the mean.}
    \label{fig:LNV}
    \end{subfigure}
    \hfill
    \begin{subfigure}[t]{0.45\linewidth}
        \includegraphics[width=1\linewidth]{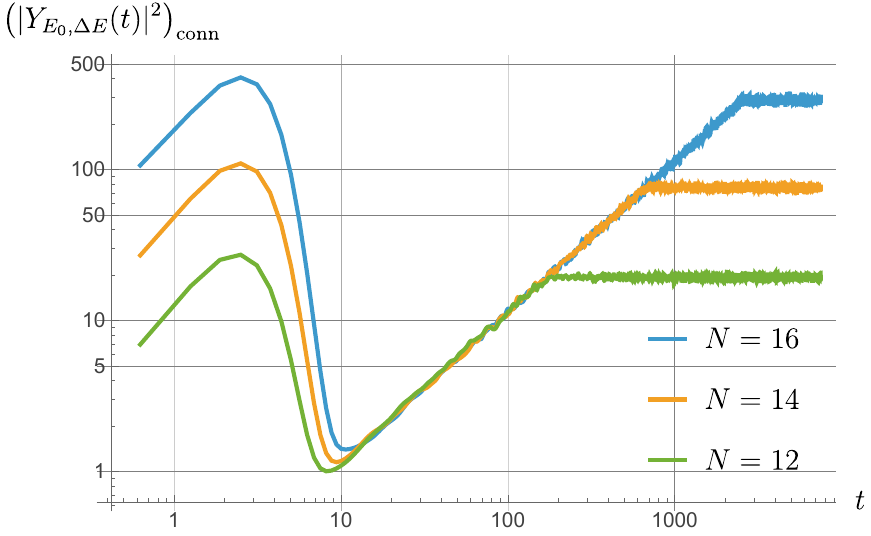}
        \caption{The connected part of $|Y_{E_0,\Delta E}|^2$ for the Berry curvature \eqref{curvSYK2} in the $r=N/2$ sector in $N=12$ (green), $N=14$ (orange), and $N=16$ (blue), averaged over an ensemble of 1000 samples. We took $\Delta E$ to be $15\%$ of the range of the (positive) spectrum, and $E_0$ to be its center.}
        \label{fig:sff_berry}  
    \end{subfigure}
    \caption{The level number variance and Gaussian-weighted spectral form factor for the Berry curvature.}
    \label{fig:LNV_SFF_berry}
\end{figure}

The ensemble-averaged level number variance and the (Gaussian-weighted) spectral form factor are plotted for the Berry curvature in Figure~\ref{fig:LNV_SFF_berry}. For the former, we find agreement in Figure \ref{fig:LNV} with the GUE level number variance over a significant fraction of the middle of the energy spectrum. The (Gaussian-weighted) spectral form factor is shown in Figure~\ref{fig:sff_berry}, and the Thouless time is consistent with being $O(1)$. In both cases, in order to definitively prove strong chaos, we need much stronger numerics which are beyond our capability.

\subsection{Berry curvature for conjugation deformations}\label{sec:SYKconj}

As reviewed in Section \ref{sec:conj}, apart from varying individual couplings, we can also consider the Berry curvature for conjugation deformations. This allows one to translate the Berry curvature into LMRS-like observables \cite{Lin:2022zxd} and also allows an easier comparison with the gravity discussion in Section \ref{sec:JTgravity}. 

Recall from the discussion in Section \ref{sec:conj} that these deformations conjugate the supercharge by $M(\lambda) = e^{\lambda \Lambda}$, which acts infinitesimally as $\partial_{\lambda} Q = [Q,\Lambda]$. If we demand that this keeps us within the moduli space $\mathcal{M}$ discussed in the previous subsection, i.e. supercharges that are cubic in the fermions, then we must have $\Lambda = K^{ij}\psi_{i}\bar{\psi}_{j}$ where $K$ is a Hermitian matrix, which changes the couplings by
\begin{equation}
    \partial_{\lambda} C_{ijk} = -C_{ajk}K_{i}^{a}-C_{iak}K_{j}^{a}-C_{ija}K_{k}^{a}\,.
\end{equation}
This makes it clear that the conjugation deformations are far from generic as their dimension is $O(N^2)$ while $\dim \mathcal{M} = O(N^3)$. One could, of course, enlarge the moduli space to allow for supercharges with higher-order interactions of the fermions, and then consider heavier deformations $\Lambda$, but we will restrict attention for now to those that keep us inside $\mathcal{M}$, and consider the more general case when we analyze $\mathcal{N}=2$ super-JT gravity in Section \ref{sec:JTgravity}.

As derived in Section \ref{sec:conj}, the expression for the Berry curvature generated by deforming with simple operators $\Lambda_\mu^T$ and $\Lambda_\nu^T$ is given by
\begin{equation}\label{conjSYK}
F_{\mu\nu} =  i[\widehat{\Lambda}_{\mu},\widehat{\Lambda}_{\nu}]- i\widehat{[\Lambda_{\mu},\Lambda_{\nu}]}\, ,
\end{equation}
where $\widehat{\Lambda} = P_{\textrm{BPS}} \,\Lambda\, P_{\textrm{BPS}}$ denote the projection of the simple operators into the BPS subspace. In particular, if we choose two simple operators that commute in the UV Hilbert space, i.e., $[\Lambda_\mu, \Lambda_\nu] = 0$, then only the first term in (\ref{conjSYK}) is non-zero. 

\begin{figure}
    \centering
    \begin{subfigure}{0.4\linewidth}
        \centering
        \includegraphics[width=1\textwidth]{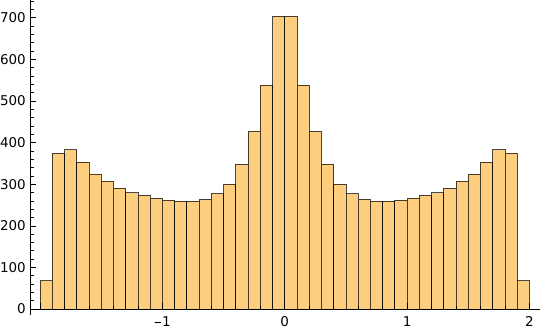}
        \subcaption{$\widehat{\Lambda}_\mu$}
    \end{subfigure}
    \hfill
    \begin{subfigure}{0.4\linewidth}
        \centering
        \includegraphics[width=1\textwidth]{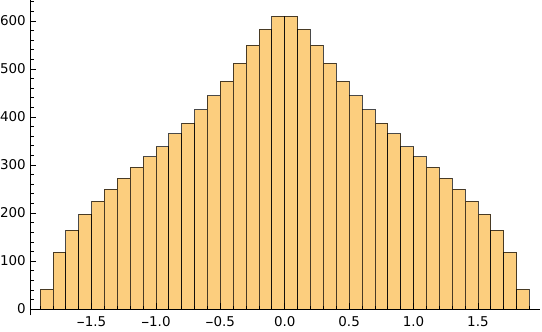}
        \subcaption{$F_{\mu\nu}$}
    \end{subfigure}
    \caption{The eigenvalue distributions of $\widehat{\Lambda}_\mu = P_{\textrm{BPS}}\Lambda_\mu P_{\textrm{BPS}}$ (same as $\widehat{\Lambda}_\nu$) (left) and $F_{\mu\nu} = i[\widehat{\Lambda}_\mu, \widehat{\Lambda}_\nu]$ (right) for the case of $N=18, r=9$. In $\widehat{\Lambda}_\mu$, it is easy to see imprints of the original spectrum of $\Lambda_\mu$, which takes value $0,\pm 2$, while such structure is washed out in the commutator.}
    \label{fig:N=18r=9}    
    \end{figure}

Following the results in \cite{Lin:2022zxd,Chang:2024lxt}, for a generic simple operator $\Lambda$, its projection into the BPS subspace $\widehat{\Lambda}$ is well approximated by a random matrix. This naturally suggests that (\ref{conjSYK}) also behaves as a random matrix. We can verify this intuition numerically. As an example, we can choose the operators $\Lambda_\mu , \Lambda_\nu$ to be
\begin{equation}
    \Lambda_\mu = \psi_1 \bar{\psi}_2 + \psi_2 \bar{\psi}_1, \quad \Lambda_\nu = \psi_3 \bar{\psi}_4 + \psi_4 \bar{\psi}_3\,.
\end{equation}
We have that $[ \Lambda_\mu , \Lambda_\nu] = 0$ before doing any projection, so only the first term in (\ref{conjSYK}) is nonzero. In Figure \ref{fig:N=18r=9}, we show the eigenvalue distribution of $\widehat{\Lambda}_\mu$ (which is the same as that of $\widehat{\Lambda}_\nu$) and that of $F_{\mu\nu} = i[\widehat{\Lambda}_\mu, \widehat{\Lambda}_\nu]$, for the case of $N=18$ and $r=9$. By studying the level statistics of $F_{\mu\nu}$, we find that it satisfies the random matrix universality of a GUE ensemble, similar to the case in Figure \ref{fig:Wignersurmise}.

\begin{figure}[t!]
\begin{center}
    \includegraphics[scale=1]{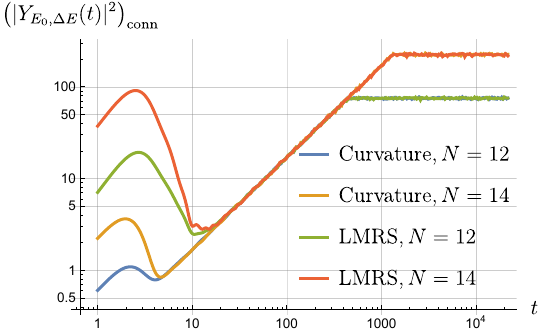}
\end{center}
\caption{We compare the connected part of the spectral form factor for the eigenvalues of $F_{\mu\nu}$ and those of $\widehat{\Lambda}_\mu$ (labeled by LMRS in the plot) for $N=12$ and $N=14$. We notice that the Thouless time, i.e., where the linear ramp kicks in, is shorter for $F_{\mu\nu}$ than $\widehat{\Lambda}_\mu$. In producing the plot, we used 5000 samples for $N=12, r=6$ and 4000 samples for $N=14,r=7$. We've chosen the same $E_0$ and $\Delta E$ as in Figure \ref{fig:sff_berry}.}
\label{fig:SFF}
\end{figure}

As a curious observation, by comparing the spectral form factors, we find that the commutator of projected operators $[\widehat{\Lambda}_\mu, \widehat{\Lambda}_\nu]$ has a shorter Thouless time compared to the projected operator itself, see Figure \ref{fig:SFF}. A potential interpretation is that neither operators  $\widehat{\Lambda}_\mu$ or $\widehat{\Lambda}_\nu$ are entirely random, but still maintain some structure in their respective preferred bases. However, there could be an additional random unitary relating the two sets of bases, and this additional randomness enters the commutator.\footnote{An example of this would be two ``banded" operators, but banded with respect to different bases.} In numerics, we also observe that the Thouless time of $[\widehat{\Lambda}_\mu, \widehat{\Lambda}_\nu]$ does not vary significantly as we vary $N$, as can be seen in Figure \ref{fig:SFF} by comparing the curves for $N=12$ and $N=14$. Since our numerics is restricted to relatively small $N$, we cannot confidently tell whether the Thouless time remains order one in the large $N$ limit. 

A natural question one can ask is that, if we approximate $\widehat{\Lambda}_\mu$, $\widehat{\Lambda}_\nu$ as two free random matrices, say drawn independently from a GUE ensemble, what would the detailed properties of the Berry curvature, i.e. the commutator, be. This problem will be analyzed in Section \ref{sec:JT Berry comp}, where we will study the properties of the Berry curvature using diagrammatic rules in $\mathcal{N}=2$ JT gravity.

\subsection{Berry curvature for monotone states}\label{sec:SYKmonotone}

In Section \ref{sec:SYKdirect} and Section \ref{sec:SYKconj}, we  studied the Berry curvature for the fortuitous states in the SYK model. Generic $\mathcal{N}=2$ SYK models do not contain monotone states, but they can exist for some non-generic choices of coupling constants. 
Some constructions were exhibited in \cite{Chang:2024lxt}, generalizing a two-flavor model of \cite{Heydeman:2022lse}. The monotones in these models are simply coupling independent, and the Berry curvature is therefore trivially zero. 

As an example where we have monotone states that nonetheless have non-zero Berry curvature, we can go to a special subspace of the moduli space (which will also play a role in Section \ref{sec:topo}), where the couplings, represented by a three-form $C_3$ are given by
\begin{equation}\label{Awedge}
    C_3 = A_1^{(1)} \wedge A_1^{(2)} \wedge A_1^{(3)} \,,
\end{equation}
where $A_1^{(a)}, a=1,2,3$ are three linearly-independent  one-forms. In such models, all the BPS states are monotones.\footnote{It should be noted that in this model, the number of monotone states with fixed $R$-charge $r>1$  grows as we take the large $N$ limit. This is different from the behavior in holographic systems, where we have a finite number of monotone states in the large $N$ limit, corresponding to the light supergravity excitations.} For example, at $r=1$, we have three BPS states
\begin{equation}
    A_{1}^{(a)} \cdot \psi \ket{0}, \quad a = 1,2,3\,,
\end{equation}
corresponding to the three one-forms.
It is easy to see that they are annihilated by $Q$ due to $C_3 \wedge A_1^{(a)} =0$. They are also annihilated by $Q^\dagger$ since $Q^\dagger$ acting on the states will decrease $r$ by three while all states have $r\geq 0$. Similarly, at $r=2$, any state of the form $A_1^{(a)} \wedge V_1$ where $V_1$ is an arbitrary one-form is also a monotone state.

One can analytically work out the Berry curvature in the special model (\ref{Awedge}) associated with variations of parameters that keep us within this class of models. Note that given (\ref{Awedge}), we can always apply a unitary transformation on the fermions, $\vec{\psi} \rightarrow U\cdot \vec{\psi}$ such that in the new basis, we have
\begin{equation}\label{ewedge}
    C_{3} = \alpha \, e^1 \wedge e^2 \wedge e^3\,, \quad \alpha > 0\,,
\end{equation}
where $e^{i}$ is the unit vector in the $i$-th direction. 
Therefore, without losing generality, the infinitesimal variations around \eqref{ewedge} that preserve \eqref{Awedge} take the form
\begin{equation}\label{varC3}
    \delta C_3 = \delta C_{123} \,e^1 \wedge e^2 \wedge e^3 + \sum_{i=4}^{N} \left( \delta C_{23i} \, e^2 \wedge e^3 \wedge e^i + \delta C_{13i} \, e^1 \wedge e^3 \wedge e^i   + \delta C_{12i} \, e^1 \wedge e^2 \wedge e^i \right) \,,
\end{equation}
Of course, if one is interested in specific changes of parameters around (\ref{Awedge}), one can easily translate those into the basis (\ref{varC3})  using the information of the unitary transformation $U$.

In Appendix \ref{app:SYKmonotone}, we compute the spectrum of the Berry curvature matrix with respect to the change of parameters in (\ref{varC3}). We find that the Berry curvature matrix is non-zero, but its spectrum only consists of three different eigenvalues $\pm 1/\alpha^2, 0$, with large degeneracies depending on the precise variations of parameters one is considering. In other words, the Berry curvature matrix, despite being non-zero, is not a random matrix. The situation here is analogous to what we found for the 1/2-BPS states of the D1/D5 CFT in Section \ref{sec:D1/D5_non-random}.

%%%%%%%%%%%
% SECTION %
%%%%%%%%%%%
\section{Berry curvature in $\mathcal{N}=2$ super-JT gravity}
\label{sec:JTgravity}

In this section we will give some evidence for the conjecture that the Berry curvature for BPS black hole microstates is strongly chaotic by analytic computations in $\mathcal{N} = 2$ super-JT gravity, a two dimensional toy model of quantum gravity which, at leading order at large $N$, is dual to the low energy limit of the $\mathcal{N} = 2$ SYK model discussed in Section \ref{sec:SYK}. To all orders in $N$, super-JT gravity is dual to an ensemble of theories \cite{Saad:2019lba,Turiaci:2023jfa}, and it captures the statistical properties of observables when averaged over the ensemble.

But $\mathcal{N}=2$ super-JT gravity is more than a toy model, as it also describes the near-horizon, low-energy dynamics around higher-dimensional supersymmetric AdS black holes \cite{Boruch:2022tno}, and so we expect it to capture the statistical properties of BPS black hole microstates in higher-dimensional theories and in their holographic duals. In particular, such black hole microstates appear in the $1/16$ BPS sector of 4d $\mathcal{N} = 4$ super Yang Mills, and the Berry curvature computed here should be contrasted with that of its more supersymmetric sectors discussed in Section \ref{sec:SYM}. The analysis also applies to the near-horizon black hole in a theory whose dual is an $\mathcal{N} = (2,2)$ 2d SCFT \cite{Boruch:2022tno}. The black holes in the theory dual to the D1/D5 system, as well as those in asymptotically flat space, have their near-horizon described by the $\mathcal{N}=4$ JT for D1/D5 \cite{Heydeman:2020hhw}, whose matter correlators are analyzed in \cite{Lin:2025wof}, and our analysis would have to be repeated for that case.

Since the theory is holographic, we can study the Berry curvature arising from conjugation deformations of the dual quantum mechanical theory, where the Hamiltonian transforms as \eqref{eqn:deltaHconjdeform}. The infinitesimal deformations are parameterized by Hermitian operators $\Lambda_{\mu,\nu}$, which we will take to commute in the UV (i.e. in the full theory). The associated Berry curvature \eqref{eqn:Fconjdef} is then given by
\begin{equation}
    \label{eq:Berry matrix UV commuting}
    F_{\mu\nu} = i[\widehat\Lambda_\mu,\widehat\Lambda_\nu] \,,
\end{equation}
where $\widehat \Lambda$ is the projection of the operator $\Lambda$ to the BPS sector. While the Berry curvature transforms in the adjoint under ``gauge transformations'' (i.e. changes of basis for the BPS subspace), its moments and polynomials made out of them, i.e.  observables of the form $\bigl\langle \Tr(F_{\mu\nu}^{k_1})\cdots\Tr(F_{\mu\nu}^{k_n})\bigr\rangle$, are ``gauge-invariant''. The brackets here denote the ensemble average. These kinds of observables are reminiscent of the LMRS criterion \cite{Lin:2022zxd}, where one studies the moments of a single operator projected to the BPS sector instead of the commutator of two projected operators. Much like it, these computations can be naturally phrased as Euclidean correlation functions, and thus calculated using the gravitational theory. 

We will work in the limit where the operators $\Lambda_{\mu,\nu}$ are very heavy, i.e. their dimension is $\Delta \to \infty$. This limit significantly simplifies our analysis (as well as those of \cite{Lin:2022zxd,Antonini:2025rmr}, for example), essentially reducing it to a problem in combinatorics. However, when our deformation comes from a marginal deformation of the dual theory, $\Delta$ is fixed by the dimension of that deformation in the effective theory that describes the near-BPS dynamics. It can be determined using holography, by starting with a massless scalar dual to a marginal deformation in the higher-dimensional background, and finding its mass (and thus its dimension $\Delta_{\rm IR}$ in the effective dual quantum mechanics) near the horizon of the black hole. For AdS$_5\times S^5$ black holes, the deformation remains marginal in the near-horizon theory, i.e. $\Delta_{\rm IR} = 1$  \cite{Ezroura:2024xba}. We leave the computation of the correlation functions for these deformations \cite{Lin:2026sli}, and thus the Berry curvature, to future work.

In order to show that $F_{\mu\nu}$ is strongly chaotic, we will compute its (connected) spectral form factor (SFF), 
\begin{equation}
    \label{eq:SFF moments}
    \text{SFF}_{\text{c}}\left(t\right) = \bigl\langle\Tr\bigl(e^{iF_{\mu\nu}t}\bigr)\Tr\bigl(e^{-iF_{\mu\nu}t}\bigr)\bigr\rangle_{\text{c}} = \!\!\sum_{m,n=0}^{\infty}\frac{(it)^{m}(-it)^{n}}{m!n!} \bigl\langle\Tr\left(F_{\mu\nu}^{m}\right)\Tr\left(F_{\mu\nu}^{n}\right)\bigr\rangle_c \,,
\end{equation}
and demonstrate that it grows linearly\footnote{Up to some oscillations around the linear trend, coming from the fact that we are not looking at the unfolded spectral form factor, i.e. we do not normalize the spectral density of $F_{\mu\nu}$ to be constant. The linear growth is still apparent when integrating over a time window.} from times of $O(1)$ in $D$, the size of the BPS subspace, up to times of order $O(D)$. The spectral form factor can thus be computed by computing usual correlation functions. We emphasize that in (\ref{eq:SFF moments}), the time variable $t$ is an auxiliary variable that is distinct from the physical time.

In Section \ref{sec:review JT} we quickly review some ingredients of $\mathcal{N}=2$ JT supergravity that are necessary for the computation of such correlation functions, and give a simple example of such computations by reviewing the LMRS computation \cite{Lin:2022zxd} of the spectrum and SFF for a single heavy operator in the BPS subspace. In Section~\ref{sec:JT Berry comp} we compute the spectrum and SFF of the Berry curvature $F_{\mu\nu}$ and show it is strongly chaotic.

\subsection{A brief review of super-JT gravity}
\label{sec:review JT}

The field content of $\mathcal{N}=2$ super-JT gravity consists of a metric $g_{\mu\nu}$, a dilaton $\phi$, a $U(1)$ $R$-charge gauge field $A$ with field strength $F = dA$, a scalar Lagrange multiplier $b$, and their fermionic superpartners, which all belong to the same supermultiplet. The Euclidean action on some manifold $M$ is schematically given by 
\begin{equation}
    I_{\mathcal{N}=2\ \text{JT}} = -S_0 \chi(M) -\frac12 \int_M \bigl[\phi(R - 2L_{\rm AdS}^{-2}) + bF + \text{(fermions)}\bigr] + \text{(boundary terms)}\,,
\end{equation}
where $\chi(M)$ is the Euler characteristic of $M$, $L_{\rm AdS}$ is the AdS radius, and $S_0$ is a free parameter in the theory, which is taken to be large. When regarded as the effective theory near the horizon of a higher-dimensional near-BPS black hole, $S_0$ is its zero-temperature entropy.  The $R$-charge is quantized\footnote{When comparing to the $\mathcal{N}=2$ SYK model discussed in the previous section, $\hat r$ is related to the number of fermions in the supercharge operator, $\hat r = \frac1q = \frac13$  and the total $R$-charge is shifted by $N/2$ compared to our SYK convention to be centered around zero, while in the reduction from supersymmetric AdS black holes $\hat r=1$ and the total $R$-charge is measured relative to that of the black hole.} in units of $\hat r$, where $1/\hat r \in \bZ$, and the smallest allowed charge for a matter field that can couple to the gauge field $A$ is $\hat r$.  For the sake of brevity in presenting the results, we will set $\hat r=1$ from now on. 

In order to compute the partition function (and other observables) of the theory, one needs to sum over all bulk geometries that obey the appropriate boundary conditions. For large $S_0$ the partition function is dominated by the disk topology, where the theory reduces to the $\mathcal{N}=2$ super-Schwarzian theory. The theory has a single dimensionful parameter, the Schwarzian coupling\footnote{When the Schwarzian theory arises in the IR dynamics of SYK, $C_S \propto N/J$ where $J$ is the variance of the SYK couplings \cite{Fu:2016vas}. When it arises in the near-horizon limit of a $D$-dimensional AdS black hole, $C_S \propto Q \propto L_{\rm AdS}^{D-1}/G_N$ \cite{Boruch:2022tno}, where $Q$ is the $R$-charge of the black hole, $G_N$ is the higher-dimensional Newton's constant, and $L_{\rm AdS}$ the AdS radius. If the dual theory is $N=4$ SYM on $S^3$ of radius $R_{S^3}$, that would imply $C_S \propto N^2 R_{S^3}$.} $C_S$. Its partition function was computed in \cite{Stanford:2017thb,Boruch:2023trc}, where it was found that the theory contains a large number of BPS states with zero energy and $R$-charges $|R| < \frac12$, with degeneracy
\begin{equation}
    n_{\text{BPS}}(R) = \cos(\pi R)\, e^{S_0} \,, \qquad  \bigl(|R|<\tfrac12\bigr)\,,
\end{equation}
thus taking large $S_0$ amounts to considering systems with a large number of BPS states. When $1/\hat r$ is odd these states are separated from the rest of the spectrum by a gap. One can then consider slow adiabatic deformations of the theory on time scales much longer than the (reciprocal of the) gap, giving rise to a well-defined Berry matrix.

As usual in (super-)JT gravity, it is convenient to regard the disc partition function not as a trace over some Hilbert space $\mathbb{H}$, but rather as an overlap between two thermofield double (TFD) states defined on a two-sided system, formally
\begin{equation}
    Z(\beta) = \Tr_{\mathbb{H}}\bigl(e^{-\beta H}\bigr) = \langle \text{TFD}_\beta| \text{TFD}_\beta\rangle = \vcenter{\hbox{
        \scalebox{0.5}{\tikzset{every picture/.style={line width=0.75pt}} %set default line width to 0.75pt        

\begin{tikzpicture}[x=0.75pt,y=0.75pt,yscale=-1,xscale=1]
%uncomment if require: \path (0,141); %set diagram left start at 0, and has height of 141

%Shape: Circle [id:dp9087914049081184] 
\draw  [line width=2.25]  (18.5,66.25) .. controls (18.5,33.25) and (45.25,6.5) .. (78.25,6.5) .. controls (111.25,6.5) and (138,33.25) .. (138,66.25) .. controls (138,99.25) and (111.25,126) .. (78.25,126) .. controls (45.25,126) and (18.5,99.25) .. (18.5,66.25) -- cycle ;
%Curve Lines [id:da23479056014115185] 
\draw [color={rgb, 255:red, 74; green, 144; blue, 226 }  ,draw opacity=1 ][line width=1.5]    (61,21) .. controls (62.11,19.01) and (63.71,18.52) .. (65.78,19.53) .. controls (68.04,20.36) and (69.56,19.68) .. (70.34,17.5) .. controls (71.27,15.3) and (72.86,14.69) .. (75.09,15.68) .. controls (76.92,16.98) and (78.56,16.76) .. (80.01,15.01) .. controls (81.88,13.5) and (83.52,13.68) .. (84.95,15.56) .. controls (86.36,17.49) and (88,17.79) .. (89.88,16.46) .. controls (91.87,15.17) and (93.53,15.53) .. (94.85,17.53) .. controls (96,19.51) and (97.59,19.89) .. (99.62,18.67) .. controls (101.63,17.46) and (103.27,17.9) .. (104.56,19.97) .. controls (105.63,22.01) and (107.21,22.48) .. (109.31,21.39) .. controls (111.6,20.51) and (113,21.31) .. (113.51,23.79) .. controls (113.34,26.08) and (114.42,27.36) .. (116.73,27.63) .. controls (119.11,28.15) and (120.03,29.54) .. (119.48,31.8) .. controls (118.92,34.14) and (119.77,35.63) .. (122.02,36.26) .. controls (124.27,36.97) and (125.01,38.42) .. (124.23,40.61) .. controls (123.4,42.79) and (124.09,44.33) .. (126.3,45.22) .. controls (128.45,46.18) and (128.95,47.75) .. (127.78,49.94) .. controls (126.16,51.43) and (125.98,53.07) .. (127.24,54.86) .. controls (128.07,57.07) and (127.39,58.58) .. (125.2,59.37) .. controls (123.13,60.78) and (122.91,62.43) .. (124.52,64.33) .. controls (126.23,65.9) and (126.28,67.58) .. (124.67,69.37) .. controls (123.06,71.1) and (123.1,72.78) .. (124.8,74.41) .. controls (126.48,76.01) and (126.48,77.69) .. (124.8,79.44) .. controls (123.09,80.99) and (123.02,82.64) .. (124.59,84.39) .. controls (126.1,86.19) and (125.9,87.84) .. (124,89.35) .. controls (122.03,90.44) and (121.53,91.99) .. (122.5,94) .. controls (122.66,96.45) and (121.49,97.67) .. (118.99,97.68) .. controls (116.67,97.37) and (115.37,98.33) .. (115.09,100.58) .. controls (114.54,102.97) and (113.15,103.89) .. (110.92,103.36) .. controls (108.57,102.89) and (107.12,103.82) .. (106.57,106.15) .. controls (106.18,108.38) and (104.79,109.28) .. (102.41,108.85) .. controls (100.09,108.42) and (98.76,109.36) .. (98.42,111.67) .. controls (98.27,113.96) and (97.1,115.12) .. (94.9,115.15) .. controls (92.94,114.46) and (91.4,115.12) .. (90.27,117.11) .. controls (88.88,118.98) and (87.22,119.16) .. (85.29,117.64) .. controls (83.67,116.03) and (82.04,116.06) .. (80.4,117.73) .. controls (78.63,119.36) and (76.89,119.31) .. (75.2,117.58) .. controls (73.74,115.82) and (72.09,115.71) .. (70.25,117.24) .. controls (68.42,118.75) and (66.79,118.58) .. (65.36,116.75) .. controls (63.97,114.88) and (62.3,114.64) .. (60.36,116.03) .. controls (58.38,117.34) and (56.8,116.95) .. (55.63,114.84) .. controls (55.32,112.68) and (54.05,111.66) .. (51.8,111.77) .. controls (49.39,111.41) and (48.39,110.07) .. (48.79,107.76) .. controls (49.18,105.41) and (48.21,104.05) .. (45.86,103.69) .. controls (43.55,103.54) and (42.42,102.31) .. (42.48,100.01) .. controls (42.37,97.66) and (41.11,96.54) .. (38.7,96.65) .. controls (36.37,96.82) and (35.15,95.73) .. (35.04,93.38) .. controls (34.99,91.01) and (33.82,89.81) .. (31.54,89.79) .. controls (29.11,89.42) and (28.14,88.05) .. (28.61,85.68) .. controls (29.54,83.65) and (29.02,82.07) .. (27.06,80.93) .. controls (25.22,79.44) and (25.11,77.81) .. (26.72,76.03) .. controls (28.42,74.44) and (28.5,72.78) .. (26.95,71.06) .. controls (25.48,69.17) and (25.7,67.48) .. (27.61,66.01) .. controls (29.56,64.68) and (29.89,63.07) .. (28.61,61.2) .. controls (27.48,59.03) and (27.98,57.4) .. (30.1,56.3) .. controls (32.24,55.45) and (32.9,53.93) .. (32.08,51.75) .. controls (31.12,49.64) and (31.69,48.07) .. (33.79,47.04) .. controls (35.86,45.93) and (36.32,44.34) .. (35.17,42.27) .. controls (34.01,40.2) and (34.46,38.58) .. (36.51,37.39) .. controls (38.56,36.3) and (39.05,34.7) .. (37.98,32.61) .. controls (36.99,30.48) and (37.61,28.98) .. (39.84,28.1) .. controls (42.07,27.65) and (43.1,26.35) .. (42.92,24.18) .. controls (43.69,21.81) and (45.19,21.16) .. (47.44,22.21) .. controls (49.23,23.68) and (50.9,23.55) .. (52.43,21.84) .. controls (54.07,20.15) and (55.74,20.08) .. (57.44,21.65) -- (61.1,20.97) ;

% Text Node
\draw (20,111) node [anchor=north west][inner sep=0.75pt]   [align=left] {$\displaystyle \beta $};

\end{tikzpicture}}}} \,, 
\end{equation}
where
\begin{equation}
    |\text{TFD}\rangle = \sum_{|n\rangle \in \mathbb{H}} e^{-\beta E_n / 2} \bigl(|n\rangle\otimes |n\rangle\bigr) = \vcenter{\hbox{
        \scalebox{0.75}{\tikzset{every picture/.style={line width=0.75pt}} %set default line width to 0.75pt        

\begin{tikzpicture}[x=0.75pt,y=0.75pt,yscale=-1,xscale=1]
%uncomment if require: \path (0,152); %set diagram left start at 0, and has height of 152

%Straight Lines [id:da8456013345650608] 
\draw  [dash pattern={on 4.5pt off 4.5pt}]  (21.5,67) -- (120.5,67) ;
%Shape: Arc [id:dp3234306626386517] 
\draw  [draw opacity=0][line width=2.25]  (134,66.25) .. controls (134,99.25) and (107.25,126) .. (74.25,126) .. controls (41.25,126) and (14.5,99.25) .. (14.5,66.25) -- (74.25,66.25) -- cycle ; \draw  [line width=2.25]  (134,66.25) .. controls (134,99.25) and (107.25,126) .. (74.25,126) .. controls (41.25,126) and (14.5,99.25) .. (14.5,66.25) ;  
%Curve Lines [id:da5175876993242863] 
\draw [color={rgb, 255:red, 74; green, 144; blue, 226 }  ,draw opacity=1 ][line width=1.5]    (120.5,67) .. controls (123.33,66.91) and (124.31,68.07) .. (123.44,70.5) .. controls (121.92,72.2) and (121.96,73.91) .. (123.55,75.62) .. controls (125.02,77.49) and (124.79,79.09) .. (122.88,80.43) .. controls (120.9,81.77) and (120.52,83.41) .. (121.75,85.35) .. controls (122.87,87.45) and (122.37,89.04) .. (120.24,90.12) .. controls (118.1,90.96) and (117.38,92.47) .. (118.07,94.64) .. controls (118.12,97.04) and (116.92,98.2) .. (114.45,98.12) .. controls (112.2,97.73) and (110.86,98.7) .. (110.43,101.04) .. controls (109.96,103.36) and (108.55,104.29) .. (106.2,103.83) .. controls (103.97,103.28) and (102.58,104.16) .. (102.03,106.49) .. controls (101.54,108.8) and (100.17,109.69) .. (97.92,109.18) .. controls (95.52,108.81) and (94.15,109.81) .. (93.8,112.16) .. controls (93.77,114.41) and (92.62,115.58) .. (90.35,115.69) .. controls (88.38,114.76) and (86.78,115.28) .. (85.53,117.23) .. controls (84.04,119.08) and (82.39,119.23) .. (80.59,117.67) .. controls (78.96,116.04) and (77.3,116.06) .. (75.63,117.72) .. controls (73.84,119.34) and (72.18,119.28) .. (70.67,117.55) .. controls (69.05,115.78) and (67.32,115.65) .. (65.49,117.17) .. controls (63.66,118.67) and (62.06,118.5) .. (60.67,116.66) .. controls (59.2,114.78) and (57.58,114.54) .. (55.82,115.94) .. controls (53.72,117.19) and (52.11,116.72) .. (50.99,114.51) .. controls (50.8,112.3) and (49.59,111.21) .. (47.34,111.22) .. controls (44.96,110.85) and (43.98,109.5) .. (44.39,107.18) .. controls (44.76,104.81) and (43.76,103.46) .. (41.4,103.12) .. controls (39.09,103.02) and (37.93,101.82) .. (37.93,99.52) .. controls (37.72,97.17) and (36.44,96.14) .. (34.07,96.43) .. controls (31.74,96.78) and (30.39,95.77) .. (30.02,93.4) .. controls (29.89,91.11) and (28.67,89.98) .. (26.35,90.01) .. controls (23.96,89.69) and (22.99,88.36) .. (23.44,86.03) .. controls (24.13,83.8) and (23.4,82.29) .. (21.25,81.51) .. controls (19.14,80.34) and (18.72,78.71) .. (20.01,76.64) .. controls (21.56,75.02) and (21.57,73.35) .. (20.03,71.63) -- (21.56,66.88) ;

% Text Node
\draw (6,102) node [anchor=north west][inner sep=0.75pt]   [align=left] {$\displaystyle \frac{\beta }{2}$};
% Text Node
\draw (45,48) node [anchor=north west][inner sep=0.75pt]   [align=left] {$\displaystyle |TFD\rangle $};

\end{tikzpicture}}}}\,.
\end{equation}  
These are the states we get by cutting open the disc diagram. The wavefunction of these states is found by canonically quantizing the $\mathcal{N}=2$ super-Schwarzian theory, which results in a supersymmetric Liouville quantum mechanics \cite{Lin:2022zxd}. That theory has a vacuum, which is the zero temperature TFD state, $|\text{BPS}\rangle = \sum_{|n\rangle \in \mathbb{H}_{\text{BPS}}} |n\rangle\otimes|n\rangle$. By definition, its amplitude counts the number of BPS states, $n_{\text{BPS}} = \langle \text{BPS}|\text{BPS}\rangle = e^{S_0}$. 

Before computing the Berry curvature \eqref{eq:Berry matrix UV commuting}, let's consider the properties of a single operator $\Lambda$ of dimension $\Delta$ when projected to the BPS subspace \cite{Lin:2022zxd}. Since the BPS subspace is the ground state of the theory, we can project $\Lambda$ down to the BPS subspace by infinite Euclidean evolution,
\begin{equation}
    \widehat \Lambda = P_{\rm BPS} \,\Lambda\, P_{\rm BPS} = \lim_{\beta\to\infty} e^{-\beta H} \, \Lambda\,  e^{-\beta H} =  \vcenter{\hbox{
        \scalebox{0.75}{\tikzset{every picture/.style={line width=0.75pt}} %set default line width to 0.75pt        

\begin{tikzpicture}[x=0.75pt,y=0.75pt,yscale=-1,xscale=1]
%uncomment if require: \path (0,152); %set diagram left start at 0, and has height of 152

%Straight Lines [id:da8456013345650608] 
\draw  [dash pattern={on 4.5pt off 4.5pt}]  (21.5,67) -- (120.5,67) ;
%Shape: Arc [id:dp3234306626386517] 
\draw  [draw opacity=0][line width=2.25]  (134,66.25) .. controls (134,99.25) and (107.25,126) .. (74.25,126) .. controls (41.25,126) and (14.5,99.25) .. (14.5,66.25) -- (74.25,66.25) -- cycle ; \draw  [line width=2.25]  (134,66.25) .. controls (134,99.25) and (107.25,126) .. (74.25,126) .. controls (41.25,126) and (14.5,99.25) .. (14.5,66.25) ;  
%Curve Lines [id:da5175876993242863] 
\draw [color={rgb, 255:red, 74; green, 144; blue, 226 }  ,draw opacity=1 ][line width=1.5]    (120.5,67) .. controls (123.33,66.91) and (124.31,68.07) .. (123.44,70.5) .. controls (121.92,72.2) and (121.96,73.91) .. (123.55,75.62) .. controls (125.02,77.49) and (124.79,79.09) .. (122.88,80.43) .. controls (120.9,81.77) and (120.52,83.41) .. (121.75,85.35) .. controls (122.87,87.45) and (122.37,89.04) .. (120.24,90.12) .. controls (118.1,90.96) and (117.38,92.47) .. (118.07,94.64) .. controls (118.12,97.04) and (116.92,98.2) .. (114.45,98.12) .. controls (112.2,97.73) and (110.86,98.7) .. (110.43,101.04) .. controls (109.96,103.36) and (108.55,104.29) .. (106.2,103.83) .. controls (103.97,103.28) and (102.58,104.16) .. (102.03,106.49) .. controls (101.54,108.8) and (100.17,109.69) .. (97.92,109.18) .. controls (95.52,108.81) and (94.15,109.81) .. (93.8,112.16) .. controls (93.77,114.41) and (92.62,115.58) .. (90.35,115.69) .. controls (88.38,114.76) and (86.78,115.28) .. (85.53,117.23) .. controls (84.04,119.08) and (82.39,119.23) .. (80.59,117.67) .. controls (78.96,116.04) and (77.3,116.06) .. (75.63,117.72) .. controls (73.84,119.34) and (72.18,119.28) .. (70.67,117.55) .. controls (69.05,115.78) and (67.32,115.65) .. (65.49,117.17) .. controls (63.66,118.67) and (62.06,118.5) .. (60.67,116.66) .. controls (59.2,114.78) and (57.58,114.54) .. (55.82,115.94) .. controls (53.72,117.19) and (52.11,116.72) .. (50.99,114.51) .. controls (50.8,112.3) and (49.59,111.21) .. (47.34,111.22) .. controls (44.96,110.85) and (43.98,109.5) .. (44.39,107.18) .. controls (44.76,104.81) and (43.76,103.46) .. (41.4,103.12) .. controls (39.09,103.02) and (37.93,101.82) .. (37.93,99.52) .. controls (37.72,97.17) and (36.44,96.14) .. (34.07,96.43) .. controls (31.74,96.78) and (30.39,95.77) .. (30.02,93.4) .. controls (29.89,91.11) and (28.67,89.98) .. (26.35,90.01) .. controls (23.96,89.69) and (22.99,88.36) .. (23.44,86.03) .. controls (24.13,83.8) and (23.4,82.29) .. (21.25,81.51) .. controls (19.14,80.34) and (18.72,78.71) .. (20.01,76.64) .. controls (21.56,75.02) and (21.57,73.35) .. (20.03,71.63) -- (21.56,66.88) ;
%Shape: Circle [id:dp127137282655171] 
\draw  [fill={rgb, 255:red, 0; green, 0; blue, 0 }  ,fill opacity=1 ] (70.25,117.25) .. controls (70.25,115.46) and (71.71,114) .. (73.5,114) .. controls (75.29,114) and (76.75,115.46) .. (76.75,117.25) .. controls (76.75,119.04) and (75.29,120.5) .. (73.5,120.5) .. controls (71.71,120.5) and (70.25,119.04) .. (70.25,117.25) -- cycle ;

% Text Node
\draw (65,128) node [anchor=north west][inner sep=0.75pt]   [align=left] {$\displaystyle \Lambda $};
% Text Node
\draw (18.09,125.57) node [anchor=north west][inner sep=0.75pt]  [rotate=-221.72,xslant=0.07] [align=left] {$\displaystyle \infty $};
% Text Node
\draw (137.63,118.07) node [anchor=north west][inner sep=0.75pt]  [rotate=-136.2,xslant=0.07] [align=left] {$\displaystyle \infty $};

\end{tikzpicture}}}} \,.
\end{equation}
We normalize $\Lambda$ such that\footnote{The two-point function in other normalizations can be calculated by inserting a full set of states along the matter worldline and evaluating the BPS Hartle-Hawking wavefunction in that basis \cite{Lin:2022zxd}.} 
\begin{equation}\label{trl2}
\bigl\langle\Tr\bigl(\widehat\Lambda^2\bigr)\bigr\rangle = C_S^{-2\Delta} e^{S_0} \,,
\end{equation} 
where the factor of $C_S^{-2\Delta}$ comes from dimensional analysis, and the factor of $e^{S_0}$ comes from that we are computing the correlator on the disc. For convenience, we will work in units where $C_S = 1$ from now on.

The spectral properties of the operator $\widehat \Lambda$ can be studied by evaluating Euclidean correlators. For example, if we denote the eigenvalue density of $\widehat \Lambda$ by $\rho_{\widehat \Lambda}$, then we can compute its moments as $\langle\Tr\bigl(\widehat{\Lambda}^k\bigr)\rangle = \int d\lambda \, \rho_{\widehat \Lambda}(\lambda) \lambda^k$. At leading order in $e^{S_0}$, this is given by the $k$-point correlation functions of $\widehat \Lambda$ on the disk. The correlator is calculated by summing over all possible worldlines in the bulk connecting pairs of operators. For example, for the four-point function
\begin{equation}
    \Tr\bigl(\widehat\Lambda^4\bigr) = \vcenter{\hbox{
        \scalebox{0.5}{\tikzset{every picture/.style={line width=0.75pt}} %set default line width to 0.75pt        

\begin{tikzpicture}[x=0.75pt,y=0.75pt,yscale=-1,xscale=1]
%uncomment if require: \path (0,180); %set diagram left start at 0, and has height of 180

%Shape: Circle [id:dp754075272165529] 
\draw  [line width=2.25]  (21.5,87.25) .. controls (21.5,54.25) and (48.25,27.5) .. (81.25,27.5) .. controls (114.25,27.5) and (141,54.25) .. (141,87.25) .. controls (141,120.25) and (114.25,147) .. (81.25,147) .. controls (48.25,147) and (21.5,120.25) .. (21.5,87.25) -- cycle ;
%Curve Lines [id:da10480930432900393] 
\draw [color={rgb, 255:red, 74; green, 144; blue, 226 }  ,draw opacity=1 ][line width=1.5]    (64,42) .. controls (65.11,40.01) and (66.71,39.52) .. (68.78,40.53) .. controls (71.04,41.36) and (72.56,40.68) .. (73.34,38.5) .. controls (74.27,36.3) and (75.86,35.69) .. (78.09,36.68) .. controls (79.92,37.98) and (81.56,37.76) .. (83.01,36.01) .. controls (84.88,34.5) and (86.52,34.68) .. (87.95,36.56) .. controls (89.36,38.49) and (91,38.79) .. (92.88,37.46) .. controls (94.87,36.17) and (96.53,36.53) .. (97.85,38.53) .. controls (99,40.51) and (100.59,40.89) .. (102.62,39.67) .. controls (104.63,38.46) and (106.27,38.9) .. (107.56,40.97) .. controls (108.63,43.01) and (110.21,43.48) .. (112.31,42.39) .. controls (114.6,41.51) and (116,42.31) .. (116.51,44.79) .. controls (116.34,47.08) and (117.42,48.36) .. (119.73,48.63) .. controls (122.11,49.15) and (123.03,50.54) .. (122.48,52.8) .. controls (121.92,55.14) and (122.77,56.63) .. (125.02,57.26) .. controls (127.27,57.97) and (128.01,59.42) .. (127.23,61.61) .. controls (126.4,63.79) and (127.09,65.33) .. (129.3,66.22) .. controls (131.45,67.18) and (131.95,68.75) .. (130.78,70.94) .. controls (129.16,72.43) and (128.98,74.07) .. (130.24,75.86) .. controls (131.07,78.07) and (130.39,79.58) .. (128.2,80.37) .. controls (126.13,81.78) and (125.91,83.43) .. (127.52,85.33) .. controls (129.23,86.9) and (129.28,88.58) .. (127.67,90.37) .. controls (126.06,92.1) and (126.1,93.78) .. (127.8,95.41) .. controls (129.48,97.01) and (129.48,98.69) .. (127.8,100.44) .. controls (126.09,101.99) and (126.02,103.64) .. (127.59,105.39) .. controls (129.1,107.19) and (128.9,108.84) .. (127,110.35) .. controls (125.03,111.44) and (124.53,112.99) .. (125.5,115) .. controls (125.66,117.45) and (124.49,118.67) .. (121.99,118.68) .. controls (119.67,118.37) and (118.37,119.33) .. (118.09,121.58) .. controls (117.54,123.97) and (116.15,124.89) .. (113.92,124.36) .. controls (111.57,123.89) and (110.12,124.82) .. (109.57,127.15) .. controls (109.18,129.38) and (107.79,130.28) .. (105.41,129.85) .. controls (103.09,129.42) and (101.76,130.36) .. (101.42,132.67) .. controls (101.27,134.96) and (100.1,136.12) .. (97.9,136.15) .. controls (95.94,135.46) and (94.4,136.12) .. (93.27,138.11) .. controls (91.88,139.98) and (90.22,140.16) .. (88.29,138.64) .. controls (86.67,137.03) and (85.04,137.06) .. (83.4,138.73) .. controls (81.63,140.36) and (79.89,140.31) .. (78.2,138.58) .. controls (76.74,136.82) and (75.09,136.71) .. (73.25,138.24) .. controls (71.42,139.75) and (69.79,139.58) .. (68.36,137.75) .. controls (66.97,135.88) and (65.3,135.64) .. (63.36,137.03) .. controls (61.38,138.34) and (59.8,137.95) .. (58.63,135.84) .. controls (58.32,133.68) and (57.05,132.66) .. (54.8,132.77) .. controls (52.39,132.41) and (51.39,131.07) .. (51.79,128.76) .. controls (52.18,126.41) and (51.21,125.05) .. (48.86,124.69) .. controls (46.55,124.54) and (45.42,123.31) .. (45.48,121.01) .. controls (45.37,118.66) and (44.11,117.54) .. (41.7,117.65) .. controls (39.37,117.82) and (38.15,116.73) .. (38.04,114.38) .. controls (37.99,112.01) and (36.82,110.81) .. (34.54,110.79) .. controls (32.11,110.42) and (31.14,109.05) .. (31.61,106.68) .. controls (32.54,104.65) and (32.02,103.07) .. (30.06,101.93) .. controls (28.22,100.44) and (28.11,98.81) .. (29.72,97.03) .. controls (31.42,95.44) and (31.5,93.78) .. (29.95,92.06) .. controls (28.48,90.17) and (28.7,88.48) .. (30.61,87.01) .. controls (32.56,85.68) and (32.89,84.07) .. (31.61,82.2) .. controls (30.48,80.03) and (30.98,78.4) .. (33.1,77.3) .. controls (35.24,76.45) and (35.9,74.93) .. (35.08,72.75) .. controls (34.12,70.64) and (34.69,69.07) .. (36.79,68.04) .. controls (38.86,66.93) and (39.32,65.34) .. (38.17,63.27) .. controls (37.01,61.2) and (37.46,59.58) .. (39.51,58.39) .. controls (41.56,57.3) and (42.05,55.7) .. (40.98,53.61) .. controls (39.99,51.48) and (40.61,49.98) .. (42.84,49.1) .. controls (45.07,48.65) and (46.1,47.35) .. (45.92,45.18) .. controls (46.69,42.81) and (48.19,42.16) .. (50.44,43.21) .. controls (52.23,44.68) and (53.9,44.55) .. (55.43,42.84) .. controls (57.07,41.15) and (58.74,41.08) .. (60.44,42.65) -- (64.1,41.97) ;
%Shape: Circle [id:dp9809289757868439] 
\draw  [fill={rgb, 255:red, 0; green, 0; blue, 0 }  ,fill opacity=1 ] (125,91.75) .. controls (125,89.96) and (126.46,88.5) .. (128.25,88.5) .. controls (130.04,88.5) and (131.5,89.96) .. (131.5,91.75) .. controls (131.5,93.54) and (130.04,95) .. (128.25,95) .. controls (126.46,95) and (125,93.54) .. (125,91.75) -- cycle ;
%Shape: Circle [id:dp13159847295207194] 
\draw  [fill={rgb, 255:red, 0; green, 0; blue, 0 }  ,fill opacity=1 ] (79,136.75) .. controls (79,134.96) and (80.46,133.5) .. (82.25,133.5) .. controls (84.04,133.5) and (85.5,134.96) .. (85.5,136.75) .. controls (85.5,138.54) and (84.04,140) .. (82.25,140) .. controls (80.46,140) and (79,138.54) .. (79,136.75) -- cycle ;
%Shape: Circle [id:dp7400833886149266] 
\draw  [fill={rgb, 255:red, 0; green, 0; blue, 0 }  ,fill opacity=1 ] (29,89.75) .. controls (29,87.96) and (30.46,86.5) .. (32.25,86.5) .. controls (34.04,86.5) and (35.5,87.96) .. (35.5,89.75) .. controls (35.5,91.54) and (34.04,93) .. (32.25,93) .. controls (30.46,93) and (29,91.54) .. (29,89.75) -- cycle ;
%Shape: Circle [id:dp046879268289208076] 
\draw  [fill={rgb, 255:red, 0; green, 0; blue, 0 }  ,fill opacity=1 ] (78.25,36.25) .. controls (78.25,34.46) and (79.71,33) .. (81.5,33) .. controls (83.29,33) and (84.75,34.46) .. (84.75,36.25) .. controls (84.75,38.04) and (83.29,39.5) .. (81.5,39.5) .. controls (79.71,39.5) and (78.25,38.04) .. (78.25,36.25) -- cycle ;

% Text Node
\draw (75,5) node [anchor=north west][inner sep=0.75pt]   [align=left] {$\displaystyle \Lambda $};
% Text Node
\draw (147,79) node [anchor=north west][inner sep=0.75pt]   [align=left] {$\displaystyle \Lambda $};
% Text Node
\draw (75,154) node [anchor=north west][inner sep=0.75pt]   [align=left] {$\displaystyle \Lambda $};
% Text Node
\draw (3,82) node [anchor=north west][inner sep=0.75pt]   [align=left] {$\displaystyle \Lambda $};
% Text Node
\draw (133.96,27.05) node [anchor=north west][inner sep=0.75pt]  [rotate=-43.36,xslant=0.07] [align=left] {$\displaystyle \infty $};
% Text Node
\draw (19.63,38.55) node [anchor=north west][inner sep=0.75pt]  [rotate=-318.14,xslant=0.07] [align=left] {$\displaystyle \infty $};
% Text Node
\draw (31.09,147.96) node [anchor=north west][inner sep=0.75pt]  [rotate=-221.72,xslant=0.07] [align=left] {$\displaystyle \infty $};
% Text Node
\draw (139.11,137.66) node [anchor=north west][inner sep=0.75pt]  [rotate=-136.2,xslant=0.07] [align=left] {$\displaystyle \infty $};

\end{tikzpicture}}}} = \vcenter{\hbox{
        \scalebox{0.5}{\begin{tikzpicture}[x=0.75pt,y=0.75pt,yscale=-1,xscale=1]
%uncomment if require: \path (0,173); %set diagram left start at 0, and has height of 173

%Shape: Circle [id:dp17091629053919355] 
\draw  [line width=2.25]  (18.5,85.86) .. controls (18.5,52.86) and (45.25,26.11) .. (78.25,26.11) .. controls (111.25,26.11) and (138,52.86) .. (138,85.86) .. controls (138,118.86) and (111.25,145.61) .. (78.25,145.61) .. controls (45.25,145.61) and (18.5,118.86) .. (18.5,85.86) -- cycle ;
%Curve Lines [id:da07545836966501895] 
\draw [color={rgb, 255:red, 74; green, 144; blue, 226 }  ,draw opacity=1 ][line width=1.5]    (61,40.61) .. controls (62.11,38.62) and (63.71,38.13) .. (65.78,39.14) .. controls (68.04,39.97) and (69.56,39.3) .. (70.34,37.12) .. controls (71.27,34.91) and (72.86,34.31) .. (75.09,35.3) .. controls (76.92,36.59) and (78.56,36.37) .. (80.01,34.62) .. controls (81.88,33.11) and (83.52,33.29) .. (84.95,35.18) .. controls (86.36,37.11) and (88,37.4) .. (89.88,36.07) .. controls (91.87,34.78) and (93.53,35.14) .. (94.85,37.14) .. controls (96,39.12) and (97.59,39.5) .. (99.62,38.29) .. controls (101.63,37.08) and (103.27,37.52) .. (104.56,39.59) .. controls (105.63,41.62) and (107.21,42.1) .. (109.31,41.01) .. controls (111.6,40.12) and (113,40.92) .. (113.51,43.41) .. controls (113.34,45.69) and (114.42,46.97) .. (116.73,47.25) .. controls (119.11,47.76) and (120.03,49.15) .. (119.48,51.42) .. controls (118.92,53.76) and (119.77,55.24) .. (122.02,55.87) .. controls (124.27,56.58) and (125.01,58.03) .. (124.23,60.23) .. controls (123.4,62.4) and (124.09,63.94) .. (126.3,64.84) .. controls (128.45,65.79) and (128.95,67.36) .. (127.78,69.55) .. controls (126.16,71.04) and (125.98,72.68) .. (127.24,74.47) .. controls (128.07,76.69) and (127.39,78.2) .. (125.2,78.99) .. controls (123.13,80.39) and (122.91,82.04) .. (124.52,83.94) .. controls (126.23,85.51) and (126.28,87.19) .. (124.67,88.98) .. controls (123.06,90.71) and (123.1,92.39) .. (124.8,94.02) .. controls (126.48,95.62) and (126.48,97.3) .. (124.8,99.05) .. controls (123.09,100.6) and (123.02,102.25) .. (124.59,104.01) .. controls (126.1,105.81) and (125.9,107.46) .. (124,108.96) .. controls (122.03,110.06) and (121.53,111.61) .. (122.5,113.61) .. controls (122.66,116.06) and (121.49,117.29) .. (118.99,117.3) .. controls (116.67,116.98) and (115.37,117.94) .. (115.09,120.19) .. controls (114.54,122.58) and (113.15,123.51) .. (110.92,122.97) .. controls (108.57,122.5) and (107.12,123.43) .. (106.57,125.76) .. controls (106.18,128) and (104.79,128.9) .. (102.41,128.46) .. controls (100.09,128.03) and (98.76,128.97) .. (98.42,131.28) .. controls (98.27,133.57) and (97.1,134.73) .. (94.9,134.76) .. controls (92.94,134.07) and (91.4,134.73) .. (90.27,136.73) .. controls (88.88,138.6) and (87.22,138.78) .. (85.29,137.25) .. controls (83.67,135.64) and (82.04,135.67) .. (80.4,137.35) .. controls (78.63,138.98) and (76.89,138.93) .. (75.2,137.19) .. controls (73.74,135.44) and (72.09,135.32) .. (70.25,136.85) .. controls (68.42,138.36) and (66.79,138.19) .. (65.36,136.36) .. controls (63.97,134.5) and (62.3,134.26) .. (60.36,135.65) .. controls (58.38,136.96) and (56.8,136.56) .. (55.63,134.45) .. controls (55.32,132.29) and (54.05,131.27) .. (51.8,131.39) .. controls (49.39,131.02) and (48.39,129.68) .. (48.79,127.37) .. controls (49.18,125.02) and (48.21,123.67) .. (45.86,123.3) .. controls (43.55,123.15) and (42.42,121.92) .. (42.48,119.63) .. controls (42.37,117.28) and (41.11,116.15) .. (38.7,116.26) .. controls (36.37,116.43) and (35.15,115.34) .. (35.04,112.99) .. controls (34.99,110.62) and (33.82,109.42) .. (31.54,109.41) .. controls (29.11,109.03) and (28.14,107.66) .. (28.61,105.29) .. controls (29.54,103.27) and (29.02,101.69) .. (27.06,100.54) .. controls (25.22,99.05) and (25.11,97.42) .. (26.72,95.64) .. controls (28.42,94.05) and (28.5,92.39) .. (26.95,90.67) .. controls (25.48,88.78) and (25.7,87.09) .. (27.61,85.62) .. controls (29.56,84.29) and (29.89,82.68) .. (28.61,80.81) .. controls (27.48,78.64) and (27.98,77.01) .. (30.1,75.91) .. controls (32.24,75.06) and (32.9,73.54) .. (32.08,71.36) .. controls (31.12,69.25) and (31.69,67.68) .. (33.79,66.65) .. controls (35.86,65.54) and (36.32,63.95) .. (35.17,61.88) .. controls (34.01,59.81) and (34.46,58.19) .. (36.51,57.01) .. controls (38.56,55.91) and (39.05,54.31) .. (37.98,52.22) .. controls (36.99,50.09) and (37.61,48.59) .. (39.84,47.71) .. controls (42.07,47.27) and (43.1,45.97) .. (42.92,43.8) .. controls (43.69,41.43) and (45.19,40.77) .. (47.44,41.83) .. controls (49.23,43.29) and (50.9,43.16) .. (52.43,41.45) .. controls (54.07,39.76) and (55.74,39.69) .. (57.44,41.26) -- (61.1,40.58) ;
%Shape: Circle [id:dp4528314006265949] 
\draw  [fill={rgb, 255:red, 0; green, 0; blue, 0 }  ,fill opacity=1 ] (122,90.36) .. controls (122,88.57) and (123.46,87.11) .. (125.25,87.11) .. controls (127.04,87.11) and (128.5,88.57) .. (128.5,90.36) .. controls (128.5,92.16) and (127.04,93.61) .. (125.25,93.61) .. controls (123.46,93.61) and (122,92.16) .. (122,90.36) -- cycle ;
%Shape: Circle [id:dp8444603120784138] 
\draw  [fill={rgb, 255:red, 0; green, 0; blue, 0 }  ,fill opacity=1 ] (76,135.36) .. controls (76,133.57) and (77.46,132.11) .. (79.25,132.11) .. controls (81.04,132.11) and (82.5,133.57) .. (82.5,135.36) .. controls (82.5,137.16) and (81.04,138.61) .. (79.25,138.61) .. controls (77.46,138.61) and (76,137.16) .. (76,135.36) -- cycle ;
%Shape: Circle [id:dp49442913653343534] 
\draw  [fill={rgb, 255:red, 0; green, 0; blue, 0 }  ,fill opacity=1 ] (26,88.36) .. controls (26,86.57) and (27.46,85.11) .. (29.25,85.11) .. controls (31.04,85.11) and (32.5,86.57) .. (32.5,88.36) .. controls (32.5,90.16) and (31.04,91.61) .. (29.25,91.61) .. controls (27.46,91.61) and (26,90.16) .. (26,88.36) -- cycle ;
%Shape: Circle [id:dp3362114799285506] 
\draw  [fill={rgb, 255:red, 0; green, 0; blue, 0 }  ,fill opacity=1 ] (75.25,34.86) .. controls (75.25,33.07) and (76.71,31.61) .. (78.5,31.61) .. controls (80.29,31.61) and (81.75,33.07) .. (81.75,34.86) .. controls (81.75,36.66) and (80.29,38.11) .. (78.5,38.11) .. controls (76.71,38.11) and (75.25,36.66) .. (75.25,34.86) -- cycle ;
%Curve Lines [id:da10117458003803836] 
\draw [line width=1.5]    (79.25,135.36) .. controls (71,105) and (52,89) .. (29.25,88.36) ;
%Curve Lines [id:da1958149956781743] 
\draw [line width=1.5]    (125.25,91.61) .. controls (102,86) and (79,70) .. (78.5,34.86) ;

% Text Node
\draw (72,3.61) node [anchor=north west][inner sep=0.75pt]   [align=left] {$\displaystyle \Lambda $};
% Text Node
\draw (144,77.61) node [anchor=north west][inner sep=0.75pt]   [align=left] {$\displaystyle \Lambda $};
% Text Node
\draw (72,152.61) node [anchor=north west][inner sep=0.75pt]   [align=left] {$\displaystyle \Lambda $};
% Text Node
\draw (0,80.61) node [anchor=north west][inner sep=0.75pt]   [align=left] {$\displaystyle \Lambda $};
% Text Node
\draw (130.96,25.67) node [anchor=north west][inner sep=0.75pt]  [rotate=-43.36,xslant=0.07] [align=left] {$\displaystyle \infty $};
% Text Node
\draw (16.63,37.16) node [anchor=north west][inner sep=0.75pt]  [rotate=-318.14,xslant=0.07] [align=left] {$\displaystyle \infty $};
% Text Node
\draw (28.09,146.57) node [anchor=north west][inner sep=0.75pt]  [rotate=-221.72,xslant=0.07] [align=left] {$\displaystyle \infty $};
% Text Node
\draw (136.11,136.28) node [anchor=north west][inner sep=0.75pt]  [rotate=-136.2,xslant=0.07] [align=left] {$\displaystyle \infty $};

\end{tikzpicture}}}} + \vcenter{\hbox{
        \scalebox{0.5}{\tikzset{every picture/.style={line width=0.75pt}} %set default line width to 0.75pt        

\begin{tikzpicture}[x=0.75pt,y=0.75pt,yscale=-1,xscale=1]
%uncomment if require: \path (0,177); %set diagram left start at 0, and has height of 177

%Shape: Circle [id:dp33309473433096026] 
\draw  [line width=2.25]  (22.5,85.86) .. controls (22.5,52.86) and (49.25,26.11) .. (82.25,26.11) .. controls (115.25,26.11) and (142,52.86) .. (142,85.86) .. controls (142,118.86) and (115.25,145.61) .. (82.25,145.61) .. controls (49.25,145.61) and (22.5,118.86) .. (22.5,85.86) -- cycle ;
%Curve Lines [id:da21861648936616196] 
\draw [color={rgb, 255:red, 74; green, 144; blue, 226 }  ,draw opacity=1 ][line width=1.5]    (65,40.61) .. controls (66.11,38.62) and (67.71,38.13) .. (69.78,39.14) .. controls (72.04,39.97) and (73.56,39.3) .. (74.34,37.12) .. controls (75.27,34.91) and (76.86,34.31) .. (79.09,35.3) .. controls (80.92,36.59) and (82.56,36.37) .. (84.01,34.62) .. controls (85.88,33.11) and (87.52,33.29) .. (88.95,35.18) .. controls (90.36,37.11) and (92,37.4) .. (93.88,36.07) .. controls (95.87,34.78) and (97.53,35.14) .. (98.85,37.14) .. controls (100,39.12) and (101.59,39.5) .. (103.62,38.29) .. controls (105.63,37.08) and (107.27,37.52) .. (108.56,39.59) .. controls (109.63,41.62) and (111.21,42.1) .. (113.31,41.01) .. controls (115.6,40.12) and (117,40.92) .. (117.51,43.41) .. controls (117.34,45.69) and (118.42,46.97) .. (120.73,47.25) .. controls (123.11,47.76) and (124.03,49.15) .. (123.48,51.42) .. controls (122.92,53.76) and (123.77,55.24) .. (126.02,55.87) .. controls (128.27,56.58) and (129.01,58.03) .. (128.23,60.23) .. controls (127.4,62.4) and (128.09,63.94) .. (130.3,64.84) .. controls (132.45,65.79) and (132.95,67.36) .. (131.78,69.55) .. controls (130.16,71.04) and (129.98,72.68) .. (131.24,74.47) .. controls (132.07,76.69) and (131.39,78.2) .. (129.2,78.99) .. controls (127.13,80.39) and (126.91,82.04) .. (128.52,83.94) .. controls (130.23,85.51) and (130.28,87.19) .. (128.67,88.98) .. controls (127.06,90.71) and (127.1,92.39) .. (128.8,94.02) .. controls (130.48,95.62) and (130.48,97.3) .. (128.8,99.05) .. controls (127.09,100.6) and (127.02,102.25) .. (128.59,104.01) .. controls (130.1,105.81) and (129.9,107.46) .. (128,108.96) .. controls (126.03,110.06) and (125.53,111.61) .. (126.5,113.61) .. controls (126.66,116.06) and (125.49,117.29) .. (122.99,117.3) .. controls (120.67,116.98) and (119.37,117.94) .. (119.09,120.19) .. controls (118.54,122.58) and (117.15,123.51) .. (114.92,122.97) .. controls (112.57,122.5) and (111.12,123.43) .. (110.57,125.76) .. controls (110.18,128) and (108.79,128.9) .. (106.41,128.46) .. controls (104.09,128.03) and (102.76,128.97) .. (102.42,131.28) .. controls (102.27,133.57) and (101.1,134.73) .. (98.9,134.76) .. controls (96.94,134.07) and (95.4,134.73) .. (94.27,136.73) .. controls (92.88,138.6) and (91.22,138.78) .. (89.29,137.25) .. controls (87.67,135.64) and (86.04,135.67) .. (84.4,137.35) .. controls (82.63,138.98) and (80.89,138.93) .. (79.2,137.19) .. controls (77.74,135.44) and (76.09,135.32) .. (74.25,136.85) .. controls (72.42,138.36) and (70.79,138.19) .. (69.36,136.36) .. controls (67.97,134.5) and (66.3,134.26) .. (64.36,135.65) .. controls (62.38,136.96) and (60.8,136.56) .. (59.63,134.45) .. controls (59.32,132.29) and (58.05,131.27) .. (55.8,131.39) .. controls (53.39,131.02) and (52.39,129.68) .. (52.79,127.37) .. controls (53.18,125.02) and (52.21,123.67) .. (49.86,123.3) .. controls (47.55,123.15) and (46.42,121.92) .. (46.48,119.63) .. controls (46.37,117.28) and (45.11,116.15) .. (42.7,116.26) .. controls (40.37,116.43) and (39.15,115.34) .. (39.04,112.99) .. controls (38.99,110.62) and (37.82,109.42) .. (35.54,109.41) .. controls (33.11,109.03) and (32.14,107.66) .. (32.61,105.29) .. controls (33.54,103.27) and (33.02,101.69) .. (31.06,100.54) .. controls (29.22,99.05) and (29.11,97.42) .. (30.72,95.64) .. controls (32.42,94.05) and (32.5,92.39) .. (30.95,90.67) .. controls (29.48,88.78) and (29.7,87.09) .. (31.61,85.62) .. controls (33.56,84.29) and (33.89,82.68) .. (32.61,80.81) .. controls (31.48,78.64) and (31.98,77.01) .. (34.1,75.91) .. controls (36.24,75.06) and (36.9,73.54) .. (36.08,71.36) .. controls (35.12,69.25) and (35.69,67.68) .. (37.79,66.65) .. controls (39.86,65.54) and (40.32,63.95) .. (39.17,61.88) .. controls (38.01,59.81) and (38.46,58.19) .. (40.51,57.01) .. controls (42.56,55.91) and (43.05,54.31) .. (41.98,52.22) .. controls (40.99,50.09) and (41.61,48.59) .. (43.84,47.71) .. controls (46.07,47.27) and (47.1,45.97) .. (46.92,43.8) .. controls (47.69,41.43) and (49.19,40.77) .. (51.44,41.83) .. controls (53.23,43.29) and (54.9,43.16) .. (56.43,41.45) .. controls (58.07,39.76) and (59.74,39.69) .. (61.44,41.26) -- (65.1,40.58) ;
%Shape: Circle [id:dp37320035840195076] 
\draw  [fill={rgb, 255:red, 0; green, 0; blue, 0 }  ,fill opacity=1 ] (126,90.36) .. controls (126,88.57) and (127.46,87.11) .. (129.25,87.11) .. controls (131.04,87.11) and (132.5,88.57) .. (132.5,90.36) .. controls (132.5,92.16) and (131.04,93.61) .. (129.25,93.61) .. controls (127.46,93.61) and (126,92.16) .. (126,90.36) -- cycle ;
%Shape: Circle [id:dp1710553151646288] 
\draw  [fill={rgb, 255:red, 0; green, 0; blue, 0 }  ,fill opacity=1 ] (80,135.36) .. controls (80,133.57) and (81.46,132.11) .. (83.25,132.11) .. controls (85.04,132.11) and (86.5,133.57) .. (86.5,135.36) .. controls (86.5,137.16) and (85.04,138.61) .. (83.25,138.61) .. controls (81.46,138.61) and (80,137.16) .. (80,135.36) -- cycle ;
%Shape: Circle [id:dp9520034904000817] 
\draw  [fill={rgb, 255:red, 0; green, 0; blue, 0 }  ,fill opacity=1 ] (30,88.36) .. controls (30,86.57) and (31.46,85.11) .. (33.25,85.11) .. controls (35.04,85.11) and (36.5,86.57) .. (36.5,88.36) .. controls (36.5,90.16) and (35.04,91.61) .. (33.25,91.61) .. controls (31.46,91.61) and (30,90.16) .. (30,88.36) -- cycle ;
%Shape: Circle [id:dp7187725233031417] 
\draw  [fill={rgb, 255:red, 0; green, 0; blue, 0 }  ,fill opacity=1 ] (79.25,34.86) .. controls (79.25,33.07) and (80.71,31.61) .. (82.5,31.61) .. controls (84.29,31.61) and (85.75,33.07) .. (85.75,34.86) .. controls (85.75,36.66) and (84.29,38.11) .. (82.5,38.11) .. controls (80.71,38.11) and (79.25,36.66) .. (79.25,34.86) -- cycle ;
%Curve Lines [id:da6240524714327924] 
\draw [line width=1.5]    (83.25,135.36) .. controls (84,106) and (100,93) .. (126,90.36) ;
%Curve Lines [id:da08351411669034081] 
\draw [line width=1.5]    (33.25,88.36) .. controls (70,90) and (84,60) .. (83.5,34.61) ;

% Text Node
\draw (76,3.61) node [anchor=north west][inner sep=0.75pt]   [align=left] {$\displaystyle \Lambda $};
% Text Node
\draw (148,77.61) node [anchor=north west][inner sep=0.75pt]   [align=left] {$\displaystyle \Lambda $};
% Text Node
\draw (76,152.61) node [anchor=north west][inner sep=0.75pt]   [align=left] {$\displaystyle \Lambda $};
% Text Node
\draw (4,80.61) node [anchor=north west][inner sep=0.75pt]   [align=left] {$\displaystyle \Lambda $};
% Text Node
\draw (134.96,25.67) node [anchor=north west][inner sep=0.75pt]  [rotate=-43.36,xslant=0.07] [align=left] {$\displaystyle \infty $};
% Text Node
\draw (20.63,37.16) node [anchor=north west][inner sep=0.75pt]  [rotate=-318.14,xslant=0.07] [align=left] {$\displaystyle \infty $};
% Text Node
\draw (32.09,146.57) node [anchor=north west][inner sep=0.75pt]  [rotate=-221.72,xslant=0.07] [align=left] {$\displaystyle \infty $};
% Text Node
\draw (140.11,136.28) node [anchor=north west][inner sep=0.75pt]  [rotate=-136.2,xslant=0.07] [align=left] {$\displaystyle \infty $};

\end{tikzpicture}}}} + \vcenter{\hbox{
        \scalebox{0.5}{\begin{tikzpicture}[x=0.75pt,y=0.75pt,yscale=-1,xscale=1]
%uncomment if require: \path (0,177); %set diagram left start at 0, and has height of 177

%Shape: Circle [id:dp7302759700567312] 
\draw  [line width=2.25]  (20.5,85.86) .. controls (20.5,52.86) and (47.25,26.11) .. (80.25,26.11) .. controls (113.25,26.11) and (140,52.86) .. (140,85.86) .. controls (140,118.86) and (113.25,145.61) .. (80.25,145.61) .. controls (47.25,145.61) and (20.5,118.86) .. (20.5,85.86) -- cycle ;
%Curve Lines [id:da3293911525126195] 
\draw [color={rgb, 255:red, 74; green, 144; blue, 226 }  ,draw opacity=1 ][line width=1.5]    (63,40.61) .. controls (64.11,38.62) and (65.71,38.13) .. (67.78,39.14) .. controls (70.04,39.97) and (71.56,39.3) .. (72.34,37.12) .. controls (73.27,34.91) and (74.86,34.31) .. (77.09,35.3) .. controls (78.92,36.59) and (80.56,36.37) .. (82.01,34.62) .. controls (83.88,33.11) and (85.52,33.29) .. (86.95,35.18) .. controls (88.36,37.11) and (90,37.4) .. (91.88,36.07) .. controls (93.87,34.78) and (95.53,35.14) .. (96.85,37.14) .. controls (98,39.12) and (99.59,39.5) .. (101.62,38.29) .. controls (103.63,37.08) and (105.27,37.52) .. (106.56,39.59) .. controls (107.63,41.62) and (109.21,42.1) .. (111.31,41.01) .. controls (113.6,40.12) and (115,40.92) .. (115.51,43.41) .. controls (115.34,45.69) and (116.42,46.97) .. (118.73,47.25) .. controls (121.11,47.76) and (122.03,49.15) .. (121.48,51.42) .. controls (120.92,53.76) and (121.77,55.24) .. (124.02,55.87) .. controls (126.27,56.58) and (127.01,58.03) .. (126.23,60.23) .. controls (125.4,62.4) and (126.09,63.94) .. (128.3,64.84) .. controls (130.45,65.79) and (130.95,67.36) .. (129.78,69.55) .. controls (128.16,71.04) and (127.98,72.68) .. (129.24,74.47) .. controls (130.07,76.69) and (129.39,78.2) .. (127.2,78.99) .. controls (125.13,80.39) and (124.91,82.04) .. (126.52,83.94) .. controls (128.23,85.51) and (128.28,87.19) .. (126.67,88.98) .. controls (125.06,90.71) and (125.1,92.39) .. (126.8,94.02) .. controls (128.48,95.62) and (128.48,97.3) .. (126.8,99.05) .. controls (125.09,100.6) and (125.02,102.25) .. (126.59,104.01) .. controls (128.1,105.81) and (127.9,107.46) .. (126,108.96) .. controls (124.03,110.06) and (123.53,111.61) .. (124.5,113.61) .. controls (124.66,116.06) and (123.49,117.29) .. (120.99,117.3) .. controls (118.67,116.98) and (117.37,117.94) .. (117.09,120.19) .. controls (116.54,122.58) and (115.15,123.51) .. (112.92,122.97) .. controls (110.57,122.5) and (109.12,123.43) .. (108.57,125.76) .. controls (108.18,128) and (106.79,128.9) .. (104.41,128.46) .. controls (102.09,128.03) and (100.76,128.97) .. (100.42,131.28) .. controls (100.27,133.57) and (99.1,134.73) .. (96.9,134.76) .. controls (94.94,134.07) and (93.4,134.73) .. (92.27,136.73) .. controls (90.88,138.6) and (89.22,138.78) .. (87.29,137.25) .. controls (85.67,135.64) and (84.04,135.67) .. (82.4,137.35) .. controls (80.63,138.98) and (78.89,138.93) .. (77.2,137.19) .. controls (75.74,135.44) and (74.09,135.32) .. (72.25,136.85) .. controls (70.42,138.36) and (68.79,138.19) .. (67.36,136.36) .. controls (65.97,134.5) and (64.3,134.26) .. (62.36,135.65) .. controls (60.38,136.96) and (58.8,136.56) .. (57.63,134.45) .. controls (57.32,132.29) and (56.05,131.27) .. (53.8,131.39) .. controls (51.39,131.02) and (50.39,129.68) .. (50.79,127.37) .. controls (51.18,125.02) and (50.21,123.67) .. (47.86,123.3) .. controls (45.55,123.15) and (44.42,121.92) .. (44.48,119.63) .. controls (44.37,117.28) and (43.11,116.15) .. (40.7,116.26) .. controls (38.37,116.43) and (37.15,115.34) .. (37.04,112.99) .. controls (36.99,110.62) and (35.82,109.42) .. (33.54,109.41) .. controls (31.11,109.03) and (30.14,107.66) .. (30.61,105.29) .. controls (31.54,103.27) and (31.02,101.69) .. (29.06,100.54) .. controls (27.22,99.05) and (27.11,97.42) .. (28.72,95.64) .. controls (30.42,94.05) and (30.5,92.39) .. (28.95,90.67) .. controls (27.48,88.78) and (27.7,87.09) .. (29.61,85.62) .. controls (31.56,84.29) and (31.89,82.68) .. (30.61,80.81) .. controls (29.48,78.64) and (29.98,77.01) .. (32.1,75.91) .. controls (34.24,75.06) and (34.9,73.54) .. (34.08,71.36) .. controls (33.12,69.25) and (33.69,67.68) .. (35.79,66.65) .. controls (37.86,65.54) and (38.32,63.95) .. (37.17,61.88) .. controls (36.01,59.81) and (36.46,58.19) .. (38.51,57.01) .. controls (40.56,55.91) and (41.05,54.31) .. (39.98,52.22) .. controls (38.99,50.09) and (39.61,48.59) .. (41.84,47.71) .. controls (44.07,47.27) and (45.1,45.97) .. (44.92,43.8) .. controls (45.69,41.43) and (47.19,40.77) .. (49.44,41.83) .. controls (51.23,43.29) and (52.9,43.16) .. (54.43,41.45) .. controls (56.07,39.76) and (57.74,39.69) .. (59.44,41.26) -- (63.1,40.58) ;
%Shape: Circle [id:dp174822296652829] 
\draw  [fill={rgb, 255:red, 0; green, 0; blue, 0 }  ,fill opacity=1 ] (124,90.36) .. controls (124,88.57) and (125.46,87.11) .. (127.25,87.11) .. controls (129.04,87.11) and (130.5,88.57) .. (130.5,90.36) .. controls (130.5,92.16) and (129.04,93.61) .. (127.25,93.61) .. controls (125.46,93.61) and (124,92.16) .. (124,90.36) -- cycle ;
%Shape: Circle [id:dp550968858118303] 
\draw  [fill={rgb, 255:red, 0; green, 0; blue, 0 }  ,fill opacity=1 ] (78,135.36) .. controls (78,133.57) and (79.46,132.11) .. (81.25,132.11) .. controls (83.04,132.11) and (84.5,133.57) .. (84.5,135.36) .. controls (84.5,137.16) and (83.04,138.61) .. (81.25,138.61) .. controls (79.46,138.61) and (78,137.16) .. (78,135.36) -- cycle ;
%Shape: Circle [id:dp9763680874748532] 
\draw  [fill={rgb, 255:red, 0; green, 0; blue, 0 }  ,fill opacity=1 ] (28,88.36) .. controls (28,86.57) and (29.46,85.11) .. (31.25,85.11) .. controls (33.04,85.11) and (34.5,86.57) .. (34.5,88.36) .. controls (34.5,90.16) and (33.04,91.61) .. (31.25,91.61) .. controls (29.46,91.61) and (28,90.16) .. (28,88.36) -- cycle ;
%Shape: Circle [id:dp7122460336823598] 
\draw  [fill={rgb, 255:red, 0; green, 0; blue, 0 }  ,fill opacity=1 ] (77.25,34.86) .. controls (77.25,33.07) and (78.71,31.61) .. (80.5,31.61) .. controls (82.29,31.61) and (83.75,33.07) .. (83.75,34.86) .. controls (83.75,36.66) and (82.29,38.11) .. (80.5,38.11) .. controls (78.71,38.11) and (77.25,36.66) .. (77.25,34.86) -- cycle ;
%Curve Lines [id:da013689319192594418] 
\draw [line width=1.5]    (34.5,88.36) .. controls (72,87) and (101,89) .. (127.25,90.36) ;
%Curve Lines [id:da42387376706853497] 
\draw [line width=1.5]    (80.5,34.86) .. controls (81,62) and (80,93) .. (81.25,135.36) ;

% Text Node
\draw (74,3.61) node [anchor=north west][inner sep=0.75pt]   [align=left] {$\displaystyle \Lambda $};
% Text Node
\draw (146,77.61) node [anchor=north west][inner sep=0.75pt]   [align=left] {$\displaystyle \Lambda $};
% Text Node
\draw (74,152.61) node [anchor=north west][inner sep=0.75pt]   [align=left] {$\displaystyle \Lambda $};
% Text Node
\draw (2,80.61) node [anchor=north west][inner sep=0.75pt]   [align=left] {$\displaystyle \Lambda $};
% Text Node
\draw (132.96,25.67) node [anchor=north west][inner sep=0.75pt]  [rotate=-43.36,xslant=0.07] [align=left] {$\displaystyle \infty $};
% Text Node
\draw (18.63,37.16) node [anchor=north west][inner sep=0.75pt]  [rotate=-318.14,xslant=0.07] [align=left] {$\displaystyle \infty $};
% Text Node
\draw (30.09,146.57) node [anchor=north west][inner sep=0.75pt]  [rotate=-221.72,xslant=0.07] [align=left] {$\displaystyle \infty $};
% Text Node
\draw (138.11,136.28) node [anchor=north west][inner sep=0.75pt]  [rotate=-136.2,xslant=0.07] [align=left] {$\displaystyle \infty $};

\end{tikzpicture}}}} \,.
\end{equation}
In the limit $\Delta\to\infty$ the sum simplifies considerably, as any crossing between two worldlines (such as the third diagram above) is exponentially suppressed in $\Delta$ (by a factor of $e^{-4\Delta\log(1+\sqrt2)}$) \cite{Lin:2022zxd}, and we can sum only over diagrams that do not contain any crossing worldlines. We note here that in the opposite limit, $\Delta \to 0$, crossed and uncrossed worldlines contribute exactly the same weight \cite{Lin:2022zxd}. The $2k^{\rm th}$ moment is then given by the number of possible Wick contractions, $(2k-1)!!$, and gives rise to a Gaussian spectral density.

Moreover, any non-crossing diagram factorizes into a product of two point functions. We see it by successively cutting the diagram between matter worldlines. The infinite Euclidean evolution on the two sides of the boundary between the matter worldlines projects the two-sided system to its ground state $|\text{BPS}\rangle$, and acts as inserting the projector $\frac{|\text{BPS}\rangle\langle\text{BPS}|}{\langle\text{BPS}|\text{BPS}\rangle} = e^{-S_0}|\text{BPS}\rangle\langle\text{BPS}|$. For example, the first diagram in the RHS above is 
\begin{equation}
         \vcenter{\hbox{
        \scalebox{0.5}{\tikzset{every picture/.style={line width=0.75pt}} %set default line width to 0.75pt        

\begin{tikzpicture}[x=0.75pt,y=0.75pt,yscale=-1,xscale=1]
%uncomment if require: \path (0,176); %set diagram left start at 0, and has height of 176

%Shape: Circle [id:dp5706539335475382] 
\draw  [line width=2.25]  (25.5,85.86) .. controls (25.5,52.86) and (52.25,26.11) .. (85.25,26.11) .. controls (118.25,26.11) and (145,52.86) .. (145,85.86) .. controls (145,118.86) and (118.25,145.61) .. (85.25,145.61) .. controls (52.25,145.61) and (25.5,118.86) .. (25.5,85.86) -- cycle ;
%Curve Lines [id:da4012794066736214] 
\draw [color={rgb, 255:red, 74; green, 144; blue, 226 }  ,draw opacity=1 ][line width=1.5]    (68,40.61) .. controls (69.11,38.62) and (70.71,38.13) .. (72.78,39.14) .. controls (75.04,39.97) and (76.56,39.3) .. (77.34,37.12) .. controls (78.27,34.91) and (79.86,34.31) .. (82.09,35.3) .. controls (83.92,36.59) and (85.56,36.37) .. (87.01,34.62) .. controls (88.88,33.11) and (90.52,33.29) .. (91.95,35.18) .. controls (93.36,37.11) and (95,37.4) .. (96.88,36.07) .. controls (98.87,34.78) and (100.53,35.14) .. (101.85,37.14) .. controls (103,39.12) and (104.59,39.5) .. (106.62,38.29) .. controls (108.63,37.08) and (110.27,37.52) .. (111.56,39.59) .. controls (112.63,41.62) and (114.21,42.1) .. (116.31,41.01) .. controls (118.6,40.12) and (120,40.92) .. (120.51,43.41) .. controls (120.34,45.69) and (121.42,46.97) .. (123.73,47.25) .. controls (126.11,47.76) and (127.03,49.15) .. (126.48,51.42) .. controls (125.92,53.76) and (126.77,55.24) .. (129.02,55.87) .. controls (131.27,56.58) and (132.01,58.03) .. (131.23,60.23) .. controls (130.4,62.4) and (131.09,63.94) .. (133.3,64.84) .. controls (135.45,65.79) and (135.95,67.36) .. (134.78,69.55) .. controls (133.16,71.04) and (132.98,72.68) .. (134.24,74.47) .. controls (135.07,76.69) and (134.39,78.2) .. (132.2,78.99) .. controls (130.13,80.39) and (129.91,82.04) .. (131.52,83.94) .. controls (133.23,85.51) and (133.28,87.19) .. (131.67,88.98) .. controls (130.06,90.71) and (130.1,92.39) .. (131.8,94.02) .. controls (133.48,95.62) and (133.48,97.3) .. (131.8,99.05) .. controls (130.09,100.6) and (130.02,102.25) .. (131.59,104.01) .. controls (133.1,105.81) and (132.9,107.46) .. (131,108.96) .. controls (129.03,110.06) and (128.53,111.61) .. (129.5,113.61) .. controls (129.66,116.06) and (128.49,117.29) .. (125.99,117.3) .. controls (123.67,116.98) and (122.37,117.94) .. (122.09,120.19) .. controls (121.54,122.58) and (120.15,123.51) .. (117.92,122.97) .. controls (115.57,122.5) and (114.12,123.43) .. (113.57,125.76) .. controls (113.18,128) and (111.79,128.9) .. (109.41,128.46) .. controls (107.09,128.03) and (105.76,128.97) .. (105.42,131.28) .. controls (105.27,133.57) and (104.1,134.73) .. (101.9,134.76) .. controls (99.94,134.07) and (98.4,134.73) .. (97.27,136.73) .. controls (95.88,138.6) and (94.22,138.78) .. (92.29,137.25) .. controls (90.67,135.64) and (89.04,135.67) .. (87.4,137.35) .. controls (85.63,138.98) and (83.89,138.93) .. (82.2,137.19) .. controls (80.74,135.44) and (79.09,135.32) .. (77.25,136.85) .. controls (75.42,138.36) and (73.79,138.19) .. (72.36,136.36) .. controls (70.97,134.5) and (69.3,134.26) .. (67.36,135.65) .. controls (65.38,136.96) and (63.8,136.56) .. (62.63,134.45) .. controls (62.32,132.29) and (61.05,131.27) .. (58.8,131.39) .. controls (56.39,131.02) and (55.39,129.68) .. (55.79,127.37) .. controls (56.18,125.02) and (55.21,123.67) .. (52.86,123.3) .. controls (50.55,123.15) and (49.42,121.92) .. (49.48,119.63) .. controls (49.37,117.28) and (48.11,116.15) .. (45.7,116.26) .. controls (43.37,116.43) and (42.15,115.34) .. (42.04,112.99) .. controls (41.99,110.62) and (40.82,109.42) .. (38.54,109.41) .. controls (36.11,109.03) and (35.14,107.66) .. (35.61,105.29) .. controls (36.54,103.27) and (36.02,101.69) .. (34.06,100.54) .. controls (32.22,99.05) and (32.11,97.42) .. (33.72,95.64) .. controls (35.42,94.05) and (35.5,92.39) .. (33.95,90.67) .. controls (32.48,88.78) and (32.7,87.09) .. (34.61,85.62) .. controls (36.56,84.29) and (36.89,82.68) .. (35.61,80.81) .. controls (34.48,78.64) and (34.98,77.01) .. (37.1,75.91) .. controls (39.24,75.06) and (39.9,73.54) .. (39.08,71.36) .. controls (38.12,69.25) and (38.69,67.68) .. (40.79,66.65) .. controls (42.86,65.54) and (43.32,63.95) .. (42.17,61.88) .. controls (41.01,59.81) and (41.46,58.19) .. (43.51,57.01) .. controls (45.56,55.91) and (46.05,54.31) .. (44.98,52.22) .. controls (43.99,50.09) and (44.61,48.59) .. (46.84,47.71) .. controls (49.07,47.27) and (50.1,45.97) .. (49.92,43.8) .. controls (50.69,41.43) and (52.19,40.77) .. (54.44,41.83) .. controls (56.23,43.29) and (57.9,43.16) .. (59.43,41.45) .. controls (61.07,39.76) and (62.74,39.69) .. (64.44,41.26) -- (68.1,40.58) ;
%Shape: Circle [id:dp9312927244560568] 
\draw  [fill={rgb, 255:red, 0; green, 0; blue, 0 }  ,fill opacity=1 ] (129,90.36) .. controls (129,88.57) and (130.46,87.11) .. (132.25,87.11) .. controls (134.04,87.11) and (135.5,88.57) .. (135.5,90.36) .. controls (135.5,92.16) and (134.04,93.61) .. (132.25,93.61) .. controls (130.46,93.61) and (129,92.16) .. (129,90.36) -- cycle ;
%Shape: Circle [id:dp14567973021368696] 
\draw  [fill={rgb, 255:red, 0; green, 0; blue, 0 }  ,fill opacity=1 ] (83,135.36) .. controls (83,133.57) and (84.46,132.11) .. (86.25,132.11) .. controls (88.04,132.11) and (89.5,133.57) .. (89.5,135.36) .. controls (89.5,137.16) and (88.04,138.61) .. (86.25,138.61) .. controls (84.46,138.61) and (83,137.16) .. (83,135.36) -- cycle ;
%Shape: Circle [id:dp9282612464130453] 
\draw  [fill={rgb, 255:red, 0; green, 0; blue, 0 }  ,fill opacity=1 ] (33,88.36) .. controls (33,86.57) and (34.46,85.11) .. (36.25,85.11) .. controls (38.04,85.11) and (39.5,86.57) .. (39.5,88.36) .. controls (39.5,90.16) and (38.04,91.61) .. (36.25,91.61) .. controls (34.46,91.61) and (33,90.16) .. (33,88.36) -- cycle ;
%Shape: Circle [id:dp18638246756634946] 
\draw  [fill={rgb, 255:red, 0; green, 0; blue, 0 }  ,fill opacity=1 ] (82.25,34.86) .. controls (82.25,33.07) and (83.71,31.61) .. (85.5,31.61) .. controls (87.29,31.61) and (88.75,33.07) .. (88.75,34.86) .. controls (88.75,36.66) and (87.29,38.11) .. (85.5,38.11) .. controls (83.71,38.11) and (82.25,36.66) .. (82.25,34.86) -- cycle ;
%Curve Lines [id:da360216708634985] 
\draw [line width=1.5]    (86.25,135.36) .. controls (78,105) and (59,89) .. (36.25,88.36) ;
%Curve Lines [id:da40704038363968986] 
\draw [line width=1.5]    (132.25,91.61) .. controls (109,86) and (86,70) .. (85.5,34.86) ;
%Straight Lines [id:da0006380163525415705] 
\draw  [dash pattern={on 4.5pt off 4.5pt}]  (26.5,27.5) -- (147.5,146.5) ;

% Text Node
\draw (79,3.61) node [anchor=north west][inner sep=0.75pt]   [align=left] {$\displaystyle \Lambda $};
% Text Node
\draw (151,77.61) node [anchor=north west][inner sep=0.75pt]   [align=left] {$\displaystyle \Lambda $};
% Text Node
\draw (79,152.61) node [anchor=north west][inner sep=0.75pt]   [align=left] {$\displaystyle \Lambda $};
% Text Node
\draw (7,80.61) node [anchor=north west][inner sep=0.75pt]   [align=left] {$\displaystyle \Lambda $};
% Text Node
\draw (137.96,25.67) node [anchor=north west][inner sep=0.75pt]  [rotate=-43.36,xslant=0.07] [align=left] {$\displaystyle \infty $};
% Text Node
\draw (41.61,17.67) node [anchor=north west][inner sep=0.75pt]  [rotate=-335.58,xslant=0.07] [align=left] {$\displaystyle \infty $};
% Text Node
\draw (35.09,146.57) node [anchor=north west][inner sep=0.75pt]  [rotate=-221.72,xslant=0.07] [align=left] {$\displaystyle \infty $};
% Text Node
\draw (160.08,117.85) node [anchor=north west][inner sep=0.75pt]  [rotate=-124.56,xslant=0.07] [align=left] {$\displaystyle \infty $};
% Text Node
\draw (8.71,58.61) node [anchor=north west][inner sep=0.75pt]  [rotate=-301.55,xslant=0.07] [align=left] {$\displaystyle \infty $};
% Text Node
\draw (126.41,151.4) node [anchor=north west][inner sep=0.75pt]  [rotate=-151.6,xslant=0.07] [align=left] {$\displaystyle \infty $};

\end{tikzpicture}}}} = e^{-S_0} \times \vcenter{\hbox{
        \scalebox{0.5}{\tikzset{every picture/.style={line width=0.75pt}} %set default line width to 0.75pt        

\begin{tikzpicture}[x=0.75pt,y=0.75pt,yscale=-1,xscale=1]
%uncomment if require: \path (0,199); %set diagram left start at 0, and has height of 199

%Shape: Circle [id:dp42243780255729046] 
\draw  [line width=2.25]  (20.5,84.86) .. controls (20.5,51.86) and (47.25,25.11) .. (80.25,25.11) .. controls (113.25,25.11) and (140,51.86) .. (140,84.86) .. controls (140,117.86) and (113.25,144.61) .. (80.25,144.61) .. controls (47.25,144.61) and (20.5,117.86) .. (20.5,84.86) -- cycle ;
%Curve Lines [id:da5780369674339896] 
\draw [color={rgb, 255:red, 74; green, 144; blue, 226 }  ,draw opacity=1 ][line width=1.5]    (63,39.61) .. controls (64.11,37.62) and (65.71,37.13) .. (67.78,38.14) .. controls (70.04,38.97) and (71.56,38.3) .. (72.34,36.12) .. controls (73.27,33.91) and (74.86,33.31) .. (77.09,34.3) .. controls (78.92,35.59) and (80.56,35.37) .. (82.01,33.62) .. controls (83.88,32.11) and (85.52,32.29) .. (86.95,34.18) .. controls (88.36,36.11) and (90,36.4) .. (91.88,35.07) .. controls (93.87,33.78) and (95.53,34.14) .. (96.85,36.14) .. controls (98,38.12) and (99.59,38.5) .. (101.62,37.29) .. controls (103.63,36.08) and (105.27,36.52) .. (106.56,38.59) .. controls (107.63,40.62) and (109.21,41.1) .. (111.31,40.01) .. controls (113.6,39.12) and (115,39.92) .. (115.51,42.41) .. controls (115.34,44.69) and (116.42,45.97) .. (118.73,46.25) .. controls (121.11,46.76) and (122.03,48.15) .. (121.48,50.42) .. controls (120.92,52.76) and (121.77,54.24) .. (124.02,54.87) .. controls (126.27,55.58) and (127.01,57.03) .. (126.23,59.23) .. controls (125.4,61.4) and (126.09,62.94) .. (128.3,63.84) .. controls (130.45,64.79) and (130.95,66.36) .. (129.78,68.55) .. controls (128.16,70.04) and (127.98,71.68) .. (129.24,73.47) .. controls (130.07,75.69) and (129.39,77.2) .. (127.2,77.99) .. controls (125.13,79.39) and (124.91,81.04) .. (126.52,82.94) .. controls (128.23,84.51) and (128.28,86.19) .. (126.67,87.98) .. controls (125.06,89.71) and (125.1,91.39) .. (126.8,93.02) .. controls (128.48,94.62) and (128.48,96.3) .. (126.8,98.05) .. controls (125.09,99.6) and (125.02,101.25) .. (126.59,103.01) .. controls (128.1,104.81) and (127.9,106.46) .. (126,107.96) .. controls (124.03,109.06) and (123.53,110.61) .. (124.5,112.61) .. controls (124.66,115.06) and (123.49,116.29) .. (120.99,116.3) .. controls (118.67,115.98) and (117.37,116.94) .. (117.09,119.19) .. controls (116.54,121.58) and (115.15,122.51) .. (112.92,121.97) .. controls (110.57,121.5) and (109.12,122.43) .. (108.57,124.76) .. controls (108.18,127) and (106.79,127.9) .. (104.41,127.46) .. controls (102.09,127.03) and (100.76,127.97) .. (100.42,130.28) .. controls (100.27,132.57) and (99.1,133.73) .. (96.9,133.76) .. controls (94.94,133.07) and (93.4,133.73) .. (92.27,135.73) .. controls (90.88,137.6) and (89.22,137.78) .. (87.29,136.25) .. controls (85.67,134.64) and (84.04,134.67) .. (82.4,136.35) .. controls (80.63,137.98) and (78.89,137.93) .. (77.2,136.19) .. controls (75.74,134.44) and (74.09,134.32) .. (72.25,135.85) .. controls (70.42,137.36) and (68.79,137.19) .. (67.36,135.36) .. controls (65.97,133.5) and (64.3,133.26) .. (62.36,134.65) .. controls (60.38,135.96) and (58.8,135.56) .. (57.63,133.45) .. controls (57.32,131.29) and (56.05,130.27) .. (53.8,130.39) .. controls (51.39,130.02) and (50.39,128.68) .. (50.79,126.37) .. controls (51.18,124.02) and (50.21,122.67) .. (47.86,122.3) .. controls (45.55,122.15) and (44.42,120.92) .. (44.48,118.63) .. controls (44.37,116.28) and (43.11,115.15) .. (40.7,115.26) .. controls (38.37,115.43) and (37.15,114.34) .. (37.04,111.99) .. controls (36.99,109.62) and (35.82,108.42) .. (33.54,108.41) .. controls (31.11,108.03) and (30.14,106.66) .. (30.61,104.29) .. controls (31.54,102.27) and (31.02,100.69) .. (29.06,99.54) .. controls (27.22,98.05) and (27.11,96.42) .. (28.72,94.64) .. controls (30.42,93.05) and (30.5,91.39) .. (28.95,89.67) .. controls (27.48,87.78) and (27.7,86.09) .. (29.61,84.62) .. controls (31.56,83.29) and (31.89,81.68) .. (30.61,79.81) .. controls (29.48,77.64) and (29.98,76.01) .. (32.1,74.91) .. controls (34.24,74.06) and (34.9,72.54) .. (34.08,70.36) .. controls (33.12,68.25) and (33.69,66.68) .. (35.79,65.65) .. controls (37.86,64.54) and (38.32,62.95) .. (37.17,60.88) .. controls (36.01,58.81) and (36.46,57.19) .. (38.51,56.01) .. controls (40.56,54.91) and (41.05,53.31) .. (39.98,51.22) .. controls (38.99,49.09) and (39.61,47.59) .. (41.84,46.71) .. controls (44.07,46.27) and (45.1,44.97) .. (44.92,42.8) .. controls (45.69,40.43) and (47.19,39.77) .. (49.44,40.83) .. controls (51.23,42.29) and (52.9,42.16) .. (54.43,40.45) .. controls (56.07,38.76) and (57.74,38.69) .. (59.44,40.26) -- (63.1,39.58) ;
%Shape: Circle [id:dp3367086450874185] 
\draw  [fill={rgb, 255:red, 0; green, 0; blue, 0 }  ,fill opacity=1 ] (78,134.36) .. controls (78,132.57) and (79.46,131.11) .. (81.25,131.11) .. controls (83.04,131.11) and (84.5,132.57) .. (84.5,134.36) .. controls (84.5,136.16) and (83.04,137.61) .. (81.25,137.61) .. controls (79.46,137.61) and (78,136.16) .. (78,134.36) -- cycle ;
%Shape: Circle [id:dp23784629642370003] 
\draw  [fill={rgb, 255:red, 0; green, 0; blue, 0 }  ,fill opacity=1 ] (28,87.36) .. controls (28,85.57) and (29.46,84.11) .. (31.25,84.11) .. controls (33.04,84.11) and (34.5,85.57) .. (34.5,87.36) .. controls (34.5,89.16) and (33.04,90.61) .. (31.25,90.61) .. controls (29.46,90.61) and (28,89.16) .. (28,87.36) -- cycle ;
%Curve Lines [id:da109810771598578] 
\draw [line width=1.5]    (81.25,134.36) .. controls (73,104) and (54,88) .. (31.25,87.36) ;

% Text Node
\draw (74,151.61) node [anchor=north west][inner sep=0.75pt]   [align=left] {$\displaystyle \Lambda $};
% Text Node
\draw (2,79.61) node [anchor=north west][inner sep=0.75pt]   [align=left] {$\displaystyle \Lambda $};
% Text Node
\draw (132.96,24.67) node [anchor=north west][inner sep=0.75pt]  [rotate=-43.36,xslant=0.07] [align=left] {$\displaystyle \infty $};
% Text Node
\draw (30.09,145.57) node [anchor=north west][inner sep=0.75pt]  [rotate=-221.72,xslant=0.07] [align=left] {$\displaystyle \infty $};

\end{tikzpicture}}}} \times \raisebox{-3.7ex}{\hbox{
        \scalebox{0.5}{\tikzset{every picture/.style={line width=0.75pt}} %set default line width to 0.75pt        

\begin{tikzpicture}[x=0.75pt,y=0.75pt,yscale=-1,xscale=1]
%uncomment if require: \path (0,175); %set diagram left start at 0, and has height of 175

%Shape: Circle [id:dp2239097503972005] 
\draw  [line width=2.25]  (23.5,86.68) .. controls (23.5,53.68) and (50.25,26.93) .. (83.25,26.93) .. controls (116.25,26.93) and (143,53.68) .. (143,86.68) .. controls (143,119.67) and (116.25,146.43) .. (83.25,146.43) .. controls (50.25,146.43) and (23.5,119.67) .. (23.5,86.68) -- cycle ;
%Curve Lines [id:da7814280758700054] 
\draw [color={rgb, 255:red, 74; green, 144; blue, 226 }  ,draw opacity=1 ][line width=1.5]    (66,41.43) .. controls (67.11,39.44) and (68.71,38.95) .. (70.78,39.96) .. controls (73.04,40.79) and (74.56,40.11) .. (75.34,37.93) .. controls (76.27,35.72) and (77.86,35.12) .. (80.09,36.11) .. controls (81.92,37.4) and (83.56,37.18) .. (85.01,35.43) .. controls (86.88,33.92) and (88.52,34.1) .. (89.95,35.99) .. controls (91.36,37.92) and (93,38.21) .. (94.88,36.88) .. controls (96.87,35.59) and (98.53,35.95) .. (99.85,37.95) .. controls (101,39.93) and (102.59,40.31) .. (104.62,39.1) .. controls (106.63,37.89) and (108.27,38.33) .. (109.56,40.4) .. controls (110.63,42.43) and (112.21,42.91) .. (114.31,41.82) .. controls (116.6,40.94) and (118,41.74) .. (118.51,44.22) .. controls (118.34,46.5) and (119.42,47.78) .. (121.73,48.06) .. controls (124.11,48.57) and (125.03,49.96) .. (124.48,52.23) .. controls (123.92,54.57) and (124.77,56.05) .. (127.02,56.68) .. controls (129.27,57.39) and (130.01,58.84) .. (129.23,61.04) .. controls (128.4,63.21) and (129.09,64.75) .. (131.3,65.65) .. controls (133.45,66.61) and (133.95,68.18) .. (132.78,70.36) .. controls (131.16,71.85) and (130.98,73.49) .. (132.24,75.28) .. controls (133.07,77.5) and (132.39,79.01) .. (130.2,79.8) .. controls (128.13,81.2) and (127.91,82.85) .. (129.52,84.75) .. controls (131.23,86.32) and (131.28,88) .. (129.67,89.79) .. controls (128.06,91.52) and (128.1,93.2) .. (129.8,94.83) .. controls (131.48,96.44) and (131.48,98.11) .. (129.8,99.86) .. controls (128.09,101.41) and (128.02,103.06) .. (129.59,104.82) .. controls (131.1,106.62) and (130.9,108.27) .. (129,109.78) .. controls (127.03,110.87) and (126.53,112.42) .. (127.5,114.43) .. controls (127.66,116.88) and (126.49,118.1) .. (123.99,118.11) .. controls (121.67,117.79) and (120.37,118.75) .. (120.09,121) .. controls (119.54,123.39) and (118.15,124.32) .. (115.92,123.79) .. controls (113.57,123.32) and (112.12,124.24) .. (111.57,126.57) .. controls (111.18,128.81) and (109.79,129.71) .. (107.41,129.27) .. controls (105.09,128.84) and (103.76,129.78) .. (103.42,132.1) .. controls (103.27,134.38) and (102.1,135.54) .. (99.9,135.57) .. controls (97.94,134.88) and (96.4,135.54) .. (95.27,137.54) .. controls (93.88,139.41) and (92.22,139.59) .. (90.29,138.06) .. controls (88.67,136.45) and (87.04,136.49) .. (85.4,138.16) .. controls (83.63,139.79) and (81.89,139.74) .. (80.2,138) .. controls (78.74,136.25) and (77.09,136.14) .. (75.25,137.67) .. controls (73.42,139.17) and (71.79,139) .. (70.36,137.17) .. controls (68.97,135.31) and (67.3,135.07) .. (65.36,136.46) .. controls (63.38,137.77) and (61.8,137.37) .. (60.63,135.27) .. controls (60.32,133.11) and (59.05,132.09) .. (56.8,132.2) .. controls (54.39,131.83) and (53.39,130.49) .. (53.79,128.18) .. controls (54.18,125.83) and (53.21,124.48) .. (50.86,124.11) .. controls (48.55,123.96) and (47.42,122.73) .. (47.48,120.44) .. controls (47.37,118.09) and (46.11,116.97) .. (43.7,117.07) .. controls (41.37,117.24) and (40.15,116.15) .. (40.04,113.8) .. controls (39.99,111.43) and (38.82,110.23) .. (36.54,110.22) .. controls (34.11,109.84) and (33.14,108.47) .. (33.61,106.1) .. controls (34.54,104.08) and (34.02,102.5) .. (32.06,101.35) .. controls (30.22,99.86) and (30.11,98.23) .. (31.72,96.45) .. controls (33.42,94.86) and (33.5,93.2) .. (31.95,91.48) .. controls (30.48,89.59) and (30.7,87.91) .. (32.61,86.44) .. controls (34.56,85.1) and (34.89,83.49) .. (33.61,81.62) .. controls (32.48,79.46) and (32.98,77.83) .. (35.1,76.72) .. controls (37.24,75.87) and (37.9,74.35) .. (37.08,72.17) .. controls (36.12,70.06) and (36.69,68.49) .. (38.79,67.46) .. controls (40.86,66.35) and (41.32,64.76) .. (40.17,62.69) .. controls (39.01,60.62) and (39.46,59) .. (41.51,57.82) .. controls (43.56,56.72) and (44.05,55.12) .. (42.98,53.03) .. controls (41.99,50.9) and (42.61,49.4) .. (44.84,48.52) .. controls (47.07,48.08) and (48.1,46.78) .. (47.92,44.61) .. controls (48.69,42.24) and (50.19,41.58) .. (52.44,42.64) .. controls (54.23,44.1) and (55.9,43.97) .. (57.43,42.26) .. controls (59.07,40.57) and (60.74,40.51) .. (62.44,42.08) -- (66.1,41.39) ;
%Shape: Circle [id:dp28428444549091725] 
\draw  [fill={rgb, 255:red, 0; green, 0; blue, 0 }  ,fill opacity=1 ] (127,91.18) .. controls (127,89.38) and (128.46,87.93) .. (130.25,87.93) .. controls (132.04,87.93) and (133.5,89.38) .. (133.5,91.18) .. controls (133.5,92.97) and (132.04,94.43) .. (130.25,94.43) .. controls (128.46,94.43) and (127,92.97) .. (127,91.18) -- cycle ;
%Shape: Circle [id:dp922828695004555] 
\draw  [fill={rgb, 255:red, 0; green, 0; blue, 0 }  ,fill opacity=1 ] (80.25,35.68) .. controls (80.25,33.88) and (81.71,32.43) .. (83.5,32.43) .. controls (85.29,32.43) and (86.75,33.88) .. (86.75,35.68) .. controls (86.75,37.47) and (85.29,38.93) .. (83.5,38.93) .. controls (81.71,38.93) and (80.25,37.47) .. (80.25,35.68) -- cycle ;
%Curve Lines [id:da3215845294045364] 
\draw [line width=1.5]    (130.25,92.43) .. controls (107,86.81) and (84,70.81) .. (83.5,35.68) ;

% Text Node
\draw (77,4.43) node [anchor=north west][inner sep=0.75pt]   [align=left] {$\displaystyle \Lambda $};
% Text Node
\draw (149,78.43) node [anchor=north west][inner sep=0.75pt]   [align=left] {$\displaystyle \Lambda $};
% Text Node
\draw (135.96,26.48) node [anchor=north west][inner sep=0.75pt]  [rotate=-43.36,xslant=0.07] [align=left] {$\displaystyle \infty $};
% Text Node
\draw (33.09,147.38) node [anchor=north west][inner sep=0.75pt]  [rotate=-221.72,xslant=0.07] [align=left] {$\displaystyle \infty $};

\end{tikzpicture}}}} \,.
\end{equation}
The diagram therefore factorizes into a product of two-point functions, and the entire correlator is reduced to a combinatorial factor, the number of diagrams with $k$ non-crossing worldlines $C_k$,
\begin{equation}
    \frac{1}{n_{\text{BPS}}} \bigl\langle\Tr\bigl(\widehat\Lambda^{2k}\bigr)\bigr\rangle = e^{-kS_0}\bigl\langle\Tr\bigl(\widehat\Lambda^{2}\bigr)\bigr\rangle^k \times C_k = C_k \,.
\end{equation}
In order to count these diagrams, pick one of the insertions $\widehat\Lambda$ in the trace as a reference point. The worldline that connects it to another insertion divides the diagram into two parts, one containing $j-1$ (non-crossing) worldlines and another containing $k-j$ worldlines, with $j$ a number between $1$ and $k$. This implies the recursion relation
\begin{equation}
    C_k = \sum_{j=1}^k C_{j-1}C_{k-j} \,, \qquad C_0 = 1 \,,
\end{equation}
which defines the Catalan numbers
\begin{equation}
    C_k = \frac{1}{k+1}\binom{2k}{k} \,.
\end{equation}
The Catalan numbers are the moments of a Gaussian random matrix whose eigenvalue density obeys the semicircle law,
\begin{equation}
    \rho_{\widehat\Lambda}(\lambda) = \frac{\sqrt{4-\lambda^2}}{2\pi} \quad \implies \quad C_k = \int_{-2}^2 d\lambda \,\rho_{\widehat\Lambda}(\lambda) \lambda^{2k} \,.
\end{equation}
The types of diagrams we considered, coming from pair-wise contractions of insertions on the boundary, are also called chord diagrams.\footnote{These are exactly the chord diagrams that appear in double-scaled SYK, where they also control the dynamics of the Hamiltonian itself \cite{Berkooz:2018jqr,Berkooz:2018qkz,Berkooz:2024lgq}. As noted by \cite{Antonini:2025rmr}, generally in double-scaled SYK each crossing of chords in a diagram is suppressed by a constant factor, and so it can provide a simplified toy model to the case of operators with finite $\Delta$.} The relation between the combinatorics of non-crossing chord diagrams and the moments of random matrices in the planar limit has been appreciated before in the literature \cite{speicher1997free,mingo2017free}. Here we simply review the implications of that for the spectral density and the spectral form factor, to introduce the technique in preparation for the novel computation for the curvature.

But in order to claim that the projected operator is chaotic, we need to show it exhibits eigenvalue repulsion, and moreover, that this repulsion ranges throughout the spectrum. To do so, we consider the double-trace moments, $\bigl\langle \Tr\bigl(\widehat\Lambda^m\bigr)\Tr\bigl(\widehat\Lambda^n\bigr) \bigr\rangle$, which will allow us to compute the spectral form factor
\begin{equation}
    \label{eq:SFF LMRS expansion}
    \text{SFF}\left(t\right) = \bigl\langle\Tr\bigl(e^{i\widehat{\Lambda}t}\bigr)\Tr\bigl(e^{-i\widehat{\Lambda}t}\bigr)\bigr\rangle = \sum_{m,n=0}^{\infty}\frac{\left(it\right)^{m}\left(-it\right)^{n}}{m!n!}\bigl\langle\Tr\bigl(\widehat{\Lambda}^{m}\bigr)\Tr\bigl(\widehat{\Lambda}^{n}\bigr)\bigr\rangle\,.
\end{equation}

The double-trace moments are correlation functions where $m$ operators are inserted on one asymptotic boundary and $n$ operators are inserted on the other. In JT gravity, one first needs to fill in the bulk geometry, and then, as above, connect in a pairwise manner all the operator insertions that belong to the same connected component. Bulk geometries in which the two boundaries are disconnected from one another give the square of the single-trace moments found above. 

The connected contribution, $\bigl\langle \Tr\bigl(\widehat\Lambda^m\bigr)\Tr\bigl(\widehat\Lambda^n\bigr) \bigr\rangle_c \equiv \bigl\langle \Tr\bigl(\widehat\Lambda^m\bigr)\Tr\bigl(\widehat\Lambda^n\bigr) \bigr\rangle -  \bigl\langle \Tr\bigl(\widehat\Lambda^m\bigr)\bigr\rangle \bigl\langle\Tr\bigl(\widehat\Lambda^n\bigr) \bigr\rangle$, is only due to connected geometries, and at large $S_0$ the leading contribution to it comes from the cylinder topology. In the limit\footnote{In the opposite limit, $\Delta\to0$, the connected piece comes from counting all possible diagrams on the cylinder where the two boundaries are connected by at least one worldline. They vanish when $m+n$ is odd, and are equal to $(m+n-1)!!$ when $m$ and $n$ are odd and to $(m+n-1)!!-(m-1)!!(n-1)!!$ when both are even. The connected SFF then behaves as $1-e^{-t^2/2}$, and plateaus after $O(1)$ time.} $\Delta \to \infty$, we again need to sum over all possible non-crossing worldlines connecting the operators. However, there are two simplifications \cite{Boruch:2023trc}: first, if there is no worldline that connects the two boundaries, the diagram vanishes. Second, we should not overcount diagrams that are related to each other by large (super-)diffeomorphisms, and so we can discard non-crossing diagrams where the worldlines (fully) wind around the cylinder. Note that relative cyclic shifts between the two boundaries are allowed. For example, the connected piece of $\langle \Tr\bigl(\widehat\Lambda^2\bigr)\Tr\bigl(\widehat\Lambda^2\bigr)\rangle$ in the limit $\Delta \to \infty$ has exactly two diagrams contributing to it (to leading order in $e^{S_0}$): 
\begin{equation}
    \bigl\langle\Tr\bigl(\widehat\Lambda^2\bigr)\Tr\bigl(\widehat\Lambda^2\bigr)\bigr\rangle_c = \vcenter{\hbox{
        \scalebox{0.5}{\input{Figures/JT_diagrams/JT_cylinder_2pt.txt}}}} + \vcenter{\hbox{
        \scalebox{0.5}{\input{Figures/JT_diagrams/JT_cylinder_2pt_2.txt}}}}\,,
\end{equation}
As before, each diagram involving non-crossing worldlines factorizes into a product of two-point functions, and we are left only with a combinatorial problem: given $m$ insertions on one boundary of the cylinder and $n$ on the other boundary, how many ways are there to pair the insertions to each other by non-crossing chords? In the math literature, these diagrams are known as \emph{annular (non-crossing) pairings} \cite{mingo2017free}. 

To solve the problem, we sum over the number of chords that connect the two boundaries to each other, denoted by $c$. For connected diagrams, $c \ge 1$. Consider now the first boundary, and label the insertions on it as $1,\dots,m$. These connecting chords divide the other $m-c$ insertions into $c$ segments. Since each such segment must connect only to itself, the lengths of the segments must be even and we denote them by $2k_1, \dots, 2k_c$, and we must also have $c \equiv m \equiv n\,\, (\textrm{mod} \,\,2)$.  As we saw above there are $C_{k_i}$ ways to draw the chords within each such segment. For each choice of $\{k_i\}$ there are $m$ ways of choosing the index of the first connecting chord, but $c$ of those ways correspond to a cyclic permutation of the segments, and therefore result in the same pairing. So each choice of $\{k_i\}$ gives rise to $\frac{m}{c}\prod_{i=1}^c C_{k_i}$ different internal pairings, and the overall number of ways to connect the chords to the insertions on the first boundary is
\begin{equation}
    \frac{m}{c}\times\sum_{\substack{k_1,\dots,k_c=0 \\ \sum_{i} k_i = \frac{m-c}{2}}}^{\frac{m-c}{2}}  \prod_{i=1}^c C_{k_i} = \binom{m}{\frac{m-c}{2}} \,,
\end{equation}
which is the Catalan convolution formula. Similarly, there are $\binom{n}{(n-c)/2}$ ways of connecting the chords to the second boundary. We now need to pair the $c$ connecting chords going out of the two boundaries, and there are $c$ ways of doing so without them crossing, due to the relative cyclic permutations between the boundaries. Putting it all together, we can evaluate the overall number of connected diagrams, and therefore the connected moment. The results are more easily presented when dividing them into the case where both $m$ and $n$ are even and to the case when both are odd (the mixed case vanishes),
\begin{align}
    \bigl\langle \Tr\bigl(\widehat\Lambda^{2a}\bigr)\Tr\bigl(\widehat\Lambda^{2b}\bigr) \bigr\rangle_{\rm conn} &= \!\!\!\sum_{r=1}^{\min(a,b)}\!\! 2r\binom{2a}{a-r}\binom{2b}{b-r} = \frac{ab}{a+b}\binom{2a}{a}\binom{2b}{b} \,,\\
    \bigl\langle \Tr\bigl(\widehat\Lambda^{2a+1}\bigr)\Tr\bigl(\widehat\Lambda^{2b+1}\bigr) \bigr\rangle_{\rm conn} &= \!\!\!\sum_{r=0}^{\min(a,b)} \!\!(2r+1)\binom{2a+1}{a-r}\binom{2b+1}{b-r} \nonumber \\
    &= \frac{(a+1)(b+1)}{a+b+1}\binom{2a+1}{a}\binom{2b+1}{b} \,.
\end{align}

\begin{figure}
    \centering
    \includegraphics[width=0.55\linewidth]{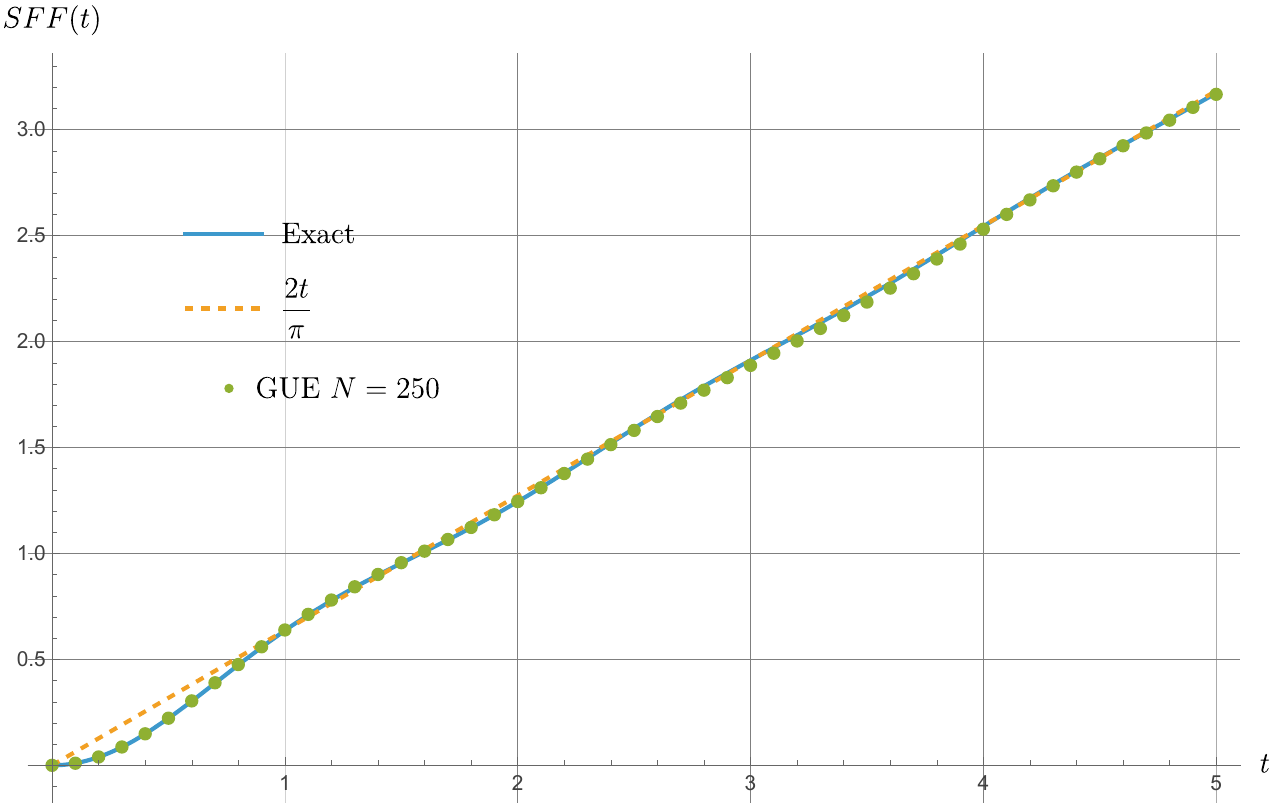}
    \caption{The SFF for an operator of dimension $\Delta\to\infty$. We plot both the exact function \eqref{eq:SFF_LMRS} (blue), the linear trend (orange), and an ensemble average over $2\times 10^4$ samples of GUE matrices of size $250 \times 250$ (green).}
    \label{fig:SFF_LMRS}
\end{figure}

The connected spectral form factor (SFF) is found by substituting the double-trace moments into \eqref{eq:SFF LMRS expansion}. The even moments contribute to the connected part of the SFF as\footnote{We used the following identities with $p = a+b$ and $p = a + b + 1$:
\begin{equation}
    \frac{1}{p} = 2\int_{0}^{1}du\,u^{2p-1} \,, \qquad -utJ_{1}\left(2ut\right)=\sum_{a=1}^{\infty}\frac{a\left(itu\right)^{2a}}{\left(a!\right)^{2}} \,, \qquad J_0(2ut) = \sum_{a=0}^{\infty}\frac{\left(iut\right)^{2a}}{\left(a!\right)^{2}} \,.
\end{equation}}
\begin{equation}
     \int_{0}^{1}\frac{2du}{u}\left(\sum_{a=1}^{\infty}\frac{a\left(itu\right)^{2a}}{\left(a!\right)^{2}}\right)\left(\sum_{b=1}^{\infty}\frac{b\left(-itu\right)^{2b}}{\left(b!\right)^{2}}\right)
    = 2t^{2}\int_{0}^{1}udu\,J_{1}^2\left(2ut\right) \,,
\end{equation}
and that of the odd moments as
\begin{equation}
    t^{2}\int_{0}^{1}2udu\left(\sum_{a=0}^{\infty}\frac{\left(iut\right)^{2a}}{\left(a!\right)^{2}}\right)\left(\sum_{b=0}^{\infty}\frac{\left(iut\right)^{2b}}{\left(b!\right)^{2}}\right) = 2t^{2}\int_0^1 udu\,J_{0}^2\left(2ut\right) \,,
\end{equation}
and so overall 
\begin{equation}
    \label{eq:SFF_LMRS}
    \text{SFF}_{\text{conn}}\left(t\right) = 2t^{2}\int_{0}^{1}udu\left(J_{0}^{2}\left(2ut\right)+J_{1}^{2}\left(2ut\right)\right) = 2t^{2}\left(J_{0}^{2}\left(2t\right)+J_{1}^{2}\left(2t\right)\right)-tJ_{0}\left(2t\right)J_{1}\left(2t\right) \,,
\end{equation}
which converges to $\frac{2t}{\pi}$ after times of $O(1)$. Both the function and the linear trend are plotted in Figure~\ref{fig:SFF_LMRS}.

\subsection{The Berry curvature}
\label{sec:JT Berry comp}
Let us now move on to the computation of the Berry curvature \eqref{eq:Berry matrix UV commuting}, i.e. its moments $\langle\Tr(F_{\mu\nu}^n)\rangle$ and double-trace moments $\langle\Tr(F_{\mu\nu}^m)\Tr(F_{\mu\nu}^n)\rangle$, which can be computed via correlation functions on the disk and on the cylinder, respectively. As before, these correlation functions would come from summing over all worldlines that connect two insertions of one of the operators. 
We assume that both operators have the same dimension $\Delta$ and take the limit\footnote{In the opposite limit $\Delta \to 0$, crossed worldlines contribute with the same weight as non-crossing ones, as mentioned above. In that case, the moments are symmetric to exchanging the order of any two operators inside a trace as, diagram by diagram, it will only change the number of crossings. But then each $F_{\mu\nu} \propto [\widehat\Lambda_\mu,\widehat\Lambda_\nu]$ contributes two terms that exactly cancel each other, and so the Berry curvature, together with all its single- and double-trace moments, precisely vanishes. Such light operators, when they exist, also do not show any chaos according to the LMRS criterion as their SFF plateaus immediately.} where $\Delta \to \infty$. As before, it allows us to consider only non-crossing worldlines, which simplifies the analysis. However, it also has a conceptual consequence: since crossed worldlines are suppressed, we have
\begin{equation}
    \frac{\Tr\bigl(\widehat\Lambda_\mu \widehat\Lambda_\nu \widehat\Lambda_\mu \widehat\Lambda_\nu\bigr)}{\Tr\bigl(\widehat\Lambda_\mu \widehat\Lambda_\mu \widehat\Lambda_\nu \widehat\Lambda_\nu\bigr)} \xrightarrow{\Delta\to\infty} 0 \,,
\end{equation}
and, in general, any configuration where there are crossed worldlines of the two types is suppressed. This means that the projected matrices $\widehat\Lambda_\mu$ and $\widehat\Lambda_\nu$ are \emph{free}, in the sense of free probability theory \cite{mingo2017free}.

\paragraph{The spectral density.}
As before, we will compute the spectral density by considering the moments
\begin{equation}
    M_{n} \equiv \frac{1}{n_{\textrm{BPS}}}\bigl\langle\Tr(F_{\mu\nu}^{n})\bigr\rangle = \frac{1}{n_{\textrm{BPS}}}i^{n} \bigl\langle\Tr\bigl(\bigl[\widehat{\Lambda}_{\mu},\widehat{\Lambda}_{\nu}\bigr]^{n}\bigr)\bigr\rangle \,.
\end{equation}
When $n$ is odd the moment vanishes due to the cyclicity of the trace. For even $n=2k$, we need to compute a sum over $2n$-point functions. On the gravity side, each correlator is given by a sum over contractions between operators of the same kind. Diagrammatically, contractions are denoted by stretching a chord between the two operators. When $\Delta\to\infty$ all crossed diagrams are exponentially suppressed as before, and only uncrossed diagrams contribute. By the same argument as above, uncrossed diagrams decompose into products of two-point functions. We thus need to count (with sign) how many uncrossed diagrams there are in $M_{2k}$.

To do so, we expand $\Tr([\widehat{\Lambda}_{\mu},\widehat{\Lambda}_{\nu}]^{2k})$ into a sum over single-traces of the form $\Tr(w_{1}\cdots w_{2k})$, where $w_{i}\in\bigl\{\widehat{\Lambda}_{\mu}\widehat{\Lambda}_{\nu},(-\widehat{\Lambda}_{\nu}\widehat{\Lambda}_{\mu})\bigr\}$. Each such single-trace contributes as the number of uncrossed diagrams, with an overall sign related to the number of $\widehat{\Lambda}_{\nu}\widehat{\Lambda}_{\mu}$ in it. Overall, we find 
\begin{equation} M_{2k}=(-1)^{k}\sum_{w_{1},\dots,w_{2k}}(-1)^{\#(\widehat{\Lambda}_{\nu}\widehat{\Lambda}_{\mu})}\#_{{\rm uncrossed}}(\Tr(w_{1}\cdots w_{2k}))\,. \end{equation} 
In any uncrossed configuration there must be an even number of operator insertions between the two ends of each chord. That means that each chord connects the first letter of one $w_{i}$ to the second letter of $w_{j}$. Thus the entire diagram must have the same number of $\widehat{\Lambda}_{\mu}\widehat{\Lambda}_{\nu}$ pairs as $\widehat{\Lambda}_{\nu}\widehat{\Lambda}_{\mu}$ pairs, giving an overall factor of $(-1)^{k}$, which cancels against the overall sign to guarantee that $M_{2k}$ is always positive. 

\begin{figure}
    \centering
    \includegraphics[width=0.5\linewidth]{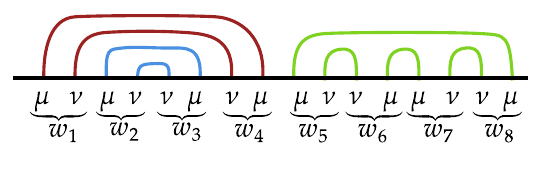}
    \caption{Example of decomposition of a diagram into cycles, each cycle drawn in a different color.}
    \label{fig:example uncrossed diagram}
\end{figure}

Now define a cycle to be a set of $\{w_{i}\}$ that are connected to each other, and to no other $w$, by chords. We demonstrate the decomposition to cycles in a single diagram with an example in Figure~\ref{fig:example uncrossed diagram}. 
Each (non-crossing) cycle contains an even number of $w$'s, and they must alternate between $\widehat{\Lambda}_{\mu}\widehat{\Lambda}_{\nu}$ and $\widehat{\Lambda}_{\nu}\widehat{\Lambda}_{\mu}$ along the cycle. Thus, each cycle gives rise to two configurations, related to each other by flipping all pairs together. Denoting by $d_{k,c}$ the number of cycles with $2k$ $w$'s and $c$ cycles, we find that the moments are
\begin{equation}
    M_{2k}=\sum_{c=1}^{k}2^{c}d_{k,c}\,.
\end{equation}
Now denote by $D\left(x,t\right)=\sum_{k,c}d_{k,c}x^{2k}t^{c}$ the generating function for $d_{k,c}$, where $x$ and $t$ are the fugacities associated with the number of pairs and with the number of cycles, respectively. The generating function for $M_{2n}$, our original moments, is simply $M\left(x\right)=\sum_{k}M_{2k}x^{2k}=D\left(x,2\right)$. We can find the generating function by summing over the length $2\ell$ of the cycle containing $w_{1}$. In that case, that cycle contributes a factor of $x^{2\ell}t$, while the elements belonging to the cycle divide the trace into $2\ell$ other segments, each of which hosts a sub-diagram of non-crossing chords, generated by the same generating function $D\left(x,t\right)$, and therefore
\begin{equation}
    D\left(x,t\right)=1+\sum_{\ell=1}^{\infty}x^{2\ell}tD^{2\ell}\left(x,t\right)=1+\frac{x^{2}tD^{2}\left(x,t\right)}{1-x^{2}D^{2}\left(x,t\right)}\,,
\end{equation}
where the $1$ represents the empty cycle. This implies that the generating function $M(x)$ satisfies
\begin{equation}
    \label{eq:M(x) implicit}
    M(x) = \frac{1 + x^2 M^2(x)}{1 - x^2 M^2(x)} \,.
\end{equation}
Now note that $d\left(x,t\right)=D\left(x,t\right)-1$ is also a generating function for $d_{k,c}$ when $k > 1$, and it satisfies
\begin{equation}
    x^{2}=\frac{d}{\left(d+1\right)^{2}\left(d+t\right)}\,.
\end{equation}
This form allows us to use the Lagrange Burmann formula, which says that if some function $u(x)$ is implicitly defined by $u = x \phi(u)$, and we want to find the coefficient of $x^n$ in the series expansion of some function $F$ of $u$, which will be denoted $[x^n]F(u)$, then
\begin{equation}
    \label{eq:Lagrange Burmann}
    [x^n]F(u) = \frac{1}{n}[u^{n-1}]\bigl(F'(u)\phi(u)^n\bigr) \,.
\end{equation}
In our case, $u = d$, $F$ is the identity, and $\phi(d) = (d+1)^2(d+t)$, and so
\begin{equation}
    \sum_{c=1}^{k}d_{k,c}t^{c}=[x^{2k}]d=\frac{1}{k}[d^{k-1}]\left(\left(d+1\right)^{2k}\left(d+t\right)^{k}\right)=\frac{1}{k}\sum_{c=1}^{k}\binom{2k}{c-1}\binom{k}{c}t^{c} \,,
\end{equation}
and finally the moments are
\begin{equation}
    M_{2k}=\frac{1}{k}\sum_{c=1}^{k}2^{c}\binom{2k}{c-1}\binom{k}{c}\,.
\end{equation}
These are the moments of the free commutator of two matrices with semi-circle spectral densities \cite{nica1998commutators}, and the spectral density of $F_{\mu\nu}$ is
\begin{equation}\label{rhocomm}
    \rho_{\rm comm}(\lambda) = \frac{\sqrt{3}}{2\pi|\lambda|}\left(h(\lambda) - \frac{3\lambda^2+1}{9h(\lambda)}\right) \,, \qquad h(\lambda) = \left(\frac{18\lambda^2+1}{27} + \sqrt{\frac{\lambda^2(1+11\lambda^2-\lambda^4)}{27}}\right)^{\frac13} \,,
\end{equation}
where the spectrum has compact support in the interval $|\lambda| \le \sqrt{\frac12\left(11+5\sqrt{5}\right)}$. Even though not very easy to see from \eqref{rhocomm}, one can verify that $\rho_{\rm comm}$ also has a square-root edge, similar to that of an ordinary random matrix. 
\begin{figure}
    \centering
    \includegraphics[width=0.55\linewidth]{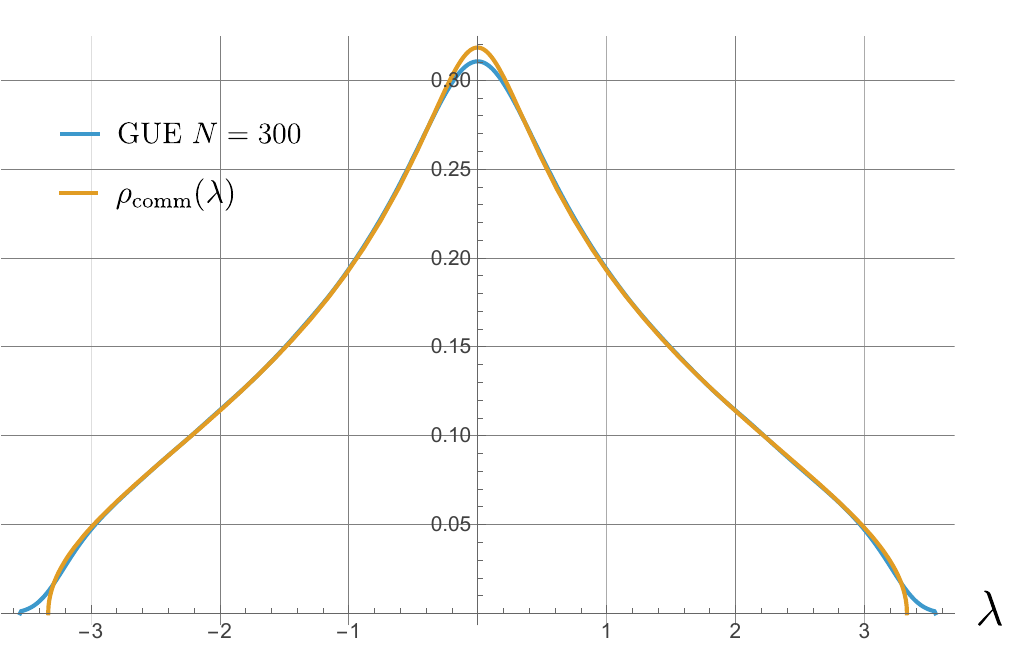}
    \caption{The spectral density of $F_{\mu\nu} = i[\widehat{\Lambda}_\mu,\widehat{\Lambda}_\nu]$, given in \eqref{rhocomm} (orange) and by averaging the commutator of two independent $300 \times 300$ GUE matrices over 200 samples (blue).}
    \label{fig:desnity_of_states_comm}
\end{figure}

\paragraph{The connected double-trace moments.}
Let us now consider the double-trace moments for the Berry curvature, $\left\langle \Tr\left(F_{\mu\nu}^{m}\right)\Tr\left(F_{\mu\nu}^{n}\right)\right\rangle = i^{m+n}\left\langle \Tr\left(\left[\widehat{\Lambda}_{\mu},\widehat{\Lambda}_{\nu}\right]^{m}\right)\Tr\left(\left[\widehat{\Lambda}_{\mu},\widehat{\Lambda}_{\nu}\right]^{n}\right)\right\rangle$. We can track the factor $i^{m+n}$ by assuming that every pair $w$ carries a factor of $-i$.  Just like before, the computation factorizes into counting the number of (connected) non-crossing diagrams on the cylinder.\footnote{We note that \cite{george2025second} seems to compute a related quantity. However, in our context, it seems not to account for possible relative cyclic shifts between the traces, such that even for $m=n=1$ where the traces identically vanish they find a non-zero answer.} Since the operators can now be different from each other, we need to track their ordering. We choose the convention that as one goes along the boundary, the interior should be to one's left, and the operators on both boundaries are ordered along that direction, i.e. 
\begin{equation}
    \label{eq:cyl ordering}
    \langle \Tr(A_1 A_2 \cdots A_n) \Tr(B_1 B_2 \cdots B_m) \rangle_{\rm cyl} = \vcenter{\hbox{
        \scalebox{0.5}{\input{Figures/JT_diagrams/JT_operator_ordering_cylinder.txt}}}} 
    \,.
\end{equation}

Let us first concentrate on the chords coming from one of the traces, i.e. from one of the boundaries of a cylinder. The chords that connect that boundary to the other separate the boundary into segments. Since the number of nodes between any two connecting chords must be even, the total number of connecting chords, $2r$, must be an even number. Each segment is bounded by two connecting chords, and there are therefore $2r$ segments. 

\begin{figure}
    \centering
    \includegraphics[width=0.5\textwidth]{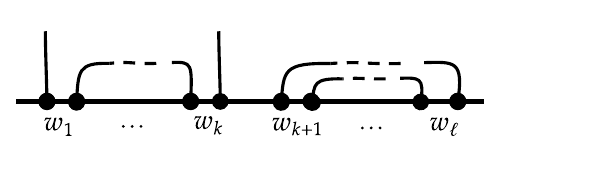}        
    \caption{Open segment (left) followed by a closed segment (right) for the half-cylinder.}
    \label{fig:open closed seg}
\end{figure}

Moreover, for each segment to have an even number of nodes, the connecting chords must alternate between the first and the second position within the pairs $w$. We will denote the pairs in a segment of length $k$ by $w_{1},\dots,w_{k}$, and the two elements in each pair as $w_{i}^{\left(1\right)}$, $w_{i}^{\left(2\right)}$. We view the pairs as ordered from left to right, such that the order of the elements is $w_{1}^{\left(1\right)}w_{1}^{\left(2\right)}\cdots w_{k}^{\left(1\right)}w_{k}^{\left(2\right)}$. There are two types of segments: open and closed, illustrated in Figure~\ref{fig:open closed seg}. In an open segment, the first connecting chord bounding the segment is attached to $w_{1}^{\left(1\right)}$, and the second connecting chord bounding the segment is attached to $w_{k}^{\left(2\right)}$. The two pairs $w_{1}$ and $w_{k}$ must therefore belong to the same cycle, which we call the outer cycle. The outer cycle (which could also contain other pairs $w_i$ with $1 < i < k$) must ultimately connect to an open segment on the other boundary of the cylinder.  In a closed segment, the connecting chords bounding it do not belong to the segment itself, but to the segments bordering it. They attach to the second element of the pair before $w_{1}$, $w_{0}^{\left(2\right)}$, and to the first element of the pair after $w_{k}$, $w_{k+1}^{\left(1\right)}$. 

The pairs within the closed segment can only connect to each other, and not with any other segment. The closed segment must then have an even number of pairs, and the generating function for all diagrams in a closed segment is exactly the same as for the disc, $M(x)$, implicitly defined in \eqref{eq:M(x) implicit}. 

The generating function for an open segment is denoted by $O(x,t)$, where $t$ is a fugacity associated with the number of pairs in the outer cycle, and $x$ with the number of pairs in the inner cycles. It counts the number of diagrams that can appear due to the inner cycles, but, since the outer cycle remains open until we glue it to the other side of the cylinder, it will only count the number of different ways to build the cycle, without summing over the possibilities of assigning values to the pairs ($w$'s) in the outer cycle. 

The outer cycle must have at least one pair in it, $w_1$ (it could happen that $w_k = w_1$). Each additional pair that belongs to the outer cycle can be separated from the previous pair by an inner cycle of some length. It would thus contribute to the generating function a factor of $t$ times the generating function for the number of diagrams in an inner segment, which is simply $M(x)$. We thus have
\begin{equation}
    \label{eq:O(x,t) def}
    O\left(x,t\right) = \sum_{j=0}^{\infty}t^{1+j}M^{j}\left(x\right)=\frac{t}{1-tM\left(x\right)}\,.
\end{equation}
We will also use the abbreviated notation\footnote{Note that $M\left(x\right)$ is an even function, $M\left(x\right)=M\left(-x\right)$, and so $O(-x) = O(-x,-x) = O(x,-x)$.} $O\left(z\right)\equiv O\left(z,z\right)$. 

Let us consider the generating function for $\left\langle \Tr\left(F_{\mu\nu}^{m}\right)\Tr\left(F_{\mu\nu}^{n}\right)\right\rangle$, i.e. for the connected diagrams on the entire cylinder, which we denote by $A\left(x,y\right)$, with the two fugacities accounting for the number of nodes in each of the two traces, by gluing the two half cylinders together and summing over the number of connecting chords $2r$. Each half of a cylinder will have $r$ open segments, each with generating function $O(x,t)$, and $r$ closed segments, each with generating function $M\left(x\right)$. 

\begin{figure}
    \centering
    \includegraphics[width=0.3\linewidth]{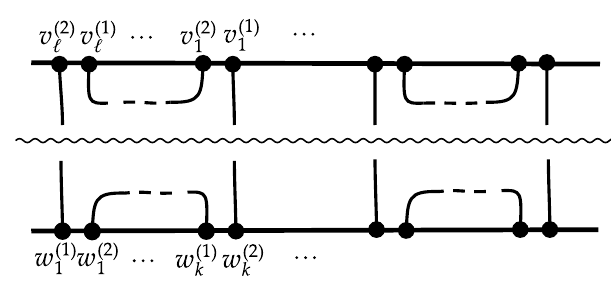}
    \hfill
    \includegraphics[width=0.3\linewidth]{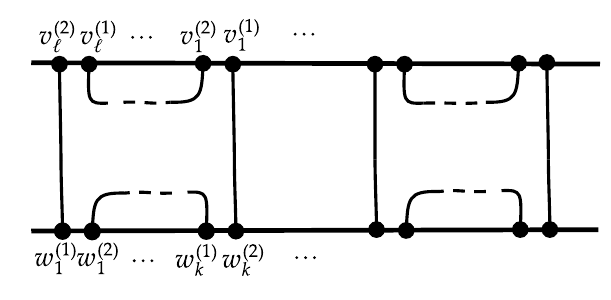}
    \hfill
    \includegraphics[width=0.3\linewidth]{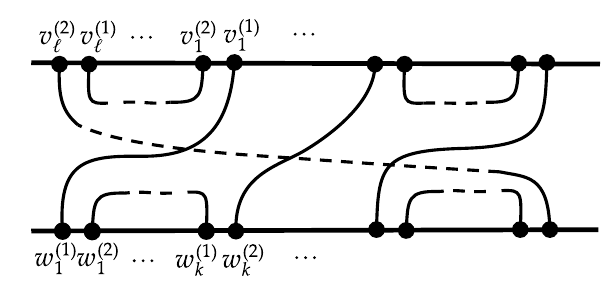}
    \caption{Gluing of two half-cylinders (left). We only draw the (two) open segments. The first type of gluing (middle) and second type of gluing (right). The dashed line on the right diagram is passing through the back of the cylinder, without crossing any other chord.}
    \label{fig:cylinder gluing}
\end{figure}

The gluing of the two cylinders can be done in two distinct ways, one way creates $r$ distinct cycles, and another way creates one cycle. Consider an open segment containing the pairs $w_{1},\dots,w_{k}$ in one half, and another open segment containing the pairs $v_{1},\dots,v_{\ell}$ in the other half. The first way of gluing them directly closes a cycle between the two open segments we considered, i.e. it glues the first element $w_{1}^{\left(1\right)}$ in one segment to the last element $v_{\ell}^{\left(2\right)}$ in the other (remember the ordering of the insertions along the two boundaries, as in \eqref{eq:cyl ordering}), and the last element $w_{k}^{\left(2\right)}$ in one segment to the first element $v_{1}^{\left(1\right)}$ in the other. The second way glues the first element $w_{1}^{\left(1\right)}$ in one segment to the first element $v_{1}^{\left(1\right)}$ in the other, and the last element $w_{k}^{\left(2\right)}$ in the first segment to the last element in the next segment on the other boundary, $v_{\ell'}^{\left(2\right)}$ on the other boundary. The gluings are illustrated in Figure~\ref{fig:cylinder gluing}.

Let us now determine the sign coming from an outer cycle (one that connects the two boundaries) with $p_{1}$ nodes on one boundary and $p_{2}$ nodes on the other, and assume $p_{1}\ge p_{2}$. First, remember that the cycle carries a factor of $i^{p_{1}+p_{2}}$ since each pair carries a factor of $i$. Once we fix the identity of one of the pairs in a cycle, say $w_{1}$, the other elements are uniquely determined. Within each boundary, the identity of the pairs in the outer cycle alternates between the two options. The sum $p_{1}+p_{2}$ must be even.\footnote{Otherwise for one choice of $w_{1}$ the number of $\widehat{\Lambda}_{\mu}\widehat{\Lambda}_{\nu}$ pairs is even and the number of $-\widehat{\Lambda}_{\nu}\widehat{\Lambda}_{\mu}$ pairs is odd, and for the other choice vice versa, so the two choices will give opposite signs and cancel one another.} Along each boundary, adjacent pairs belonging to the same cycle alternate. In the first way of gluing, the pairs on opposite sides of the connecting chords are the opposite of each other, and so they also alternate between boundaries. Since there are even number of pairs in the cycle, half of them come with a minus sign, and the overall sign is $i^{p_{1}+p_{2}}\left(-1\right)^{\frac{p_{1}+p_{2}}{2}}=1$. For the second type of gluing, the pairs on opposite sides of the connecting chords are the same, and therefore the ordering of the pairs is the same on both boundaries, and the first $p_{2}$ terms just cancel each other's sign. The sign must come from the last $p_{1}-p_{2}$ pairs on the first boundary. Since they alternate between one another, and there is an even number of them, the sign is $i^{p_{1}+p_{2}}\left(-1\right)^{\frac{p_{1}-p_{2}}{2}}=\left(-1\right)^{p_{1}}$. 

The generating function with $2r$ connecting chords that implements these gluings is as follows. First, each half cylinder has $r$ closed segments, whose generating functions give $M^{r}(x)M^{r}(y)$. Each half cylinder also has $r$ open segments, each with a generating function of $O(x)$. In order to account for a possible sign of $\left(-1\right)^{p}$ where $p$ is the number of pairs in the outer cycle we will need to replace it by $O(-x)$. The first way of gluing the two half cylinders pairs these open segments into $r$ additional cycles. Each such cycle has an even number of pairs overall. The generating function for two such glued open cycles is thus $\frac{O\left(x\right)O\left(y\right)+O\left(-x\right)O\left(-y\right)}{2}$, which projects out the cases where the sum of the powers of $x$ and $y$ is odd. There is also an extra factor of $2$ for the two possible choices of the identity of the pairs in the open cycle. Overall, the first way of gluing is generated by
\begin{equation}
    \left(2\times\frac{O\left(x\right)O\left(y\right)+O\left(-x\right)O\left(-y\right)}{2}\right)^{r}M^{r}\left(x\right)M^{r}\left(y\right)\,.
\end{equation}
In the second way of gluing all $2r$ open segments connect into a single cycle (with no additional sign). Accounting for the sign, this gluing is generated by $O^{r}(-x)O^{r}(y)$. Projecting out the cases where the outer cycle has an odd number of elements and accounting for the two options to choose the pairs in the cycle, the second way of gluing is generated by
\begin{equation}
    \left(O^{r}\left(-x\right)O^{r}\left(y\right)+O^{r}\left(x\right)O^{r}\left(-y\right)\right)M^{r}\left(x\right)M^{r}\left(y\right)\,.
\end{equation}
When putting it all together, we must remember that so far we have only counted the diagrams up to cyclic permutations of each trace. In other words, we assumed that there are $2r$ connecting chords, and summed over the number of pairs that could separate the first and the second connecting chords, the second and the third, and so on, but we haven't decided which pair $w_{i}$ has the first connecting chord. When computing $\left\langle \Tr\left(F_{\mu\nu}^{m}\right)\Tr\left(F_{\mu\nu}^{n}\right)\right\rangle$, there are $m$ such choices for the first trace, multiplied by $n$ such choices for the second trace. These extra factors are incorporated by acting on the generating function with the operator $x\partial_{x}y\partial_{y}$. However, there are exactly $r$ cases where the shifts are synchronized such that there is no relative shift, and they amount to the same division of open and closed segments on both traces. These $r$ cases do not amount to a new pairing, and so we should explicitly divide by $r$ to avoid overcounting. Finally, we can put it all together and write
\begin{multline}
    \label{eq:A sum over r}
    A\left(x,y\right) = x\partial_{x}y\partial_{y} \sum_{r=1}^{\infty}\frac{M^{r}\left(x\right)M^{r}\left(y\right)}{r}\Big[\bigl(O\left(x\right)O\left(y\right) + O\left(x\right)O\left(y\right)\bigr)^{r} \\
    + O^{r}\left(-x\right)O^{r}\left(y\right) + O^{r}\left(x\right)O^{r}\left(-y\right)\Big]\,.
\end{multline}
After performing the sum over $r$, denoting $u(x) \equiv M(x)O(x)$, $v(y) \equiv M(y)O(y)$, and expressing the derivatives as $\partial_x = u'(x)\partial_u$, $\partial_y = v'(y)\partial_v$, we find
\begin{equation}
    A(x,y) = 2xyu'v'\left[-\tfrac{1-u^{2}v^{2}+2uv+\left(u+v\right)^{2}}{\left(1-\left(u+v\right)^{2}-u^{2}v^{2}\right)^{2}}+\tfrac{1+\left(u-v\right)^{2}}{\left(1-\left(u-v\right)^{2}\right)^{2}}\right] \,.
\end{equation}
This equips us with all the necessary information to compute the spectral form factor of the Berry curvature, which we now turn to.

\begin{figure}
    \centering
    \includegraphics[width=0.35\linewidth]{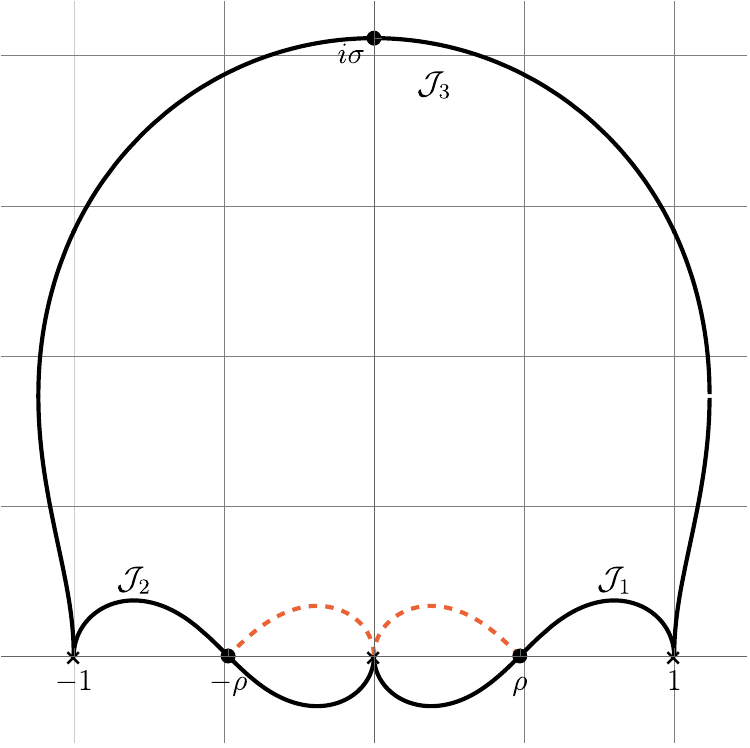}
    \caption{The Lefschetz thimbles needed for the saddle point approximation of \eqref{eq:SFF comm integral}. In dotted red line, the poles \eqref{eq:SFF poles} that are inside the integration contour.}
    \label{fig:Lefschetz SFF}
\end{figure}

\paragraph{The spectral form factor.}
Remember that $A(x,y) = \sum_{n,m=0}^{\infty} a_{m,n} x^{m}y^{n}$ is the generating function of the (connected) double-trace moments ${a_{m,n} = \left\langle \Tr\left(F_{\mu\nu}^{m}\right)\Tr\left(F_{\mu\nu}^{n}\right)\right\rangle_\text{c}}$, but we are interested in the spectral form factor \eqref{eq:SFF moments}. Fortunately, the latter can be related to $A$ via the integral transform\footnote{Around $x=y=0$ the functions $u$ and $v$ are analytic and  $u = v = 0$ at the origin, as can be inferred from \eqref{eq:M(x) implicit} and \eqref{eq:O(x,t) def}. Therefore $A(x,y)$ is also analytic there, and we can use Cauchy's residue formula to write
\begin{equation*}
    a_{m,n} = \frac{1}{\left(2\pi i\right)^{2}}\oint\frac{dx}{x}\oint\frac{dy}{y}\,\frac{A\left(x,y\right)}{x^{m}y^{n}} \,,
\end{equation*}
where the contours encircle the origin and contain no other poles. Substituting this into the definition of $\mathrm{SFF}_c$ and summing the series gives the integral formula in \eqref{eq:SFF integral xy}.}
\begin{equation}
    \label{eq:SFF integral xy}
    \text{SFF}_{\text{c}}\left(t\right) = \frac{1}{\left(2\pi i\right)^{2}}\oint\frac{dx}{x}\oint\frac{dy}{y}A\left(x,y\right)e^{it \left(\frac1x - \frac1y\right)} \,,
\end{equation}
where the two contours encircle the origin and contain no other singularities of the integrand. We would like to show that the spectral form factor grows linearly with $t$. First, we bring it to a more convenient form by changing the integration variables from $x,y$ to $u,v$, 
\begin{equation}
    \label{eq:SFF comm integral}
    \text{SFF}_{\text{c}}\!\left(t\right) =\ \frac{1}{2\pi^{2}}\oint\!du\oint\!dv\left( \tfrac{1+\left(u-v\right)^{2}}{\left(1-\left(u-v\right)^{2}\right)^{2}} - \tfrac{1-u^{2}v^{2}+2uv+\left(u+v\right)^{2}}{\left(1-\left(u+v\right)^{2}-u^{2}v^{2}\right)^{2}} \right) \exp\left[it\left(\tfrac{1+u^{2}}{u\left(1-u^{2}\right)} - \tfrac{1+v^{2}}{v\left(1-v^{2}\right)}\right)\right] \,,
\end{equation}
where it is more amenable to a saddle point approximation. The saddle points for both $u$ and $v$ are at 
\begin{equation}
    \pm \rho = \pm \sqrt{\sqrt{5}-2} \quad \text{and}\quad \pm i \sigma = \pm i \sqrt{2+\sqrt{5}} \,.
\end{equation}
We thus need to deform the integration contour, which starts as a small counterclockwise contour around the origin in each variable, into the Lefschetz thimbles associated with these saddles. Since the exponent gives rise to essential singularities at $u = 0,\pm 1$, and the thimbles connect these singularities, we should deform the contour such that it approaches these singularities from a direction where the integrand is exponentially suppressed. This determines the three thimbles for the $u$ variable, $\mathcal{J}_{1,2,3}$, to be the ones shown in Figure~\ref{fig:Lefschetz SFF}. The thimbles for $v$ are the mirror images of those with respect to the real axis. 

\begin{figure}
    \centering
    \includegraphics[width=0.55\linewidth]{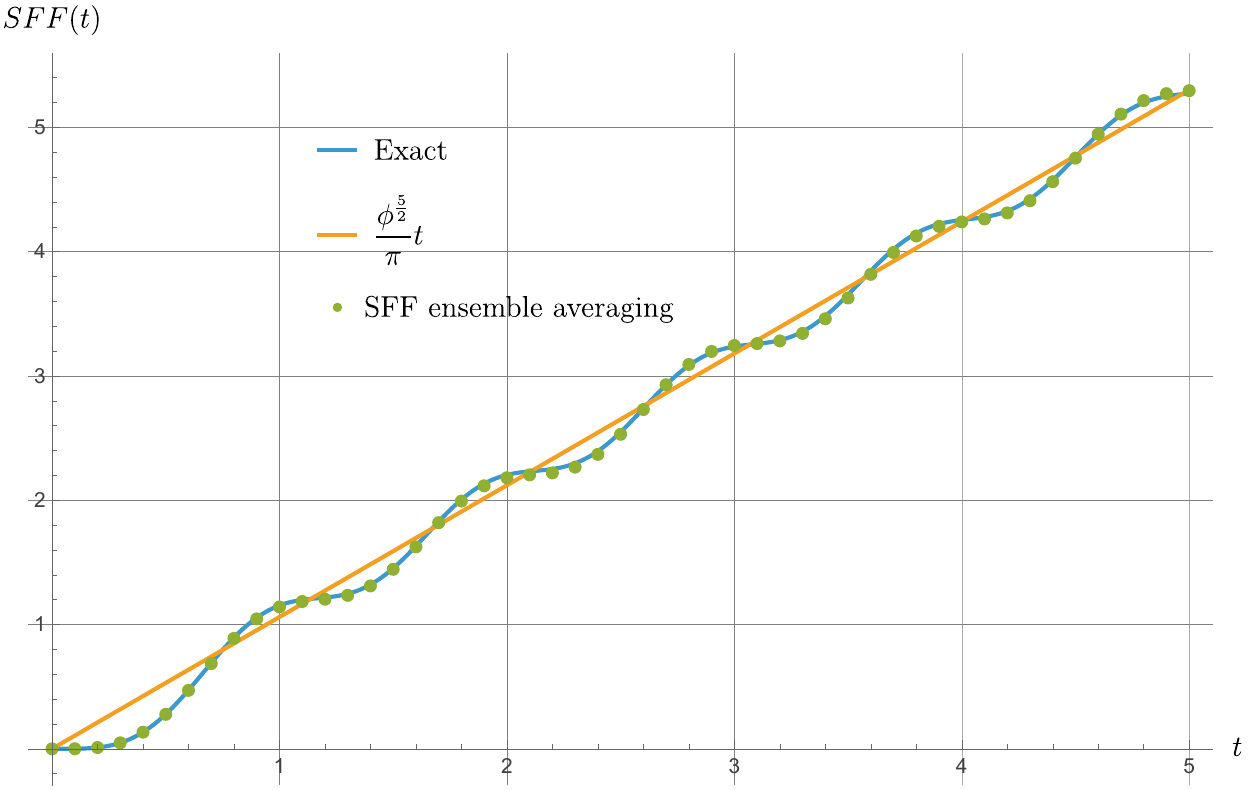}
    \caption{The spectral form factor of the Berry curvature $F_{\mu\nu}$, via numerical evaluation of \eqref{eq:SFF comm integral} (blue), the linear contribution \eqref{eq:SFF comm linear} (orange), and by direct averaging over $2\times 10^4$ samples of the commutator of two independently chosen $200 \times 200$ GUE matrices (green).}
    \label{fig:SFF Berry curvature JT}
\end{figure}

But one needs to be careful when deforming the contours, as the integrand also has poles. In particular, for any fixed $v$, the poles of the second term are at
\begin{equation}
    \label{eq:SFF poles}
    u_{\pm} = \frac{-v\pm\sqrt{1+v^{2}-v^{4}}}{1+v^{2}} \,,
\end{equation}
and they intersect the Lefschetz thimbles precisely at $(u,v) = (\rho,\rho)$ and at $(-\rho,-\rho)$. For any fixed $v$, when deforming the $u$ contour past a pole, we need to add a small clockwise contour around it to cancel its contribution to our original contour. Overall, this creates a small tube that encircles the singular curve, whose endpoints are at $v = \rho$ and $v = -\rho$. Computing the residue due to the $u$ integral and integrating along the curve, we find that the contribution of these poles to the SFF is
\begin{equation}
    \label{eq:SFF comm linear}
    \int_{-\rho}^{\rho}dv\frac{1-4v^{2}-v^{4}}{2\pi v^{2}\left(1-v^{2}\right)^{2}}t = \frac{\phi^{5/2}}{\pi} t \,,
\end{equation}
where $\phi = \frac12(\sqrt{5}+1)$ is the golden ratio. This indeed gives the linear trend of the SFF, as shown\footnote{The numerical evaluation of the integral \eqref{eq:SFF comm integral} also precisely agrees with numerically extracting the first 100 coefficients of $A(x,y)$ from \eqref{eq:A sum over r} using the Lagrange Burmann theorem \eqref{eq:Lagrange Burmann} and computing the SFF using them.} in Figure~\ref{fig:SFF Berry curvature JT}. The contributions from the other saddles are subleading. The spectral form factor thus exhibits linear growth from times that are $O(1)$, and the Berry curvature is strongly chaotic.

We would like to highlight that what we have derived in this subsection is of independent interest in contexts outside the Berry curvature and gravity, particularly for random matrix theory and free probability. What we have achieved, essentially, is to compute the connected correlators of the moments of the commutators of two independent (free) random Hermitian matrices $M_1, M_2$ (drawn from a Gaussian potential), i.e. $\langle \textrm{Tr}([M_1, M_2]^n)\textrm{Tr}([M_1, M_2]^m) \rangle$ as well as the connected spectral form factor of the commutator $[M_1, M_2]$ at times shorter than the plateau time. In Figure~\ref{fig:SFF Berry curvature JT}, we compare our analytic results for the connected spectral form factor with the numerically computed result from averaging over a sample of $2\times10^4$ pairs of independent Gaussian Hermitian random matrices of sizes $200 \times 200$.

%%%%%%%%%%%
% SECTION %
%%%%%%%%%%%
\section{Topological aspects of Berry curvature in $\mathcal{N} = 2$ SYK}
\label{sec:topo}

So far, we have been focusing on the local properties of the Berry curvature $F_{\mu\nu}$, and contrasting its chaotic properties for states of supersymmetric black holes versus those that are not. However, even though a bit tangent to our main discussion, it is also interesting to study the global/topological aspects of Berry curvature, characterized by the quantized Chern numbers when integrating the curvature over non-contractible spheres in the moduli space, see Section \ref{sec:general} for a brief review \cite{Kiritsis:1986re, KAUFMANN2020103892}. This quantization is rather amusing because we found in Section \ref{sec:SYK} that the eigenvalues of the curvature for BPS black hole microstates exhibit random matrix statistics, nonetheless the integral of its trace on closed manifolds must be simply ($2\pi$ times) an integer. 

\subsection{Special loci in the moduli space}
\label{sec:specialloci}

The $\mathcal{N}=2$ SYK model provides an interesting set up to study the topological aspects. Recall that the moduli space is 
\begin{equation}
    \mathcal{M} = \mathbb{C}^{\binom{N}{3}}/U(1) \cong \mathbb{C}\mathbb{P}^{\binom{N}{3}-1} \times (0,\infty)
\end{equation}
up to subtleties described below, whose real dimension is $2\binom{N}{3} -1$.

Inside the moduli space parametrized by the complex three form (modulo a phase) $\{C_3\}$, there are many interesting loci where the BPS spectra differ from those at generic couplings.  As a trivial example, for any value of $N$, there always exists a point
\begin{equation}\label{sing1}
  \mathcal{M}_0: \quad  C_{3} = 0 \,,
\end{equation}
where the Hamiltonian is simply zero and all the states are BPS (zero energy). As we approach this point from a generic direction, the eigenvalues of the Berry curvature, defined for the set of states that are BPS at generic couplings, exhibits a diverging behavior due to the additional zero energy states. The subspace $\mathcal{M}_0$ given by (\ref{sing1}) has zero dimension. There exist some more interesting loci with larger dimensions that one can easily identify. One, which we already mentioned in Section \ref{sec:SYKmonotone}, is when the three form $C_3$ can be decomposed into the wedge product of three one-forms $A_1^{(a)}, a= 1,2,3$, i.e.
\begin{equation}\label{3oneform}
  \mathcal{M}_1 : \quad   C_3 = A_1^{(1)} \wedge A_1^{(2)} \wedge A_1^{(3)}\,. 
\end{equation}
Clearly, we have $\mathcal{M}_0 \subset \mathcal{M}_1$. Away from $\mathcal{M}_0$, the real dimension of $\mathcal{M}_1$ is given by $6N-17$. To see this, we notice that given a $C_3$ of the form (\ref{3oneform}), we can always use a $U(N)$ transformation on the fermions to go to a frame where $C_3 = \alpha e^{1}\wedge e^{2} \wedge e^{3}$, where $\alpha \in \mathbb{R}$ and $e^{1,2,3}$ are the unit vectors in the $1,2,3$ directions. Now, in the neighborhood of this point, any $C_3$ that takes the form (\ref{3oneform}) can be written as $C_3 = (\alpha + \delta \alpha) (e^1 +\delta v^1)\wedge (e^2 + \delta v^2)\wedge (e^3 + \delta v^3)$, where $\delta \alpha \in \mathbb{C}$ and $v^1,v^2,v^3$ are complex vectors that are orthogonal to the hyperplane spanned by $e^{1,2,3}$, as in \eqref{varC3}.  This gives $2 + 3 \times 2( N- 3) =2 (3N -8)$ independent real parameters.  After quotienting by an overall phase rotation, we are left with\footnote{As a simple sanity check, when $N=3$, the coupling always has the form (\ref{3oneform}), and we have $6N-17 = 1$, matching with the single parameter in $|C_{123}|$.}\textsuperscript{,}\footnote{Alternatively, one can derive this dimension formula using the fact that the space of three-forms decomposable into a wedge product of three one-forms (up to an overall complex scaling) is the Pl\"ucker embedding of the Grassmannian $\mathrm{Gr}(3,N)$ inside the projectivized space of three-forms $\mathbb{P}(\Lambda^{3}(\mathbb{C}))$. Using $\dim\mathrm{Gr}(3,N) = 3(N-3)$, multiplying by $2$ to convert from complex to real dimensions, and adding $1$ since we only want to quotient by a phase not an overall complex scaling, we obtain \eqref{eqn:dim_decomposable3form}.} 
\begin{equation}\label{eqn:dim_decomposable3form}
    \dim \mathcal{M}_{1} = 6N -17 \,.
\end{equation}
Along $\mathcal{M}_1$, the system also has more BPS states compared to a generic point in $\mathcal{M}$. Whereas at generic points on the moduli space there is $R$-charge concentration and BPS states are restricted to those in Table \ref{tab:N=2SYK_BPSdegeneracy}, on $\mathcal{M}_1$ there are BPS states at any $R$-charge $r$, like those built from $A^{(a)}_1\wedge V^{(r-1)}$ where $V^{(r-1)}$ is an $r-1$-form orthogonal to $e^{1,2,3}$.

The subspace $\mathcal{M}_1$ can be further embedded in a larger subspace, given by
\begin{equation}\label{2form1form}
  \mathcal{M}_2: \quad   C_3 = B_2 \wedge A_1 \,,
\end{equation}
where $B_2$ is a two form and $A_1$ is a one form.  We can similarly count the dimension of the subspace at a generic point on it by first using a $U(N)$ transformation to bring it to the form $C_3 = e^1 \wedge (\sum_{1 < i<j \leq N} \alpha_{ij} e^{i}\wedge e^j)$ . In the vicinity of this point, we can perturb $C_3$ as $(e^1 + \delta v^1)\wedge (\sum_{1 < i<j \leq N} (\alpha_{ij} + \delta \alpha_{ij}) e^{i}\wedge e^j)$ while preserving the form (\ref{2form1form}), where $\delta v^1$ is orthogonal to $e^1$. This brings $2(N-1) + 2 \times \frac{(N-1)(N-2)}{2} = N(N-1)$ real parameters, and after subtracting one corresponding to the overall phase, we find the dimension to be\footnote{For $N=3$ and $N=4$, the counting needs to be modified. At $N=3$, we have $\mathcal{M}_2 = \mathcal{M}_1 = \mathcal{M}$ and the dimension is one. At $N=4$,  the $3$ components of $\delta v^1$ leads to variations that are proportional to each other, so we need to subtract $2\times 2$ from the general counting, which gives $4^2-4-1-4 = 7$. For $N=4$, we also have $\mathcal{M}_1 = \mathcal{M}_2 = \mathcal{M}$, as we show in Appendix \ref{app:forms}.}  
\begin{equation}
    \dim \mathcal{M}_{2} = N^2 - N -1 \,.
\end{equation}
Similar to $\mathcal{M}_1$, there are also additional BPS states when we are in $\mathcal{M}_2$, such as $A_1 \cdot \psi \ket{0}$ at $r=1$.

Apart from the subspaces $\mathcal{M}_0$, $\mathcal{M}_1$, $\mathcal{M}_2$, that are easy to identify, there could be other loci in $\mathcal{M}$ where the BPS spectra differ from those at generic couplings. Some of these loci could in principle exist not for general $N$, but only when $N$ satisfies some conditions. For instance, when $N = \tilde{N}^2 - 1$, $\tilde{N}\in \mathbb{Z}$, we could choose $C_3$ to be the structure constant of $\textrm{su}(\tilde{N})$, and the model defined this way indeed has a different BPS spectra \cite{Chen:2025sum} (see also \cite{Biggs:2026zir} for another such example). A complete classification of all such loci is an interesting but likely difficult problem that goes beyond the scope of this work.

However, for $N$ even we can give the following criterion, in each charge sector, for the largest submanifolds in which extra BPS states can appear \cite{Chang:2024lxt}. The Hilbert space $\mathcal{H}^{r}$ of fermion number $r$ is isomorphic to constant holomorphic forms of degree $r$ so $\mathcal{H}^{r} \cong \bigwedge^{r}(\mathbb{C})$ with dimension $D_{r} = {N \choose r}$. Then the supercharge acting on the $r$-th fermion sector is a map $Q_{r} : \mathcal{H}^{r} \to \mathcal{H}^{r+3}$, which can be represented by a $D_{r+3} \times D_{r}$ matrix, denoted $M_{r}(C_{3})$. Define
\begin{equation}\label{eqn:Dpdef}
  \tilde{D}_{r} = \sum_{n=0}^{\floor{\frac{r}{3}}}(-1)^{n}D_{r-3n}  
\end{equation}
Then, using the definition of $Q$-cohomology along with the rank-nullity theorem, it is straightforward to prove the following theorem.

\begin{theorem}%\textbf{Theorem.} 
There are no $Q$-cohomology classes for $r < \frac{N-3}{2} \Leftrightarrow \rank(M_{r}(C_{3})) = \tilde{D}_{r}$.
\end{theorem}
\noindent Thus, the condition that there exist BPS states in an $R$-charge de-concentrated sector $r < \frac{N-3}{2}$ is equivalent to $\rank(M_{r}(C)) < \tilde{D}_{r}$, which implies that all $\tilde{D}_{r}$-minors of $M_{r}(C)$ vanish. This gives a complicated set of polynomial equations in the $C_{3}$ that defines, in that charge sector, the largest subspaces of $\mathcal{M}$ in which extra BPS states can appear, illustrating why a deeper understanding of the topology of such spaces is likely difficult. Nevertheless, this immediately implies that the real codimension of these spaces must be at least $4$ since the von Neumann-Wigner theorem discussed in Section \ref{sec:basics} implies that it must be at least $3$, but the polynomial equations defining it are complex so it must have even real codimension. Note, however, that this theorem does not give any information on the non-maximal singular subspaces, i.e., those sub-subspaces where even more BPS states can appear, such as $\mathcal{M}_{1}$ inside of $\mathcal{M}_{2}$.

By particle-hole symmetry, this covers all of the other $R$-charge de-concentrated sectors. Furthermore, the $R$-charge concentrated sectors $r =  \frac{N}{2},\frac{N}{2} \pm 1$ can only have extra BPS states when the de-concentrated sectors  $r = \frac{N}{2} \pm 2$ or $\frac{N}{2} \pm 3$ do because long multiplets must descend to the BPS subspace and states in these multiplets have $R$-charge differing by $3$. Hence it reduces to the analysis above.\footnote{For $N$ odd, $R$-charge de-concentration works differently because, even for generic couplings, there are ``exceptional'' BPS states in the sectors $r = \frac{N \pm 3}{2}$. So new BPS states can appear in these sectors without any $R$-charge de-concentration. This is why the argument only works for $N$ even.} We denote the union of all such singular subspaces by $\mathcal{M}_{\mathrm{sing}}$ and the non-singular moduli space as
\begin{equation}
    \mathcal{M}_{\mathrm{non-sing}} = \left(\mathbb{C}^{{N \choose 3}} \backslash \mathcal{M}_{\mathrm{sing}}\right)/U(1)\,.
\end{equation}

\subsection{Enhanced moduli redundancies and stratification}
\label{sec:enhancedsym}

The removal of these singular moduli is not the end of the story because beyond the global $U(1)$ transformation of the moduli space that rotates all of the couplings by the same phase, there can be subspaces of moduli space where additional $U(1)$ transformations exist so the symmetry group is enhanced to a larger group. We emphasize that these are not symmetries of the theory with a fixed set of couplings, rather they are transformations in moduli space that preserve the Hamiltonian. These transformations, which we will call redundancies, must also be quotiented out to obtain the correct moduli space and leads to a complicated topological structure. Indeed, the resulting space is no longer a manifold but rather a stratified space. We now demonstrate that these are the only possible extra continuous moduli redundancies and determine for which submanifolds they exist along with the redundancy group for each such submanifold.\footnote{Recall that we defined moduli space to mean the space of distinct Hamiltonians $H$, up to discrete symmetries. That is, we quotient by all continuous symmetries of moduli space that leave $H$ invariant, but not discrete ones.} 

First, let us see why the only global redundancy is $U(1)$. Denote the indices appearing in the supercharge by $I = (ijk)$ and write the Hamiltonian as
\begin{equation}
    H = \sum_{I,J}C_{I}^{\dagger}\Omega_{IJ}C_{J}\,, \qquad \Omega_{IJ} = \{(\psi_{i}\psi_{j}\psi_{k})^{\dagger},\psi_{l}\psi_{m}\psi_{n}\}\,.
\end{equation}
To preserve the fact that $H$ is quadratic in $C_{I}$, we assume that any redundancy group $\mathcal{G}$ of $H$ must be a matrix Lie group acting linearly on the $C_{I}$ via matrix multiplication. Taking the trace over the fermion Hilbert space to obtain $\omega_{IJ} = \Tr_{\mathbb{H}_{\mathrm{SYK}}}\Omega_{IJ}$, we see that $\omega_{IJ}$ is a positive Hermitian form which defines a Hilbert space inner product on $\mathbb{C}^{\mathcal{D}}$ where $\mathcal{D} = {N \choose 3}$. For any $U \in \mathcal{G}$ to preserve this inner product for all $C_{I}$, meaning it is a global redundancy, it must be unitary. Hence $\mathcal{G} \subset U(\mathcal{D})$.  Moreover, $U$ is diagonal: because $C_{I}$ can be anything, we have an operator equality $\sum_{K,L}U^{\dagger}_{IK}\Omega_{KL}U_{LI} = \Omega_{II}$ for any $I$, but off-diagonal elements in $U$ would imply some linear combination of the $\Omega_{KL}$ is equal to $\Omega_{II}$, which can never happen since all the $\Omega_{IJ}$ are linearly independent.

We conclude that $U = \diag(e^{i\theta_{1}} \ldots e^{i\theta_{\mathcal{D}}})$. This transforms $H$ as
\begin{equation}
    H \to \sum_{I,J}e^{i (\theta_{J}-\theta_{I})}C_{I}^{\dagger}\Omega_{IJ}C_{J}\,.
\end{equation}
Observe that $\Omega_{IJ} = 0$ if and only if $I \cap J = \emptyset$. So invariance of $H$ requires $\theta_{I}=\theta_{J}$ whenever $I \cap J \neq \emptyset$. But this implies that all $\theta_{I}$ are equal since, even if $I \cap J = \emptyset$, there always exists $K$ such that $I \cap K \neq \emptyset$ and $J \cap K \neq \emptyset$ so $\theta_{I}=\theta_{J}=\theta_{K}$. This proves that $U(1)$ is the only global redundancy. 

\begin{figure}
\begin{center}
\begin{tikzpicture}[
  % diamond-friendly node style
  vertex/.style={circle,draw=black,minimum size=10mm,inner sep=0pt},
  line/.style={draw=black}
]

% ---------- Component 1: subsets of {1,2,6,7} (diamond layout)
\node[vertex] (A) at (0, 2) {$1 \; 2 \; 6$};  % top
\node[vertex] (B) at (2, 0) {$1 \; 2 \; 7$};  % right
\node[vertex] (C) at (0,-2) {$1 \; 6 \; 7$};  % bottom
\node[vertex] (D) at (-2,0) {$2 \; 6 \; 7$};  % left

% edges (K4) — explicit
\draw[line] (A)--(B);
\draw[line] (A)--(C);
\draw[line] (A)--(D);
\draw[line] (B)--(C);
\draw[line] (B)--(D);
\draw[line] (C)--(D);

% ---------- Component 2: subsets of {3,4,5,8} (diamond layout, shifted right)
\node[vertex] (E) at (6, 2) {$3 \; 4 \; 5$};  % top
\node[vertex] (F) at (8, 0) {$3 \; 4 \; 8$};  % right
\node[vertex] (G) at (6,-2) {$3 \; 5 \; 8$};  % bottom
\node[vertex] (H) at (4, 0) {$4 \; 5 \; 8$};  % left

% edges (K4) — explicit
\draw[line] (E)--(F);
\draw[line] (E)--(G);
\draw[line] (E)--(H);
\draw[line] (F)--(G);
\draw[line] (F)--(H);
\draw[line] (G)--(H);

\end{tikzpicture}
\end{center}
\caption{Graph of indices for $N=8$ with two connected components $\mathcal{I}_{1}$ and $\mathcal{I}_{2}$ consisting of all subsets of size $3$ of $\{1,2,6,7\}$ and $\{3,4,5,8\}$, respectively.}
\label{fig:graph}
\end{figure}
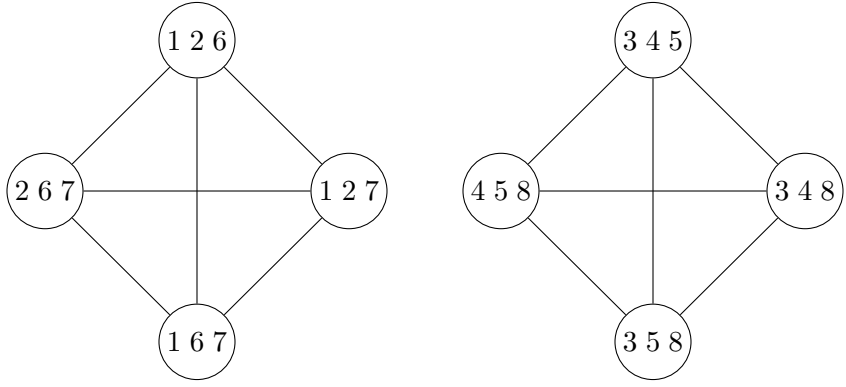

Now, suppose there is a subspace $\mathcal{N} \subset \mathcal{M}$ on which this global redundancy is enhanced to a larger compact Lie group $\tilde{\mathcal{G}} \supset \mathcal{G}$. Then it must have $\mathrm{rank}(\tilde{\mathcal{G}}) \geq 2$ so its maximal torus is at least two-dimensional. Take a $U(1)$ in this maximal torus not equal to the global $U(1)$. Then under this additional $U(1)$, $C_{I} \to e^{i q_{I}\theta}C_{I}$ for $q_{I} \in \mathbb{Z}$ with not all $q_{I}$ equal. Take $C_I$ and $C_J$ that transform with different charges. Then, by the argument above, for any $K$ such that $|K \cap I| \neq \emptyset$ and $|K \cap J|\neq\emptyset$, we must have $C_K = 0$. The fermion indices appearing in the non-vanishing $C$'s must decompose into disjoint components, $\mathcal{I}_1,\dots,\mathcal{I}_k$, and the Hamiltonian can be thought of as a sum over Hamiltonians acting on separate sectors,
\begin{equation}
    H = \sum_{i=1}^{m}H_{(i)} \,, \qquad H_{(i)} = \sum_{I,J \in \mathcal{I}_{i}}C_{I}^{\dagger}\Omega_{IJ}C_{J} \,.
\end{equation}
These connected components of indices $\mathcal{I}_{1},\ldots,\mathcal{I}_{m}$ can be visualized via a graph where we assign a vertex to each index $I$ for which $C_{I} \neq 0$ and attach an edge connecting $C_{I}$ to $C_{J}$ if $I \cap J \neq \emptyset$. Then the connected components of the graph are the $\mathcal{I}_{i}$, as illustrated in Figure \ref{fig:graph}.

We want to determine whether these subspaces $\mathcal{N} \subset \mathcal{M}$ have extra BPS states because if they do, then they are part of the singular subspaces discussed in the previous subsection, which get removed. Consider first the case where $m=2$ so there are only two connected components of indices with $k$ and $N-k$ indices, respectively. Since the two Hamiltonians are decoupled, the BPS states of the full theory are the tensor product of the BPS states from each decoupled theory. We divide the analysis into $N$ even or odd.

\noindent $\boxed{N\;\mathrm{even}}$ If $N$ is even, then $k$ and $N-k$ are both even or both odd. When both are even, we get BPS states with $R$-charge $r=(\frac{(N-k)}{2}-1)+(\frac{k}{2}-1) = \frac{N}{2}-2$ so $R$-charge de-concentration occurs. When both are odd, if either $k$ or $N-k$ is congruent to $1 \pmod 4$, then there are BPS states with $R$-charge $r = (\frac{(N-k-1)}{2})+(\frac{k-3}{2}) = \frac{N}{2}-2$, again giving $R$-charge de-concentration. However, when both $k$ and $N-k$ are congruent to $3 \pmod 4$, then no $R$-charge de-concentration occurs, and hence there are no extra BPS states.

\noindent $\boxed{N\;\mathrm{odd}}$ For $N$ odd, take $k$ odd and $N-k$ even, without loss of generality. If $N \equiv 3 \pmod 4$, then the BPS states with $R$-charge $r=(\frac{(N-k)}{2}-1)+(\frac{k-1}{2}) = \frac{N-3}{2}$ give $R$-charge de-concentration. If $N \equiv 1 \pmod 4$, then we have two cases. For $k \equiv 1 \pmod 4$, then we find $R$-charge de-concentration from BPS states with $R$-charge $r=(\frac{(N-k)}{2}-1)+\frac{k-3}{2} = \frac{N-5}{2}$. For $k \equiv 3 \pmod 4$, there is no $R$-charge de-concentration, but there are additional BPS states in the $r=\frac{N-3}{2}$ sector. From Table \ref{tab:N=2SYK_BPSdegeneracy}, the number of such states is $3^{\frac{(N-k)}{2}-1} \times 3^{\frac{(k-1)}{2}} = 3^{\frac{(N-3)}{2}}> 3$ since $N \geq 9$ in this case.

To summarize, when there are two connected components of indices, for $N \equiv 2 \pmod 4$ and $k \equiv 3 \pmod 4$ there are no new BPS states on the subspace $\mathcal{N}$ so they do not get removed when deleting the singular moduli. In all other cases one always finds extra BPS states so after removing singular loci there are no subspaces with enhanced symmetries. Let us thus focus on the former case to see what enhanced symmetries exist. Since $N-k$ and $k$ are odd, any submanifolds $\tilde{\mathcal{N}} \subset \mathcal{N}$ corresponding to more than two connected components of indices would necessarily have extra BPS states so those are ruled out. By treating each Hamiltonian $H_{(1)}$ and $H_{(2)}$ separately and applying the argument from the global case, the only redundancy acting on each set of couplings separately is $U(1)$, giving a total of $U(1)^{2}$. What about symmetries that mix the couplings between the two Hamiltonians? They involve disjoint sets of fermions so at most there could be subspaces $\tilde{\mathcal{N}} \subset \mathcal{N}$ where the couplings between the two sets are related and the $U(1)$ action gets twisted by $\Gamma \subset S_{\mathcal{D}}$. For example, for $N=6$ with $H_{(1)} = |C_{123}|^{2}\{\psi_{1}\psi_{2}\psi_{3},(\psi_{1}\psi_{2}\psi_{3})^{\dagger}\}$ and $H_{(2)} = |C_{456}|^{2}\{\psi_{4}\psi_{5}\psi_{6},(\psi_{4}\psi_{5}\psi_{6})^{\dagger}\}$, then generically there is only $U(1)^{2}$ redundancy, but for $|C_{123}|=|C_{456}|$, the redundancy is enhanced to the freely generated group $\tilde{\mathcal{G}} = \langle \left(\begin{smallmatrix}e^{i\theta} & 0 \\ 0 & e^{i\theta'}\end{smallmatrix}\right), \left(\begin{smallmatrix}0 & e^{i\gamma} \\ e^{i\gamma'} & 0 \end{smallmatrix}\right)\rangle \cong U(1)^{2} \rtimes \mathbb{Z}_{2}$.

\begin{figure}
\begin{center}
\hspace*{-2cm}
\begin{tikzpicture}

\shade[ball color = green!50, opacity = 0.4] (-4,0) circle (2cm);
\draw (-4,0) circle (2cm);
\draw[very thick,red] (-6,0) arc (180:360:2 and 0.6);
\draw[dashed, very thick, red] (-2,0) arc (0:180:2 and 0.6);

\draw[->] (-1.2,0) -- (1.2,0);
\node at (0,0.5) {$(\frac{\pi}{2},\varphi) \sim (\frac{\pi}{2},0)$};

\shade[ball color = green!50, opacity = 0.4] (3,1) circle (1cm);
\draw (3,1) circle (1cm);
\draw (2,1) arc (180:360:1 and 0.3);
\draw[dashed] (4,1) arc (0:180:1 and 0.3);

\shade[ball color = green!50, opacity = 0.4] (3,-1) circle (1cm);
\draw (3,-1) circle (1cm);
\draw (2,-1) arc (180:360:1 and 0.3);
\draw[dashed] (4,-1) arc (0:180:1 and 0.3);

\fill[red] (3,0) circle (0.1cm);

\end{tikzpicture}
\caption{Quotient of $S^{2}$ by $U(1)$ rotation of the equator $(\theta=\frac{\pi}{2},\varphi) \sim (\theta=\frac{\pi}{2}, 0)$ leading to the stratified space $S^{2} \vee S^{2}$.}
\label{fig:equatorquotientsphere}
\end{center}
\end{figure}
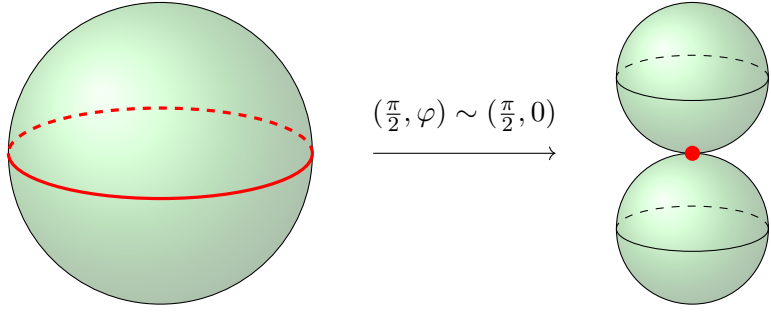

This leads to an intricate topological structure for the moduli space for $N \equiv 2 \pmod 4$ as we must quotient by all enhanced symmetries on any subspaces. In general, performing the quotient by a group action on a submanifold of a manifold will lead to a space that is no longer a manifold. As the simplest example, consider the sphere $S^{2}$ and quotient by the $U(1)$ action of azimuthal rotations only along the equator: $(\theta=\frac{\pi}{2},\varphi) \sim (\theta=\frac{\pi}{2}, 0)$. This cannot be a manifold because any neighborhood of the quotiented equator will not be diffeomorphic to an open subset of $\mathbb{R}^{2}$. The resulting space is topologically the wedge product $S^{2} \vee S^{2}$ given by gluing two $S^{2}$'s at a single point, as shown in Figure \ref{fig:equatorquotientsphere}. The topology of this space is drastically different from the topology of a single $S^{2}$ since all of their homotopy groups $\pi_{k}$ are different for $k \geq 2$ \cite{MR658304}. Spaces constructed in this way from quotients of submanifolds are known as stratified spaces since they can be constructed by gluing together manifolds of various dimensions which comprise of the strata.\footnote{More precisely, our construction gives a Whitney stratified space, which satisfies some extra niceness conditions on how it is glued together \cite{Nicolaescu2011}.}

The moduli space of $\mathcal{N}=2$ SYK is thus obtained as follows. First, we remove the singular moduli and then quotient the resulting manifold by the global $U(1)$ moduli redundancy. For $N \equiv 2 \pmod 4$, we need to additionally find all possible disconnected graphs constructed from the set of indices that have two connected components and then quotient each of the corresponding subspaces $\mathcal{N}$ by its redundancy group $U(1)^{2} \rtimes \Gamma$ for $\Gamma \subset S_{\mathcal{D}}$. This leads to a stratified space, which we write as
\begin{equation}\label{eqn:SYKmodulispace_final}
\mathcal{M} = \left(\left(\mathbb{C}^{{N \choose 3}} \backslash \mathcal{M}_{\mathrm{sing}}\right)/U(1)\right)_{\mathrm{strat.}}.
\end{equation}

\subsection{Analyzing the moduli spaces for small $N$}
\label{sec:modulismallN}

Even though a full analysis of the moduli space and its associated topological properties appears to be a difficult task, for small values of $N$, such as $N=4,5,6$ we can perform a more thorough study.

\paragraph{$\boxed{N=4}$} We start with the simplest non-trivial case, $N=4$. One can show (see Appendix \ref{app:forms}) that at $N=4$, one can always decompose $C_3$ as the wedge product of three one-forms, and therefore $\mathcal{M}_1 = \mathcal{M}_2 = \mathcal{M}$. We can summarize the BPS spectra as Table \ref{tab:N=4}. 
\begin{table}[h]
\centering
{\renewcommand{\arraystretch}{1.4}
\begin{tabular}{|c|c|c|c|c|c|}
\hline
$R$-charge & 0 & 1 & 2 & 3 & 4 \\
\hline
$\mathcal{M}\backslash \mathcal{M}_0$ & 0 & 3 & 6 & 3 & 0 \\
\hline
$\mathcal{M}_0$ & 1 &  4 & 6 & 4  & 1 \\
\hline
\end{tabular}}
\caption{BPS spectra at $N=4$}\label{tab:N=4}
\end{table}

\begin{figure}[t!]
\begin{center}
    \includegraphics[scale=0.25]{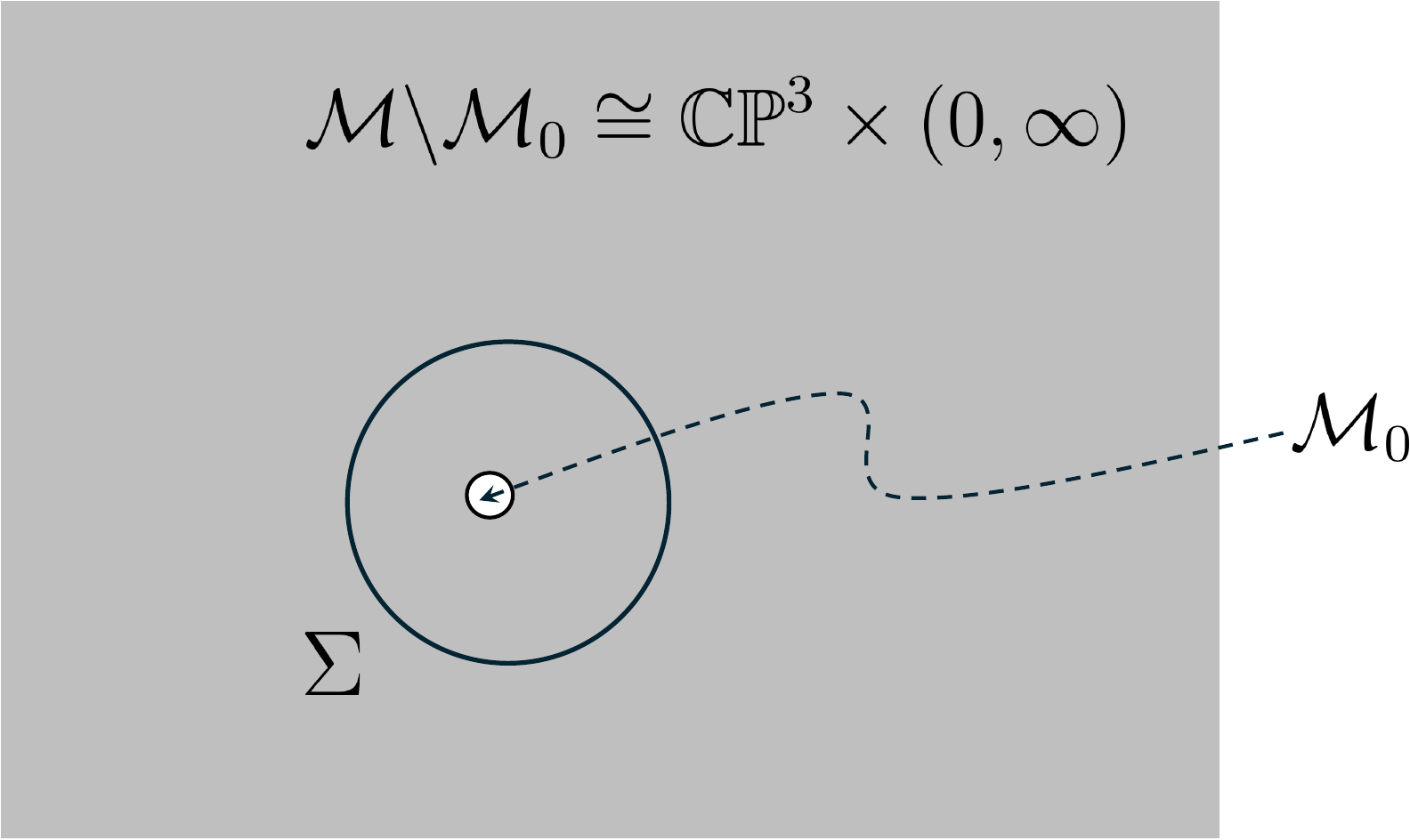}
\end{center}
\caption{A schematic illustration of the moduli space at $N=4$. $\Sigma$, which is topologically a two-sphere, becomes non-contractible after $\mathcal{M}_0$ is carved out from the moduli space.}
\label{fig:topoNeq4}
\end{figure}

We have $\mathcal{M}\cong \mathbb{C}^{3} \times [0,\infty)$, which does not contain any non-contractible $S^2$. However, once we carve out the point $\mathcal{M}_0$ where the BPS spectra changes, we have
\begin{equation}
    \mathcal{M} \backslash \mathcal{M}_0 \cong \mathbb{C}\mathbb{P}^{3} \times (0,\infty)\,.
\end{equation}
Since $(0,\infty)$ deformation retracts to a point, we have\footnote{We have homotopy groups $\pi_2(\mathbb{C}\mathbb{P}^n) \cong \mathbb{Z}$ for $n\geq 1$.}
\begin{equation}
    \pi_2 ( \mathcal{M} \backslash \mathcal{M}_0 ) =\pi_2 (\mathbb{C}\mathbb{P}^{3} \times (0,\infty))= \pi_2 (\mathbb{C}\mathbb{P}^{3} ) \cong \mathbb{Z}\,,
\end{equation}
while all higher homotopy groups vanish. This tells us that in the carved out moduli space $\mathcal{M} \backslash \mathcal{M}_0$, there are now non-contractible $2$-spheres, as can be detected by a quantized Chern number. An explicit construction of such a non-contractible sphere is given by\footnote{This can be seen by noting that the phase of $C_{234}$ winds non-trivially on an $S^1$ near the south pole of $\Sigma$, i.e. $\theta \sim \pi$, which forbids a lift of a smooth map from $S^2 \rightarrow \mathbb{C}\mathbb{P}^3$ into a smooth map from $S^2 \rightarrow S^7$, which implies the non-contractibility of the $S^2$ in $\mathbb{C}\mathbb{P}^3 \cong S^7/ U(1)$.}
\begin{equation}
  \Sigma: \,\,   \{C_{123}, C_{134},  C_{124}, C_{234} \} = \left\{\cos \frac{\theta}{2},  \cos \frac{\theta}{2}, \cos \frac{\theta}{2}, \sin \frac{\theta}{2} e^{in\varphi}\right\}, \,\, \theta \in [0,\pi], \, \varphi \in [0,2\pi), \,\, n \in \mathbb{Z}_{\neq 0}\,. 
\end{equation}
In Figure \ref{fig:topoNeq4}, we illustrate (in a highly schematic way) the moduli space and the chosen two-sphere $\Sigma$.
Computing the Berry curvature for the $3$ BPS states at $r=1,3$, or the $6$ BPS states at $r=2$ and integrating over $\Sigma$, one finds
\begin{equation}\label{intTrF}
    c_{1} = \frac{1}{2\pi}\int_{\Sigma} \textrm{Tr} F = \left\{ \begin{aligned} 
    n, \quad & r = 1,\\
    0, \quad & r= 2, \\
    -n, \quad & r = 3\,.
    \end{aligned} \right. 
\end{equation}
The answer (\ref{intTrF}) is easy to interpret.  Let's focus on $n=1$ and look at the $R$-charge  sector $r=1$. From Table \ref{tab:N=4}, we see that as we approach $\mathcal{M}_0$, a single long multiplet with $(r,r+3) = (1,4)$ joins the BPS sector. This long multiplet contributes Chern number $+1$, which is picked up by the sphere $\Sigma$ $n$ times. On the other hand, at $r=3$, we have a long multiplet $(r-3,r) = (0,3)$ that becomes zero energy. This long multiplet contributes a Chern number $-1$ instead. The fact that the long multiplets $(r-3,r)$ and $(r,r+3)$  contribute oppositely to the Chern number can be seen explicitly from (\ref{FxSYK}), as they enter the first and the second term, respectively, which differ by a minus sign. For the $R$-charge sector $r=2$, the BPS spectra does not change as we approach $\mathcal{M}_0$, and as a consequence we get zero quantized Chern number as in (\ref{intTrF}). 

\paragraph{$\boxed{N=5}$} For $N=5$, one can show (see Appendix \ref{app:forms}) that every three-form can be decomposed into the wedge product of a one-form with a two-form so $\mathcal{M}=\mathcal{M}_{2}$. Writing $C_{3} = B_{2} \wedge A_{1}$, this explains why the $r=1$ sector always has at least $1$ BPS state, which is $A_{1} \cdot \psi \ket{0}$. For $r=3$, new BPS states only occur at the origin, i.e. $\mathcal{M}_0$, since it is in the same cochain complex as $r=0$, and by particle-hole symmetry the same is true for $r=2$. Thus, the only possible sector where we can find new BPS states away from the origin is $r=1$ (and by particle-hole symmetry, the $r=4$ sector). To see that this only happens on $\mathcal{M}_{1}$, observe that all such states are determined by a one-form $\alpha$ and the BPS condition is $C \wedge \alpha = 0$. Suppose such an $\alpha \in V=\mathrm{span}(A_{1})^{\perp}$ exists, i.e., there is a state different from the ubiquitous one. It suffices to consider the case where $B_{2} \in \Lambda^{2}V$. Then $C \wedge \alpha = 0$ implies $B_{2} \wedge \alpha = 0$. Now, let us further decompose $V=U \oplus U^{\perp}$ where $U = \mathrm{span}(\alpha)$. Note $B_{2}$ can always be written as
\begin{equation}
    B_{2} = \alpha \wedge w + W\,, \qquad w \in U^{\perp}\,, \; W \in \Lambda^{2}U^{\perp},
\end{equation}
and hence $B_{2} \wedge \alpha = W \wedge \alpha=0$. But $W$ is in the orthogonal subspace to $\alpha$ so $W=0$, and thus $B_{2}$ is a wedge product of two one-forms, showing that we are in subspace $\mathcal{M}_1$. 
\begin{table}[h]
\centering
{\renewcommand{\arraystretch}{1.4}
\begin{tabular}{|c|c|c|c|c|c|c|}
\hline
$R$-charge & 0 & 1 & 2 & 3 & 4 & 5 \\
\hline
$\mathcal{M}\backslash \mathcal{M}_{1}$ & 0 & 1 & 9 & 9 & 1 & 0 \\
\hline
$\mathcal{M}_{1} \backslash \mathcal{M}_{0}$ & 0 & 3 & 9 & 9 & 3 & 0 \\
\hline
$\mathcal{M}_0$ & 1 & 5 & 10 & 10 & 5  & 1 \\
\hline
\end{tabular}}
\caption{BPS spectra at $N=5$}\label{tab:N=5}
\end{table}

The theorem in Appendix \ref{sec:homotopy} then gives the following results for the homotopy groups of $\mathcal{M}_{\mathrm{non-sing}}$:
\begin{equation}\label{eqn:homotopyN=5}
    N=5: \;\; \pi_{1}(\mathcal{M}_{\mathrm{non-sing}}) = 0\,, \qquad \pi_{2}(\mathcal{M}_{\mathrm{non-sing}}) \cong \mathbb{Z}\,, \qquad \pi_{3}(\mathcal{M}_{\mathrm{non-sing}}) = \pi_{4}(\mathcal{M}_{\mathrm{non-sing}}) = 0
\end{equation} 
and thus the first Chern number $c_{1}$ can be non-trivial, but $c_{2}=0$.

\paragraph{$\boxed{N=6}$} The case $N=6$ is more interesting because the space is stratified and generically $C_{3}$ does not decompose into a product of low degree forms. Nevertheless, one can still characterize all the submanifolds where new BPS states appear. Since $N$ is even, as discussed at the end of Section \ref{sec:specialloci}, the appearance of new BPS states is equivalent to $R$-charge de-concentration. The only sector that can de-concentrate is $r=1$ (and $r=5$ by particle-hole symmetry). That is, we want to find all $C$ for which there exists a one-form $\alpha \neq 0$ such that $C \wedge \alpha = 0$. The argument is similar to the one for $N=5$ above. For a given $C$, suppose there exists such an $\alpha \neq 0$. Define the covector space $V=\mathrm{span}(\alpha)$ and decompose $\mathbb{C}^{N} = V \oplus V^{\perp}$. We can always write $C$ as
\begin{equation}
    C = \alpha \wedge w + W\,, \qquad w \in \Lambda^{2}(V^{\perp})\,,\; W \in \Lambda^{3}(V^{\perp}) \implies C \wedge \alpha = W \wedge \alpha.
\end{equation}
Imposing $W \wedge \alpha = 0$ implies $W = 0$ since they live in orthogonal covector spaces, and hence $C = \alpha \wedge w$. Therefore, $R$-charge de-concentration occurs if and only if $C = A_{1} \wedge B_{2}$ for a $1$-form $A_{1}$ and $2$-form $B_{2}$. 

By the same argument as for $N=5$, the only way to get even more BPS states is when $B_{2}$ is a wedge product of two one-forms. Thus, we have completely determined all possible enhancements in the number of BPS states in the moduli space for $N=6$, with the resulting structure
\begin{equation}
    \mathcal{M}_{0} \subsetneq \mathcal{M}_{1} \subsetneq \mathcal{M}_{2} \subsetneq \mathcal{M}.
\end{equation}
The corresponding numbers of BPS states are given in Table \ref{tab:N=6}.

\begin{table}[h]
\centering
{\renewcommand{\arraystretch}{1.4}
\begin{tabular}{|c|c|c|c|c|c|c|c|}
\hline
$R$-charge & 0 & 1 & 2 & 3 & 4 & 5 & 6 \\
\hline
$\mathcal{M}\backslash \mathcal{M}_{2}$ & 0 & 0 & 9 & 18 & 9 & 0 & 0 \\
\hline
$\mathcal{M}_{2} \backslash \mathcal{M}_{1}$  & 0 & 1 & 10 & 18 & 10 & 1 & 0 \\
\hline
$\mathcal{M}_{1} \backslash \mathcal{M}_{0}$ & 0 & 3 & 12 & 18 & 12 & 3 & 0 \\
\hline
$\mathcal{M}_0$ & 1 & 6 & 15 & 20 & 15 & 6  & 1 \\
\hline
\end{tabular}}
\caption{BPS spectra at $N=6$}\label{tab:N=6}
\end{table}

The stratification of this space is simple: there are ${6 \choose 3} = 15$ many graphs $\Gamma_{A}$ that have two connected components, each consisting of a single vertex and no edges, obtained by setting all of the $C$s to zero except $C_{ijk}$ and $C_{lmn}$ with $\{i,j,k\} \cap \{l,m,n\} = \emptyset$. The corresponding two-dimensional subspace $\mathcal{N} \cong \mathbb{C}^{2}$ has a $U(1)^2$ redundancy that acts by a phase rotation on each copy of $\mathbb{C}$ so $\mathcal{N}/U(1)^{2} \cong \mathbb{R}_{+}^{2}$. There is then a subspace $\tilde{\mathcal{N}} \subset \mathcal{N}$ defined by $|C_{ijk}|=|C_{lmn}|$ where the redundancy is further enhanced to $U(1)^{2} \rtimes \mathbb{Z}_{2}$. However, the extra twisting by $\mathbb{Z}_{2}$ does not affect the quotient geometry so the resulting space is $\tilde{\mathcal{N}}/(U(1)^{2} \rtimes \mathbb{Z}_{2}) \cong \mathbb{R}_{+}$.

To understand the topology of $\mathcal{M} \backslash \mathcal{M}_{2}$, one can compute its homotopy groups. Excising $\mathcal{M}_{2}$ from $\mathcal{M}$ could lead to higher homotopy groups being non-trivial. Using the theorem in Appendix \ref{sec:homotopy}, we find that many are trivial while the second one is non-trivial, viz.,
\begin{equation}\label{eqn:homotopyN=6}
    N=6: \;\; \pi_{1}(\mathcal{M}_{\mathrm{non-sing}}) = 0\,, \qquad \pi_{2}(\mathcal{M}_{\mathrm{non-sing}}) \cong \mathbb{Z}, \qquad \pi_{i}(\mathcal{M}_{\mathrm{non-sing}}) = 0\,, \quad 3 \leq i \leq 8\,.
\end{equation} 
Therefore, the Chern numbers must be $c_{2}=c_{3}=c_{4}=0$, while $c_{1}$ could be non-zero. Moreover, for any $N > 6$ even, the same theorem gives
\begin{equation}\label{eqn:homotopyNeven}
    N>6,\mathrm{\;even}: \;\; \pi_{1}(\mathcal{M}_{\mathrm{non-sing}}) = 0\,, \qquad \pi_{2}(\mathcal{M}_{\mathrm{non-sing}}) \cong \mathbb{Z}\,.
\end{equation} 
This motivates us to study the first Chern numbers even for larger $N$, which we shall now do. 

\subsection{Exponentially large Chern numbers at large $N$}
\label{sec:Chernnumbers}

For larger values of $N$, we have not performed a systematic study of the moduli space. However, we have explored numerically the Berry phase integrated on candidate $S^2$'s in the moduli space and found that indeed there are quantized Chern numbers associated to them. In fact, as $N$ becomes larger, the Chern numbers grow exponentially with $N$. Intuitively, this is because  the number of additional BPS states on the singular subspaces such as $\mathcal{M}_0, \mathcal{M}_1, \mathcal{M}_2$, which can be thought of as sourcing the Berry flux, grow exponentially with $N$. 

An interesting aspect is that as $N$ gets larger, the Berry curvature matrix $F$ at a generic point in the moduli space $\mathcal{M}$ closely resembles a random matrix, as we have demonstrated in Section \ref{sec:SYK}. However, the fact that the integral of $\textrm{Tr}F$ on a non-contractible $S^2$ equals an integer reflects the fact that the underlying $F$ are  not completely random, but rather are subjected to topological constraints.\footnote{It would be interesting to see whether one could define ensembles of random matrix ``fields" on moduli spaces with such topological constraints built in, and study the interplay of randomness and topology. We leave this for future study.}

We can demonstrate the existence of large Chern numbers for a large family of $S^2$'s inside the moduli space, parameterized by 
\begin{equation}\label{couplingS2}
  \Sigma : \quad   \vec{C} = U \cdot \left( \cos \frac{\theta}{2}\,, \sin \frac{\theta}{2} e^{i \varphi}, 0,\ldots,0 \right)^t \,, \,\,\theta \in [0,\pi]\,, \, \varphi \in [0,2\pi) \,.
\end{equation}
Here $\vec{C} = (C_{123}, C_{124}, \ldots)^t$ are the couplings and $U$ is some unitary transformation one can choose. For generic choices of $U$, the $S^2$ as parameterized by (\ref{couplingS2}) does not intersect any special subspaces such as $\mathcal{M}_{0,1,2}$ so it is properly included in the moduli space where the special subspaces are carved out.

\begin{figure}[t!]
\begin{center}
    \includegraphics[scale=0.5]{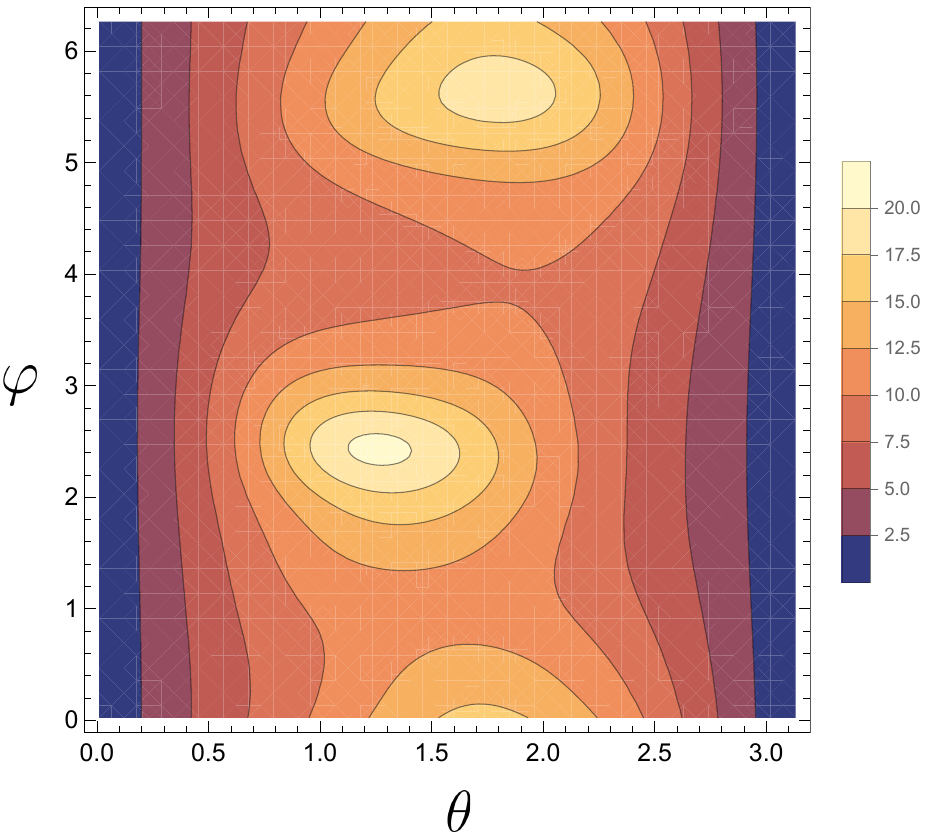}
\end{center}
\caption{$\textrm{Tr}F$ as a function of $\theta$ and $\varphi$, for a typical $\Sigma$ in the family (\ref{couplingS2}) with $N=8, r = 3$. We have $\int_{\Sigma} \textrm{Tr}F = 27 \times 2\pi$ up to small numerical errors. To obtain this plot, we computed the local Berry curvature on a $300 \times 300$ grid and then performed a smooth interpolation.}
\label{fig:Berrycontour}
\end{figure}

For a typical $\Sigma$ in the family (\ref{couplingS2}), numerically, we find that
\begin{equation}\label{bigtrF}
    c_{1}^{(r)} = \frac{1}{2\pi}\int_{\Sigma} \textrm{Tr}F^{(r)} = \binom{N}{r+3} - \binom{N}{r-3}\,,
\end{equation}
where $r = \frac{N}{2}, \frac{N}{2} \pm 1$ for $N$ even, and $r = \frac{N\pm 1}{2}$ for $N$ odd.\footnote{As seen in Table \ref{tab:N=2SYK_BPSdegeneracy}, when $N = 1 \mod 4$, there are $1$ or $3$ BPS states in the sectors $r =\frac{N\pm 3}{2} $. 
We have not found a clear pattern for their $\int_{\Sigma} \textrm{Tr}F$.} We have checked the formula (\ref{bigtrF}) up to $N=10$ and we expect it to be true generally. Note that the two terms on the right hand side of (\ref{bigtrF}) has the interpretation of the number of long multiplets $(r,r+3)$ and $(r,r-3)$ that become zero energy when $C_3$ approaches zero. In Figure \ref{fig:Berrycontour}, we show an example  at $N=8, r=3$ of how $\textrm{Tr}F$ typically varies as a function of $\theta,\varphi$. 

At large $N$, except the case $r= N/2$ for $N$ even, for which the Chern number is zero, we have an exponentially large Chern number. For example, at $r=N/2-1$ for $N$ even, we can use the Stirling approximation and find
\begin{equation}
\label{eqn:ChernnumbergeneralN}
   c_{1}^{(r)} = \frac{1}{2\pi}\int_{\Sigma} \textrm{Tr}F = \binom{N}{\frac{N}{2}  -1 +3} - \binom{N}{\frac{N}{2} - 1 -3} \sim 24 \sqrt{\frac{2}{\pi}} \frac{1}{N^{\frac{3}{2}}} 2^{N}\,.
\end{equation}

Note that so far our exploration has been limited to the first Chern number. Given the richness of the moduli space, we expect that there should also exist surfaces carrying the higher Chern numbers. We leave the identification of such surfaces and their associated Chern numbers to future study.

%%%%%%%%%%%
% SECTION %
%%%%%%%%%%%
\section{Discussion}
\label{sec:disc}

This work investigated whether the Berry curvature matrix of the BPS states at fixed charges can be used as a diagnostic for chaos through the comparison of its eigenvalue repulsion to random matrix theory. We studied this question in many different holographic examples, including the D1/D5 CFT, $\mathcal{N}=4$ super-Yang Mills theory, and the $\mathcal{N}=2$ supersymmetric SYK model, as well as in $\mathcal{N}=2$ super-JT gravity. In all such cases, we found that the monotone states dual to smooth, horizonless supergravity solutions have a Berry curvature matrix that is not a random matrix, while that of the fortuitous states does have random matrix statistics.

In light of the results of \cite{Chen:2024oqv} using the LMRS criterion, the fact that the Berry curvature of the monotones has no chaos, not even weak chaos, may seem surprising. We speculate that this is due to two effects. One is that we heavily used that the monotonous states considered preserved a large amount of supersymmetry, as compared to sectors containing fortuitous states.\footnote{Except for the particular 1/4-BPS states in D1/D5 and the monotones in $\mathcal{N}=2$ SYK that we considered, which were still perhaps ``too simple".} This allowed us in certain cases to derive the $\tt$  equations and in others, it simplified the integrated four-point function for the curvature. The second is that the 1/2-BPS operator associated with the marginal deformations seems to behave more simply than heavier, generic 1/2-BPS operators.  The fact that we do not find random matrix behavior for sectors with large amount of supersymmetries aligns with a no-go argument for a Schwarzian description in such situations \cite{Heydeman:2025vcc}. However, it is important to understand if the Berry curvature of monotonous states with less supersymmetry, such as the 1/8-BPS states in $\mathcal{N}=4$ SYM, is chaotic or not. It would also be curious if the curvature could be computed from the bulk point of view, as the dual geometries are known.\footnote{Belin et al. \cite{Belin:2018fxe} also studied a Berry curvature in a holographic setting, but it is of a different nature than ours. They compute an abelian Berry curvature by varying the parameters of a particular state or wavefunction, and relate it holographically to the bulk symplectic form on the moduli space of semiclassical bulk states, whereas we compute the non-abelian Berry curvature describing the mixing of an entire degenerate subspace under deformations of the theory.}

Another possible generalization is to compute the curvature for lighter operators with less supersymmetry below the black hole threshold, in order to see if they are weakly chaotic or not chaotic at all, and if the amount of chaos depends on their dimension. Examples for such states include 1/16-BPS operators with dimension $\Delta\sim O(N^0)$, where integrability methods could be used, or perhaps those with $\Delta \sim O(N)$ whose bulk dual contains branes.

Our gravitational analysis of Section~\ref{sec:JTgravity} could also be improved. For example, one could consider a conjugation deformation by an exactly marginal operator, instead of a heavy operator. That would require a better understanding of the effect of crossed Witten diagrams \cite{Lin:2026sli}. Another interesting generalization is the effect of more general deformations, not just conjugation deformations, for example using \eqref{eqn:curvformula2}.

We conclude with a comparison to the literature and several open questions.

\paragraph{Comparison to other chaos diagnostics.}

We have proposed a new measure of chaos for the BPS sector of supersymmetric theories, but there exist two other measures that have previously been studied in the literature. The first is the LMRS criterion, discussed throughout this work, which considers the random matrix theory statistics of the eigenvalues of a ``simple" operator projected to the BPS subspace \cite{Lin:2022zxd}. The second is the information entropy, which measures the spread of an energy eigenvector over a ``simple" basis for the Hilbert space \cite{ZELEVINSKY199685,IZRAILEV1990299}.\footnote{See also recent work on ``quantum magic" for fortuitous states \cite{Malvimat:2026oqf}.}  When the subspace is degenerate, one can act with a Haar random unitary on a fixed state in the subspace and average the entropy over an ensemble of such unitaries. For chaotic systems, the information entropy is expected to be the same as that of a random state, which has been found to be true for typical fortuitous states in $\mathcal{N}=2$ SYK, but not for typical monotonous ones \cite{Chang:2024lxt}.\footnote{The average information entropy has also been studied for all non-multi-graviton states, including non-BPS ones, that have the same charges as the lowest energy 1/16-BPS black hole state in $SU(2)$ $\mathcal{N}=4$ super-Yang Mills theory. The averaged entropy was found to be very close to that of a random state \cite{Budzik:2023vtr}.}  

One nice feature of our proposal is that it does not require a definition of ``simple". Given a $d$-dimensional superconformal field theory purely in terms of its primary operators $\{\mathcal{O}_{i}\}$ and their OPE data $\{\Delta_{i},C_{ijk}\}$, the Berry curvature is an intrinsic object in the SCFT with no ambiguities in its definition: one finds all of the marginal operators, namely those with $\Delta=d$ that are neutral under all global symmetries that one wishes to preserve, and then one computes the integrated four-point function in \eqref{berry4point} for the BPS states. In fact, an explicit expression in terms of OPE data can be derived \cite{Ranganathan:1993vj, Balthazar:2022hzb}.

\paragraph{Can fortuitous and monotone mix?} In situations such as the 1/16-BPS sector of $\mathcal{N}=4$ SYM theory, we could have monotone states and fortuitous states residing in the same charge sector. An interesting question is whether fortuitous and monotone states mix under adiabatic change of the couplings. We are not aware of an abstract argument that forbids such mixings from happening. However, if indeed the monotone states are dual to horizonless geometries while the fortuitous states are dual to macroscopic black holes, then one might expect the mixing between them to be non-perturbatively small in the large $N$ limit, as they are governed by different bulk saddle points. This would imply an almost block-diagonal structure in the Berry curvature matrix in the boundary theory. Since the size of the fortuitous block is exponentially larger than the monotone block, we naively expect that this will make the full Berry curvature matrix exhibit strong chaos. It would be nice to understand this structure better.

\paragraph{Other moduli spaces.} 

There are other interesting moduli spaces that are relevant for holography, besides the ones studied in this work, and here we give a survey of some of them.\footnote{This discussion is by no means exhaustive and, in particular, does not include many interesting examples of $3d$ $\mathcal{N}=2$ holographic SCFTs with conformal manifolds.} One that has gained renewed interest is the Type IIB string theory in $\mathrm{AdS}_{3} \times S^{3} \times S^{3} \times S^{1}$ \cite{Gukov:2004ym,Eberhardt:2017pty, Witten:2024yod}. Although the dual CFT is not known explicitly, the BPS spectrum computed from supergravity and the worldsheet exhibits a very non-trivial structure \cite{Eberhardt:2017pty, Eberhardt:2017fsi, Heydeman:2025fde, Murthy:2025moj}. There also exists a whole host of holographic four-dimensional $\mathcal{N}=1$ SCFTs dual to Type IIB string theory on $\mathrm{AdS}_{5} \times Y_{p,q}$ where $Y_{p,q}$ is a five-dimensional toric Sasaki-Einstein manifold \cite{Benvenuti:2004dy}, with the simplest being the conifold theory $Y_{1,0}$ \cite{Klebanov:1998hh}. Their moduli spaces were studied in \cite{Benvenuti:2005wi} and found to generically have three complex dimensions, except for the conifold theory which has five complex dimensions.

If one is willing to break some amount of supersymmetry, then the moduli space of $\mathcal{N}=4$ SYM can be enlarged. The space of marginal deformations that preserve $\mathcal{N}=1$ supersymmetry is three complex-dimensional \cite{Leigh:1995ep}. The dual supergravity picture for one of them, the $\beta$ deformation, corresponds to a particular $SL(2,\mathbb{R})$ transformation of a complexified combination of the $B$-field and metric on a two-torus inside the $S^{5}$ of $\mathrm{AdS}_{5} \times S^{5}$ \cite{Lunin:2005jy}. Under such deformations, it has been found that many 1/4-BPS states get lifted (see e.g. \cite{McLoughlin:2020siu}). An interesting feature of the $\beta$-deformed theories is that one can show analytically that new BPS states appear whenever $\mathrm{Re}(\beta)$ is rational \cite{Lunin:2005jy}, so these points in moduli space could lead to non-trivial sources for the Berry curvature.

 One nice property of all of these various $\mathcal{N}=1$ theories, not present for $\mathcal{N}=4$ SYM, is that their moduli spaces have real dimension greater than two so Chern numbers of the Berry curvature can be computed and may lead to interesting topological constraints. On the other hand, a computation of Berry curvature for ${\cal N}=1$ would lead to similar difficulties as that of 1/8-and 1/16-BPS sectors in ${\cal N}=4$.

\paragraph{Thermofield double state and breakdown of geometry.}

Consider the CFT thermofield state (TFD) dual to an ordinary eternal AdS black hole $\ket{\mathrm{TFD}(t=0)}_{\beta} = \frac{1}{\sqrt{Z(\beta)}}\sum_{n}e^{-\frac{\beta E_{n}}{2}}\ket{n}_{L}\ket{n}_{R}$ at inverse temperature $\beta$ and at boundary time $t=0$. As one evolves this state in time with the boundary Hamiltonian, each energy eigenstate in the superposition obtains a phase $e^{-iE_{n}t}$. Since the energy differences are $O(e^{-S})$, for times $t \sim O(e^{S})$ these phases will look like random phases and the TFD state at such times will look like the state obtained from a random unitary acting on $\ket{\mathrm{TFD}(t=0)}_{\beta}$. The corresponding bulk picture is that non-perturbative effects become important and topology change can occur, as explicitly demonstrated in JT gravity \cite{Saad:2019lba, Saad:2019pqd, Iliesiu:2021ari, Stanford:2022fdt}. The same will be true if one instead considers the microcanonical TFD. 

However, this cannot be the case for the microcanonical TFD of a BPS black hole, defined as the maximally entangled state in the BPS subspace at some fixed large charges $q_{i} \sim N^{2}$,
\begin{equation}\label{eqn:BPS_TFD}
    \ket{\mathrm{BPS\;TFD},\{q_{i}\}} = \frac{1}{\sqrt{d}}\sum_{a \in \mathbb{H}_{\mathrm{BPS}}^{\{q_{i}\}}}\ket{\mathrm{BPS},\{q_{i}\},a}_{L}\ket{\mathrm{BPS},\{q_{i}\},a}_{R}
\end{equation}
because all states in the superposition have the same energy, fixed in terms of the charges, and hence all obtain the same phase. Although the geometry already has some peculiar features, such as a classically infinitely long throat with large quantum fluctuations that make it finite \cite{Lin:2022zxd}, nothing non-trivial can happen to the state, and hence to the dual geometry, as a function of boundary time. The Berry phase provides an alternate way to obtain a random state: if we vary the couplings along a closed loop in moduli space, then since we have argued that the Berry curvature is a random matrix, the resulting unitary acting on the BPS TFD state will transform it into a random state. It would be very interesting to understand what the dual geometric picture is. When the deformation is viewed perturbatively, long time evolution would involve long BPS wormholes with a large number of operator insertions.\footnote{See also related discussions of geometric phases, wormholes and operator algebras \cite{Nogueira:2021ngh, Banerjee:2022jnv, Banerjee:2023eew}.}

\paragraph{Non-BPS black holes and splitting degeneracy.}

One could ask whether the Berry curvature can still be used as a probe of chaos for non-BPS black holes. As a first step, suppose we start with a supersymmetric Hamiltonian and infinitesimally deform the theory to break supersymmetry. This will lift the BPS degeneracy, but the formerly BPS states (fBPS) will still be separated from the rest of the spectrum by a gap if the supersymmetry-breaking deformation is small. The Berry curvature is no longer non-Abelian, but instead we have a direct sum of roughly $e^{S(E_{0})}$ many $U(1)$-bundles each with their own curvature. There are then two natural objects to consider to measure RMT-like repulsion:
\begin{enumerate}[label=(\arabic*)]
    \item The Berry curvatures of each of the fBPS states, treated as the eigenvalues of a diagonal matrix. An important difference here compared to the supersymmetric case is that the Berry curvature, written as a sum over intermediate states in \eqref{eqn:curvformula}, now receives contributions from other formerly BPS states, which give a large contribution since their energy difference is small.
    \item A modified ``Berry curvature'' for two formerly BPS states $\ket{a}$, $\ket{b}$
            \begin{equation}
                \left(F_{\mu\nu}^{(n)}\right)_{ab} = \sum_{m \not \in \mathbb{H}_{\rm fBPS}}\frac{i}{(E_{a}-E_{m})(E_{b}-E_{m})}\bra{b}\partial_{\mu}H\ket{m}\bra{m}\partial_{\nu}H\ket{a} - (\mu \leftrightarrow \nu)\,.
            \end{equation}
        This becomes the BPS Berry curvature as we take the supersymmetry-breaking deformation parameter to zero.
\end{enumerate}
Which of these two quantities is the relevant one is a matter of how the couplings are varied such that the adiabatic theorem holds. Suppose the energies of the fBPS states range from $0$ to $\epsilon$ for some $\epsilon \ll E_{\mathrm{gap}} \sim 1/S$ so that they are still well below the gap to the non-fBPS states. If the couplings are varied sufficiently slowly so that changes in the matrix elements of the Hamiltonian are smaller than the energy differences $\Delta E_{\mathrm{fBPS}}$ between fBPS states, then quantity (1) is the natural object since they will not mix. On the other hand, changes in the couplings slow enough that the fBPS states do not mix with non-fBPS states but fast enough that they do mix with each other, i.e., if one evolves for times $S \ll t \ll e^{S}/\epsilon$, then the modified Berry curvature in (2) is the appropriate quantity to consider.

Now, let us return to non-BPS black holes. The analog of the BPS sector is to consider a microcanonical window centered around some energy $E_{0} \sim N^{2}$. Evolving for an exponentially long time $t \gg 1/\Delta E \sim e^{S(E_{0})}$ puts us in the realm of option (1). A small numerical experiment can be made in the non-supersymmetric SYK model\footnote{We used a system with $N=26$ Majorana fermions, restricted to the even parity sector.} to understand the behavior in that case. Looking at any small energy window and computing its curvature under deformation of the couplings, we found that the repulsion of nearest-neighbor Berry curvatures is consistent with Poisson statistics, not RMT statistics, which might be expected given that we are examining individual states separately.  Perhaps option (2) could give RMT statistics, but it is more confusing for non-BPS black holes because there is no gap between the microcanonical window and all of the other states in the theory and so there is no time scale in which the states mix only within our window, and not with states outside of it.

There is an alternative way to use the Berry curvature to see that non-BPS black holes are chaotic. One can ask whether the Berry curvature of a state in the microcanonical window looks like the Berry curvature of an eigenstate of a random matrix. To be more precise, Berry and Shukla \cite{Berry_2018, Berry_2019} considered families of random matrix Hamiltonians, such as ones of the form $H = H_{0}+xH_{x}+yH_{y}$, where $H_{0}$, $H_{x}$, and $H_{y}$ are all random matrices from the GUE ensemble, and analyzed $F_{xy}$ for a generic state in the Hilbert space. They estimated that there is the following universal behavior in the probability that the Berry curvature of an eigenstate can be large: $P(|F|) \sim |F|^{-\frac{5}{2}}$ as $|F| \to \infty$. This simple result, which they also verified numerically, was obtained by assuming that the dominant contribution to the curvature comes from nearest-neighbor terms in the sum over intermediate states \eqref{eqn:curvformula}, and then using the quadratic level repulsion of nearest-neighbors for GUE random matrices. Hence it probes whether the Hamiltonian has RMT-like level repulsion. We have tested this numerically for the non-supersymmetric SYK model with $N=18$ Majorana fermions, finding very good agreement: $P(|F|) \sim |F|^{-2.48}$.

\paragraph{Berry curvature for operators.} Up to now we had considered the Berry curvature arising from the mixing of some degenerate subspace under deformations by some parameter $\lambda$. But we can also ask what happens to the space of operators acting on the subspace, which is perhaps more accessible in gravitational computations, where one can naturally construct states in the two-sided Hilbert space,\footnote{The two-sided Hilbert space is isomorphic to the space of operators, not necessarily local operators, acting on the Hilbert space.} such as the thermofield double state and states of the form $e^{-\beta H/4} O e^{-\beta H / 4}$ where $O$ is some other local operator. This formulation might also be more suitable for generalization to the non-BPS case, as perhaps it captures some property of the relevant operator algebra. 

The construction is as follows. Suppose we have a family of Hamiltonians parameterized by some parameter $\lambda$, with a basis $|n(\lambda)\rangle$ for our subspace $\mathbb{H}_{0}$ at each point in parameter space. Operators can be written explicitly in this basis as
\begin{equation}
    O(\lambda) = \sum_{n,m}  O_{mn}\ket{m(\lambda)}\bra{n(\lambda)} \,.
\end{equation}
The operators form a Hilbert space, and so we will sometimes denote the operators as states, $|O)$, with the natural inner product induced by the trace over the original Hilbert space, $(A|B) = \Tr(A^\dagger B)$ and a resolution of the identity operator, $\sum_O |O)(O| = \1$. The operators transform under an arbitrary change of basis as
    \begin{equation}
    \ket{n} \to U\ket{n} \quad \implies \quad |O) \to \cU |O) \equiv |U O U^\dagger) \,.
\end{equation}
They are sensitive only to $SU(\dim \mathbb{H}_0)$ transformations, and not to an overall $U(1)$ phase factor. We can now define a connection $\mathcal{A}_\mu$ on the space of operators. It turns out to be related to the Berry connection $A_\mu$ on the original Hilbert space, and locally, using a basis $\{|O^\alpha)\}$ for the operators, it has the form
\begin{equation}
    \mathcal{A}_\mu = \sum_{\alpha,\beta} \Tr\bigl(A_\mu \bigl[O^\beta, O^{\alpha\dagger}\bigr]\bigr)  \, \bigl|O^\alpha\bigr)\bigl( O^\beta\bigr| \,,
\end{equation}
where the trace is over the original Hilbert space. The Berry curvature in operator space\footnote{Note that $\mathcal{F}$ is Hermitian, and transforms in the adjoint action of $\mathcal{U}$ under changes of basis.} is also related to the original Berry curvature, 
\begin{equation}
    \mathcal{F}_{\mu\nu} = \sum_{\alpha,\beta} \Tr\bigl(F_{\mu\nu}[O^\beta,O^{\alpha\dagger}]\bigr)  \, \bigl|O^\alpha\bigr)\bigl( O^\beta\bigr| \,.
\end{equation}
As seen from here, even if we choose our subspace $\mathbb{H}_0$ to contain non-degenerate states, such that the original Berry curvature $F_{\mu\nu}$ is diagonal in the energy basis, the operator Berry curvature $\mathcal{F}_{\mu\nu}$ might remain complicated when it is hard to find a basis of operators $\{|O^\alpha)\}$ with a simple form in the energy basis.

\paragraph{Berry curvature in $N$.} We have considered the Berry curvature of states with respect to continuous physical parameters of the theory. As a curiosity, one might wonder whether one can define a notion of Berry curvature in $N$, to the extent that $N$ can be treated as a continuous parameter. There are two issues with this idea, one is that the analytic continuation in $N$ is naively highly non-unique, and the second being that given such a continuation, the resulting theory at non-integer $N$ could be non-unitary. Nonetheless, explicit prescriptions for analytic continuing in $N$ have been explored for the $\mathcal{N}=4$ SYM theory in \cite{Budzik:2023vtr} and for the $\mathcal{N}=2$ SUSY SYK model in \cite{Chang:2024lxt}, where the latter discussion in fact preserves unitarity even for non-integer $N$. In these contexts, we can study the properties of the Berry curvature and we expect to see interesting features associated with fortuity, that due to additional fortuitous states becoming BPS at integer $N$, they should ``source" a non-trivial Berry curvature in the complex $N$ plane.  More generally,  we expect that if we study the Berry phase for fortuitous BPS states, or for non-BPS black hole microstates, we shall find very different behavior compared to the monotone states.

\section*{Acknowledgments}
The authors would like to thank Ofer Aharony, Alex Belin, Francesco Benini, Micha Berkooz, Suzanne Bintanja, Chi-Ming Chang, Jan de Boer, Alessandra Gnecchi, Monica Guica, Luca Iliesiu, Shota Komatsu, Wolfgang Lerche, Guanda Lin, Henry Lin, Henry Maxfield, Shiraz Minwalla, Damian Musk, Eliezer Rabinovici, Stephen Shenker, Mykhaylo Usatyuk, Spenta Wadia and Haoyu Zhang for useful discussions. We also acknowledge some useful conversations with ChatGPT, Claude, and Gemini. YC and SC-E thank the Aspen Center for Physics for hospitality during the program ``Recent Developments in String Theory'', which is supported by National Science Foundation grant PHY-2210452. SC-E would like to thank the programs ``Quantum Gravity, Holography, and Quantum Information'' at the
International Institute of Physics, Natal and ``Gravity – New quantum and string perspectives'' at the
Centro de Ciencas de Benasque for hospitality during this work. OM would also like to thank the Leinweber Institute for Theoretical Physics at UC Berkeley for its hospitality during the initiation of the project. YC acknowledges support from DE-SC0021085, DE-SC0026143 and Simons Foundation 926198, DS. SC-E was supported in part by the Department of Energy, Office of Science, Office of High Energy Physics under QuantISED Award DE-SC0019380. OM is supported by the ERC-COG grant NP-QFT No. 864583 “Non-perturbative dynamics of quantum fields: from new deconfined phases of matter to quantum black holes”, by the MUR-FARE2020 grant No. R20E8NR3HX “The Emergence of Quantum Gravity from Strong Coupling Dynamics”, and by the INFN ``Iniziativa Specifica"  GAST.

\appendix

%%%%%%%%%%%
% SECTION %
%%%%%%%%%%%
\section{Quantum metric tensor}
\label{sec:quantummmetric}

In this appendix, we discuss the quantum metric tensor, which is the symmetrized cousin of the Berry curvature, and the two can be combined to form the quantum geometric tensor (QGT). Positive semi-definiteness of the QGT leads to an interesting upper bound on the first Chern number for the Hilbert bundle \eqref{eqn:firstChernnumber} in terms of the ``quantum volume" of the $2$-sphere.

\subsection{Construction of quantum metric tensor}

We present an intuitive construction of the quantum metric tensor \cite{Provost:1980nc}. Suppose we want to define a metric for the Hilbert bundle of states $\ket{\psi_{n}(\lambda)}$ at energy $E=E_{n}(\lambda)$, which we assume for the moment is non-degenerate. The most naive definition would be to inspect
\begin{equation}
\big|\ket{\psi_{n}(\lambda+d\lambda)}-\ket{\psi_{n}(\lambda)}\big|^{2} = |\ket{\partial_{\mu}\psi_{n}(\lambda)}d\lambda^{\mu}|^{2} = \underbrace{\langle\partial_{\mu}\psi_{n}(\lambda)|\partial_{\nu}\psi_{n}(\lambda)\rangle}_{g_{\mu\nu}^{(n)\mathrm{naive}}(\lambda)} \, d\lambda^{\mu}d\lambda^{\nu}.
\end{equation}
Observe that, since $\mathrm{Im}\,\langle\partial_{\mu}\psi_{n}(\lambda)|\partial_{\nu}\psi_{n}(\lambda)\rangle$ is anti-symmetric, it does not contribute above so we can take $g_{\mu\nu}^{\mathrm{naive}}$ to be the real part. The real part is symmetric as expected for a metric. The naivety of this metric tensor is that it is not gauge-invariant. Under a change of phase 
\begin{equation}
    \ket{\psi_{n}(\lambda)} \to \ket{\tilde{\psi}_{n}(\lambda)} = e^{i\alpha(\lambda)}\ket{\psi_{n}(\lambda)},
\end{equation} 
the naive metric transforms as
\begin{equation}
    g_{\mu\nu}^{(n)\mathrm{naive}} \to g_{\mu\nu}^{\mathrm{naive}}-i\partial_{\mu}\alpha(\lambda)\langle\psi_{n}|\partial_{\nu}\psi_{n}\rangle-i\partial_{\nu}\alpha(\lambda)\langle\psi_{n}|\partial_{\mu}\psi_{n}\rangle + \partial_{\mu}\alpha(\lambda)\partial_{\nu}\alpha(\lambda).
\end{equation}
From this gauge transformation, we can see how to improve it to a gauge-invariant metric tensor
\begin{equation}
    g_{\mu\nu}^{(n)} = g_{\mu\nu}^{(n)\mathrm{naive}}+\langle\psi_{n}|\partial_{\mu}\psi_{n}\rangle\langle\psi_{n}|\partial_{\nu}\psi_{n}\rangle.
\end{equation}
This metric is positive semi-definite so it is not technically a metric. To see this, we rewrite the metric as
\begin{equation}
    d\ell^{2} = g_{\mu\nu}^{(n)}d\lambda^{\mu}d\lambda^{\nu} = \big|\ket{\psi_{n}(\lambda+d\lambda)}-\ket{\psi_{n}(\lambda)}\big|^{2}-\big|\langle\psi_{n}(\lambda)|\psi_{n}(\lambda+d\lambda)\rangle|^{2}
\end{equation}
and then use the Cauchy-Schwarz inequality (C-S). Therefore, it has zero directions whenever the C-S inequality is saturated, namely when $\ket{\psi_{n}(\lambda+d\lambda)} \propto \ket{\psi_{n}(\lambda)}$.

This can be generalized to a degenerate subspace $\mathbb{H}_{n}$ where the metric is now
\begin{equation}
    (g_{\mu\nu}^{(n)})_{ab} = \frac{1}{2}\left(\langle\partial_{\mu}(n,b)|\partial_{\nu}(n,a)\rangle - \sum_{c \in \mathbb{H}_{n}}\langle\partial_{\mu}(n,b)|(n,c)\rangle\langle(n,c)|\partial_{\nu}(n,a)\rangle\right) + (\mu \leftrightarrow \nu).
\end{equation}
Notice that the trace of the quantum metric over $ab$ indices is positive semi-definite by the same reasoning as the non-degenerate case.

\subsection{Quantum geometric tensor}

The Berry curvature and quantum metric tensor can be nicely packaged into a single tensor, known as the quantum geometric tensor \cite{PhysRevB.81.245129}, defined by 
\begin{equation}
    (\chi_{\mu\nu}^{(n)})_{ab} = \bra{\partial_{\mu} (n,b)}(1-P_{n})\ket{\partial_{\nu} (n,a)},
\end{equation}
which has
\begin{equation}
    g_{\mu\nu}^{(n)} = \frac{1}{2}\left(\chi_{\mu\nu}^{(n)}+\chi_{\mu\nu}^{(n)\dagger}\right), \qquad F_{\mu\nu}^{(n)} = i\left(\chi_{\mu\nu}^{(n)}-\chi_{\mu\nu}^{(n)\dagger}\right).
\end{equation}

The quantum geometric tensor has the property that, after tracing over $ab$ indices, it is a positive semi-definite matrix in the $\mu\nu$ indices. This will allow us to derive inequalities between different $\mu\nu$ components for the trace over $ab$, the most simple of which gives an upper bound on the Berry curvature in terms of the quantum metric. When the subspace is non-degenerate, positive semi-definiteness is trivial since $(1-P_{n})$ has this property. To prove it in the degenerate case for $\Tr(\chi_{\mu\nu}^{(n)})$, we rewrite it in terms of projectors
\begin{equation}\label{eqn:trchi}
    \Tr(\chi_{\mu\nu}^{(n)}) = \Tr\left(P_{n}(\partial_{\mu}P_{n})(1-P_{n})(\partial_{\nu}P_{n})P_{n}\right).
\end{equation}
Now, since $P_{n}^{2}=P_{n}$ and $P_{n}^{\dagger}=P_{n}$, we can define the operators
\begin{equation}
    B_{\mu}^{(n)} = (1-P_{n})(\partial_{\mu}P_{n})P_{n}
\end{equation}
so \eqref{eqn:trchi} becomes
\begin{equation}
    \Tr(\chi_{\mu\nu}^{(n)}) = \Tr\left(B_{\mu}^{(n)\dagger}B_{\nu}^{(n)}\right),
\end{equation}
which is simply the standard inner product on operators $\langle B_{\mu}^{(n)},B_{\nu}^{(n)} \rangle$. Therefore, $\Tr(\chi_{\mu\nu}^{(n)})$ is the Gram matrix, which is indeed positive semi-definite.

\subsection{Bound on first Chern number}

The positive semi-definiteness of the QGT is powerful because Sylvester's criterion implies that every $k \times k$ principal minor is non-negative. For the $1 \times 1$ principal minors, this is simply positive semi-definiteness of the quantum metric. But the $2 \times 2$ principal minors lead to the interesting bound for $\mu \neq \nu$
\begin{equation}\label{eqn:Berrycurvbound}
    \Tr(\chi_{\mu\mu}^{(n)})\Tr(\chi_{\nu\nu}^{(n)}) \geq \left|\Tr(\chi_{\mu\nu}^{(n)})\right|^{2} \implies \Tr(g_{\mu\mu}^{(n)})\Tr(g_{\nu\nu}^{(n)})-\Tr(g_{\mu\nu}^{(n)})^{2} \geq \frac{1}{4}\Tr(F_{\mu\nu}^{(n)})^{2}.
\end{equation}
One can analyze this bound numerically for the $\mathcal{N}=2$ SYK model analyzed in Section \ref{sec:SYK} and we found that it is far from saturated. 

It is not clear how to define volumes when the metric is matrix-valued, but it seems reasonable to use the trace of the metric. Then we recognize the lefthand side of the bound in \eqref{eqn:Berrycurvbound} as the determinant of a 2d metric. Thus, the integral of the trace of the curvature over $2$-spheres is bounded by its ``quantum volume"
\begin{equation}\label{eqn{volbound}}
    \mathrm{qvol}(S^{2}) \equiv \int_{S^{2}} \sqrt{\Tr\left(g^{(n)}\right)} \geq \frac{1}{2}\left|\int_{S^{2}} \Tr F^{(n)}\right|,
\end{equation}
where we used that $\int_{S^{2}} |\Tr F^{(n)}| \geq |\int_{S^{2}} \Tr F^{(n)}|$. The righthand side is the absolute value of the first Chern number introduced in \eqref{eqn:firstChernnumber}. This inequality has been studied extensively in models of topological insulators \cite{PhysRevB.90.165139, Peotta, Mera_2021, PhysRevB.104.045103}, for which it gives a lower bound on superfluidity weight in terms of a topological invariant of the system. Perhaps the larger principal minors also give interesting bounds.

%%%%%%%%%%%
% SECTION %
%%%%%%%%%%%
\section{Details of D1/D5 Berry curvature}
\label{sec:D1/D5details}    

In this Appendix, we provide some of the different details needed in Section \ref{sec:D1/D5}. The small and medium superconformal algebras in two dimensions are given in \ref{sec:N=(4,4)SCA}. A detailed explanation of the mathematical background for vector bundles over symmetric base spaces is expounded in \ref{sec:D1/D5math}. Then \ref{app:realchiral} gives a convenient basis for the chiral primaries in $\mathcal{N}=(4,4)$ theories. In \ref{sec:CRCidentity}, we provide a proof of \eqref{identitychiral} using crossing symmetry. \ref{sec:desccurv} shows why the curvature of a level-$1$ superconformal descendant is the sum of the curvature of the chiral primary plus the curvature of the superconformal generator. The rest of the Appendix is dedicated to further explaining the Berry curvature for the $\mathrm{Sym}^{N}(T^{4})$ CFT, including the $N=1$ case in \ref{sec:T4}, the general $N$ case at generic points of the conformal manifold in \ref{sec:detailssymt4}, some explicit computations at the orbifold point in \ref{app:orbifold}, and a proof of factorization of the 1/2-BPS partition function into that of a single $T^4$ and that of primitive chiral primaries in \ref{app:factorization}.

%%%%%%%%%%%
% SECTION %
%%%%%%%%%%%
\subsection{Two-dimensional $\mathcal{N}=(4,4)$ superconformal algebras}
\label{sec:N=(4,4)SCA}

There are three different types of $\mathcal{N}=(4,4)$ superconformal algebras in two dimensions, known as small, medium or large (contracted), and large. They can be distinguished by their different Kac-Moody algebras, which are $\mathfrak{su}(2)$, $\mathfrak{su}(2) \oplus \mathfrak{u}(1)^{4}$, and $\mathfrak{su}(2) \oplus \mathfrak{su}(2) \oplus \mathfrak{u}(1)$, respectively \cite{Ali:2003aa}. We will only need the small and medium $\mathcal{N}=(4,4)$ SCAs in this work so we focus on those.

\subsubsection{Small $\mathcal{N}=(4,4)$ SCA}
\label{sec:smallN=(4,4)SCA}

The left-moving generators of the small $\mathcal{N}=(4,4)$ superconformal algebra are the stress-tensor $T(z)$, four supercharges $G^{a A}(z)$, $\alpha,A=\pm$, and the $SU(2)_{R}$ $R$-symmetry generators $J^{i}(z)$, $i=1,2,3$. The supercharges transform as a doublet under the $SU(2)_{R}$ symmetry acting on the first superscript. The algebra also has an outer automorphism $SU(2)_{\mathrm{out}}$ which only acts on the supercharges, transforming as a doublet via rotation on the second superscript.

The commutation relations are given by
\begin{align}\label{eqn:smallN=(4,4)SCA}
\begin{split}
    [L_{m},L_{n}] &= (m-n)L_{m+n}+\frac{c}{12}n(n^{2}-1)\delta_{n+m,0}
    \\ [L_{m},J_{n}^{i}] &= -nJ_{m+n}^{i}
    \\ [J_{m}^{i},J_{n}^{j}] &= m\frac{\hat{k}}{2}\delta^{ij}\delta_{n+m,0}+i{\epsilon^{ij}}_{k}J_{m+n}^{k}
    \\ [L_{m},G_{r}^{a A}] &= \left(\frac{m}{2}-r\right)G_{m+r}^{a A}
    \\ [J_{m}^{i},G_{r}^{a A}] &= \frac{1}{2}(\sigma^{iT})^{a}_{b}G_{m+r}^{b A}
    \\ \{G_{r}^{a A},G_{s}^{b B}\} &= \epsilon^{AB}\left(\epsilon^{ab}\frac{c}{6}\left(r^{2}-\frac{1}{4}\right)\delta_{r+s,0}+(r-s)(\sigma^{iT})^{a}_{c}\epsilon^{cb}J_{r+s}^{i}+\epsilon^{ab}L_{r+s}\right) 
\end{split}
\end{align}
where $n,m \in \mathbb{Z}$, $r,s \in \mathbb{Z}+\frac{1}{2}$ and $\hat{k}=c/6$ is the level of the Kac-Moody algebra. We use the convention $\epsilon^{+-} = -1$. The same story holds for the right-moving generators.

\subsubsection{Medium $\mathcal{N}=(4,4)$ SCA}
\label{sec:mediumN=(4,4)SCA}

The left-moving medium (or large (contracted)) $\mathcal{N}=(4,4)$ SCA contains the same generators as the small algebra, as well as $8$ additional generators given by four additional bosonic currents $U^{i}(z)$ forming a $\mathfrak{u}(1)^{4}$ Kac-algebra and four fermionic spin-$\frac{1}{2}$ operators $\Psi^{a A}(z)$ \cite{Ali:1993sd}. It is convenient to repackage the bosonic currents as $U^{kl} = U^{i}(\sigma^{i})^{kl}$.

We have the same commutation relations as \eqref{eqn:smallN=(4,4)SCA} along with the following new ones
\begin{align}\label{eqn:mediumN=(4,4)SCA}
\begin{split}
    [L_{m},U_{n}^{kl}] &= -nU_{n+m}^{kl}
    \\ [U_{m}^{kl},U_{n}^{k'l'}] &= -m\epsilon^{kk'}\epsilon^{ll'}\delta_{n+m,0}
    \\ [G_{r}^{a A},U_{m}^{kl}] &= -im\epsilon^{Al}\epsilon^{Bk}\Psi_{r+m}^{a B}
    \\ [L_{m},\Psi_{r}^{a A}] &= -\left(\frac{m}{2}+r\right)\Psi_{m+r}^{a A}
    \\ [J_{m}^{i},\Psi_{r}^{a A}] &= \frac{1}{2}(\sigma^{iT})^{a}_{b}\Psi_{m+r}^{b A}
    \\ \{G_{r}^{a A},\Psi_{s}^{b B}\} &= i\epsilon^{ab}\epsilon^{kB}U_{r+s}^{kA}
    \\ \{\Psi_{r}^{a A},\Psi_{s}^{b B}\} &= -\epsilon^{AB}\epsilon^{ab}\delta_{r+s,0}
\end{split}
\end{align}
with all others equal to zero. These two-index SU(2) doublet fermions are related to the single-index complex fermions in Section \ref{sec:symT4} by $\Psi^{a+} = \Psi^{\alpha}$ and $\Psi^{\alpha-}=\bar{\Psi}^{3-\alpha}$, where $a=(-1)^{\alpha-1}$. The same story holds for the right-moving generators.

%%%%%%%%%%%
% SECTION %
%%%%%%%%%%%
\subsection{Mathematical background for curvature}
\label{sec:D1/D5math}

Consider the D1/D5 moduli space $\mathcal{M}$ in \eqref{eqn:D1/D5modulispace_ST}
\begin{equation}
    \mathcal{M} \cong \frac{SO_{0}(4,n)}{SO(4) \times SO(n)}\,.
\end{equation}
It has the nice property that it is \textit{locally symmetric} defined as follows.

\noindent \textbf{Definition.} A Riemannian manifold $W$ is \textit{locally symmetric} if its Riemann curvature is covariantly constant and it is \textit{globally symmetric} if each $p \in W$ is an isolated fixed point of an involutive isometry.\footnote{An involutive isometry is one which squares to the identity.}

One can prove that every globally symmetric manifold is locally symmetric. Cartan classified all globally symmetric (irreducible) spaces of which $\mathcal{M}$ is one (see Table II in Chapter IX of \cite{Helgason1962}), and hence it is also locally symmetric. Consider the vector bundle $E^{(r,\widetilde{r})}$ of chiral primaries with charges $(r,\widetilde{r})$ over base $\mathcal{M}$. Since $\mathcal{M}$ is locally symmetric, we can apply Lemma 2.3 in \cite{guijarro2001parallel} and the discussion directly below it to determine the Riemann tensor for our vector bundle as follows.\footnote{The Lemma also needs a condition they call `Condition A' and $\mathcal{M}$ satisfies this condition by their Theorem 1.1.} 

Let the vector fibre $\mathbb{H}_{\mathrm{chiral}}^{(r,\widetilde{r})}$ have dimension $m$. Then the Riemann tensor on $E^{(r,\widetilde{r})}$ can be obtained from the Riemann tensor on $\mathcal{M}$ by a representation $\rho$ of the holonomy group $SO(4) \times SO(n)$ on the fibre:
\begin{equation}\label{eqn:holonomyrep}
    \rho: SO(4) \times SO(n) \to SO(m)\,.
\end{equation}
More precisely, consider the Riemann curvature tensor ${R^{\mathcal{M};\sigma}}_{\mu\nu\rho}$ on $\mathcal{M}$, which we can think of as defining, at each point $p \in \mathcal{M}$, a map
\begin{align}
\begin{split}
    R^{\mathcal{M}}: \Lambda^{2}\mathcal{M}_{p} &\to \mathfrak{hol}(\mathcal{M}_{p})
    \\ (u \wedge v) &\mapsto {{(R_{\mu\nu}^{\mathcal{M}})_{\rho}}^{\sigma}}u^{\mu}v^{\nu}
\end{split}
\end{align}
where $\mathfrak{hol}(\mathcal{M}_{p})=\mathfrak{so}(4) \times \mathfrak{so}(n)$ is the Lie algebra for the holonomy group $\mathrm{Hol}(\mathcal{M}_{p})=SO(4) \times SO(n)$. For the bundle $E^{(r,\widetilde{r})}$ over $\mathcal{M}$, the Riemann curvature on $E^{(r,\widetilde{r})}$ with respect to a connection $\nabla^{E}$ is a map from 2-forms on the base space $\mathcal{M}$ into the holonomy algebra of the fibre
\begin{align}
\begin{split}
    R^{E}: \Lambda^{2}\mathcal{M}_{p} &\to \mathfrak{hol}\left(\mathbb{H}_{\mathrm{chiral},\, p}^{(r,\widetilde{r})}\right)
    \\ (u \wedge v) &\mapsto {{(R_{\mu\nu}^E)}_{a}}^{b}u^{\mu}v^{\nu}
\end{split}
\end{align}
where $\mathbb{H}_{\mathrm{chiral},\, p}^{(r,\widetilde{r})}$ is the fibre at $p$ and $\mu,\nu$ label coordinates on $\mathcal{M}$ and $a,b$ label elements of $\mathbb{H}_{\mathrm{chiral},\, p}^{(r,\widetilde{r})}$. The key mathematical result is that if $\nabla^{E}R^{E}=0$, then there exists a representation $\rho$ of the holonomy group such that the corresponding map $\hat{\rho}$ on Lie algebras gives
\begin{equation}
    R^{E} = \hat{\rho} \circ R^{\mathcal{M}}\,.
\end{equation}
Notice that the Berry curvature \eqref{eqn:D1D5_1/2BPS_tildeF_final_N=4} always depends trivially on the $SO(4)$ directions so the representation is only nontrivial for the $SO(n)$ part. We emphasize that finding this representation $\rho$ for a given vector bundle is in general very non-trivial.

Thus, to compute the curvature $R^{E}$, we first need to compute the Riemann tensor on the moduli space $R^{\mathcal{M}}$, which can be done by leveraging the fact that $\mathcal{M}$ is a symmetric space, using some mathematical results in \cite{eschenburg1997lecture}. The isometry group is $SO_{0}(4,n)$ which acts transitively so the tangent space at a point $p$ is the set of Killing vectors that are non-zero at $p$. Every symmetric space has a Cartan decomposition of the Lie algebra $\mathfrak{g}$ of isometries as
\begin{equation}\label{eqn:Cartandecomp}
    \mathfrak{g} = \mathfrak{t} \oplus \mathfrak{p}
\end{equation}
where $\mathfrak{t}$ is the isotropy algebra given by
\begin{equation}
    \mathfrak{t} = \{X\;\mathrm{Killing}\,|\,X|_{p}=0\}
\end{equation}
which can be proven to be the same as the holonomy algebra (see Theorem 7.2 in \cite{eschenburg1997lecture}) so in our case $\mathfrak{t} = \mathfrak{so}(4) \oplus \mathfrak{so}(n)$, and $\mathfrak{p}$ is defined as
\begin{equation}\label{eqn:pdef}
    \mathfrak{p} = \{X\;\mathrm{Killing}\,|\,\nabla_{\mu}X_{\nu}|_{p}=0\}\,,
\end{equation}
which must be the tangent space $\mathfrak{p} \cong T_{p}\mathcal{M} \cong \mathbb{R}^{4} \otimes \mathbb{R}^{n}$ since it gives all non-zero tangent vectors at $p$. It is easy to show using \eqref{eqn:pdef} that $[\mathfrak{p},\mathfrak{p}] \subset \mathfrak{t}$ and one can prove with more work that $[\mathfrak{t},\mathfrak{p}] \subset \mathfrak{p}$. From this, \cite{eschenburg1997lecture} proves that the Riemann curvature, as a linear map on the tangent space acts via the adjoint action, viz.,
\begin{equation}\label{eqn:Riemann_symmspaces}
    R^{\mathcal{M}}(u,v) \cdot w = {{(R_{\mu\nu}^{\mathcal{M}})_{\rho}}^{\sigma}}u^{\mu}v^{\nu}w^{\rho} = -[[u,v],w]^{\sigma}, \qquad u,v,w \in \mathfrak{p}\,.
\end{equation}

This is all rather abstract so let's write down some explicit matrices. Recall that the Lie algebra $\mathfrak{so}(4,n)$ is defined by
\begin{equation}
    \mathfrak{so}(4,n) = \{A \in M_{n+4}(\mathbb{R})\,|\,A^{T}\eta+\eta A = 0\}, \qquad \eta \equiv \left(\begin{matrix} \mathbbm{1}_{4} & 0 \\ 0 & -\mathbbm{1}_{n} \end{matrix}\right)\,.
\end{equation}
Write a general element as 
\begin{equation}
    S = \left(\begin{matrix} A_{4 \times 4} & B_{n \times n} \\ C_{n \times n} & D_{4 \times 4} \end{matrix} \right) \in \mathfrak{so}(4,n)\,.
\end{equation}
Then the condition $S^{T}\eta+\eta S = 0$ implies
\begin{equation}
    A = -A^{T}, \quad D = -D^{T}, \quad B = C^{T} \implies A \in \mathfrak{so}(4), \quad D \in \mathfrak{so}(n)\,,
\end{equation}
and thus
\begin{equation}
    S = \left(\begin{matrix}
        A & 0 \\ 0 & D
    \end{matrix}\right) + \left(\begin{matrix}
        0 & C \\ C^{T} & 0
    \end{matrix}\right) \in \underbrace{(\mathfrak{so}(n) \oplus \mathfrak{so}(4))}_{\mathfrak{t}} \oplus \underbrace{\mathbb{R}^{4n}}_{\mathfrak{p}}\,.
\end{equation}
Note that the inner product on tangent vectors is given by the Zamolodchikov metric $g_{\mu\nu} = \hat{k} \delta_{ab}\delta_{ij}$ on $\mathcal{M}$ where the overall scale $\hat{k}=c/6$ is the Kac-Moody level determined by the $R$-current two-point function. Choose an orthonormal basis $\{e_{ai}=\frac{1}{\sqrt{\hat{k}}}e_{a}^{(4)}e_{i}^{(n)T}\}$ of rank-1 matrices for $\mathfrak{p}$ where $e_{b}^{(l)}$ is an orthonormal basis for $\mathbb{R}^{l}$ with respect to the standard inner product in that vector space. We embed this basis into $\mathfrak{so}(4,n)$ in the way it appears above:
\begin{equation}
    \hat{e}_{ai} = \left(\begin{matrix} 0 & e_{ai} \\ e_{ai}^{T} & 0 \end{matrix}\right)\,.
\end{equation}
Computing the commutator of two such basis elements gives
\begin{equation}
    [\hat{e}_{ai},\hat{e}_{bj}] = \frac{1}{\hat{k}}\left(\begin{matrix} \delta_{ij}\mathcal{T}_{ab}^{(4)} & 0 \\ 0 & \delta_{ab}\mathcal{T}_{ij}^{(n)} \end{matrix}\right)
\end{equation}
where $\mathcal{T}_{cd}^{(l)} = \left(e_{c}^{(l)}e_{d}^{(l)T}-e_{d}^{(l)}e_{c}^{(l)T}\right)$ are the standard generators of $\mathfrak{so}(l)$. To find the Riemann curvature \eqref{eqn:Riemann_symmspaces}, we take the commutator with a third basis element
\begin{equation}\label{eqn:Riemann_symmspaces_explicit}
    R^{\mathcal{M}}(\hat{e}_{ai},\hat{e}_{bj}) \cdot \hat{e}_{ck} = -[[\hat{e}_{ai},\hat{e}_{bj}],\hat{e}_{ck}] = \frac{1}{\hat{k}}\Big(\delta_{ab}\left(\delta_{ik}\hat{e}_{cj}-\delta_{jk}\hat{e}_{ci}\right)+\delta_{ij}\left(\delta_{ac}\hat{e}_{bk}-\delta_{bc}\hat{e}_{ak}\right)\Big)\,.
\end{equation}
The Riemann tensor in component form can be obtained by projecting the vector on the RHS of \eqref{eqn:Riemann_symmspaces_explicit} onto a basis vector $e_{\mathfrak{d}l}$ and using orthonormality of the basis
\begin{equation}\label{eqn:Riemann_symmspaces_final}
    {R_{(ai)(bj)(ck)}^{\mathcal{M}}}^{(dl)} = \frac{1}{\hat{k}}\Big(\delta_{ab}\delta_{c}^{d}\left(\delta_{ik}\delta_{j}^{l}-\delta_{jk}\delta_{i}^{l}\right)+\delta_{ij}\delta_{k}^{l}\left(\delta_{ac}\delta_{b}^{d}-\delta_{bc}\delta_{a}^{d}\right)\Big) = \frac{1}{k}\Big(\delta_{ab}\delta_{c}^{d}(\mathcal{T}_{ij}^{(n)})_{k}^{l}+\delta_{ij}\delta_{k}^{l}(\mathcal{T}_{ab}^{(4)})_{cd}\Big)\,.
\end{equation}
Now, to find the Riemann tensor on the vector bundle, we simply apply our representation $\hat{\rho}$ for the chiral primaries at some fixed charge which will in general reside in a reducible representation $\mathcal{R}$ of the holonomy algebra $\mathfrak{so}(4) \oplus \mathfrak{so}(n)$. Recall that $\hat{\rho}^{\mathcal{R}}$ is always trivial for the $\mathfrak{so}(4)$ factor, i.e., $\hat{\rho}^{\mathcal{R}}(\mathcal{T}_{ab}^{(4)})=0$, so the representation is determined by representation matrices
\begin{equation}
    \Sigma_{ij}^{\mathcal{R}} = \hat{\rho}^{\mathcal{R}}(\mathcal{T}_{ij}^{(n)})\,,
\end{equation}
and hence the curvature is
\begin{equation}
    R_{(ai)(bj)KL}^{E} = \frac{1}{\hat{k}}\delta_{ab}(\Sigma_{ij}^{\mathcal{R}})_{KL}\,.
\end{equation}
%

%%%%%%%%%%%
% SECTION %
%%%%%%%%%%%

\subsection{Reality properties of the ${\cal N}=(4,4)$ chiral ring}
\label{app:realchiral}

Consider the vector space $V_{r,\widetilde{r}}$ of all chiral primaries of dimension $(r,\widetilde{r})$. By acting with $SU(2)$ R-symmetry lowering operators as $(J^-)^{2r}(\widetilde{J}^-)^{2\widetilde{r}}$, we have a linear map $\lowerj:V_{r,\widetilde{r}}\rightarrow V_{-r,-\widetilde{r}}$ into the space of anti-chiral primaries. Hermitian conjugation maps the space $V_{-r,-\widetilde{r}}$ back into $V_{r,\widetilde{r}}$. Composing the two maps, we get an {\it antilinear} map $\Theta:V_{r,\widetilde{r}}\rightarrow V_{r,\widetilde{r}}$ defined by $\Theta(\varphi)=[\lowerj (\varphi)]^\dagger$. With appropriate normalization of ${\cal J}^-$, this map is anti-unitary. Now we distinguish two cases:
\begin{enumerate}
    \item If the chiral primaries under consideration are bosonic ($r+\widetilde{r}$ is an integer), then we have that $\Theta^2=1$. This follows from using the $SU(2)$ algebra and the fact that $\lowerj$ involves an even number of lowering operators. Since we have an antiunitary map $\Theta$ which also satisfies $\Theta^2=1$, we can select a basis $\varphi_K$ of chiral primaries in $V_{r,\widetilde{r}}$ such that $\Theta (\varphi_K) =\varphi_K$, that is
    $
    {1\over \sqrt{(2r)! (2\widetilde{r})!}}(J^-)^{2r} (\overline{J}^-)^{2\bar{r}}\cdot \varphi_K = \varphi_K^\dagger$. 
In this basis the OPE coefficients between three bosonic chiral primaries $\varphi_K,\varphi_L,\varphi_M$ are real
\be
(C_{KL}^M)^* =C_{KL}^M\,.
\ee
\item If the chiral primaries are fermionic ($r+\widetilde{r}$ is half-integer), then we have that $\Theta^2=-1$. This implies that ${\rm dim}\,V_{r,\widetilde{r}}= 2n$, for some integer $n$. In this case, it is possible to find a basis of $V_{r,\widetilde{r}}$ consisting of two sets of vectors, $\varphi_A,A=1,\ldots,n$ and $\widetilde{\varphi}_A,A=1,\ldots,n$ such that $\Theta(\varphi_A) =\widetilde{\varphi}_A$ and $\Theta(\widetilde{\varphi}_A)=-\varphi_A$. Now we can consider the OPE coefficient between bosonic $\varphi_K$ and fermionic $\varphi_A$ chiral primaries, resulting into another fermionic chiral primary $\varphi_B$, and we have the reality condition
\be
\label{realityferm}
C_{K\,\,\Theta(A)}^{\Theta(B)} = (C_{KA}^B)^*
\ee
\end{enumerate}
where we used that $\Theta(K)=K$, since $K$ is assumed to be bosonic.

\subsection{A proof of \eqref{identitychiral} based on crossing symmetry}
\label{sec:CRCidentity}

Here we provide a direct proof of \eqref{identitychiral} based on Ward identities and crossing symmetry, which does not make use of conformal perturbation theory. We want to show that the chiral ring coefficients in a (small) ${\cal N}=(4,4)$ SCFT obey
\be
\label{identitychiral2}
[C_i,C_i^\dagger]_{KL}= \delta_{KL}\left(1-{1 \over \hat{k}}(r+\widetilde{r})\right)
\ee
where $i$ is any chiral primary of dimension $({1\over 2},{1\over 2})$ and $K,L$ are general chiral primaries. The following proof makes use of the $SU(2)$ R-symmetry of the ${\cal N}=(4,4)$ theory and generally we do not expect \eqref{identitychiral2} to hold in a ${\cal N}=(2,2)$ theory.

We denote by $\varphi_{K},\varphi_L$  general chiral primary operators of $SU(2)_R$ charge equal to $(r,\widetilde{r})$. We denote by $\varphi_i$ a chiral primary of dimension $({1\over 2},{1\over 2})$. Using \eqref{n4ward} twice and for $j=i$, we find
\be
\label{anothereq4}
\,\,\Box_x  \langle \varphi_L^\dagger(\infty) \,\varphi_i(x)\,\, \varphi_i^\dagger(y) \,\,\varphi_K(0)\rangle = \,\,\Box_y  \langle \varphi_L^\dagger(\infty) \,\varphi_i(x)^\dagger\,\, \varphi_i(y) \,\,\varphi_K(0)\rangle\,.
\ee
Further using a conformal Ward identity, the RHS can be rewritten as $\Box_x ({|x|^2 \over |y|^2}\langle \varphi_L^\dagger(\infty) \,\varphi_i(x)^\dagger\,\, \varphi_i(y) \,\,\varphi_K(0)\rangle)$. So all in all we have that \eqref{anothereq4} can be expressed as
\be
\label{eq3}
\Box_x G(x,y) =0
\ee
where we defined
\be
\label{defg2}
G(x,y) \equiv \langle \, \varphi_L^\dagger(\infty)\,  \varphi_i (x)\,  \varphi_i^\dagger(y)\,  \varphi_K(0)\rangle - {|x|^2 \over |y|^2} \langle \varphi_L^\dagger(\infty) \, \varphi_i^\dagger (x) \, \varphi_i(y) \, \varphi_K(0)\rangle\,.
\ee
Let us keep $y$ fixed and examine $G(x,y)$ as a function of $x$. First, from \eqref{eq3} we have that $G(x,y)$ is a harmonic function in $\mathbb{R}^2-\{0,y\}$. Second, we notice that the function is smooth as $x\rightarrow 0$ (due to the structure of the OPE and BPS bounds),  so it can be smoothly extended into a harmonic function on $\mathbb{R}^2-\{y\}$.

Now we will use the fact that if we have a harmonic function in ${\mathbb R}^2-\{y\}$, which has no logarithmic singularities at $y$ as guaranteed by the OPE, then the average of this function on a circle of radius $r$ centered around $y$ is independent of $r$. This will allow us to relate the asymptotic value of $G$ at $x\rightarrow \infty$ to the short distance OPE as $x\rightarrow  y$. To proceed, we define
\be
M(r) = {1\over 2\pi r} \int_{C_r} G(x,y)
\ee
where $C_r$ is a circle in $x$ of radius $r$ centered around $y$. In the equation above we keep $y$ fixed. The fact that $G(x,y)$ is harmonic in $x$ in the domain ${\mathbb R}^2-\{y\}$ implies that
\be
\label{twolims}
\lim_{r\rightarrow \infty} M(r) = \lim_{r\rightarrow 0} M(r)
\ee
provided that both limits exist.

By a double OPE expansion as $|x|\rightarrow \infty$ we find that
 \be
\label{liminf}
 \lim_{r\rightarrow \infty} M(r) =- {[C_i,C_i^\dagger]_{KL} \over |y|^2}\,.
 \ee

For the limit as $r\rightarrow 0$, we will use the OPE between the operator at $x$ and $y$. Suppose we have an operator of dimension $(h,\widetilde{h})$ showing up in the OPE of $\varphi_i(x)$ with $\varphi^\dagger_i(y)$. It could be a primary or descendant. The OPE looks like
\be
\label{opeII}
\varphi_i(x) \varphi_i^\dagger(y) =\ldots+  A{{\cal O}(y) \over (x-y)^{1-h}(\overline{x}-\overline{y})^{1-\widetilde{h}}}+\ldots
\ee
This gives the following contribution to the expression $G$
\be
A' \cdot {1 \over (x-y)^{1-h}(\overline{x}-\overline{y})^{1-\widetilde{h}}}\cdot \left[{1\over y^h \bar{y}^{\bar{h}}}-{1\over |y|^2 x^{h-1}\overline{x}^{\widetilde{h}-1}}\right] 
\ee
where $A'\equiv A\cdot \langle \varphi_K^\dagger(\infty)\,\,{\cal O}(1) \varphi_L(0)\rangle$. Now we write $x=y+re^{i\theta}$. Suppose first that $h\geq \widetilde{h}$, in which case we have $h=\widetilde{h}+s$ where the spin $s$ is a positive integer or zero (the operator ${\cal O}$ cannot be fermionic). Then the expression above takes the form
\be
A' \cdot {r^s e^{i\theta s} \over r^{2-2\overline{h}}}\cdot \left[{y^{-s}\over|y|^{2\widetilde{h}}}-{x^{-s}\over |y|^2 |x|^{2\widetilde{h}-2}}\right]\,.
\ee
It is easy to see that if $s\geq 2$, then the expression above vanishes after integrating over $\theta$ and then taking the limit $r\rightarrow 0$. So we only need to examine the cases $s=0$ and $s=1$. 

For $s=0$, the expression above simplifies to
\be
A' {1\over r^{2-2\widetilde{h}}} \left[{1\over |y|^{2\widetilde{h}}}-{1\over |y|^2 (y\overline{y}+ \overline{y} r  e^{i\theta}+ y r e^{-i\theta}+r^2)^{\widetilde{h}-1}}\right] \,.
\ee
Expanding the brackets in a Taylor series in $r$ around $r=0$, we find that the $O(r^0)$ term vanishes. The $O(r^1)$ terms vanish due to the integration over $\theta$. The $O(r^2)$ and higher terms do not contribute in the limit $r\rightarrow 0$ as long as $\widetilde{h}>0$. Finally if $\widetilde{h}=0$, corresponding to the case where ${\cal O}$ is the identity operator, we find that we get a finite limit and equal to
\be
-{\delta_{KL}\over |y|^2}
\ee
where we assume we work in a basis where the chiral primaries $\varphi_i$ have unit-normalized 2-point functions.

Performing a similar analysis for the case $s=1$, we find that only $\widetilde{h}=0$ operators give a finite contribution. These are $(1,0)$ conserved currents, which in our case contributes 
\be
{r\over \hat{k}} \delta_{KL} {1\over |y|^2}\,.
\ee
We find analogous results for the $(0,1)$ current. So all in all we find that
\be
\label{limzero}
\lim_{r\rightarrow 0} M(r) = -{\delta_{KL}\over |y|^2}\left(1 - {1 \over \hat{k}}(r+\widetilde{r})\right)\,.
\ee
Finally, replacing \eqref{liminf} and \eqref{limzero} in equation \eqref{twolims} we get the  desired identity
\be
\label{finalidentity}
[C_i,C_i^\dagger]_{KL} =  \delta_{KL}\left(1-{1\over \hat   {k}} (r+\widetilde{r})\right)\,.
\ee

\subsection{On the curvature of descendants}
\label{sec:desccurv}

In this appendix we consider the relative Berry phase of primaries and descendants in the same (short) supermultiplet. We start with a quick review of the curvature of the generators of the ${\cal N}=4$ SCA. 

First, we show that the curvature of the stress tensor is zero. Consider two marginal operators, normalized so that
\be
\langle {\cal O}_\mu(x) \,\,{\cal O}_\nu(y)\rangle = {\delta_{\mu\nu} \over |x-y|^4}\,.
\ee
Then, using Ward identites, we find the following 4-point function
\be
\langle T(\infty) \,\, {\cal O}_\mu(x) \,\,{\cal O}_\nu(y) \,\, T(0)\rangle =\delta_{\mu\nu}\left[ {c\over 2}{1\over |x-y|^4} + {x^2+y^2 \over x^2 y^2(\overline{x}-\overline{y})^2 }\right]\,.
\ee
We see that the antisymmetric combination relevant for the computation of the curvature \eqref{berry4point} vanishes. Hence the curvature of $T(z)$ is zero.

Similarly, we find the curvature of the $SU(2)$ R-currents vanishes, which follows from the fact that the marginal operators are R-neutral, hence a correlator of the form $\langle J J  O O\rangle$ factorizes in the obvious way and then we get a vanishing result upon antisymmetrization \eqref{berry4point}.

Finally, the curvature of the supercurrents is more interesting. It was explicitly computed for ${\cal N}=(2,2)$ SCFTs in \cite{deBoer:2008ss} and for ${\cal N}=(4,4)$ SCFTs in \cite{Gomis:2016sab}. The result is that the curvature for the left-moving supercharges corresponds to an $SU(2)$ factor inside the $SO(4)$ part of the $SO(4)\times SO(n)$ holonomy of the conformal manifold \eqref{eqn:D1/D5modulispace_ST}, while the right-moving supercharges correspond to the second of the $SU(2)$ factors inside the $SO(4)$.

Now, we consider the relative Berry phase between superconformal primaries and certain descendants in the same supermultiplet. Suppose we denote by ${\cal V}_p$ the bundle of the primaries of a given charge, ${\cal V}_s$ the bundle of a symmetry current (which could be either of $J^i(z), G^{aA}(z), T(z)$ or one of right-moving generators) and ${\cal V}_d$ the bundle of the descendants that we get by acting with the corresponding symmetry generators on the primaries. Each of these bundles has its own connection and curvature which can be computed by \eqref{berry4point}. We want to understand to what extent the curvature of the descendant is equal to  the sum of the curvature of the corresponding primary plus that of the symmetry current. We can think of the OPE between the primary and the current as a point-wise multiplication mapping from ${\cal V}_p \otimes {\cal V}_s \rightarrow {\cal V}_d$. A sufficient condition for the curvatures to be additive is that this mapping is covariantly constant. Equivalently, it is sufficient that the 3-point functions between the primaries, descendants and symmetry current is covariantly constant. 

Consider for example the superconformal primaries to be chiral primary operators $\varphi_i$ and the descendant to be $[G^{-C}_{-r},\varphi_i]$. We expect that the curvature of the descendants is equal to the sum of the curvature of the supercurrents plus that of the primaries $\varphi_i$. For this it would be sufficient that the following 3-point function is covariantly constant
\be
\langle [G_{-r}^{-D},\varphi_j]^\dagger(x)\,\, G^{-C}(y)\,\,\varphi_i(z)\rangle\,.
\ee
This follows if we can show that
\be
\label{addholon}
\langle [G_{-r}^{-B},\varphi_j]^\dagger(x)\,\, G^{-A}(y)\,\,\varphi_i(z)\,\,{\cal O}_{a,k}(w)\rangle = 0
\ee
for any marginal operator ${\cal O}_{a,k}(w))$. To prove \eqref{addholon} we write the marginal operator as in \eqref{marginal2}, i.e.
\be
{\cal O}_{a,k} = {1\over \sqrt{2}}\sigma_{a,{A\dot{B}}} \{G_{-{1\over2}}^{+A},[\widetilde{G}_{-{1\over 2}}^{+\dot{B}},\varphi_k^\dagger]\} \,.
\ee
We use a superconformal Ward identity with a Killing spinor vanishing at $x$ to move the $\widetilde{G}^{+\dot{B}}_{-{1\over 2}}$ supercharge away from the operator at $w$. We do not get a contribution from $\varphi_i(z)$ because it was assumed to be chiral primary, we do not get a contribution from $G^{-C}(y)$ since it has a regular OPE with $\widetilde{G}^{+\dot{B}}$, and we do not get a contribution from the operator at $x$ because it is a superconformal primary on the right-moving sector and the Killing spinor was selected to vanish at $x$. Hence the correlator \eqref{addholon} vanishes. 

All in all we have shown that the curvature of descendants of the form $[G_{-r}^{-C},\varphi_i]$, with $\varphi_i$ chiral primary, is the sum of curvature of the chiral primaries plus the curvature of the supercurrents.

A similar argument like the one above can be used to show that descendants of the form $[L_{-m},\varphi_i]$ or of the form $[J_{-m}^{a},\varphi_i]$ have the same Berry curvature as that of $\varphi_i$, since the curvature of $T(z)$ and $J^{a}(z)$ vanishes.

We expect a similar pattern for more general descendants of $\varphi_i$, but we will not provide a complete analysis in this paper. As a special case, in section \ref{sec:D1/D5_1/4BPS} we discuss the case of descendants generated by multiple $J_{-m}^{a}$ insertions.

%%%%%%%%%%%
% SECTION %
%%%%%%%%%%%
\subsection{The case of a single ${T^4}$} 
\label{sec:T4}

\paragraph{Field content and symmetries.}

We consider a 2d CFT of 4 compactified bosons $X^I$ and 4 real left-moving fermions $\psi^I$ plus 4 right-moving fermions $\widetilde{\psi}^I$. The theory has ${\cal N}=(4,4)$ symmetry. Due to the free fermions, the symmetry algebra is larger than the small ${\cal N}=(4,4)$ SCA and is sometimes called contracted large $N=4$ SCA (or medium SCA) with $c=6$ \cite{Ali:1993sd}.

The conformal manifold of this theory is the Narain moduli space which locally looks like
\be
\label{narainspace}
SO_{0}(4,4) \over SO(4) \times SO(4)
\ee
corresponding to choices of the metric $G_{IJ}$ of the $T^4$ and the B-field $B_{IJ}$. Here we only indicate the local structure. The Zamolodchikov metric is the classical metric on this coset.

At a generic point on the conformal manifold, the left-moving fermions of the theory generate a left-moving $SO(4)$ symmetry, whose currents are fermion bilinears $\psi^I \psi^J$. We can decompose this as $SO(4) = SU(2)_R \otimes SU(2)_{\rm flavor}$, where the first factor will play the role of the left-moving R-symmetry of the ${\cal N}=4$ SCA. The theory also has a $U(1)^4$ left-moving global symmetry with currents $\partial X^I$. We have the same symmetries in the right-moving sector, see \cite{Volpato:2014zla} for a review.

We define a complex basis of fermions $\psi^1,\psi^2$ and their conjugates. In this basis, the R-symmetry currents can be chosen as
\be
\label{t4rsym}
J_{R}^+ = - \psi^1 \psi^2 \qquad J_{R}^- = \psi^{1\dagger} \psi^{2\dagger} \qquad J_R^3 = {1\over 2}\left(:\psi^1 \psi^{1\dagger}:+:\psi^2 \psi^{2\dagger}:\right)
\ee
and the flavor symmetry currents as
\be
\label{t4flav}
J_{{\rm flavor}}^+ = - \psi^1 \psi^{2\dagger} \qquad J_{{\rm flavor}}^- = \psi^{1\dagger} \psi^2
\qquad J_{{\rm flavor}}^3 = {1\over 2}\left(-:\psi^1 \psi^{1\dagger}:+:\psi^2 \psi^{2\dagger}:\right)\,.
\ee
These are all conserved everywhere on the conformal manifold \eqref{narainspace}.

We move on to the supercharges of the theory. In the case of $T^4$, we have the medium ${\cal N}=4$ SCA with $c=6$. In this case, the supercurrents are constructed as linear combinations of products of the fermionic currents and U(1)$^4$ currents, i.e. the 4 real supercurrents $G^a,a=1,..,4$ are of the form
\be
G^a = K^a_{IJ}\,\, \psi^I \, \partial X^{J}
\ee
where $K^a_{IJ}$ are fixed real matrices and $\psi^I$ fermions in the real basis.\footnote{There are $4\times 4=16$ operators of the form $\psi^I \partial X^J$, all of which have dimension $(\frac{3}{2},0)$ all over the Narain moduli space. But only certain linear combinations, determined by the choice of the matrices $K^a_{IJ}$ above, have the expected mutual OPE structure to be interpreted as supercurrents, as we review.}

There are certain conditions that we need to impose on the candidate supercurrents, see for instance \cite{Alvarez-Gaume:1981exa}. Let us start more generally by assuming that we have $a=1,\ldots,{\cal N}$ such real and mutually compatible supercurrents. The first condition is that the OPE of each supercurrent with itself must have the form
\be
G^a(z) G^a(0) = {2c/3 \over z^3}+{T(0) \over z} +\ldots
\ee
where $T(z)$ is the full stress tensor of the theory. 
This requires
\be
\label{condition1}
(K^a)^T K^a = K^a (K^a)^T =1\qquad \forall a\,.
\ee
The second condition is that if we take two such supercurrents $G^a,G^b$,with $a\neq b$, then their OPE must only contain R-currents. This leads to conditions of the form
\be
\label{condition2a}
K^a\, (K^b)^T + K^b\,(K^a)^T = 0\qquad a\neq b
\ee
and
\be
\label{condition2b}
(K^a)^T\, K^b + (K^b)^T\,K^a = 0 \qquad a\neq b\,.
\ee
One can check that the conditions \eqref{condition1}, \eqref{condition2a}, \eqref{condition2b} have a solution for $4\times 4$ matrices $K^a$, provided that ${\cal N}\leq 4$.
A convenient basis is to take the first of them to be the identity $K^1={\mathbbm{1}}$. Then the conditions \eqref{condition2a}, \eqref{condition2b} imply that the remaining matrices $K^a$ must be antisymmetric and they must obey
\be
\label{cliff}
\{K^a , K^b\} =-2\delta^{ab}\,.
\ee
This is a real Clifford algebra with negative signature. 
If the matrices $K^a$ are $4 \times 4$, then the maximal range of $a,b$ for which \eqref{cliff} has solutions is equal to 3. Then taking the 3 $K$'s and solving \eqref{cliff} together with the first matrix $K^1={\mathbbm{1}}$, we find that the maximum number of supercharges of the theory is ${\cal N}=4$, as expected.

However, the solution is not unique: suppose we take $F_{IJ},B_{IJ}$ to be two independent $SO(4)$ transformations, which can be thought of as acting on the fermions/bosons respectively. Suppose we start with a choice of $K^{a}_{IJ}$ that obey the conditions above and suppose that we define a new set of matrices by
\be
\widetilde{K}^a_{IJ} = F_{IK} K^a_{KL} B_{LJ}\,.
\ee
Then one can see that conditions \eqref{condition1}, \eqref{condition2a}, \eqref{condition2b} are also satisfied by $\widetilde{K}^a_{IJ}$. This means that the choice of the supercurrents of the ${\cal N}=4$ algebra for $T^4$ can be rotated by $SO(4)_F\otimes SO(4)_B$, though certain combinations may leave the set of supercharges invariant.

As we will see later, the $SO(4)_F$ part does not have any curvature on the conformal manifold. On the other hand, the $SO(4)_B$ part has non-trivial $SO(4)$ holonomy, the same as that of the $\partial X^I$ operators. This means that we need to associate an $SO(4)_{\rm left} \otimes SO(4)_{\rm right}$ holonomy to the superalgebra of $T^4$, unlike the $SU(2)_{\rm left} \otimes SU(2)_{\rm right}$ holonomy of the superalgebra of $K3$.

\paragraph{The chiral ring.}

The 1/2-BPS states of the theory transform in the $(r,\widetilde{r})$ representation of $SU(2)_{\rm left}\otimes SU(2)_{\rm right}$ R-symmetry and have dimensions $(h,\widetilde{h})=(r,\widetilde{r})$. We label these states by the highest-weight state of $(J_{R}^{3,\rm left},J_{R}^{3,\rm right})$. We have the following 1/2-BPS states, which correspond to the Hodge diamond of a $T^4$ 
\[
\renewcommand{\arraystretch}{0.9}
\setlength{\arraycolsep}{10pt}
\begin{array}{ccccccccc}
&&&&
\begin{array}{@{}c@{}}
\color{red!70!black}{1}\\[-2pt]
\color{blue!70!black}{(1,1)}\\[-2pt]
\color{green!50!black}{\psi^{1}\psi^{2}\widetilde{\psi}^{1}\widetilde{\psi}^{2}}
\end{array}
&&&&
\\[8pt]

&&&
\begin{array}{@{}c@{}}
\color{red!70!black}{2}\\[-2pt]
\color{blue!70!black}{\left(1,\frac{1}{2}\right)}\\[-2pt]
\color{green!50!black}{\psi^{1}\psi^{2}\widetilde{\psi}^{j},\ j=1,2}
\end{array}
&&
\begin{array}{@{}c@{}}
\color{red!70!black}{2}\\[-2pt]
\color{blue!70!black}{\left(\frac{1}{2},1\right)}\\[-2pt]
\color{green!50!black}{\psi^{i}\widetilde{\psi}^{1}\widetilde{\psi}^{2},\ i=1,2}
\end{array}
&&&
\\[8pt]

&&
\begin{array}{@{}c@{}}
\color{red!70!black}{1}\\[-2pt]
\color{blue!70!black}{(1,0)}\\[-2pt]
\color{green!50!black}{\psi^{1}\psi^{2}}
\end{array}
&&
\begin{array}{@{}c@{}}
\color{red!70!black}{4}\\[-2pt]
\color{blue!70!black}{\left(\frac{1}{2},\frac{1}{2}\right)}\\[-2pt]
\color{green!50!black}{\psi^{a}\widetilde{\psi}^{a'},\ a,a'=1,2}
\end{array}
&&
\begin{array}{@{}c@{}}
\color{red!70!black}{1}\\[-2pt]
\color{blue!70!black}{(0,1)}\\[-2pt]
\color{green!50!black}{\widetilde{\psi}^{1}\widetilde{\psi}^{2}}
\end{array}
&&
\\[8pt]

&&&
\begin{array}{@{}c@{}}
\color{red!70!black}{2}\\[-2pt]
\color{blue!70!black}{\left(\frac{1}{2},0\right)}\\[-2pt]
\color{green!50!black}{\psi^{a},\ a=1,2}
\end{array}
&&
\begin{array}{@{}c@{}}
\color{red!70!black}{2}\\[-2pt]
\color{blue!70!black}{\left(0,\frac{1}{2}\right)}\\[-2pt]
\color{green!50!black}{\widetilde{\psi}^{j},\ j=1,2}
\end{array}
&&&
\\[8pt]

&&&&
\begin{array}{@{}c@{}}
\color{red!70!black}{1}\\[-2pt]
\color{blue!70!black}{(0,0)}\\[-2pt]
\color{green!50!black}{\mathbf{1}}
\end{array}
&&&&
\end{array}
\]
Notice that we can go to a slightly different basis where the $({1\over 2},{1\over 2})$ chiral primaries satisfy the reality condition \eqref{hermbasis}, which is of the form
\begin{align}
\begin{split}
   \phi_{1} &= \frac{1}{\sqrt{2}}\left(\psi^{1}\widetilde{\psi}^{1}+\psi^{2}\widetilde{\psi}^{2}\right)
   \\   \phi_{2} &= \frac{i}{\sqrt{2}}\left(\psi^{1}\widetilde{\psi}^{1}-\psi^{2}\widetilde{\psi}^{2}\right)
   \\   \phi_{3} &= \frac{1}{\sqrt{2}}\left(\psi^{1}\widetilde{\psi}^{2}-\psi^{2}\widetilde{\psi}^{1}\right)
   \\   \phi_{4} &= \frac{i}{\sqrt{2}}\left(\psi^{1}\widetilde{\psi}^{2}+\psi^{2}\widetilde{\psi}^{1}\right)\,.
\end{split}
\end{align}

\paragraph{Berry curvature on Narain moduli space.}

There are $4\times 4=16$ marginal operators, which are super-descendants of the $({1\over 2},{1\over 2})$ chiral primaries. The marginal operators are
\be
\label{marginalt4}
{\cal O}_{(II')}=\partial X^I \overline{\partial}X^{I'}\,.
\ee
They correspond to tangent vectors on the Narain moduli space \eqref{narainspace}, so we expect that they have an $SO(4)\times SO(4)$ holonomy. To compute the curvature of the marginal operators we start with the following 4-point function
\be
\langle {\cal O}_{(BB')}(\infty)\,\,  {\cal O}_{(II')}(x)\,\, {\cal O}_{(JJ')}(y)\,\,{\cal O}_{(AA')}(0)\rangle
\ee
$$
=\left({\delta^{IJ}\delta^{AB}\over(x-y)^2}+{\delta^{IA}\delta^{JB}\over x^2}+{\delta^{IB}\delta^{AJ}\over y^2}\right)
\left({\delta^{I'J'}\delta^{A'B'}\over (\overline{x}-\overline{y})^2}+{\delta^{I'A'}\delta^{J'B'}\over \overline{x}^2}+{\delta^{I'B'}\delta^{J'A'}\over \overline{y}^2}\right)\,.
$$
Taking the antisymmetric combination and performing the necessary integral \eqref{berry4point}, we recover that the curvature is that corresponding to the tangent vectors of the coset
\be
(F_{(II')(JJ')})_{(AA')(BB')}=i\, \delta^{I'J'}\delta^{A'B'}\left(\delta^{IA} \delta^{JB} -\delta^{IB}\delta^{JA}\right) +i \,\delta^{IJ}\delta^{AB}\left(\delta^{I'A'} \delta^{J'B'} -\delta^{I'B'}\delta^{J'A'}\right)\,.
\ee
Let us now consider the currents $\partial X^A$ and compute their curvature. We need
\be
\langle \partial X^B (\infty)\,\,{\cal O}_{(II')}(x)\,\,{\cal O}_{(JJ')}(y)\,\, \partial X^A(0)\rangle
=\left({\delta^{IJ}\delta^{AB}\over(x-y)^2}+{\delta^{IA}\delta^{JB}\over x^2}+{\delta^{IB}\delta^{JA}\over y^2}\right){\delta^{I'J'}\over (\overline{x}-\overline{y})^2}\,.
\ee
Performing the integrals, we find that the left-moving currents $\partial X^A$ have curvature 
\be
(F_{(II'),(JJ')})_{BA} = i\, \delta^{I'J'}\left(\delta^{IA} \delta^{JB} -\delta^{IB}\delta^{JA}\right)
\ee
which corresponds to the left part of the $SO(4)\times SO(4)$ holonomy group of the coset.
By a similar computation, we find that the holonomy of the right-moving currents $\overline{\partial}X^{A'}$ corresponds to the right $SO(4)$ factor of the $SO(4)\times SO(4)$ holonomy group
\be
(F_{(II'),(JJ')})_{B'A'} = i\, \delta^{IJ}\left(\delta^{I'A'} \delta^{J'B'} -\delta^{I'B'}\delta^{J'A'}\right)\,.
\ee

\paragraph{Curvature of fermions and of the chiral ring.}

Since the theory is free everywhere on the conformal manifold, we can directly compute the curvature of chiral primaries over the conformal manifold  by the doubly-integrated 4-point function of two marginal operators and two chiral primaries \eqref{berry4point}. Let us denote by $\phi_{K,L}$ two general chiral primaries in the $T^4$ theory, which is a product of a number of left- and right- moving fermions. Since the marginal operators are made out of bosons and the theory is free, we find that the 4-point function factorizes
\be
\label{factort4}
\langle \phi_L^\dagger(\infty)\,\, {\cal O}_{(II')}(x)\,\,{\cal O}_{(JJ')}(y)\,\,\phi_K(0)\rangle = \langle \phi_L^\dagger(\infty) \,\,\phi_K(0)\rangle\,\cdot\,\langle \, {\cal O}_{(II')}(x)\,\,{\cal O}_{(JJ')}(y)\rangle\,.
\ee
Upon anti-symmetrization in $x\leftrightarrow y$ we get a vanishing result for \eqref{berry4point}, which means that the Berry curvature for chiral primary states vanishes.

It is straightforward to get  the same result  using the general form of the $\tt$ equations \eqref{ttgenj}, where all currents of the theory \eqref{t4rsym} and \eqref{t4flav} have to be carefully taken into account. \footnote{Depending on how one selects the basis of $({1\over 2},{1\over 2})$ chiral primaries the contribution of the currents can be different, though of course the final result for the curvature is basis-independent once expressed covariantly.}

\paragraph{A clarification.}

In general, we can think of the tangent bundle of the conformal manifold as the product of the bundle of $({1\over 2},{1\over2})$ chiral primaries times the bundle of the supercurrents.
In theories with the small ${\cal N}=4$ SCA, such as the case of $K3$ or $\mathrm{Sym}^N(K3)$, the $({1\over 2},{1\over2})$ chiral primaries have a non-trivial $SO(n)$ holonomy (where $n=20$ or $n=21$ respectively). The supercharges of the small ${\cal N}=(4,4)$ SCA also have an $SU(2)_{\mathrm{left}} \otimes SU(2)_{\mathrm{right}}= SO(4)$ holonomy  \cite{deBoer:2008ss,Gomis:2016sab}. So all in all the marginal operators have $SO(4)\otimes SO(n)$ holonomy, as expected for the tangent bundle since the conformal manifold is the symmetric space ${SO_{0}(4,n)\over SO(4)\times SO(n)}$ \eqref{eqn:D1/D5modulispace_ST}. 

In the case of a single $T^4$ sigma model, the conformal manifold is again of the same form ${SO_{0}(4,4)\over SO(4)\times SO(4)}$, so the marginal operators have $SO(4)\times SO(4)$ holonomy. However, in this case, and unlike the cases discussed above, the $SO(4)\times SO(4)$ holonomy of the marginal operators comes entirely from the holonomy of the supercharges --- since the chiral primaries for $T^4$ are flat. As discussed earlier in this subsection, the supercurrents for $T^4$ do indeed have enhanced $SO(4)\times SO(4)$ holonomy, as opposed to the smaller $SU(2)\times SU(2)$ holonomy of the supercurrents in theories with the small ${\cal N}=4$ SCA.

\subsection{Details on curvature for ${\rm Sym}^N(T^4)$}
\label{sec:detailssymt4}

Here we discuss some details of the curvature of chiral primaries for ${\rm Sym}^N(T^4)$ at a generic point on the conformal manifold.

{\bf Case 1: both deformations along ${\cal M}_4$}

In this case the curvature of chiral primaries vanishes. To see this, start with the 4-point function relevant for computing the curvature
\be
\langle \phi_L^\dagger(\infty) \,\, {\cal O}_\mu(x) \,\,{\cal O}_\nu(y)\,\,\phi_K(0)\rangle\,.
\ee
For deformations along ${\cal M}_4$, the marginal deformations are of $J\widetilde{J}$ form, where $J,\widetilde{J}$ are one of the $U(1)^4_L,U(1)^4_R$ bosonic currents. The chiral primaries are neutral under these currents, hence the 4-point function above factorizes
\be
\langle \phi_L^\dagger(\infty) \,\, {\cal O}_\mu(x) \,\,{\cal O}_\nu(y)\,\,\phi_K(0)\rangle = 
\langle \phi_L^\dagger(\infty) \,\,\phi_K(0)\rangle
\cdot
\langle {\cal O}_\mu(x) \,\,{\cal O}_\nu(y)\rangle\,,
\ee
which implies
\be
(F_{\mu\nu})_{KL}=0\,.
\ee
We can reproduce the same result using the $\tt$ equations \eqref{ttgenj}. Let us start by considering the case where the chiral primaries are primitive (in the sense of \eqref{primaryfermionic}). Then we have the 4-point function
\be
\langle \phi_L^\dagger(\infty)\,\, \Psi^\alpha \widetilde{\Psi}^{\alpha'}(x)\,\,(\Psi^\beta \widetilde{\Psi}^{\beta'})^\dagger(y)\,\,\phi_K(0)\rangle = {\delta_{KL}\delta^{\alpha \beta}\delta^{\alpha'\beta'}\over |x-y|^2}\,.
\ee
From this we can extract $[C_{\alpha\alpha'},C_{\beta\beta'}^\dagger]_{KL}=g_{KL}\delta_{\alpha\beta}\delta_{\alpha'\beta'}$. This cancels the identity contribution on the RHS of \eqref{ttgenj} and there is no current contribution,\footnote{This follows from the fact that $\phi_K,\phi_L$ are neutral under double-trace R-current \eqref{doubletracer} and flavor currents \eqref{doubletracef}, since they are primitive, and that $\Psi^\alpha \widetilde{\Psi}^{\alpha'},\Psi^\beta \widetilde{\Psi}^{\beta'}$ are neutral under the shifted R-symmetry current \eqref{shiftedcurrent}.} so indeed the $\tt$ equations \eqref{ttgenj} predict vanishing curvature. A similar analysis of the $\tt$ equations can be done for chiral primaries which are decorated with fermions.

{\bf Case 2: mixed deformations}

Let us take ${\cal O}_\mu$ to be a marginal operator along ${\cal M}_4$, which means it is of $J\widetilde{J}$ form, and ${\cal O}_\nu$ a marginal operator along ${\cal M}_5$. Then we find that
\be
\langle \phi_L^\dagger(\infty)\,\, {\cal O}_\mu(x)\,\, {\cal O}_\nu(y)\,\,\phi_K(0) \rangle = 0\,.
\ee
This follows by thinking of the 4-point function above as the coincident limit of a 5-point function where the two currents in ${\cal O}_\mu$ are separated. This 5-point function vanishes  from holomorphicity of the current in combination with the fact that the other operators are neutral under them. So again we find
\be
(F_{\mu\nu})_{KL}=0
\ee
along mixed directions.

We can also see this in terms of the $\tt$ equations. Again, consider the case where $\phi_K,\phi_L$ are primitive. Then we consider the correlator
\be
\langle \phi_L^\dagger(\infty)\,\, \Psi^\alpha \widetilde{\Psi}^{\alpha'}(x)\,\,\phi_j^\dagger(y) \,\, \phi_K(0)\rangle=0
\ee
where $\phi_j$ is a $({1\over 2},{1\over 2})$ chiral primary 
which is also primitive (and hence the corresponding marginal operator is along ${\cal M}_5$). The vanishing of this 4-point function follows from holomorphicity of the fermions and the fact that all other operators are primitive and hence obey \eqref{primaryfermionic}. The vanishing of the correlator above implies  $[C_{\alpha\alpha'},C_j^\dagger]=0$. There is no identity contribution or current contribution in \eqref{ttgenj}, so again the curvature computed by the $\tt$ equations \eqref{ttgenj} is zero. Similarly, we can analyze decorated chiral primaries.

{\bf Case 3: both deformations along ${\cal M}_5$}

The first thing to notice is that the 4-point function computing the curvature
\be
\langle \phi_K^\dagger(\infty) \,\,{\cal O}_\mu(x)\,\,{\cal O}_\nu(y) \,\, \phi_L(0)\rangle
\ee
is zero if the external chiral primaries have a different number of free fermion decorations. This means that we can examine the operator mixing order-by-order in the fermion number.

Let us start with the case where the external states are primitive chiral primaries. These states have vanishing double-trace charges and their shifted-$SU(2)_R$ charge \eqref{shiftedcharges} is equal to their usual $SU(2)_R$ charge, i.e $\widehat{r}=r$. Hence the only difference in the $\tt$ equations is the shift $\hat{k}\rightarrow \hat{k}-1$ due to the normalization of the current $\widehat{J}_{R,3}$ \eqref{shiftedcurrent}. Hence we arrive at equation \eqref{curvaturet4main}.

Then, it is straightforward to check that for chiral primaries decorated by fermions the curvature turns out to be the same as that of the corresponding primitive chiral primary. This happens because the $[C,C^\dagger]$ term is not modified and also the current contributions are not modified: 
we use the properly-shifted single-trace R-charge \eqref{shiftedcharges}, which does not change via decorating with fermions and is still the same as that of the corresponding fermionic primary, 
and we do not get any contributions from the double-trace currents since the chiral primaries corresponding to deformations along ${\cal M}_5$ are fermionic primaries and have vanishing double-trace charge.

\subsection{Orbifold checks}
\label{app:orbifold}

\subsubsection{Curvature computations in orbifold CFT}

In this section, we perform various checks of the curvature at the orbifold point, focusing on chiral primaries of dimension $({1 \over 2},{1 \over 2})$. There are 9 such states. Four of them are ``double trace" (and hence not primitive), we denote those simply as  $\Psi^\alpha \widetilde{\Psi}^{\alpha'}$ with $\alpha,\alpha'=1,2$. The remaining 5 are primitive chiral primaries. One of these 5 primaries is the twist-$2$ operator $\Sigma_2$. The other 4 are defined as follows.

First we introduce some notation. We denote the basic fermions of the theory as
\be
\Psi^\alpha(z) = {1\over \sqrt{N}} \sum_m \psi^\alpha_m(z) \qquad \alpha=1,2
\ee
where the sum is over the $N$ strands of the orbifold theory. We denote multi-trace products of such fermions by simply multiplying together the letters, for example $\Psi^\alpha \widetilde{\Psi}^{\alpha'}$ are the aforementioned double-trace operators.

We also have the single-trace operators
\be
\label{naivest}
 {1\over \sqrt{N}} \sum_m \psi_m^\alpha\widetilde{\psi}_m^{\alpha'}(z)\,.
\ee
Both of the above are normalized to have unit-normalized 2-point functions. 

Notice that single- and multi-trace operators are not orthogonal at finite $N$. It is convenient to perform a change of basis, to somewhat shifted single-trace operators, which we denote by hats and which are defined as
\be
\label{defhat}
\widehat{\Psi^\alpha \widetilde{\Psi}^{\alpha'}} = {1\over \sqrt{ N-1}} \sum_m \psi_m^\alpha \widetilde{\psi}_m^{\alpha'}(z) - {1\over \sqrt{N-1}} \Psi^\alpha \widetilde{\Psi}^{\alpha'}\,.
\ee
At large $N$ this is mostly single trace. The relative normalization is selected so that the hatted operators are orthogonal to the double-trace ones and the overall normalization is selected so that the hatted operators are unit-normalized. The hatted operators are also primitive in the sense of \eqref{primaryfermionic}.

To summarize, the 9 chiral primaries of dimension $({1\over 2},{1\over 2})$ are the double-trace operators $\Psi^\alpha \widetilde{\Psi}^{\alpha'}$, the hatted operators $\widehat{\Psi^\alpha \widetilde{\Psi}^{\alpha'}}$, and the twist-2 operator $\Sigma_2$. They are all mutually orthogonal.

We will now verify the general picture for the curvature of chiral primaries presented in Subsection \ref{sec:symT4} by computing the curvature of the $({1\over 2},{1\over 2})$ chiral primaries at the orbifold locus by direct orbifold computations and using equation \eqref{ttgenj}. First, we enumerate the conserved currents of the theory and construct orthogonal combinations so that we can apply \eqref{ttgenj}. We have the following $SU(2)$ multiplets of currents
\begin{enumerate}
\item R-symmetry 
\be
J_{R}^+ = -\sum_m \psi_m^1 \psi^2_m \qquad J_{R}^- = \sum_m \psi_m^{1\, \dagger} \psi^{2\, \dagger}_m\qquad J_{R}^3= {1\over 2} \sum_m (:\psi^1_m  \psi_m^{1\, \dagger}: + :\psi^2_m  \psi_m^{2\, \dagger}:)\,.
\ee
\item Accidental flavor symmetry
\be
J_{{\rm flavor}}^+ = -\sum_m \psi_m^1 \psi^{2\,\dagger}_m  \qquad J_{{\rm flavor}}^- =  \sum_m \psi_m^{1\, \dagger} \psi^{2}_m \qquad J_{{\rm flavor}}^3 =  {1\over 2} \sum_m (-:\psi^1_m  \psi_m^{1\, \dagger}: + :\psi^2_m  \psi_m^{2\, \dagger}:)\,.
\ee
\item Double-trace R-symmetry 
\be
J^{D,+}_{R} = -\Psi^1 \Psi^2 \qquad J^{D,-}_{R} = \Psi^{1\,\dagger} \Psi^{2\,\dagger} \qquad 
J^{D,3}_{R}= {1\over 2} (:\Psi^1 \Psi^{1\,\dagger}:+:\Psi^2 \Psi^{2\,\dagger}:)\,.
\ee
\item Double-trace flavor symmetry
\be
J^{D,+}_{{\rm flavor}} = -\Psi^1 \Psi^{2\,\dagger}\qquad J^{D,-}_{{\rm flavor}} = \Psi^{1\,\dagger}\Psi^2 \qquad J^{D,3}_{{\rm flavor}} ={1\over 2} (-:\Psi^1 \Psi^{1\,\dagger}:+:\Psi^2 \Psi^{2\,\dagger}:)\,.
\ee
\end{enumerate}

We construct orthogonal combinations of the currents. We find
\begin{equation}
    \langle J_{R} \cdot J_{R} \rangle = \frac{N}{2}, \quad \langle J_{R} \cdot J^D_{R} \rangle = {1\over 2}, \quad \langle J_{R}^D \cdot J_R^D \rangle = \frac{1}{2}\,.
\end{equation}
Therefore, we define
\begin{equation}
\label{shiftedcurrent2}
    \widehat{J_{R,3}} = J_{R,3}-J_{R,3}^{D}
\end{equation}
and consider the orthogonal currents $\widehat{J_R}$ and $J_R^D$.

In the case where both deformations are along ${\cal M}_4$, we expect that the curvature of chiral primaries will vanish. We already provided a general argument in Appendix \ref{sec:detailssymt4}, which continues to hold on the orbifold locus. Similarly, the general arguments in Appendix \ref{sec:detailssymt4} guarantee the vanishing of the curvature along mixed directions between ${\cal M}_4$ and ${\cal M}_5$.

We then come to the more interesting case where both deformations are along ${\cal M}_5$. Here we expect that the 5 primitive chiral primaries of dimension $({1\over 2},{1\over 2})$, namely $\widehat{\Psi^\alpha \widetilde{\Psi}^{\alpha'}}$ and $\Sigma_2$ will have a curvature corresponding to the fundamental representation of $SO(5)$, while the remaining 4 double-trace chiral primaries $\Psi^\alpha \widetilde{\Psi}^{\alpha'}$ will have vanishing curvature. We will confirm this result by using the $\tt$ equations \eqref{ttgenj} for all possible deformations along ${\cal M}_5$.

\paragraph{Both deformations untwisted.} 

We take the two deformations to be descendants of $\widehat{\Psi^\gamma \widetilde{\Psi}^{\gamma'}}$ and $\widehat{\Psi^\delta \widetilde{\Psi}^{\delta'}}$, so both are untwisted and along the part of the orbifold locus ${\cal M}_0$ inside ${\cal M}_5$.

First, we consider the 4-point function
\be
G(x,y)=\langle \widehat{\Psi^\beta\widetilde{\Psi}^{\beta'}}^\dagger (\infty) \,\, \widehat{\Psi^\gamma \widetilde{\Psi}^{\gamma'}}(x)\,\,
\widehat{\Psi^\delta \widetilde{\Psi}^{\delta'}}^\dagger(y)\,\,\widehat{\Psi^\alpha \widetilde{\Psi}^{\alpha'}} (0)\rangle\,.
\ee
Using \eqref{defhat} and doing Wick contractions, we find
\be
\label{basic4}
G(x,y) = {\delta^{\alpha\beta} \delta^{\alpha'\beta'} \delta^{\gamma\delta} \delta^{\gamma'\delta'} \over |x-y|^2}  + {\delta^{\beta \gamma}\delta^{\beta'\gamma'}\delta^{\alpha\delta}\delta^{\alpha'\delta'}\over |y|^2}+{1\over N-1}\left(  {\delta^{\gamma \delta} \delta^{\alpha\beta} \delta^{\beta'\gamma'}\delta^{\alpha'\delta'} \over (x-y) \overline{y}} 
+ {\delta^{\beta\gamma}\delta^{\alpha\delta} \delta^{\gamma'\delta'}\delta^{\alpha'\beta'} \over y (\overline{x}-\overline{y})} \right)\,.
\ee

We now switch to a ``real basis" of chiral primaries by defining
\be
\phi_i \equiv A_{i,\alpha\alpha'} \widehat{\Psi^\alpha \widetilde{\Psi}^{\alpha'}}\,, \qquad i = 1,2,3,4
\ee
and the matrices $A_{i,\alpha \alpha'}$ can be taken to be $A_i={1\over \sqrt{2}}(\sigma_1,\sigma_2,\sigma_3,i {\mathbbm{1}})$.  Remember, this basis was defined in order to satisfy
\be
[J_{R}^-,[\widetilde{J}_{R}^-, \varphi_i]]= \varphi_i^\dagger\,.
\ee
Notice that the 2-point function in this basis is 
\be
\langle \varphi_i (\infty) \varphi_j^\dagger(0)\rangle = A_{i,\alpha\alpha'} A^*_{j,\beta\beta'} \langle  \widehat{\Psi^\alpha \widetilde{\Psi}^{\alpha'}}(\infty)  \widehat{\Psi^\beta \widetilde{\Psi}^{\beta'}}(0)\rangle =  A_{i,\alpha\alpha'} A^*_{j,\beta\beta'} \delta_{\alpha\beta} \delta_{\alpha'\beta'} = {\rm Tr}[A_i A_j^\dagger] = \delta_{ij}
\ee
where the trace is over $2\times2$ matrices.

Using \eqref{basic4}, we rewrite the 4-point function in the new basis
\be
G'(x,y) = \langle \varphi_l^\dagger(\infty) \,\varphi_i(x) \, \varphi_j^\dagger(y)\, \varphi_k(0)\rangle  = {\delta_{ij}\delta_{kl} \over |x-y|^2} + {\delta_{il}\delta_{jk} \over |y|^2} + {1\over N-1} \left({{\rm Tr}[A_l^*\, A_i^T\, A_j^* A_k^T]\over (x-y)\overline{y}} + {{\rm Tr}[A_l^\dagger A_i A_j^\dagger A_k] \over y(\overline{x}-\overline{y})}\right)\,.
\ee
One can see that the numerators in the last two terms are both equal and the expression above can be rewritten as
\be
G'(x,y)={\delta_{ij}\delta_{kl} \over |x-y|^2} + {\delta_{il}\delta_{kj} \over |y|^2} + {1\over N-1} ({1\over 2} \delta_{ij}\delta_{kl} + {1\over 2} \delta_{il} \delta_{jk} - {1 \over 2} \delta_{jl} \delta_{ik})\left({1\over (x-y)\overline{y}} + {1 \over y(\overline{x}-\overline{y})}\right)\,.
\ee
Now, by taking limits of this 4-point function, we can extract the desired terms. First, let us take the $\lim_{x\rightarrow 0,y\rightarrow \infty} (|y|^2 G'(x,y))$ to extract the ``raising term" in the commutator $[C,C^\dagger]$. We find
\be
\label{identitycrt4}
C_{ik}^M C_{jl}^M = \delta_{ij}\delta_{kl} + \delta_{il} \delta_{jk} -{1\over N-1}\left(\delta_{ij}\delta_{kl} + \delta_{il} \delta_{jk} - \delta_{lj} \delta_{ik} \right)
\ee
where the sum over $M$ is over chiral primaries of dimension $(1,1)$ and where we used the reality of the chiral ring coefficients for bosonic primaries. This is precisely the identity \eqref{chiralringidentity} discussed in \cite{Gomis:2016sab} with $\hat{k}=N-1$, as expected due to the general discussion in Section \ref{sec:symT4}.

For the full commutator, we find
\be
[C_i, C_j^\dagger]_{kl} =   \delta_{ij}\delta_{kl} -{1\over N-1}\left(\delta_{ij}\delta_{kl} + \delta_{il} \delta_{kj} - \delta_{jl} \delta_{ik} \right)\,.
\ee
Finally, for the current term contribution we get the identity and the contribution from the currents.\footnote{Naively, we would use \eqref{eqn:4ptfn_currentscontr} to isolate the {\it total} contribution of all currents, including the accidental one, and we would find that the current term contribution is
\be
 \delta_{IJ}\delta_{KL} -{1\over N-1}\left(\delta_{IJ}\delta_{KL} + \delta_{IK} \delta_{JL} - \delta_{KJ} \delta_{IL} \right)\,.   
\ee
This seems to precisely cancel the $[C,C^\dagger]$ term and seems to predict vanishing curvature along these deformations, which contradicts our expectation. To get the correct result we need to remove the contribution of the accidental current ``by hand".} It is important to keep only the currents that remain conserved in a neighborhood of the orbifold point, namely the shifted single-trace R-symmetry current $\widehat{J_{R,3}}$ \eqref{shiftedcurrent2}  with contribution ${1\over N-1}\delta_{ij}\delta_{kl}$. Putting everything together, let us evaluate the curvature using the $\tt$ equations, using again the notation for the marginal operators $\mu = (a,i),\nu=(b,j)$. After some basic algebra, we find
\be
(F_{\mu\nu})_{kl} = i\, \delta_{ab}{1\over N-1}\left(\delta_{il} \delta_{jk} - \delta_{jl} \delta_{ik} \right)
\ee
which is precisely what we expect for a generator in the fundamental of $SO(4)\in SO(5)$. Notice that if $i=j$ this vanishes, as expected.

Finally, we consider the case where the external states involve the twisted chiral primary $\Sigma_2$. Since we are considering two untwisted deformations, and using twist conservation, there is no mixing between $\Sigma_2$ and $\varphi_i$ since the relevant 4-point functions vanish. Then we consider
\be
\langle \Sigma_2^\dagger(\infty)\,\,\varphi_i(x) \,\, \varphi_j^\dagger(y) \,\, \Sigma_2(0)\rangle ={ \delta_{ij} \over |x-y|^2}\left[{N-2 \over N-1} + {1\over N-1} {|x| \over |y|}\right]\,.
\ee
From this, we extract
\be
[C_i,C_j^\dagger]_{22}=\delta_{ij}{N-2 \over N-1}\,.
\ee
From the identity and shifted R-current contribution, we get
\be
\delta_{ij} - {1\over N-1}\delta_{ij}
\ee
which cancels the term above, as expected for this particular component of the curvature.

\paragraph{Both deformations along ${\cal M}_5$, twisted-untwisted.}

Now we consider one deformation along an untwisted $\varphi_i$ and the other along $\Sigma_2$. By twist conservation, the only nontrivial matrix elements are those mixing $\Sigma_2$ with other untwisted chiral primaries $\varphi_J$. Let us consider the orbifold 4-point function
\be
\langle \varphi_l^\dagger(\infty)\,\,\varphi_i(x)\,\,\Sigma_2^\dagger(y) \,\, \Sigma_2(0)\rangle = {\delta_{il}\over |y|^2}\left[{N-2 \over N-1} + {1\over N-1} {|x| \over |x-y|}\right]
\ee
From this, we extract that
\be
[C_i,C_{\Sigma_2}^\dagger]_{\Sigma_2\,l} = \delta_{il} \left[{N-2 \over N-1}-1\right]= -{\delta_{il}\over N-1}\,.
\ee
There is no identity or current contribution and we find that the curvature is
\be
(F_{\mu\nu})_{\Sigma_2\,l}= i\,\delta_{ab}{1\over {N-1}}\delta_{il}
\ee
along the directions $\mu=(a,i)$ and $\nu=(b,\Sigma_2)$.

Combining this with the previous results, we reproduce the expected statement that the 5 chiral primaries $\varphi_i,\Sigma_2$ have a curvature in the fundamental representation of $SO(5)$. The prefactor ${1\over N-1}$ is fixed by the scaling of the metric of the conformal manifold ${\cal M}_5$.

\paragraph{Both deformations along ${\cal M}_5$, both twisted.}
\label{orbcheckssigma2}

Finally, we consider the case where both deformations are associated to $\Sigma_2$. As we can see from the general formula \eqref{eqn:D1D5_1/2BPS_tildeF_final_N=4}, we expect the curvature of all chiral primaries to vanish along these directions: suppose two marginal operators of the form \eqref{marginal1} have $i=j$. If they also have the same index $a=b$, then they are exactly the same operator and then the curvature vanishes by anti-symmetry. If $a\neq b$ the curvature vanishes again because of the $\delta_{ab}$ in the formula \eqref{eqn:D1D5_1/2BPS_tildeF_final_N=4}.

\subsubsection{Some orbifold checks of identity \eqref{identitychiral}}

We now provide some checks of the identity \eqref{identitychiral} at the orbifold point, where we have to be careful that in the case of ${\rm Sym}^N(T^4)$ we need to shift $N\rightarrow N-1$, as discussed in Section \ref{sec:symT4}. Then the identity \eqref{identitychiral} takes the form
\be
\label{identitychiral2_twist2}
[C_{\Sigma_2},C_{\Sigma_2}^\dagger]_{KL} = \delta_{KL}\left(1-{1\over N-1}(r+\widetilde{r})\right)\,.
\ee
We will first try to check the identity \eqref{identitychiral2_twist2} for the case where the external states are primitive $({1\over 2},{1\over 2})$ chiral primaries, so in this case $r=\widetilde{r}={1\over 2}$. From twist-conservation, we notice that $[C_{\Sigma_2}, C_{\Sigma_2}^\dagger]_{kl}$ can be non-zero only if $k,l$ are both twisted or both untwisted. We consider these two cases separately.

Suppose that both external operators are untwisted. In this case, we need the orbifold 4-point function
\be
G_3(x,y) = \langle\varphi_l^\dagger(\infty)\,\,\Sigma_2(x)\,\,\Sigma_2^\dagger(y) \,\,\varphi_k(0)\rangle ={\delta_{kl}\over |x-y|^2}\left({N-2 \over N-1} + {1\over N-1}{|x|\over |y|}\right)\,.
\ee
From this, we extract
\be
\label{ccdrhat}
[C_{\Sigma_2}, C_{\Sigma_2}^\dagger]_{kl} ={\delta_{kl}} {N-2 \over N-1} \,.
\ee
From the identity and current contribution we get
\be
\delta_{kl}\left(1 -{1\over N-1}\right) 
\ee
which agrees with the one above, so \eqref{identitychiral2_twist2} is obeyed since $r=\widetilde{r}={1\over 2}$.

Finally, suppose both external operators are the twist-$2$ operator $\Sigma_2$.  This case is somewhat more complicated. For a direct check like the ones above one could start with the 4-point function
\be
\langle (\Sigma_2)^\dagger(\infty) \,\,\Sigma_2(x)\,\,\Sigma_2^\dagger(y) \,\,\Sigma_2(0)\rangle
\ee
and try to extract $[C,C^\dagger]$ by taking various limits. However, we were not able to find a simple expression for this 4-point function in the literature. Alternatively, one can try to compute $[C,C^\dagger]$ by computing all possible 3-point functions of the form $\langle \Sigma_2 \Sigma_2 \varphi^\dagger\rangle$ where $\varphi$ is a chiral primary of dimension $(1,1)$ For $N=2$ we find
\be
[C_{\Sigma_2},C_{\Sigma_2}^\dagger]_{{\Sigma_2} {\Sigma_2}} = 0\,.
\ee
For the RHS of \eqref{identitychiral2_twist2} we have $r=\widetilde{r}=1/2$. Using that $N=2$ we find that also the RHS is equal to $0$ and again \eqref{identitychiral2_twist2} works.

We can also work out $[C_{\Sigma_2}, C_{\Sigma_2}^\dagger]_{KL}$ for general 1/2-BPS external states and in principle at general $N$ using the results in \cite{Avery:2010hs,Carson:2014ena}. In these references, it was worked out what states one can get by acting $\Sigma_2$ (or $\Sigma_2^\dagger$) on a 1/2-BPS state, from which we can then compute the matrices $C_{\Sigma_2},  C_{\Sigma_2}^\dagger$. Using this data, we have done the computation explicitly up to $N=6$ and can fully reproduce the results in Section \ref{ccdagertt} from this independent method.

\subsection{On the factorization of the $1/2$-BPS partition function for ${\rm Sym}^N(T^4)$}\label{app:factorization}

We start with the grand-canonical partition function for 1/2-BPS states \eqref{goettsche} which reads
\be
\label{goettschet4}
\sum_{N=0}^\infty Q^N Z_N(y,\widetilde{y}) = \prod_{m=1}^\infty Z^{(m)}(Q,y,\widetilde{y}) \,,
\ee
where 
\begin{align}
\begin{split}
&Z^{(m)}(Q,y,\widetilde{y})
\\  &={(1+Q^m y^{m}\widetilde{y}^{m-1})^2(1+Q^m y^{m-1}\widetilde{y}^{m})^2(1+Q^m y^{m}\widetilde{y}^{m+1})^2(1+Q^m y^{m+1}\widetilde{y}^{m})^2\over (1-Q^m y^{m-1}\widetilde{y}^{m-1})(1-Q^m y^{m+1}\widetilde{y}^{m-1})(1-Q^m y^{m-1}\widetilde{y}^{m+1})(1-Q^m y^{m+1}\widetilde{y}^{m+1}) (1-Q^m y^{m}\widetilde{y}^{m})^4}\,.
\end{split}
\end{align}
We focus on a particular $m$ and redefine $q\equiv Q^m y^{m-1}\widetilde{y}^{m-1}$, so we have
\be
Z^{(m)}={(1+q y)^2(1+q \widetilde{y})^2(1+q y^2 \widetilde{y})^2(1+q y \widetilde{y}^2)^2\over (1-q)(1- q y^2)(1-q \widetilde{y}^2)(1-q y^2 \widetilde{y}^2) (1-q  y \widetilde{y})^4}\,.
\ee
It is then straightforward to check that this can also be written as
\be
\label{splitting}
Z^{(m)} = 1 + q (1+y)^2(1+\widetilde{y})^2X \,,
\ee
where
\be
\label{splitting2}
X=X_1+X_2+X_3+X_4+X_5 \,,
\ee
and
\be
X_1 = \frac{q^3 y^4 \widetilde{y}^4}{\left(1-q y^2\right) \left(1-q \widetilde{y}^2\right) (1-q y \widetilde{y})^4 \left(1-q y^2 \widetilde{y}^2\right)} \,,
\ee
\be
X_2 = \frac{2 q^2 y \widetilde{y}^2(1+y \widetilde{y})    }{(1-q) \left(1-q \widetilde{y}^2\right) (1-q y \widetilde{y})^4 \left(1-q y^2 \widetilde{y}^2\right)} \,,
\ee
\be
X_3 = \frac{2 q^2 y^2 \widetilde{y} (1+y \widetilde{y})}{(1-q) \left(1-q y^2\right) (1-q y \widetilde{y})^4 \left(1-q y^2 \widetilde{y}^2\right)} \,,
\ee
\be
X_4= \frac{1+q y \widetilde{y}}{(1-q) \left(1-q y^2\right) \left(1-q \widetilde{y}^2\right) (1-q y \widetilde{y})^3 \left(1-q y^2 \widetilde{y}^2\right)} \,,
\ee
\be
X_5=\frac{q^3 y^2 \widetilde{y}^2 \left(q^3 y^4 \widetilde{y}^4+2 q y^3 \widetilde{y}^2+y^2 \left(2 q \left(\widetilde{y}^3+\widetilde{y}\right)+1\right)+2 y \widetilde{y} (q \widetilde{y}+2)+\widetilde{y}^2+1\right)}{(1-q) \left(1-q y^2\right) \left(1-q \widetilde{y}^2\right) (1-q y \widetilde{y})^4 \left(1-q y^2 \widetilde{y}^2\right)} \,.
\ee
Now, we notice that in a power series expansion in $q$ of the factor $X$ in \eqref{splitting}, each power $q^n$ comes multiplied by a polynomial in $y,\widetilde{y}$ with positive integer coefficients. This follows by direct observation of the $q$-expansion of each of the $X_i$'s presented above. Since each of $Z^{(m)}$ can be written as \eqref{splitting} with the corresponding $X$ having positive, integer coefficients in its $q-$expansion, it follows that
\be
Z_N(y,\widetilde{y}) = (1+y)^2(1+\widetilde{y})^2 \widehat{Z}_N(y,\widetilde{y})\,,\quad N\geq 2 \,,
\ee
with $\widehat{Z}_N(y,\widetilde{y})$ being a polynomial in $y,\widetilde{y}$, with positive integer coefficients.

\section{Details for ${\cal N}=4$ SYM computations}
\label{sec:SYMdetails}

This appendix contains the technical details needed in Section \ref{sec:SYM}, including a proof using crossing symmetry in \ref{app:4dcrossing} that the Berry curvature vanishes for 1/2-BPS states and a careful analysis in \ref{app:n4bd1} and \ref{app:n4bd2} of the OPE contributions to the integrated four-point function in the Berry curvature of 1/4-BPS states, showing that they are all zero.

\subsection{A crossing symmetry proof of \eqref{identitychiral4d}}
\label{app:4dcrossing}

Here we provide a proof of \eqref{identitychiral4d} at general values of the coupling, based on crossing symmetry. The proof is similar to the one in Appendix \ref{sec:CRCidentity} for 2d ${\cal N}=(4,4)$ SCFTs.

We denote by $\varphi^{\pm}$ the highest/lowest weight states of a 1/2-BPS multiplet. Using the fact that marginal operators are neutral under $SU(4)$, we have that
\be
\langle \phi_L^{-}(\infty) \,\, {\cal O}_\tau(x)\,\, {\cal O}_\tau^\dagger (y) \,\, \phi_K^+(0)\rangle =\langle \phi_L^{+}(\infty) \,\, {\cal O}_\tau(x)\,\, {\cal O}_\tau^\dagger (y) \,\, \phi_K^-(0)\rangle \,.
\ee
Using superconformal Ward identities we can rewrite this as
\be
\Box_x \Box_x \langle \phi_L^{-}(\infty) \,\, \phi_2(x)\,\, \phi_2^\dagger (y) \,\, \phi_K^+(0)\rangle = \Box_y \Box_y \langle \phi_L^{+}(\infty) \,\, \phi_2(x) \,\, \phi_2^\dagger(y)\,\,\phi_K^-(0)\rangle \,,
\ee
where we use the notation $\phi_2 ={\rm Tr}(Z^2)$. 
Using a conformal Ward identity\footnote{The Ward identity we use holds for the 4-point function of two scalar, primary operators at $x,y$ each with $\Delta=2$ and two operators at $0,\infty$ which are scalar and have arbitrary (same) dimension.} to slightly rewrite the RHS, the condition above takes the form
\be
\label{bifunction}
\Box_x \Box_x G(x,y) = 0 \,,
\ee
with
\be
\label{defgapp}
G(x,y) \equiv  \langle \phi_L^{-}(\infty) \,\, \phi_2(x)\,\, \phi_2^\dagger (y) \,\, \phi_K^+(0)\rangle - {|x|^4\over |y|^4} \langle \phi_L^{+}(\infty) \,\, \phi_2(x) \,\, \phi_2^\dagger(y)\,\,\phi_K^-(0)\rangle \,.
\ee
We notice that $G$ is regular as $x\rightarrow 0$. The function $G$ is ``bi-harmonic" \eqref{bifunction} on ${\mathbb R}^4-\{y\}$. Let us define the sphere-average as
\be
M(r) \equiv {1\over 2\pi^2 r^3}\int_{S_r} G(x,y) \,,
\ee
where here we take $y$ to be fixed and we integrate over $x$ on a sphere of radius $r$ centered around $y$. For a bi-harmonic function, such sphere-averages must be of the form 
\be
M(r) = \alpha + \beta r^2+ \gamma/r^2 \,,
\ee
for some constants $\alpha,\beta,\gamma$.\footnote{One way to see it is: expand $G$ in spherical harmonics (as a function of $x$, with $y$ fixed and spherical expansion centered around $y$). Only the s-wave contributes to the sphere-average. A priori the s-wave is an arbitrary spherically symmetric function $f(r)$. Demanding that the $\Box_x \Box_x$ kills $f(r)$, we get that $f(r)=\alpha+\beta r^2 +\gamma/r^2$.} Since the expression \eqref{defgapp} does not blow up as $x\rightarrow \infty$, we have that $\beta=0$. Hence we get 
\be
\label{biharmrel}
\lim_{r\rightarrow \infty} M(r) = \lim_{r\rightarrow 0} (M(r)-\gamma/r^2) \,.
\ee

First, it is clear by taking the $x\rightarrow \infty$ limit of \eqref{defgapp} that 
\be
\label{limminf}
\lim_{r\rightarrow\infty} M(r) =-{1\over |y|^4} [C_2,C_2^\dagger]_{KL} \,.
\ee

Second, we examine the limit $r\rightarrow 0$. 
We proceed by considering the contribution of scalar primaries appearing in the $x,y$ OPE. Suppose such an operator has dimension $\Delta$. Then we get a contribution of the form
\be
{1\over |y|^{4+\Delta}}{|y|^4-|x|^4\over |x-y|^{4-\Delta}} \,.
\ee
Now, remember that in ${\cal N}=4$ SYM we have a unitarity bound $\Delta\geq 2$ for scalar operators (except for the identity). For $\Delta\geq 2$ the contribution above does not behave like ${1\over r^2}$, where $r=|x-y|$, so it does not contribute to $\gamma$ (but could potentially contribute to $\alpha$). So for determining $\gamma$ we only need to consider the identity operator. From the identity operator we get 
\be
-\left({1\over |y|^4} + {3 \over r^2 |y|^2}\right)\delta_{KL} \,,
\ee
where we assume that we are in a basis where $\langle \phi_L^\dagger(\infty) \phi_K(0)\rangle = \delta_{KL}$. So we find
\be
\gamma=-{3 \over |y|^2}\delta_{KL} \,,
\ee
since no other operators (of any spin) can potentially contribute to $\gamma$.

The next step is to compute the constant $\alpha$. This can be computed by taking the $r\rightarrow 0$ limit of $M(r)-\gamma/r^2$. From the identity operator we get the contribution
\be
-{1 \over |y|^4}\delta_{KL} \,.
\ee
For operators of dimension $\Delta=2$ we find that the primary, as well as the first descendant (but not higher descendants), contribute to $\alpha$ a factor of 
\be
-{1\over |y|^4} {2 \,p\over 3c}\delta_{KL} \,,
\ee
where $p$ is the $SU(4)$ R-charge $[0,p,0]$ of the external states and $c$ one of the conformal anomalies of the CFT. Scalar operators of dimension $\Delta>2$ do not contribute. For spin-1 currents, we find the contribution to $\alpha$ equal to 
\be
{1\over |y|^4}\cdot \left({p\over 6c}\right) \delta_{KL} \,.
\ee
Higher spin operators (or fermionic operators) do not contribute.

Now, combining the contribution of the identity, the scalar of $\Delta=2$ (and its first descendant) and the currents we find that the constant $\alpha$ is
\be
\alpha= -{1\over |y|^4} \left(1+{p \over 2c}\right) \delta_{KL} \,,
\ee
or
\be
\lim_{r\rightarrow 0}(M(r)-\gamma/r^2)=-  {1\over |y|^4} \left(1+{p \over 2c}\right) \delta_{KL} \,.
\ee
Combining this with \eqref{limminf} and using \eqref{biharmrel}, we get the desired identity \eqref{identitychiral4d}
\be
[C_2,C_2^\dagger]_{KL}= \left(1+{p\over 2 c}\right)\delta_{KL} \,,
\ee
which guarantees the vanishing of the curvature of 1/2-BPS operators via \eqref{maintt4d}.

\subsection{Boundary terms as $x,y\rightarrow 0$}
\label{app:n4bd1}

We want to study the OPE of ${\rm Tr}(Y^2)$ and of ${\rm Tr}(\overline{Y}^2)$ with $\phi_K^+$, where the latter is either a 1/2- or 1/4-BPS highest-weight state and determine whether there are any singular terms that could contribute to the boundary terms. First, let us remember that if we have a scalar operator in an $SU(4)$ representation $[k,p,r]$ then we have two different unitarity bounds: one from the left-chiral supercharges, which reads $\Delta \geq {k + 2p +3r \over 2}$, and one from the right-chiral supercharges, which reads $\Delta\geq {3k + 2p + r \over 2}$. For a given scalar operator the overall bound is $\Delta \geq \mathrm{max}\{{k + 2p +3r \over 2},{3k + 2p + r \over 2}\}$.

We work in conventions where the three simple roots of $SU(4)$ are
\be
\alpha_1 = (2,-1,0) \,, \qquad \alpha_2 = (-1,2,-1)\,, \qquad \alpha_3 = (0,-1,2) \,.
\ee
We start with highest-weight vectors and act with lowering operators $E_i^\dagger$, which change the weight vector by minus the vectors of the roots mentioned above.

We denote by $\phi_2={\rm Tr}(Z^2)$ the highest-weight state of the representation $[0,2,0]$. 
 We want to identify what types of states could potentially appear in the OPE between ${\rm Tr}(Y^2)= \kappa E_3^\dagger E_2^\dagger E_3^\dagger E_2^\dagger \phi_2$ and $\phi_K^+$. In principle, we could consider the Clebsch-Gordan decomposition, but for our purposes it is simpler to enumerate all possible representations that could {\it potentially} appear in this product of representations and check that they would all lead to regular terms in the OPE. To identify those, we consider the full set of states in the product of the representations $\phi_2$ and $\phi_K^+$. All these states are of the general form $(E_{i_1}^\dagger \ldots E_{i_m}^\dagger \phi_2) \cdot (E_{j_1}^\dagger \ldots E_{j_n}^\dagger \phi_K^+)$. Certain linear combinations of these states are $SU(4)$ highest weight states and others are $SU(4)$ descendants. The $SU(4)$ quantum numbers of the primaries provide unitarity bounds on the conformal dimension of the operators that can appear in the OPE. From charge conservation, any state that appears on the RHS of the OPE we are considering must contain two $E_2^\dagger$ roots and two $E_3^\dagger$ roots acting, in some combination, on $\phi_2$ and $\phi_K^+$. Since the roots are linearly independent, the root $\alpha_1$ does not participate and we can ignore it for now. For our purposes it is then sufficient to classify the states according to the number (which we call ``level" below) of $E_2^\dagger,E_3^\dagger$ roots that we need to act with on $\phi_2$ and $\phi_K^+$ to get the corresponding highest weight state of the representation.

\paragraph{Level 0:} Here we only have the highest weight state $\phi_2 \cdot \phi_K^+$ with $SU(4)$ charges and unitarity bound
\be
[k,p+2,k] \qquad \Delta \geq p+2k+2 \,.
\ee
Given the constraint on $\Delta$, any $SU(4)$ descendant of this state that could potentially appear in our OPE would be regular.

\paragraph{Level 1:} Here we have states where we have used only one root, either $E_2^\dagger$ or $E_3^\dagger$. Some of these states will be descendants of the previous one, while others may be highest weights of genuinely new representations. We do not care about the details of this decomposition, since as stated above we want to imagine the largest possible set of representations that could potentially appear (even if they do not actually appear). Hence at this level we get two new weights with corresponding unitarity bounds
\begin{align}
[k+1,p,k+1]_2 & \qquad \Delta \geq p+2k+2 \,, \\
[k,p+3,k-2]_3 & \qquad \Delta \geq p+2k+2\,.
\end{align}
The subscript indicates which roots we have acted with. So again the OPE would be regular. 

\paragraph{Level 2:} Here we have states with two roots selected from 2, 3. The possible highest-weight states are
\begin{alignat}{3}
 &[k+2,p-2,k+2]_{22} & \qquad \Delta \geq p+2k+2 \,, \\
&[k+1,p+1,k-1]_{23} & \qquad \Delta \geq p+2k+2 \,, \\
&[k,p+4,k-4]_{33} & \qquad \Delta \geq p+2k+2 \,.
\end{alignat}
The OPE would be regular.\footnote{Notice that the way we are doing this analysis (i.e. not a complete Clebsch-Gordan decomposition, but only classifying the possible highest weights that could in-principle appear), we do not care about the ordering of the roots in the subscripts.}

\paragraph{Level 3:} States with 3 roots
\begin{alignat}{2}
&[k+2,p-1,k]_{223} &\qquad \Delta \geq p+2k +2 \,, \\ 
&[k+1,p+2,k-3]_{233} &\qquad \Delta \geq p+2k+2 \,.
\end{alignat}
The OPE is regular.

\paragraph{Level 4:} States with 4 roots
\be
[k+2,p,k-2]_{2233} \qquad \Delta \geq p+2k +2 \,.
\ee
The OPE is regular.

We can do a similar analysis for the OPE of ${\rm Tr}(\overline{Y}^2)$ with $\phi_K^+$ and we find again that the OPE is regular. 

So all in all we have found that there are no boundary terms from $x,y \rightarrow 0$ when we integrate by parts in \eqref{integralcurv2}.

\subsection{Boundary terms as $|x|,|y|\rightarrow 1$}
\label{app:n4bd2}

Since there are no boundary terms at $x,y=0$, integrating by parts in \eqref{integralcurv2}, we can write the curvature as 
\begin{align}
\label{boundary1}
&(F_{\tau\overline{\tau}})_{KL}= \lim_{r\rightarrow 1^-} \, {i\over (2\pi)^4}\int_{|x|=r} d^3\Omega_x \int_{|y|=1} d^3 \Omega_y\,\,|x|^2 |y|^2 \,(x\cdot\partial_x)\,(y\cdot\partial_y) \times\cr &\times\left[\langle \phi_L^-(\infty)\,\,{\rm Tr}(Y^2)(x)\,\, {\rm Tr}(\overline{Y}^2)(y)\,\, \phi_K^+(0) \rangle -\langle \phi_L^-(\infty) \,\,{\rm Tr}(\overline{Y}^2)(x)\,\,{\rm Tr}(Y^2)(y) \,\,\phi_K^+(0) \rangle\right].
\end{align}
As explained in \cite{Papadodimas:2009eu}, for regions of the integral where $\Omega_x\neq \Omega_y$, we notice that due to the antisymmetrization, the regions related by $(\Omega_x,\Omega_y)\leftrightarrow (\Omega_y,\Omega_x)$ cancel exactly in the limit $r\rightarrow 1^-$. The only subtlety is along the diagonal $\Omega_x=\Omega_y$ in the limit $r\rightarrow 1^-$. For that limit, we can use the OPE between the operators at $x,y$.

We classify the contributions according to spin. Only representations with $l=0,1$ can contribute in the limit (since operators with $l\geq 2$ would have $\Delta\geq 4$ and then the singularity as $x\rightarrow y$ is not hard enough to give a finite piece in the limit $r\rightarrow 1$).

\paragraph{Scalar contribution.}

Suppose we have a scalar operator ${\cal O}$ of dimension $\Delta$ in the OPE. As we will see, it is important to also keep the first conformal descendant (higher descendants do not contribute). So we have
\be
{\rm Tr}(Y^2)(x)\cdot {\rm Tr}(\overline{Y}^2)(0) = \ldots+ {C^{\cal O}_{Y^2\,\overline{Y}^2} \over |x|^{4-\Delta}}\left({\cal O}(0)+ {1\over 2} x^\mu \partial_\mu {\cal O}(0) + \ldots \right) + \ldots \,.
\ee
Then in the 4-point functions we will have
\begin{center}
\begin{minipage}{\linewidth}
    \begin{multline}
    \langle \phi_L^-(\infty)\,\,{\rm Tr}(Y^2)(x)\,\, {\rm Tr}(\overline{Y}^2)(y)\,\,\phi_K^+(0)\rangle = \\
    C^{\cal O}_{Y^2\,\overline{Y}^2}\, C_{{\cal O} K L} \left({1\over |x-y|^{4-\Delta}|y|^\Delta} -{\Delta \over 2}{(x-y)\cdot y\over |x-y|^{4-\Delta}|y|^{\Delta+2}} +\ldots\right)+\ldots \,,
\end{multline}
\end{minipage}
\end{center}
where dots indicate other conformal descendants of ${\cal O}$, which however will not contribute, as well as other conformal families.

For simplicity of notation, let us define
\be
\Xi_\Delta(x,y)\equiv \left({1\over |x-y|^{4-\Delta}|y|^\Delta} -{\Delta \over 2}{(x-y)\cdot y\over |x-y|^{4-\Delta}|y|^{\Delta+2}}\right) \,.
\ee
Hence the contribution from the scalar ${\cal O}$ to the curvature \eqref{boundary1} is
\be
\label{scalarintegral}
{i\over(2\pi)^4} C^{\cal O}_{Y^2\,\overline{Y}^2}\, C_{{\cal O} K L}\,\lim_{r\rightarrow 1^-} \int_{|x|=r} d^3\Omega_x \int_{|y|=1} d^3 \Omega_y\,\,|x|^2 |y|^2 \,(x\cdot\partial_x)\,(y\cdot\partial_y) \left[\Xi_\Delta(x,y) - \Xi_\Delta(y,x) \right] \,.
\ee
Here it is important to remember that for neutral operators there is a unitarity restriction $\Delta \geq 2$ from the norm of a second super-descendant \cite{Dolan:2002zh}. Of course, we can also have $\Delta=0$, but then the integrand above is identically zero. Hence we only need to examine this formula for $\Delta \geq 2$.

Carefully evaluating the integral \eqref{scalarintegral} in Mathematica and taking the limit $r\rightarrow 1^-$, we find that \eqref{scalarintegral} is zero for any $\Delta \geq 2$. Hence we do not have any scalar contribution.\footnote{Notice that if we look at exactly $\Delta=2$ we find a non-vanishing contribution from the primary which precisely cancels with the contribution of the first descendant.}

\paragraph{Current contribution.}

A scaling argument of the singularity implies that when $l=1$, only operators with $\Delta=3,l=1$ can potentially contribute. The contribution from all such conserved spin-1 currents to the 4-point function can be encoded in a coefficient $D_{KL}$, in an expansion of the 4-point function where we have kept only the spin-1 primaries
\be
\label{yetanothercor}
\langle \phi_L^-(\infty)\,\,{\rm Tr}(Y^2)(x)\,\, {\rm Tr}(\overline{Y}^2)(y)\,\,\phi_K^+(0)\rangle= \ldots+ D_{KL} {(x-y)\cdot y \over |x-y|^2|y|^4} +\ldots
\ee
Then the contribution from currents to \eqref{boundary1} is
\be
{i\over (2\pi)^4} D_{KL}\,\lim_{r\rightarrow 1^-} \int_{|x|=r} d^3\Omega_x \int_{|y|=1} d^3 \Omega_y\,\,|x|^2 |y|^2 \,(x\cdot\partial_x)\,(y\cdot\partial_y) \left[{(x-y)\cdot y \over |x-y|^2|y|^4}- {(y-x)\cdot x \over |x-y|^2|x|^4}  \right].
\ee
After some algebra, we find that the contribution is
\be
\label{currentcont}
-i  D_{KL} \,,
\ee
so conserved currents could potentially contribute. However, we will now provide an argument that $D_{KL}=0$ and hence there is no current contribution. 

To get some intuition about why we might expect that $D_{KL}=0$, consider the following argument: in the free theory the external operators $\phi_A^-,\phi_B^+$ only contain the letters $\overline{Z},\overline{X}$ and  $Z,X$ respectively. Hence in the free theory the correlator \eqref{yetanothercor} factorizes and as a result $D_{KL}=0$ at zero coupling. In fact, $D_{KL}=0$ continues to be true at all values of the coupling since the operators ${\rm Tr}(Y^2)$ and $\phi_K^+$ are charged under mutually orthogonal generators of $SO(6)$ and hence there is no common R-current connecting them in the 4-point function above.

In the free theory, this argument is not complete because we have additional accidental currents besides the $SU(4)$ R-currents. The argument just outlined implies that the {\it total} contribution $D_{KL}$ from the currents vanishes. It could logically be that the $SU(4)$ R-currents give a non-trivial contribution which is canceled by the accidental currents. Let us see why this is not a possibility. These accidental currents are in the Konishi multiplet and they are linear combinations of $J_{\rm scalars}$ and $J_{\rm fermions}$, in free-field language. Notice that  $J_{\rm scalars} +J_{\rm fermions}$ is the protected $SU(4)$ R-current. The orthogonal combination is in the Konishi multiplet and is an accidental current. However, the fermions do not click with the external operators $\phi_K^+,\phi_L^-$ in the free theory. Hence the contribution in the free theory from genuine SU(4) R-currents and from Konishi currents must be proportional to each other and with a positive coefficient of proportionality (since in both it is $J_{\rm scalar}$ that clicks). Since we know that the sum of the two vanishes (due to factorization of the full correlator in the free theory) they should both vanish individually. Moreover from Ward identities we know that $D_{KL}$ from the genuine R-currents is protected, and is constant so it should vanish everywhere on the conformal manifold.

\section{Details of $\mathcal{N}=2$ SYK Berry curvature}
\label{sec:SYKdetails}

This appendix contains all of the technical details from Section \ref{sec:SYK}. We prove the decomposition of three-forms into products for $N=4,5$ in Section \ref{app:forms}. Then, in Section \ref{app:SYKmonotone}, we compute the Berry curvature of the BPS states in an integrable variant of the model where the coupling is a wedge product of one-forms. Finally, Section \ref{sec:homotopy} contains our proof of a theorem on low-degree homotopy groups of the non-singular moduli space $\mathcal{M}_{\mathrm{non-sing}}$ applicable to $N=5$ and all $N \geq 6$ even.

\subsection{Decomposition of three forms}\label{app:forms}

In this appendix, we discuss some basic properties of three forms which were used in the discussion of Section \ref{sec:topo}. 

First, we can show that a $3$-form $C_3$ can always be decomposed into the wedge product of three $1$-forms in $4$ dimensions, which is consistent with the general counting argument in Section \ref{sec:specialloci}. To see this, we note that $*C_3$ is a one form, which without losing generality we can choose to be in the direction of $e^1$ with $e^1$ being a unit vector. Now, consider an orthonormal frame formed by $\{e^1,e^2,e^3,e^4\}$, it is apparent that the only component $C_3$ can have is $e^2 \wedge e^3 \wedge e^4$, since otherwise $*C_3$ would contain components other than $e^1$. Therefore, $C_3$ is decomposable into three one-forms.

The second statement we would like to show is that a $3$-form $C_3$ can always be decomposed into the wedge product of a $1$-form and a $2$-form in 5 dimensions. Again, we consider the Hodge dual $B_{2} = \ast C_{3}$. This gives an anti-symmetric matrix $(B_{2})_{ij}$ in $5$ dimensions, which must have rank at most $4$ since any anti-symmetric matrix in odd dimension $n$ has rank at most $n-1$. So $B_{2}$ has a non-trivial kernel, implying $\iota_{v}B_{2} = 0$ for some vector $v$. Then $\ast(C_{3} \wedge \hat{v}) = \ast(\iota_{v}B_{2})=0$, where $\hat{v}$ is the one-form corresponding to $v$. Hence $C_{3} \wedge \hat{v}=0$, which implies $C_{3} = \tilde{B}_{2} \wedge \hat{v}$ for some $2$-form $\tilde{B}_{2}$.

\subsection{Calculating Berry curvature for $C_3 = A_1^{(1)} \wedge A_1^{(2)} \wedge A_1^{(3)}$}\label{app:SYKmonotone}

In this appendix, we include some details on the computation of the Berry curvature for the model where $C_3 = A_1^{(1)} \wedge A_1^{(2)} \wedge A_1^{(3)}$, as discussed in Section \ref{sec:SYKmonotone}. Recall from the main text, we can go to a basis of the fermions where \begin{equation}\label{e1e2e3}
    C_{3} = \alpha \, e^1 \wedge e^2 \wedge e^3\,, \quad \alpha > 0\,,
\end{equation}
where $e^{i}$ is the unit vector in the $i$-th direction. Infinitesimal variations can be parameterized as 
\begin{equation}\label{Cvars}
     \delta C_3 = \delta C_{123} \,e^1 \wedge e^2 \wedge e^3 + \sum_{i=4}^{N} \left( \delta C_{23i} \, e^2 \wedge e^3 \wedge e^i + \delta C_{13i} \, e^1 \wedge e^3 \wedge e^i   + \delta C_{12i} \, e^1 \wedge e^2 \wedge e^i \right) \,.
\end{equation}
Below,
 we denote the directions of variations as $(ijk)$. 

At the point (\ref{e1e2e3}), the BPS subspace with $R$-charge $R=r$ is given by
\begin{equation}
   \mathcal{H}_{\textrm{BPS}}^{r} \cong \left(\textrm{span} \{  \ket{1}, \ket{2 } ,\ket{3 } \} \otimes \mathcal{H}_{\textrm{rest}}^{r-1} \right) \oplus \left(\textrm{span} \{  \ket{12}, \ket{13 } ,\ket{23 } \} \otimes \mathcal{H}_{\textrm{rest}}^{r-2} \right) \,,
\end{equation}
where $\ket{1},\ldots, \ket{23}$ are states in the Hilbert space of $\psi_1, \psi_2,\psi_3$ with the specified fermions excited, while $\mathcal{H}_{\textrm{rest}}^{r}$ denotes the subspace in the Hilbert space of $\psi_4,\ldots, \psi_N$ with fermion occupation number $r$. The BPS Hilbert spaces have dimension
\begin{equation}
    \dim \mathcal{H}_{\textrm{BPS}}^{r} = 3{N-2 \choose r-1}\,. 
\end{equation}

We can calculate the Berry curvature using 
\begin{equation}
    \left( F_{\mu \bar{\nu}} \right)_{ab} (x)  =  i\bra{b} \psi_i \psi_j \psi_k   \frac{H}{(H-x)^2} \bar{\psi}_n \bar{\psi}_m \bar{\psi}_l \ket{a} -  i\bra{b}  \bar{\psi}_n \bar{\psi}_m \bar{\psi}_l    \frac{H}{(H-x)^2} \psi_i \psi_j \psi_k \ket{a},
\end{equation}
with respect to $\mu = (ijk)$ and $\nu = (lmn)$ and taking the limit $x\rightarrow 0$ in the end. Since the Hamiltonian is just a constant $\alpha^2$ in the non-BPS sector, the $x\rightarrow 0$ limit is given by
\begin{equation}\label{Fxapp}
   (  F_{\mu \bar{\nu}} )_{ab} = \frac{i}{\alpha^2} \bra{b} \psi_i \psi_j \psi_k   P_{\textrm{non-BPS}} \bar{\psi}_n \bar{\psi}_m \bar{\psi}_l    \ket{a} - \frac{i}{\alpha^2} \bra{b}  \bar{\psi}_n \bar{\psi}_m \bar{\psi}_l  P_{\textrm{non-BPS}} \psi_i \psi_j \psi_k \ket{a}.
\end{equation}
Notice that when either $\mu$ or $\nu$ is equal to $(123)$, the Berry curvature vanishes since either the bra or the ket states in (\ref{Fxapp}) are annihilated by $\psi_{1} \psi_2 \psi_3$ or $\bar{\psi}_{1} \bar{\psi}_2 \bar{\psi}_3$. Therefore, we only need to consider the cases where
$\mu,\nu$ are chosen from the second term in (\ref{Cvars}). 

\paragraph{Diagonal variations.} Consider the example of diagonal Berry curvature $\mu = \nu = (12k)$, where $k \in \{4,\ldots,N\}$. Then for the first term in (\ref{Fxapp}) to be nonzero, we must have the bra and ket states to contain the factor $\ket{12}$. On the other hand, for the second term to be non-zero, we need the bra and ket states to contain the factor $\ket{3}$. 

Let's focus on the first possibility, where the bra state must be $\ket{12}\otimes \ket{r-2}$ and the ket state must be $\ket{12}\otimes \ket{r'-2}$, where $\ket{r-2},\ket{r'-2}$ are arbitrary states in $\mathcal{H}_{\textrm{rest}}^{r-2}$. We then have 
\begin{equation}\label{12r}
 \left(  \bra{12} \otimes \bra{r-2} \right) \psi_1 \psi_2 \psi_k   P_{\textrm{non-BPS}}\bar{\psi}_k \bar{\psi}_2 \bar{\psi}_1   \left( \ket{12} \otimes \ket{r'-2} \right) = \bra{r-2} \psi_k  \bar{\psi}_k  \ket{r'-2}\,.
\end{equation}
Therefore, the problem reduces to that of finding the eigenvalues of $\psi_k  \bar{\psi}_k$ in the subspace spanned by $\ket{12} \otimes \mathcal{H}_{\textrm{rest}}^{r-2}$, which has dimension $\binom{N-3}{r-2}$. Since the operator is simply the occupation number operator for fermion $\psi_k$, we get eigenvalue $1\times 1/\alpha^2$ with multiplicity $\binom{N-4}{r-3}$ and eigenvalue $0\times 1/\alpha^2$ with multiplicity $\binom{N-4}{r-2}$.

Now let's consider the second possibility for (\ref{Fxapp}) to be nonzero, where we have both the bra and ket states to be in $\ket{3} \otimes \mathcal{H}_{\textrm{rest}}^{r-1}$. Similar to (\ref{12r}), we now have
\begin{equation}
   - \left(  \bra{3} \otimes \bra{r-1} \right)   \bar{\psi}_k \bar{\psi}_2 \bar{\psi}_1 P_{\textrm{non-BPS}}  \psi_1 \psi_2 \psi_k  \left( \ket{3} \otimes \ket{r'-1} \right) = -      \bra{r-1}  (1- \psi_k \bar{\psi}_k   ) \ket{r'-1} .
\end{equation}
So we find that in the subspace spanned by $\ket{3} \otimes \mathcal{H}_{\textrm{rest}}^{r-1}$, which has dimension $\binom{N-3}{r-1}$, we have eigenvalue $-1 \times \frac{1}{\alpha^2}$ with multiplicity $\binom{N-4}{r-1}$ and eigenvalue $0 \times 1/\alpha^2$ with multiplicity $\binom{N-4}{r-2}$. 

To summarize, we find that $F_{\mu \bar{\mu}}$ with $\mu = (12k)$ has eigenvalues\footnote{Note that $F_{\mu\bar{\mu}}$ is anti-Hermitian, so we are really looking at the eigenvalues of $-iF_{\mu \bar{\mu}}$.} 
\begin{equation}
\begin{aligned}
   \frac{1}{\alpha^2}  \,\, \textrm{with multiplicity} \,\, \binom{N-4}{r-3}\,, \quad \quad
   -\frac{1}{\alpha^2}  \,\, \textrm{with multiplicity} \,\, \binom{N-4}{r-1} \,,
\end{aligned}
\end{equation}
and the rest of the eigenvalues being zero.

\paragraph{Off-diagonal variations.} As another example, we can consider an off-diagonal variation of the couplings, say $\mu = (12k)$ and $\nu = (13l)$, with $l \neq k$ For the first term in (\ref{Fxapp}) to be non-zero, we need $\ket{a}$ to include a $\ket{13}$ factor, and $\bra{b}$ to include a $\bra{12}$ factor. On the other hand, for the second term in (\ref{Fxapp}) to be non-zero, we need $\ket{a}$ to include a $\ket{3}$ factor and $\bra{b}$ to include a $\bra{2}$ factor. Focusing on the first term, we then have
\begin{equation}\label{psikl}
    (\bra{12} \otimes \bra{r-2}) \psi_1 \psi_2 \psi_k P_{\textrm{non-BPS}} \bar{\psi}_l \bar{\psi}_3 \bar{\psi}_1 ( \ket{13} \otimes \ket{r'-2}) = (\bra{12} \otimes \bra{r-2}) \psi_2 \psi_k \bar{\psi}_l  \bar{\psi}_3 (\ket{13}  \otimes \ket{r'-2})\,.
\end{equation}
We need to recall from Footnote \ref{footnoteholomorphic} that $F_{\mu \bar{\nu}}$ is non-Hermitian, and we can form a Hermitian combination by considering $F_{\mu \bar{\nu}} - F_{\nu\bar{\mu}}$. From the $- F_{\nu \bar{\mu}}$ piece, instead of (\ref{psikl}), we get
\begin{equation}
    - (\bra{13} \otimes \bra{r-2}) \psi_3 \psi_l \bar{\psi}_k  \bar{\psi}_2 (\ket{12}  \otimes \ket{r'-2}) \,.
\end{equation}
Therefore, in the subspace spanned by $\ket{12} \otimes \mathcal{H}_{\textrm{rest}}^{r-2}$ and $\ket{13} \otimes \mathcal{H}_{\textrm{rest}}^{r-2}$, the eigenvalues of $F_{\mu \bar{\nu}} - F_{\nu\bar{\mu}}$ are equivalent to $1/\alpha^2$ multiplying the eigenvalues of $i (\psi_2 \psi_k \bar{\psi}_l \bar{\psi}_3 - \psi_3 \psi_l \bar{\psi}_k \bar{\psi}_2)$, which results in  eigenvalues $1/\alpha^2, - 1/\alpha^2$ both with multiplicity $\binom{N-5}{r-3}$, and the rest of the eigenvalues being zero. 

For the second term in (\ref{Fxapp}), following a similar discussion, we have eigenvalues $1/\alpha^2, - 1/\alpha^2$ both with multiplicity $\binom{N-5}{r-2}$, and the rest of the eigenvalues being zero. 

Adding the contribution from the two terms together, we find that $F_{\mu \bar{\nu}} - F_{\nu\bar{\mu}}$ with $\mu = (12k)$ and $\nu = (13l)$ with $k\neq l$ has eigenvalues
\begin{equation}
     \frac{1}{\alpha^2}  \,\, \textrm{with multiplicity} \,\, \binom{N-4}{r-2} \,, \quad \quad
   -\frac{1}{\alpha^2}  \,\, \textrm{with multiplicity} \,\, \binom{N-4}{r-2} \,,
\end{equation}
and the rest of the eigenvalues being zero.

One can of course further consider Berry curvature associated with  more general variations of couplings. We have verified numerically that the results are qualitatively the same, namely the eigenvalues are $\pm 1/\alpha^2,0$ with large degeneracies.

\subsection{Homotopy groups of $\mathcal{M}_{\mathrm{non-sing}}$}
\label{sec:homotopy}

In this appendix, we prove a theorem that gives the low-degree homotopy groups of the non-singular moduli space $\mathcal{M}_{\mathrm{non-sing}}$ for $N=5$ and even $N \geq 6$, which were stated in \eqref{eqn:homotopyN=5}, \eqref{eqn:homotopyN=6}, and \eqref{eqn:homotopyNeven}.
Our goal is to understand the topology of the non-singular moduli space
\begin{equation}
    \mathcal{M}_{\mathrm{non-sing}} = (\mathbb{C}^{{N \choose 3}} \backslash \mathcal{M}_{\mathrm{sing}})/U(1) \cong (\mathbb{C}\mathbb{P}^{{N \choose 3}-1} \backslash \mathbb{P}\mathcal{M}_{\mathrm{sing}}) \times \mathbb{R}_{+} \,,
\end{equation}
where $\cong$ means homeomorphic. In particular, we would like to understand the homotopic properties of $\mathcal{M}_{\mathrm{non-sing}}$. By a trivial deformation retract, we have the homotopy equivalence
\begin{equation}
    \mathcal{M}_{\mathrm{non-sing}} \simeq \mathbb{C}\mathbb{P}^{{N \choose 3}-1} \backslash \mathbb{P}\mathcal{M}_{\mathrm{sing}}\,.
\end{equation}
Therefore, the problem reduces to studying $\mathbb{C}\mathbb{P}^{{N \choose 3}-1} \backslash \mathbb{P}\mathcal{M}_{\mathrm{sing}}$. We will always assume that $\mathbb{P}\mathcal{M}_{\mathrm{sing}}$ is a projective variety. This is true for $N=5$ because the space $\mathcal{M}_{1}$ is determined by the Pl\"ucker relations, which are a set of homogeneous polynomials in the $C_{ijk}$, and it is true for $N \geq 6$ even since, by the Theorem in Section \ref{sec:specialloci}, $\mathcal{M}_{\mathrm{sing}}$ is defined by $\mathrm{rank}\, M_{\frac{N}{2}-2}(C) < \widetilde{D}_{\frac{N}{2}-2}$ so all $\tilde{D}_{\frac{N}{2}-2}$-minors of $M_{\frac{N}{2}-2}(C)$ vanish, which gives a set of homogeneous polynomials in the $C_{ijk}$. From the arguments in Section \ref{sec:specialloci}, we find the projective dimensions\footnote{Recall that the projective dimension is equal to the complex dimension minus $1$ since projectivization removes one complex dimension.} for $\mathbb{P}\mathcal{M}_{\mathrm{sing}}$ displayed in Table \ref{tab:projectivedimMsing}.

\begin{table}[h]
\centering
\caption{Projective dimensions of $\mathbb{P}\mathcal{M}_{\mathrm{sing}}$ for different $N$.}
\label{tab:projectivedimMsing}
{\renewcommand{\arraystretch}{1.4}
\begin{tabular}{|c|c|}
\hline
$N$ & $\mathrm{Proj.\;dim.\;of\;\mathbb{P}\mathcal{M}_{\mathrm{sing}}}$ \\
\hline
$5$ & $6$ \\
\hline
$6$ & $14$ \\
\hline
$N > 6, \mathrm{\;even}$ & $\leq {N \choose 3}-3$ \\
\hline
\end{tabular}}
\end{table}

We will prove the following:

\begin{theorem}
    For $N=5$ and $N \geq 6$ even, the space $\mathcal{M}_{\mathrm{non-sing}}$ has homotopy groups 
\begin{equation}
    \pi_{1}(\mathcal{M}_{\mathrm{non-sing}}) = 0, \quad \pi_{2}(\mathcal{M}_{\mathrm{non-sing}}) \cong \mathbb{Z}, \quad \pi_{i}(\mathcal{M}_{\mathrm{non-sing}}) = 0, \quad 3 \leq i \leq 2\left({N \choose 3}-d-2\right) \,,
\end{equation} 
where $d$ is the projective dimension of $\mathbb{P}\mathcal{M}_{\mathrm{sing}}$.
\end{theorem}

Plugging in the dimensions in Table \ref{tab:projectivedimMsing}, we obtain the results in \eqref{eqn:homotopyN=5}, \eqref{eqn:homotopyN=6}, and \eqref{eqn:homotopyNeven}. Note that it is possible that higher-degree homotopy groups are non-trivial, but we have not been able to determine them.

The proof of the theorem is very long so we will break it into several lemmas. It uses the following properties of $\mathbb{P}\mathcal{M}_{\mathrm{sing}}$: (1) it has projective dimension $d$ and (2) it is a projective variety.

Let $Y \subset \mathbb{C}\mathbb{P}^{n}$ be a projective variety with projective dimension $d$. Define $\mathrm{Cone}(Y) = \{v \in \mathbb{C}^{n+1}\,|\,[v] \in Y\}$ and let $\Sigma = \mathrm{Cone}(Y) \cap S^{2n+1}$. The proof of the Theorem boils down to showing that the corresponding homotopy groups of $S^{2n+1}\backslash\Sigma$ are trivial when $Y=\mathbb{P}\mathcal{M}_{\mathrm{sing}}$.

\noindent \textbf{Definition.} A topological space $\mathcal{T}$ is locally contractible if for any $x \in \mathcal{T}$, every neighborhood $U$ of $x$ contains a neighborhood $x \in V \subset U$ such that $V$ is contractible.

\begin{lemma} \label{lem:contractible}
$\Sigma$ is locally contractible.
\end{lemma}

\begin{proof}
Observe that $\Sigma$ is a complex algebraic variety in $\mathbb{C}^{n+1}$ defined by the homogeneous polynomials defining $Y$ along with the (non-homogeneous) polynomial $\sum_{i=1}^{n+1}|z_{i}|^{2}-1=0$. By Hironaka's theorem \cite{Hironaka1974TriangulationsOA}, every complex algebraic variety (even a singular one) has a triangulation, i.e., it is a simplicial complex. Hence, $\Sigma$ is a CW complex.\footnote{Note that $\Sigma$ is not necessarily a manifold, e.g. $\mathcal{M}_{2}$ is singular along the decomposable submanifold $\mathcal{M}_{1}$.}$^{,}$\footnote{The original proof that complex algebraic varieties are CW complexes is actually much older than Hironaka's theorem and was proved by Koopman and Brown in the 1930s \cite{KoopmanBrown1949}, but Hironaka's proof is much clearer. See \cite{Goresky2012IntroThomMather} for a discussion of the history of the subject.} By Proposition A.4 in Hatcher \cite{Hatcher:478079}, every CW complex is locally contractible.
\end{proof}

\begin{lemma}
\label{lem:pi1}
If $\mathrm{codim}_{\mathbb{R}}\Sigma \geq 3$ inside of $S^{2n+1}$, then $\pi_{1}(S^{2n+1}\backslash\Sigma) = 0$.
\end{lemma}

\noindent \textit{Intuition}: 
Removing a codimension-$3$ subspace from a simply-connected space means you can always slide loops around the resulting hole to contract them. While this intuition is roughly correct, it can fail. For example, if the subspace is not locally flat inside the larger space, known as a ``wild'' embedding \cite{FERRY1989197}. In fact, in $\mathbb{R}^{n}$ for any $n$, there exist complicated zero-dimensional subspaces whose complement is not simply-connected \cite{c808a5f1-9785-3607-8b40-9d6b5459ff4c}. It can also fail if the ambient space is not a manifold because it is singular. We will use the fact that (1) the ambient space $S^{2n+1}$ is a manifold and (2) $\Sigma$ is triangulable.

The crux of the proof of Lemma~\ref{lem:pi1} is to apply the following theorem:
\begin{theorem} 
\label{thm:positioning}
\textbf{(General positioning theorem for embeddings).}
Given simplicial subcomplexes $P \supset P_{0}$ and $Q$ of an $n$-dimensional piecewise linear manifold $M$ with $\dim (P-P_{0})=p$ and $\dim Q = q$, there exists a homotopy $h$ of $P$ in $M$, relative to $P_{0}$, such that $\dim\left(h(P-P_{0},1) \cap Q\right) \leq p+q-n$.
\end{theorem}

See \cite{BryantPLTopologyNotes} for the necessary background and a proof of this theorem.\footnote{The theorem as stated there actually proves the stronger statement of $\epsilon$-isotopic, which includes homotopic as a special case (we will only need homotopic).} The basic idea is that one can smoothly deform $P$ inside $M$, while holding $P_{0}$ inside of it fixed, in order to minimize its intersection with $Q$, and the general positioning theorem for embeddings gives a precise bound on how small this intersection can be. The proof of Lemma \ref{lem:pi1} then goes as follows.

\begin{proof}[Proof of Lemma~\ref{lem:pi1}.]
We will show that $\pi_{1}(S^{2n+1}\backslash\Sigma) \cong \pi_{1}(S^{2n+1})$. Consider the inclusion map $i : S^{2n+1}\backslash\Sigma \to S^{2n+1}$. This induces a map on fundamental groups
\begin{equation}
    i_{\ast} : \pi_{1}(S^{2n+1}\backslash\Sigma) \to \pi_{1}(S^{2n+1})\,,
\end{equation}
which is trivially a group homomorphism, so we need to show that it is a bijection. 

For surjectivity, we will show that every loop $\gamma : S^{1} \to S^{2n+1}$ can be continuously deformed to a $\gamma'$ such that $\mathrm{im}(\gamma') \cap \Sigma = \emptyset$. This requires the general position theorem for embeddings. Observe that $\mathrm{im}(\gamma)$ is not necessarily a simplicial complex, even though $S^{1}$ is a simplicial complex. Nevertheless, since $\gamma$ is continuous and $S^{2n+1}$ is a simplicial complex, by the simplicial approximation theorem (Hatcher Thm 2C.1 \cite{Hatcher:478079}), there exists $\tilde{\gamma}$ homotopic to $\gamma$ such that $\mathrm{im}(\tilde{\gamma})$ is a simplicial subcomplex of $S^{2n+1}$. Recall from the proof of Lemma~\ref{lem:contractible} that $\Sigma$ is triangulable and hence is a simplicial complex. Furthermore, since $\Sigma$ is closed in $S^{2n+1}$ and $S^{2n+1}$ is compact, $\Sigma$ is compact. These properties allow us to apply the general position theorem for embeddings to $\Sigma$ and $\mathrm{im}(\tilde{\gamma})$ (with $P_{0}=\emptyset$), which implies that there exists a homotopy $h: S^{2n+1} \times I \to S^{2n+1}$ with $h(x,0) = x$ such that
\begin{equation}
    \dim\left(h(\mathrm{im}(\tilde{\gamma}),1) \cap \Sigma\right) \leq \dim \left(h(\mathrm{im}(\tilde{\gamma}),1)\right) - \mathrm{codim}_{\mathbb{R}}\Sigma \leq -2 \,,
\end{equation}
where the last inequality used $\mathrm{codim}_{\mathbb{R}}\Sigma \geq 3$, and hence $h(\mathrm{im}(\tilde{\gamma}),1) \cap \Sigma = \emptyset$. Thus, we obtain the desired surjectivity.

For injectivity, consider any $\gamma : S^{1} \to S^{2n+1}\backslash\Sigma$ such that $i_{\ast}([\gamma]) = [0]$. Then $i(\gamma)$ can be deformed in $S^{2n+1}$ to a point and we need to show that this deformation can be done without intersecting $\Sigma$. First, we use the simplicial approximation theorem again. Since $S^{2n+1}$ can always be triangulated such that $\Sigma$ is a subcomplex, $S^{2n+1}\backslash\Sigma$ is a simplicial complex. So we obtain $\tilde{\gamma}$ homotopic to $\gamma$ with $\mathrm{im}(\tilde{\gamma})$ a simplicial complex. Since $i(\tilde{\gamma})$ is null-homotopic, it can be extended to a map $f : D^{2} \to S^{2n+1}$ such that $f|_{\partial D^{2}} = i(\tilde{\gamma})$. Next, use the relative simplicial approximation theorem \cite{Zeeman1964RelativeSimplicialApproximation} to obtain $\tilde{f} : D^{2} \to S^{2n+1}$ homotopic to $f$, with $\mathrm{im}(\tilde{f})$ a simplicial subcomplex, that fixes $\partial D^{2}$, i.e., $f|_{\partial D^{2}} = \tilde{f}|_{\partial D^{2}}$. Then the general position theorem for embeddings with $P_{0}=\mathrm{im}(\tilde{\gamma})$ gives a homotopy $H$ of $\mathrm{im}(\tilde{f})$ in $S^{2n+1}$ fixing $P_{0}$ such that
\begin{equation}
    \dim\left(H(\mathrm{im}(\tilde{f})-\mathrm{im}(\tilde{\gamma}),1) \cap \Sigma\right) \leq \dim\left(H(\mathrm{im}(\tilde{f})-\mathrm{im}(\tilde{\gamma}),1)\right) - \mathrm{codim}_{\mathbb{R}}\Sigma \leq -1\,,
\end{equation}
and hence $H(\mathrm{im}(\tilde{f})-\mathrm{im}(\tilde{\gamma}),1) \cap \Sigma = \emptyset$. Thus, $\gamma$ is null-homotopic.
\end{proof}

\begin{lemma}
\label{lem:pii}
If $\Sigma$ satisfies the assumptions of Lemma~\ref{lem:pi1} and $Y$ has projective dimension $d<n$, then $\pi_{i}(S^{2n+1}\backslash\Sigma) = 0$ for $2 \leq i \leq 2(n-d-1)$.
\end{lemma}

\begin{proof}
Recall from the proof of Lemma~\ref{lem:pi1} that $\Sigma$ is a compact subset of $S^{2n+1}$. Then, by Lemma~\ref{lem:contractible}, we can use Alexander duality given by Corollary 3.45 in Hatcher \cite{Hatcher:478079} to conclude that the homology and cohomology groups are related by
\begin{equation}
    H_{i}(S^{2n+1}\backslash\Sigma) \cong H^{2n-i}(\Sigma) \,,
\end{equation}
for $i \geq 1$.\footnote{Hatcher \cite{Hatcher:478079} uses the reduced homology groups $\tilde{H}_{i}$ but these are isomorphic to the homology groups for $i \geq 1$ (same for cohomology).} Since $Y$ has projective dimension $d$, the real dimension of $\Sigma$ satisfies $\mathrm{dim}_{\mathbb{R}}\Sigma \leq 2d+1$. In general, $H^{m}(\mathcal{S}) = 0$ when $m > \mathrm{dim}_{\mathbb{R}}(\mathcal{S})$ so
\begin{equation}
    H_{i}(S^{2n+1}\backslash\Sigma) = 0 \quad \mathrm{for} \quad i \leq 2(n-d-1)\,.
\end{equation}

To relate these homology groups to homotopy groups, we use the Hurewicz Theorem. First, observe that $S^{2n+1}\backslash\Sigma$ is path-connected since $S^{2n+1}$ is path-connected and $\Sigma$ has $\mathrm{codim}_{\mathbb{R}}\Sigma \geq 2$ so any path connecting two points in $S^{2n+1}$ can always be deformed to avoid $\Sigma$. Now, let $k \geq 2$ be the smallest positive integer such that $\pi_{k}(S^{2n+1}\backslash\Sigma) \neq 0$ ($k$ cannot be less than $2$ by Lemma \ref{lem:pi1}). Then the Hurewicz Theorem states that $\pi_{k}(S^{2n+1}\backslash\Sigma) \cong H_{k}(S^{2n+1}\backslash\Sigma)$, but we showed that $H_{i}(S^{2n+1}\backslash\Sigma) = 0$ for $i \leq 2(n-d-1)$ so we must have $k > 2(n-d-1)$, thus proving the lemma.
\end{proof}

We are now ready to prove the Theorem.

\begin{proof}[Proof of Theorem~\ref{thm:positioning}.]
Since $\mathbb{P}\mathcal{M}_{\mathrm{sing}}$ is a projective variety and it satisfies the assumptions of Lemma \ref{lem:pi1}, we can apply the Lemmas. It has projective dimension $d$ so by Lemmas 2 and 3, $\pi_{i}(S^{2n+1}\backslash\Sigma) = 0$ for $1 \leq i \leq 2(n-d-1)$ where $n = {N \choose 3}-1$. Now, consider the standard principal $U(1)$ bundle
\begin{equation}
    S^{1} \xrightarrow{i} S^{2n+1} \xrightarrow{p} \mathbb{C}\mathbb{P}^{n}\,.
\end{equation}
By Remark 1.1.2 in \cite{Neeb:DiffTopologyFiberBundles:2024}, for any open set $U \subset \mathbb{C}\mathbb{P}^{n}$, we have a principal $U(1)$ bundle $S^{1} \rightarrow p^{-1}(U) \rightarrow U$ so taking $U = \mathbb{C}\mathbb{P}^{n}\backslash\mathbb{P}\mathcal{M}_{\mathrm{sing}}$, we obtain
\begin{equation}
    S^{1} \xrightarrow{i} S^{2n+1}\backslash\Sigma \xrightarrow{p} \mathbb{C}\mathbb{P}^{n}\backslash\mathbb{P}\mathcal{M}_{\mathrm{sing}}\,.
\end{equation}
This leads to the following long exact sequence of homotopy groups:
\begin{equation}
    \ldots \rightarrow \pi_{i}(S^{1}) \xrightarrow{i_{\ast}} \pi_{i}(S^{2n+1}\backslash\Sigma) \xrightarrow{p_{\ast}} \pi_{i}(\mathbb{C}\mathbb{P}^{n}\backslash\mathbb{P}\mathcal{M}_{\mathrm{sing}}) \xrightarrow{\partial_{\ast}} \pi_{i-1}(S^{1}) \rightarrow \ldots
\end{equation}
We now apply this case-by-case to compute $\pi_{i}(\mathbb{C}\mathbb{P}^{n}\backslash\mathbb{P}\mathcal{M}_{\mathrm{sing}})$.

\underline{$i \neq 2$}: Since $\pi_{i}(S^{2n+1}\backslash\Sigma)=\pi_{i-1}(S^{1})=0$, we have
\begin{equation}
    \pi_{i}(\mathbb{C}\mathbb{P}^{n}\backslash\mathbb{P}\mathcal{M}_{\mathrm{sing}}) = \ker\partial_{\ast}= \mathrm{im}\, p_{\ast} = 0\,.
\end{equation}

\underline{$i = 2$}: Since $\pi_{1}(S^{1}) \cong \mathbb{Z}$, we obtain 
\begin{equation}
    0 \xrightarrow{p_{\ast}} \pi_{2}(\mathbb{C}\mathbb{P}^{n}\backslash\mathbb{P}\mathcal{M}_{\mathrm{sing}}) \xrightarrow{\partial_{\ast}} \mathbb{Z} \xrightarrow{p_{\ast}} 0\,,
\end{equation}
and hence $\pi_{2}(\mathbb{C}\mathbb{P}^{n}\backslash\mathbb{P}\mathcal{M}_{\mathrm{sing}}) \cong \mathbb{Z}$.
\end{proof}

\bibliographystyle{ourbst}
\bibliography{Refs}
\end{document}